\documentclass[10pt,journal,compsoc]{IEEEtran}
\usepackage{amsmath}  
\usepackage{amssymb}  
\usepackage[table]{xcolor}
\usepackage{hyperref}
\usepackage{prettyref}
\usepackage[normalem]{ulem}
\usepackage{booktabs}
\usepackage{caption}

\newcommand\scalemath[2]{\scalebox{#1}{\mbox{\ensuremath{\displaystyle #2}}}}

\newcommand{\set}[1]{\left\{{#1}\right\}}

\usepackage{rotating}
\newcommand{\rulesep}{\unskip\ \vrule\ }

\newcommand{\pp}{\mathbf{p}}

\renewcommand{\tt}{\mathbf{t}}

\newcommand\restr[2]{{\left.\kern-\nulldelimiterspace{}#1\right|_{#2}}}

\usepackage[switch]{lineno} 

\begin{document}

\newcommand{\mytitle}[1]{Visual-Guided Mesh Repair}
\title{\mytitle{\\*}}

\title{Visual-Guided Mesh Repair}
\author{Zhongtian~Zheng,
        Xifeng~Gao,
        Zherong~Pan,
        Wei~Li,
        Peng-Shuai~Wang,
        Guoping~Wang,
        Kui~Wu
\IEEEcompsocitemizethanks{\IEEEcompsocthanksitem This work was done when Z. Zheng was an intern at LightSpeed Studios.
\IEEEcompsocthanksitem Z. Zheng, P. Wang, and G. Wang are with Peking University.
\IEEEcompsocthanksitem X. Gao, Z. Pan, W. Li, and K. Wu are with LightSpeed Studios.
\IEEEcompsocthanksitem Corresponding Author: Kui Wu, E-mail: kwwu@global.tencent.com
}
\thanks{Manuscript received XXX, 2021; revised XXX.}}

\markboth{\emph{Zheng \MakeLowercase{\textit{et al.}}}: \mytitle{ }}%
{\emph{Zheng \MakeLowercase{\textit{et al.}}}: \mytitle{ }}

\IEEEpubid{0000--0000/00\$00.00~\copyright~2021 IEEE}

\IEEEtitleabstractindextext{%
\begin{abstract}
Mesh repair is a long-standing challenge in computer graphics and related fields. Converting defective meshes into watertight manifold meshes can greatly benefit downstream applications such as geometric processing, simulation, fabrication, learning, and synthesis. In this work, we first introduce three visual measures for visibility, orientation, and openness, based on ray-tracing. We then present a novel mesh repair framework that incorporates visual measures with several critical steps, i.e., open surface closing, face reorientation, and global optimization, to effectively repair defective meshes, including gaps, holes, self-intersections, degenerate elements, and inconsistent orientations. Our method reduces unnecessary mesh complexity without compromising geometric accuracy or visual quality while preserving input attributes such as UV coordinates for rendering. We evaluate our approach on hundreds of models randomly selected from ShapeNet and Thingi10K, demonstrating its effectiveness and robustness compared to existing approaches.
\end{abstract}

\begin{IEEEkeywords}
Mesh repairing, geometry processing.
\end{IEEEkeywords}}
\maketitle

%

\section{Introduction} \label{sec:intro}
Meshes in games created by modelers often prioritize visual appearance over geometric and topological correctness, leading to various defects like gaps, holes, self-intersections, singular elements, and inconsistent orientations~\cite{Attene2013survey}. Additionally, raw data from online repositories like ShapeNet~\cite{Chang2015shapenet} may contain quality issues like duplicated faces, self-intersections, and non-manifold elements. Due to these issues, meshes with such defects become invalid for downstream applications. Therefore, developing a robust mesh repair pipeline is essential. 
Three key properties should be pertained during mesh repair, so it can facilitate the downstream applications. First, the mesh should be \textit{manifold} to calculate differential quantities such as normals and curvatures. Second, the mesh should be \textit{watertight} to have a well-defined interior and exterior volume.
Finally, the mesh should be repaired with \textit{minimal modification}, preserving the sharp features and UVs of the input mesh as much as possible. These properties are crucial for simulation, 3D printing, geometric Boolean operators, and learning-related applications such as shape analysis and synthesis~\cite{hanocka2019meshcnn,hu2022subdivision}. Besides the three features, \textit{robustness} and \textit{efficiency} are desirable features that must be considered in mesh repair algorithms. 

\begin{figure*}[ht]
 \centering
\includegraphics[trim=450 300 500 320,clip,width=0.19\linewidth]{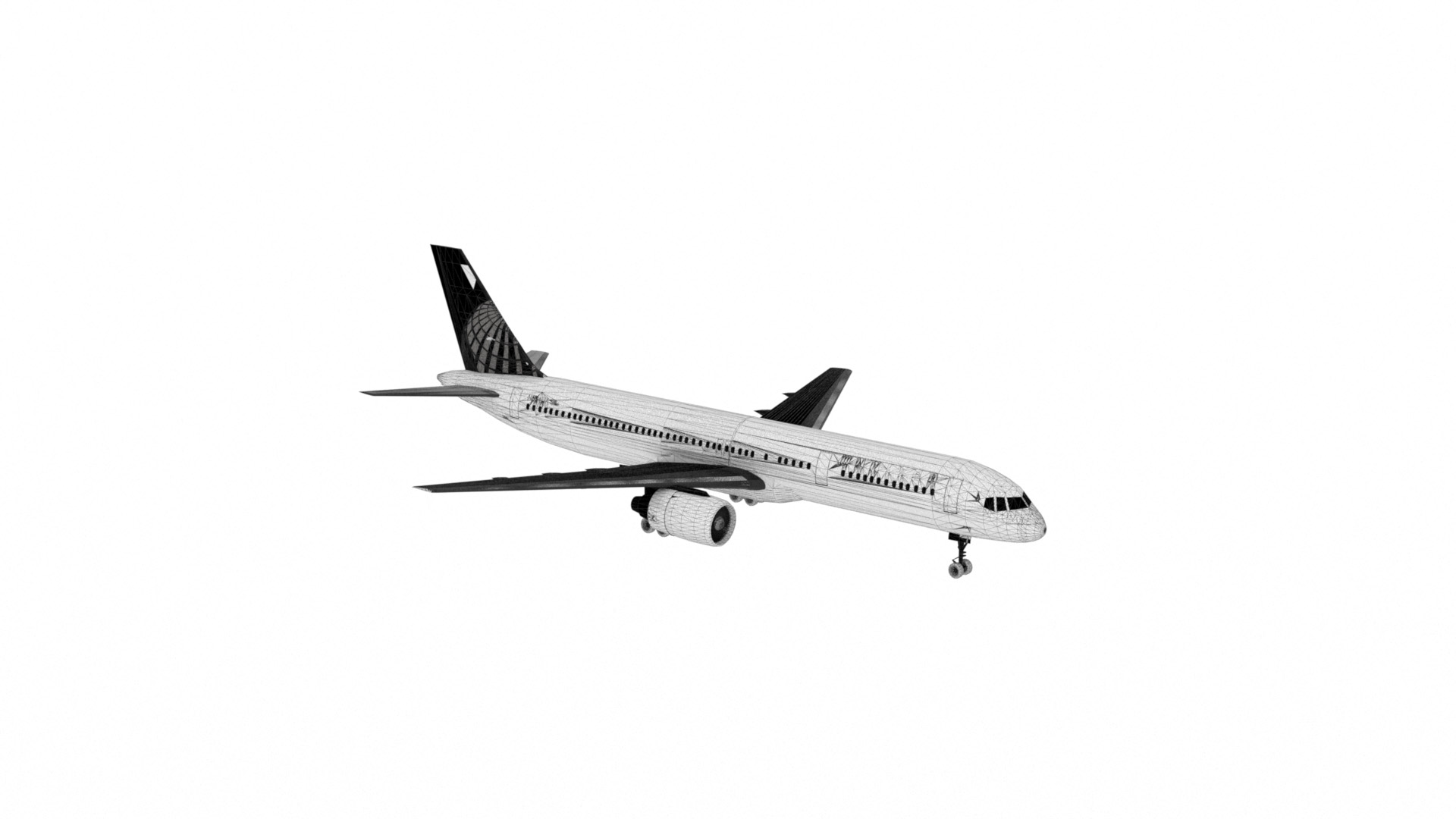}\hfill
\frame{\includegraphics[trim=300 100 600 400,clip,width=0.14\linewidth]{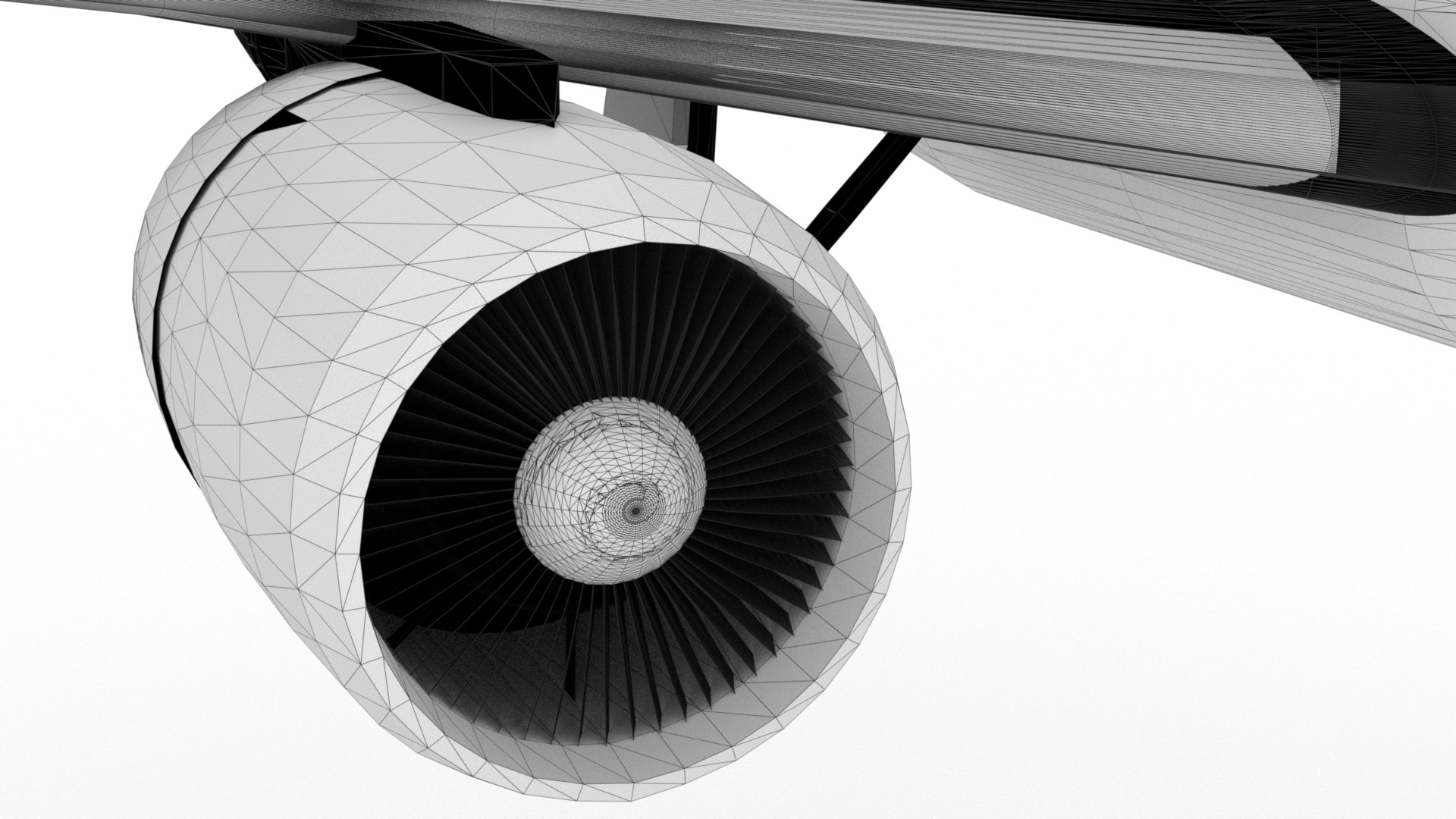}}\hfill
\includegraphics[trim=450 300 500 320,clip,width=0.19\linewidth]{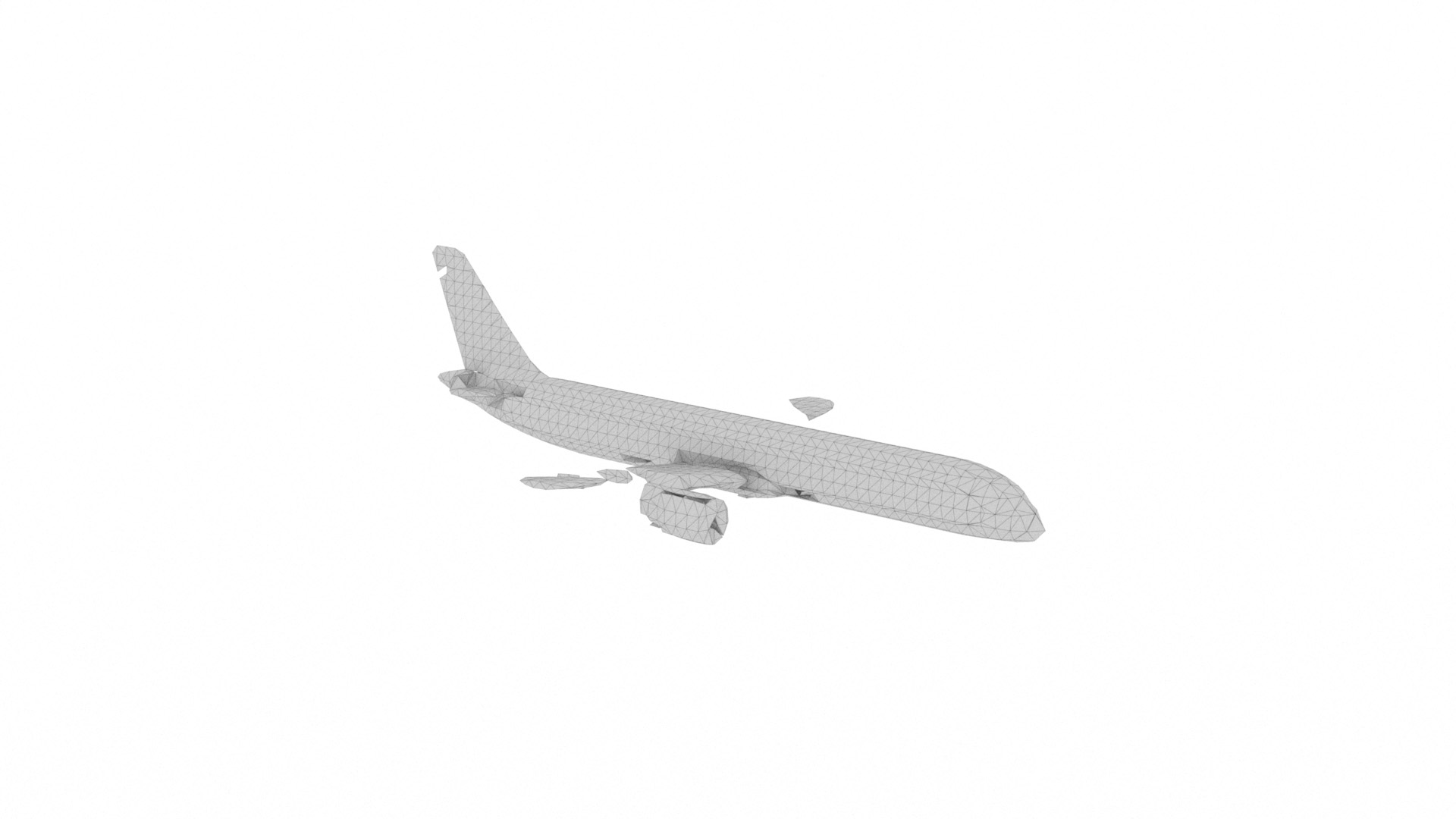}\hfill
\frame{\includegraphics[trim=300 100 600 400,clip,width=0.14\linewidth]{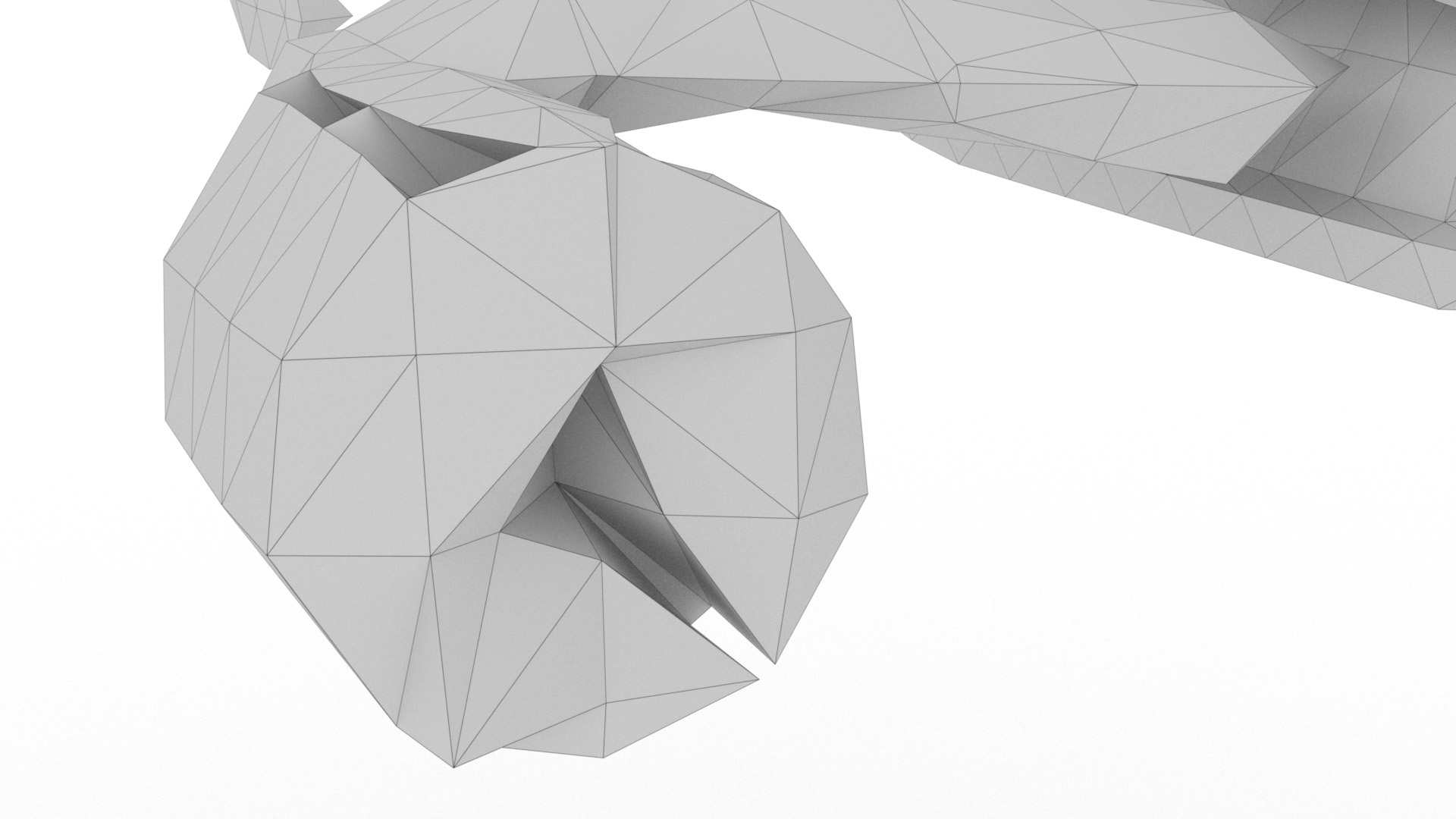}}\hfill
\includegraphics[trim=450 300 500 320,clip,width=0.19\linewidth]{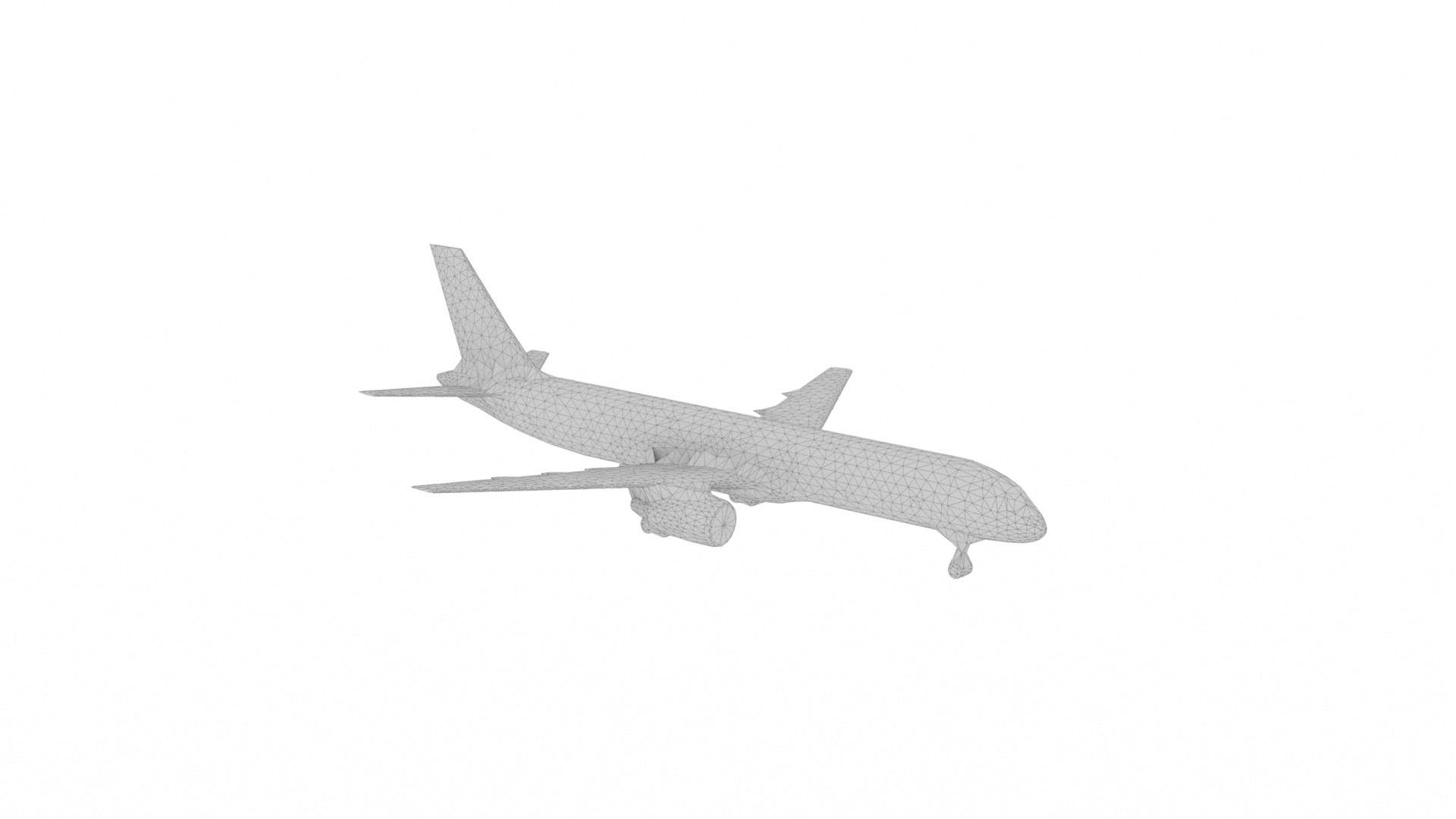}\hfill
\frame{\includegraphics[trim=300 100 600 400,clip,width=0.14\linewidth]{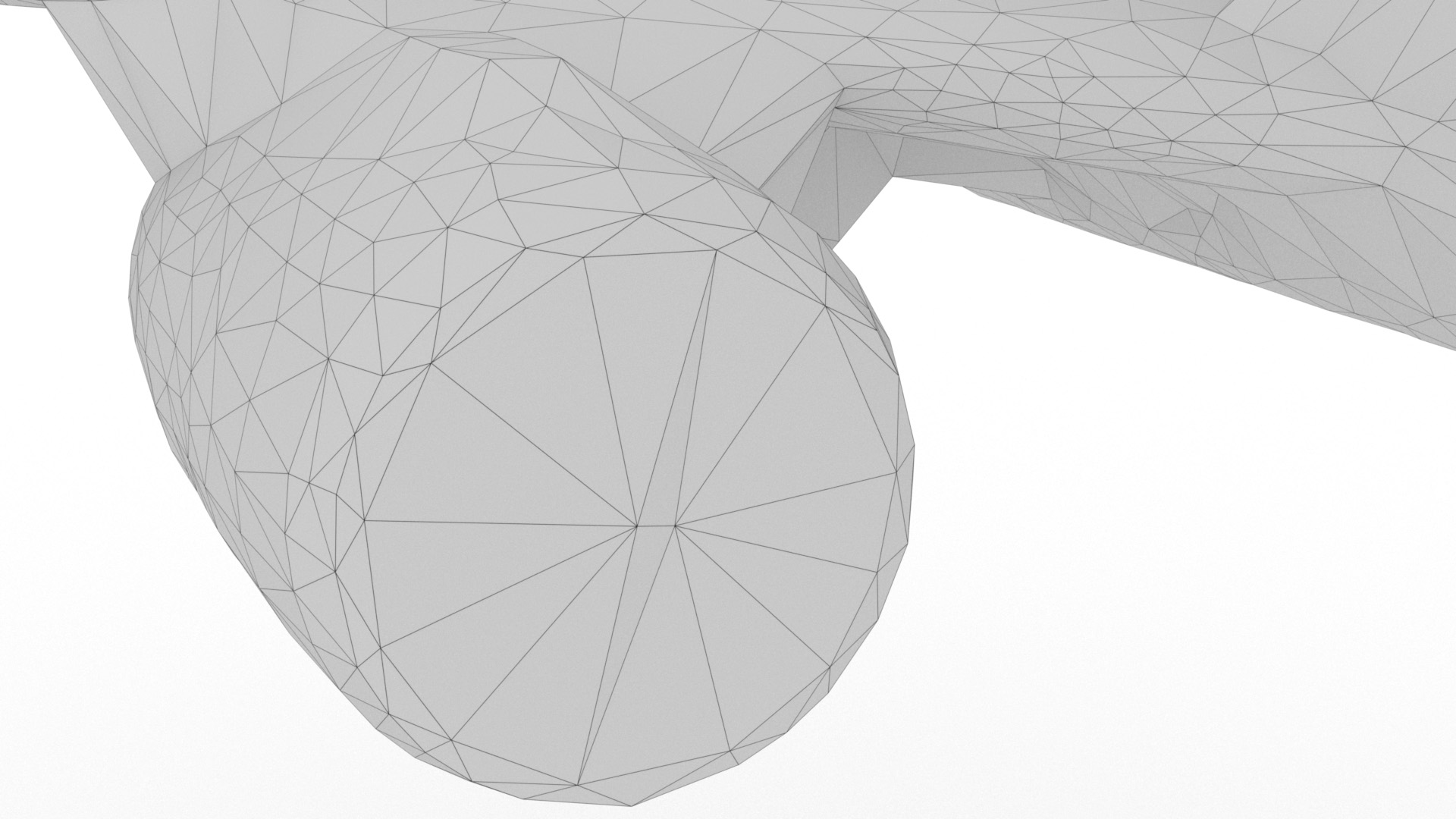}}\\
\begin{minipage}{0.33\linewidth}\centering\footnotesize Input with texture\end{minipage}\hfill%
\begin{minipage}{0.33\linewidth}\centering\footnotesize PolyMender~\cite{Ju2004polymender}\end{minipage}\hfill%
\begin{minipage}{0.33\linewidth}\centering\footnotesize AlphaWrapping~\cite{Portaneri2022AlphaWrapping}\end{minipage}\\\vspace{0.5em}

\includegraphics[trim=450 300 500 320,clip,width=0.19\linewidth]{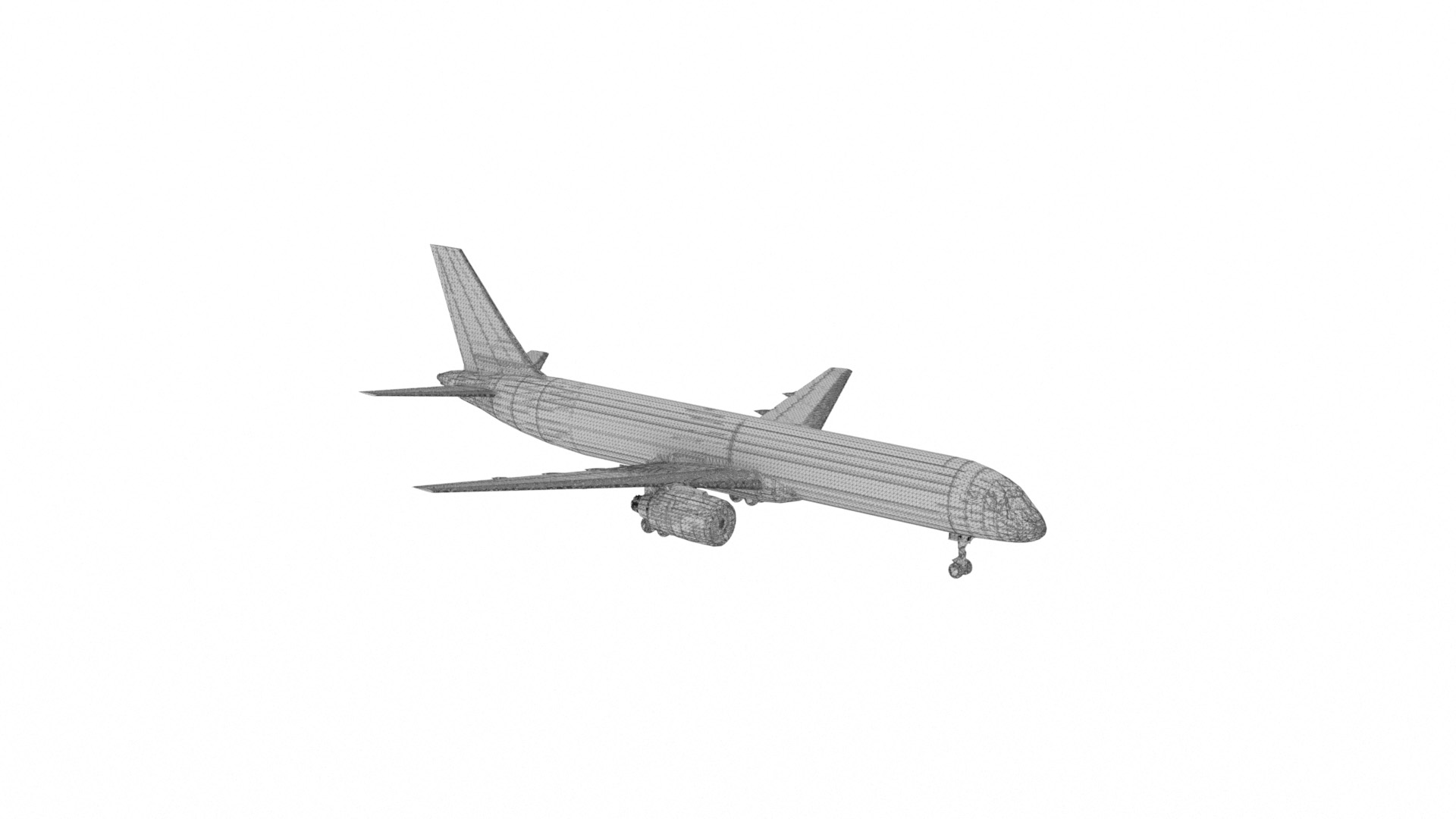}\hfill
\frame{\includegraphics[trim=300 100 600 400,clip,width=0.14\linewidth]{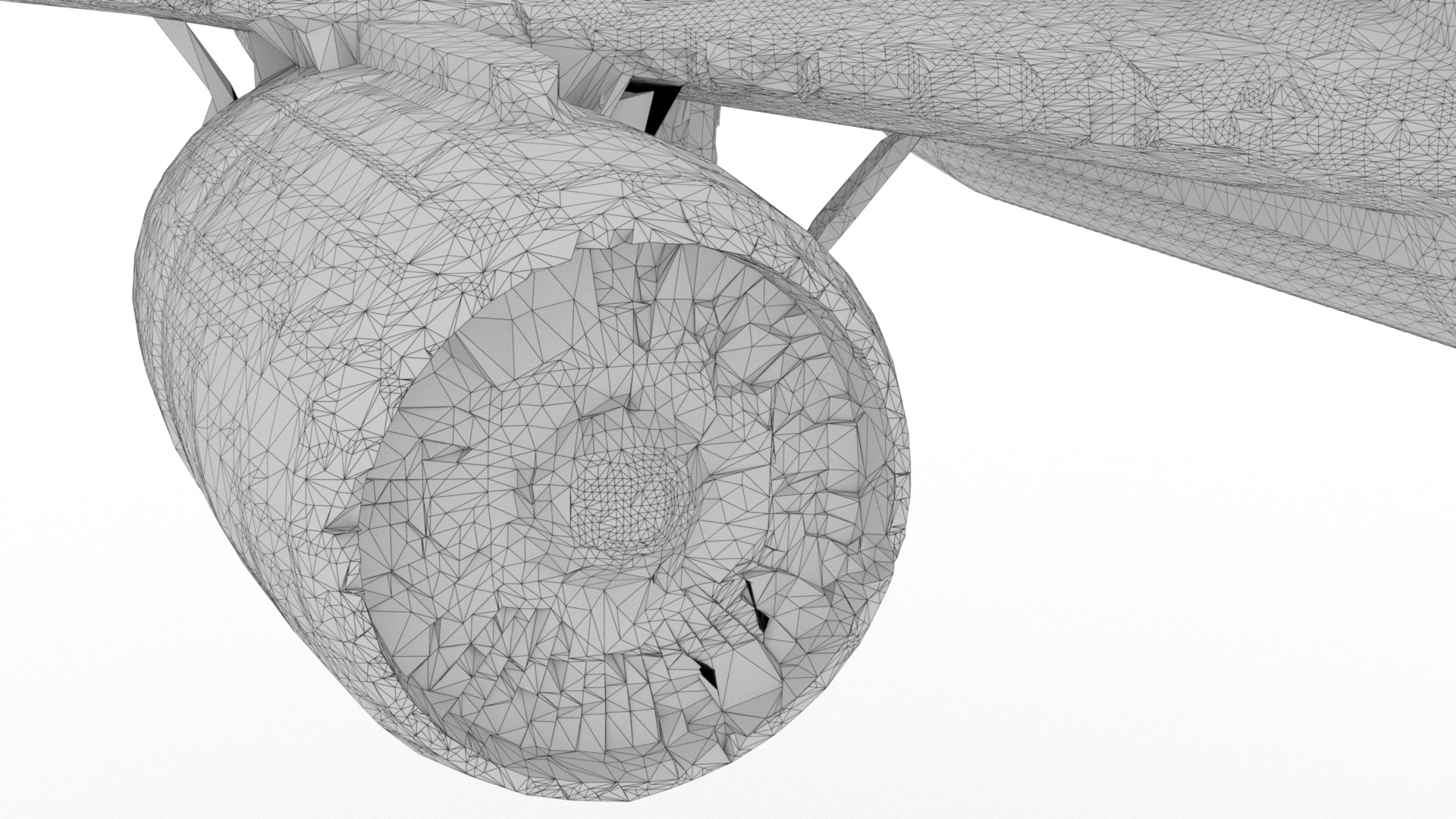}}\hfill
\includegraphics[trim=450 300 500 320,clip,width=0.19\linewidth]{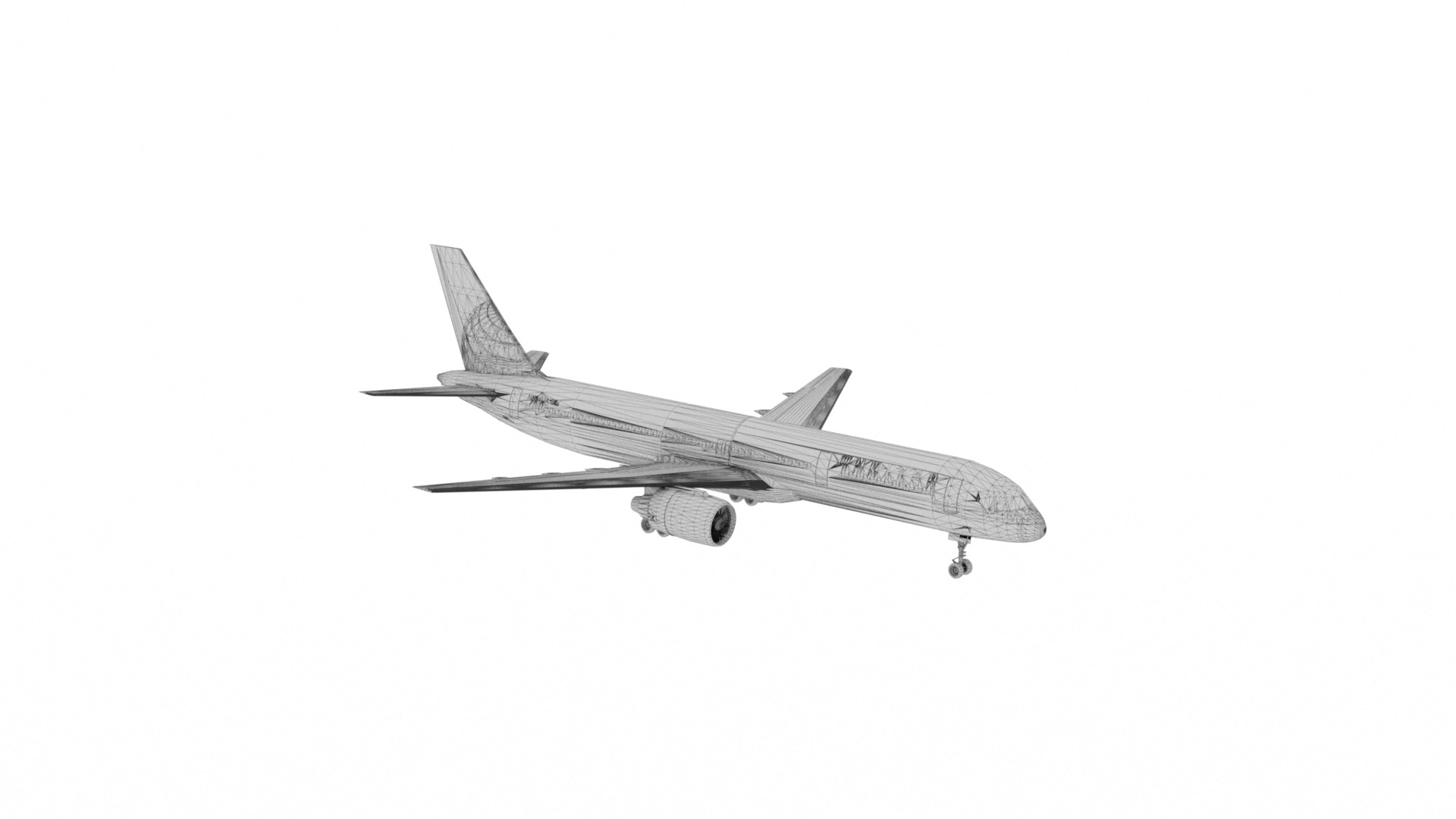}\hfill
\frame{\includegraphics[trim=300 100 600 400,clip,width=0.14\linewidth]{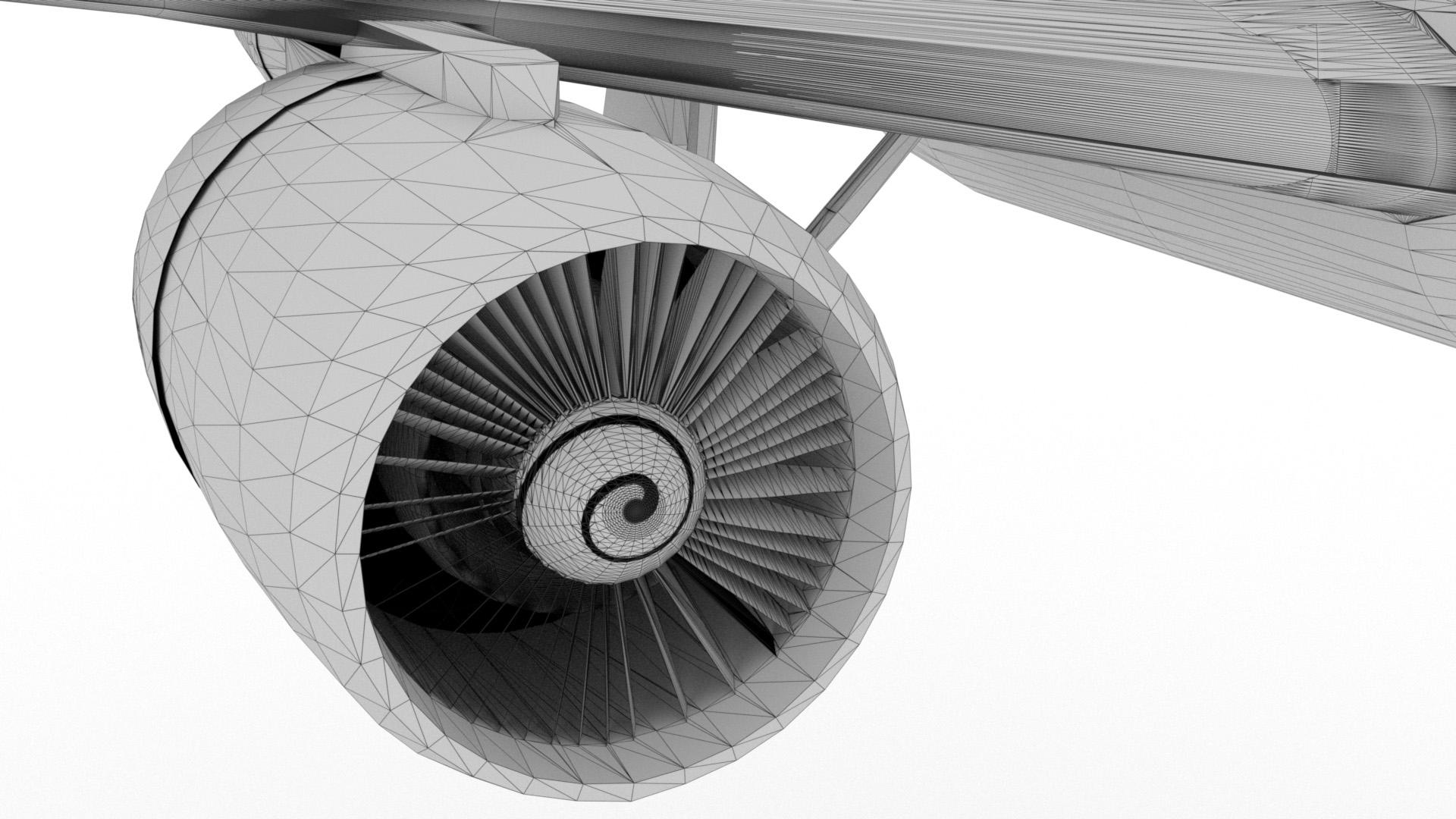}}\hfill
\includegraphics[trim=450 300 500 320,clip,width=0.19\linewidth]{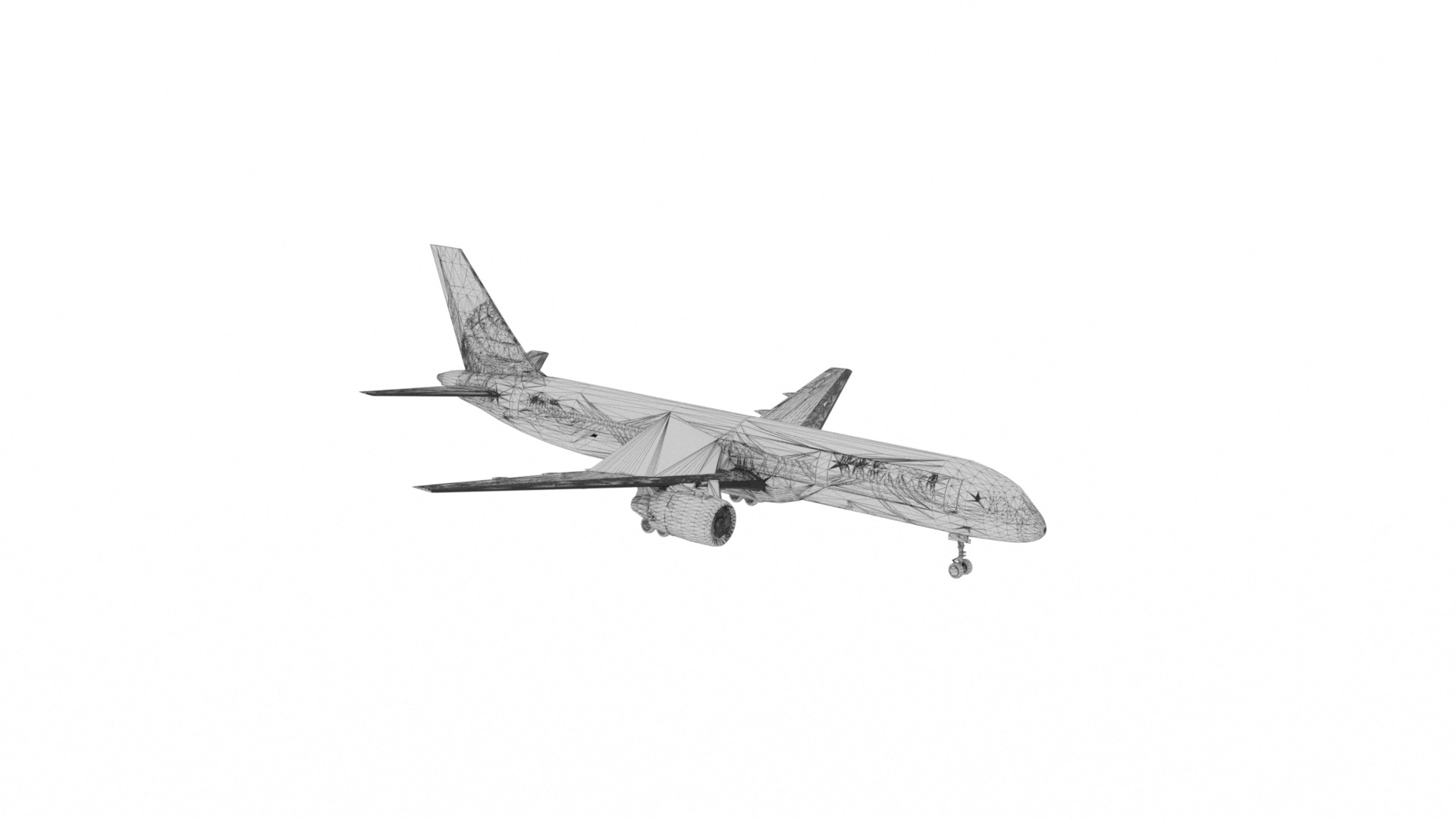}\hfill
\frame{\includegraphics[trim=300 100 600 400,clip,width=0.14\linewidth]{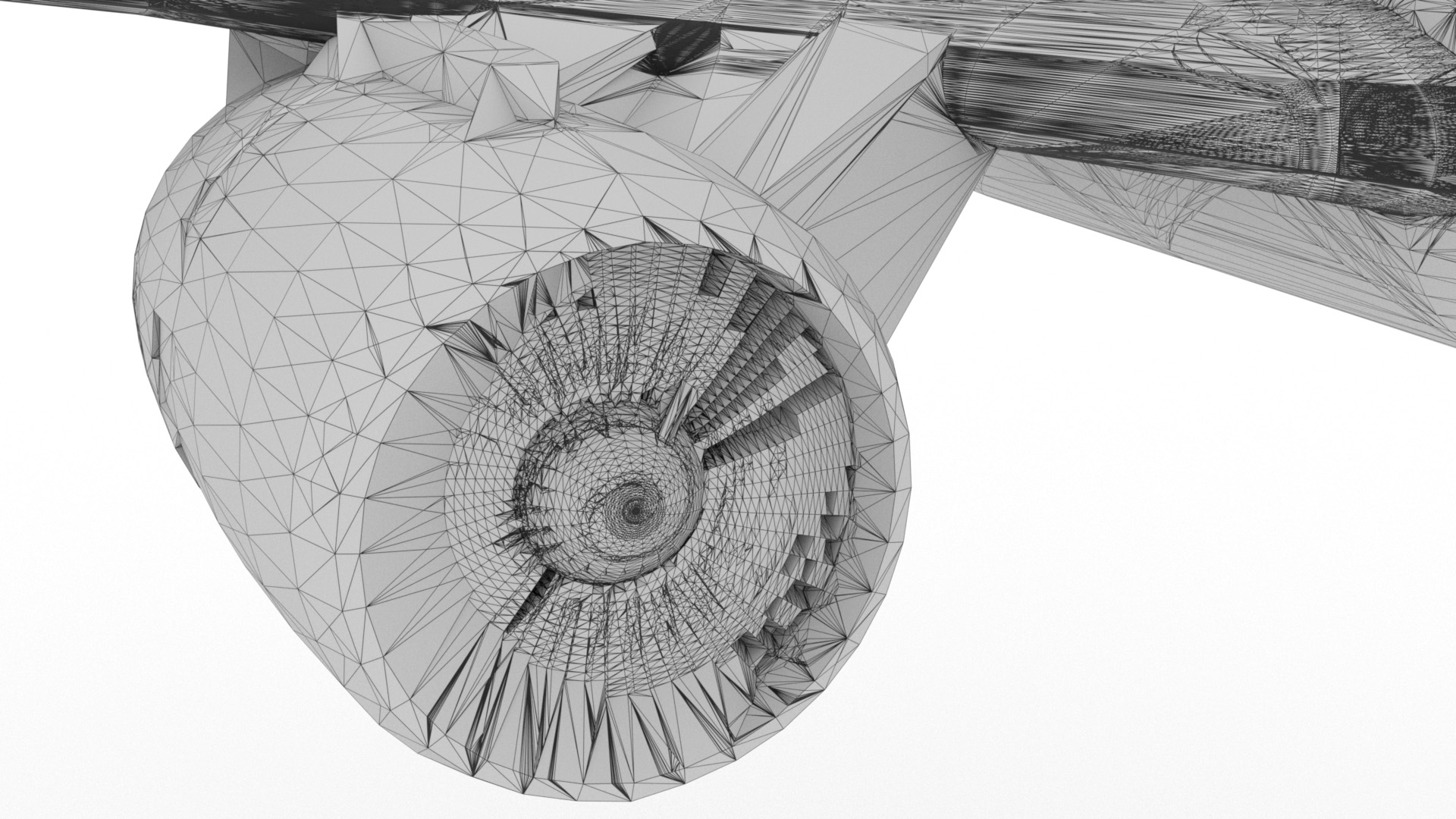}}\\
\begin{minipage}{0.33\linewidth}\centering\footnotesize ManifoldPlus~\cite{huang2020manifoldplus}\end{minipage}\hfill%
\begin{minipage}{0.33\linewidth}\centering\footnotesize VisualRepair~\cite{Chu2019visualrepair}\end{minipage}\hfill%
\begin{minipage}{0.33\linewidth}\centering\footnotesize VolumeMesher~\cite{Diazzi2021volumemesher}\end{minipage}\\\vspace{0.5em}

\includegraphics[trim=450 300 500 320,clip,width=0.19\linewidth]{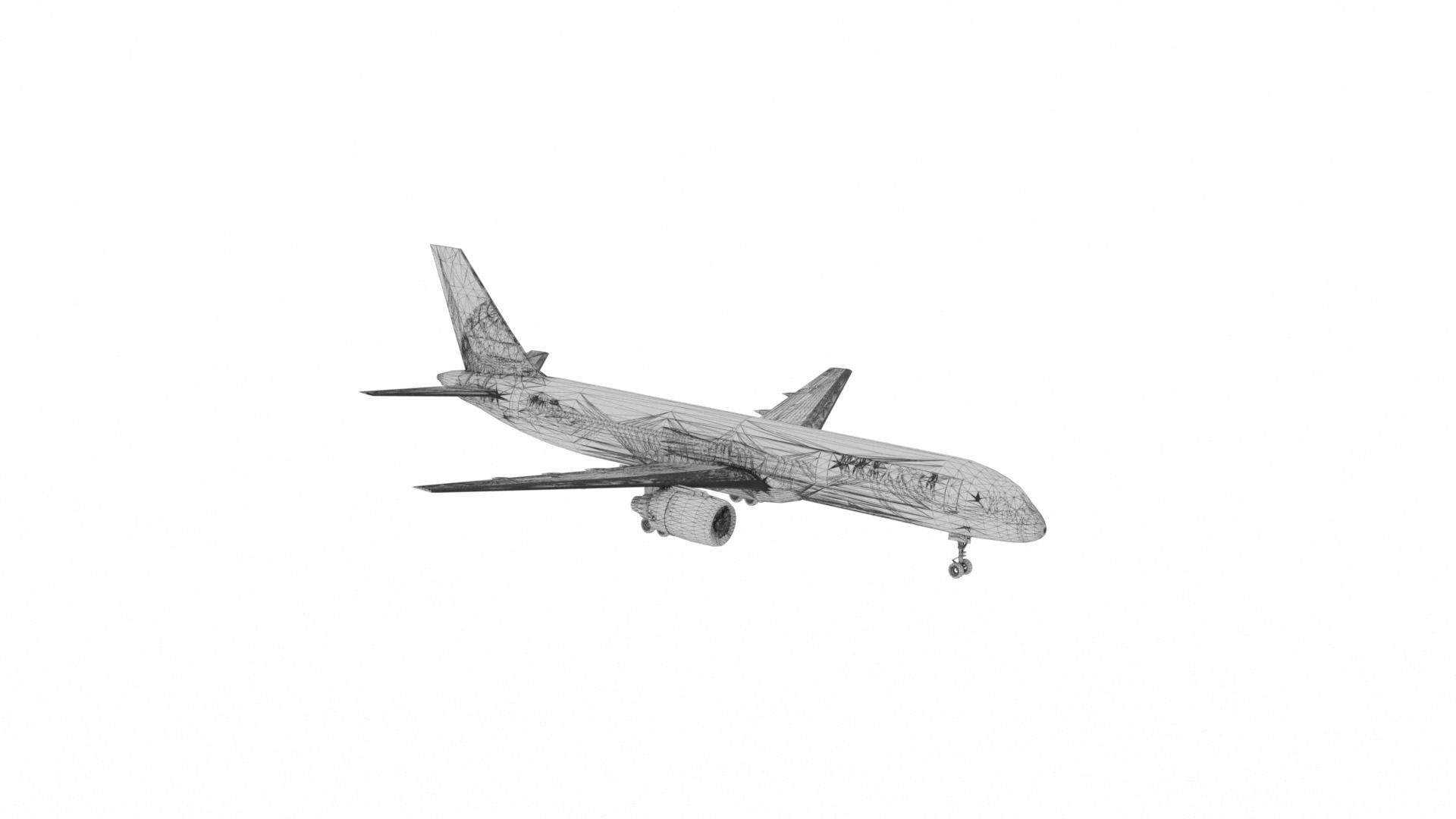}\hfill
\frame{\includegraphics[trim=300 100 600 400,clip,width=0.14\linewidth]{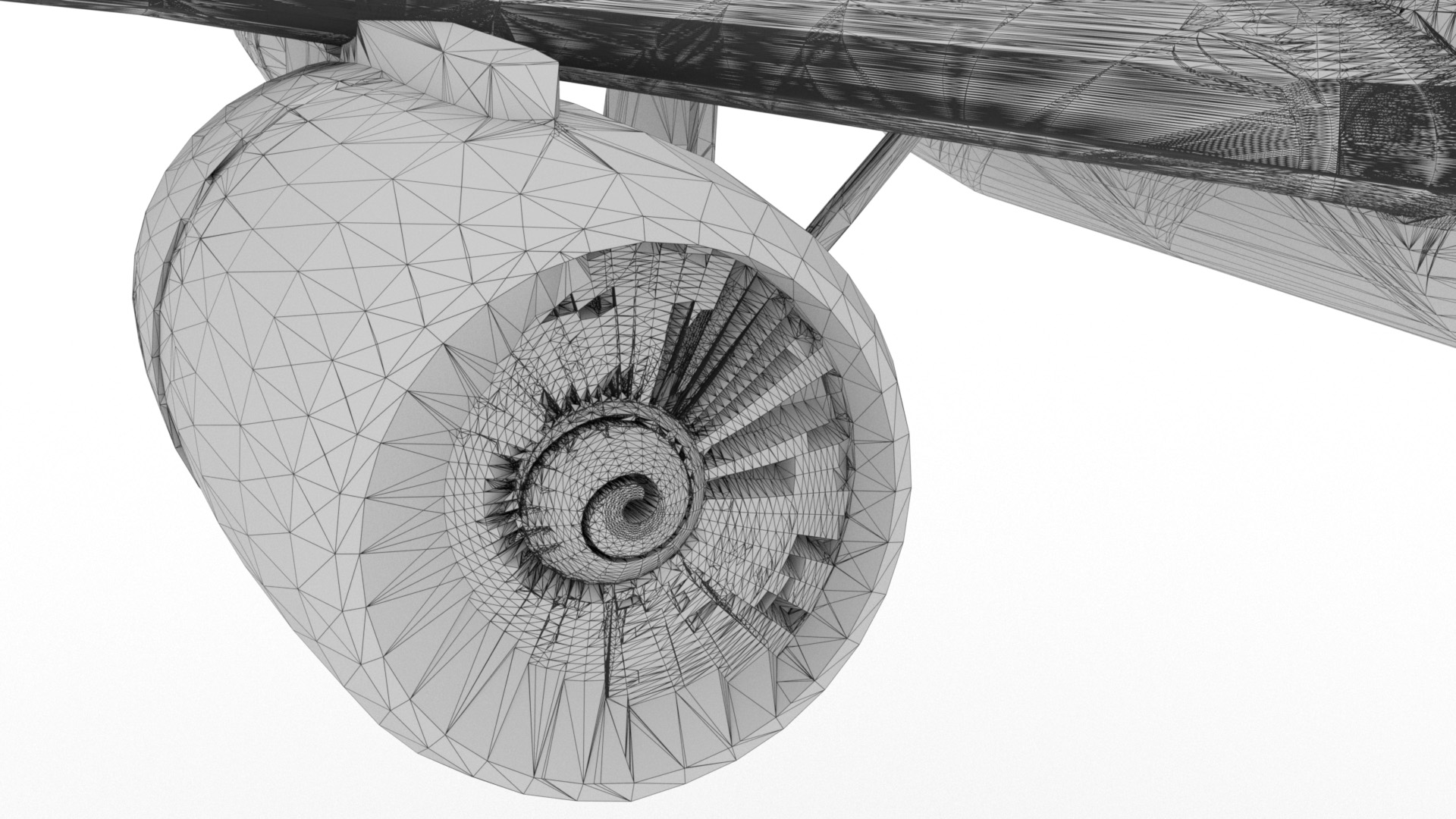}}\hfill
\includegraphics[trim=450 300 500 320,clip,width=0.19\linewidth]{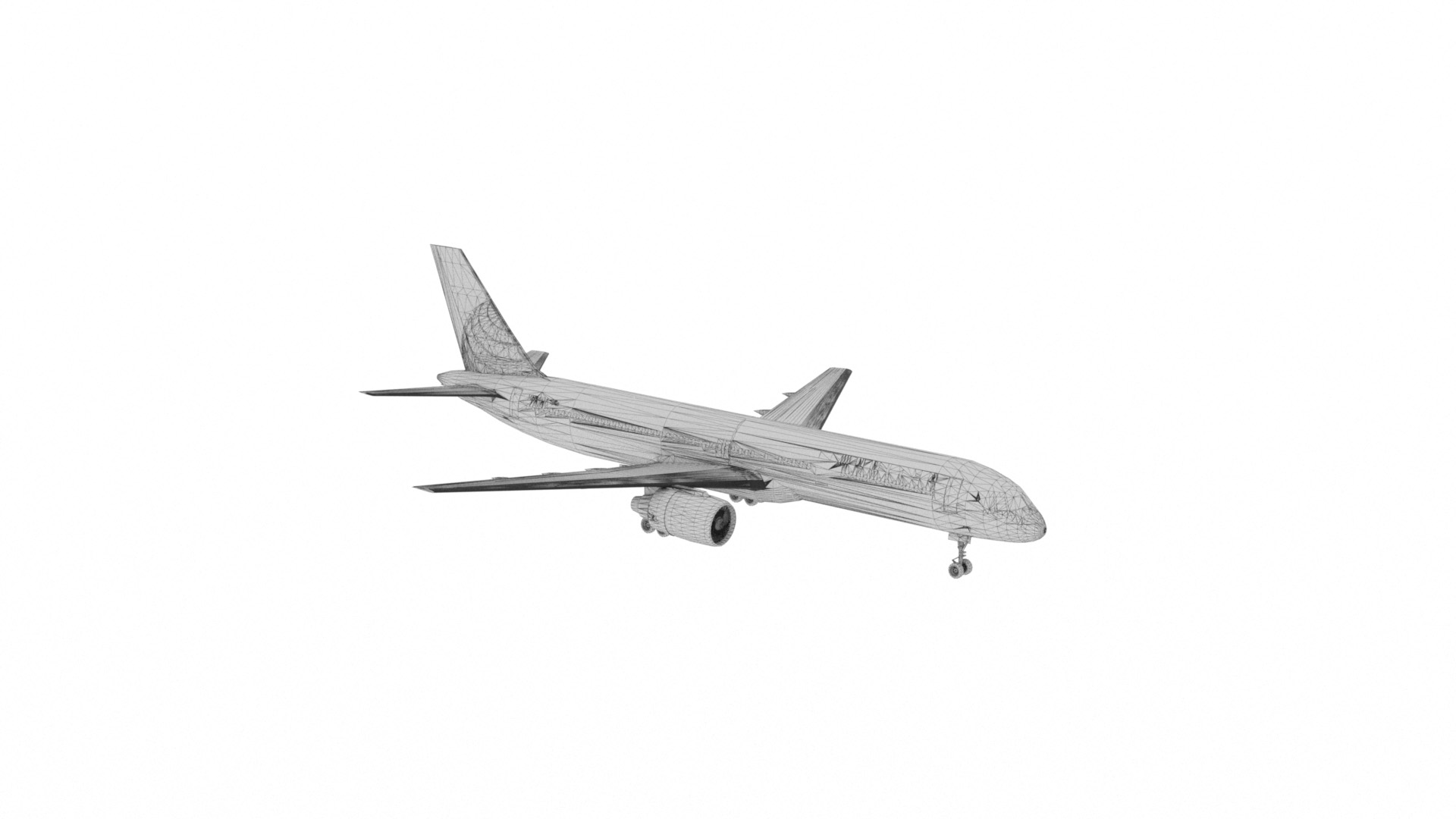}\hfill
\frame{\includegraphics[trim=300 100 600 400,clip,width=0.14\linewidth]{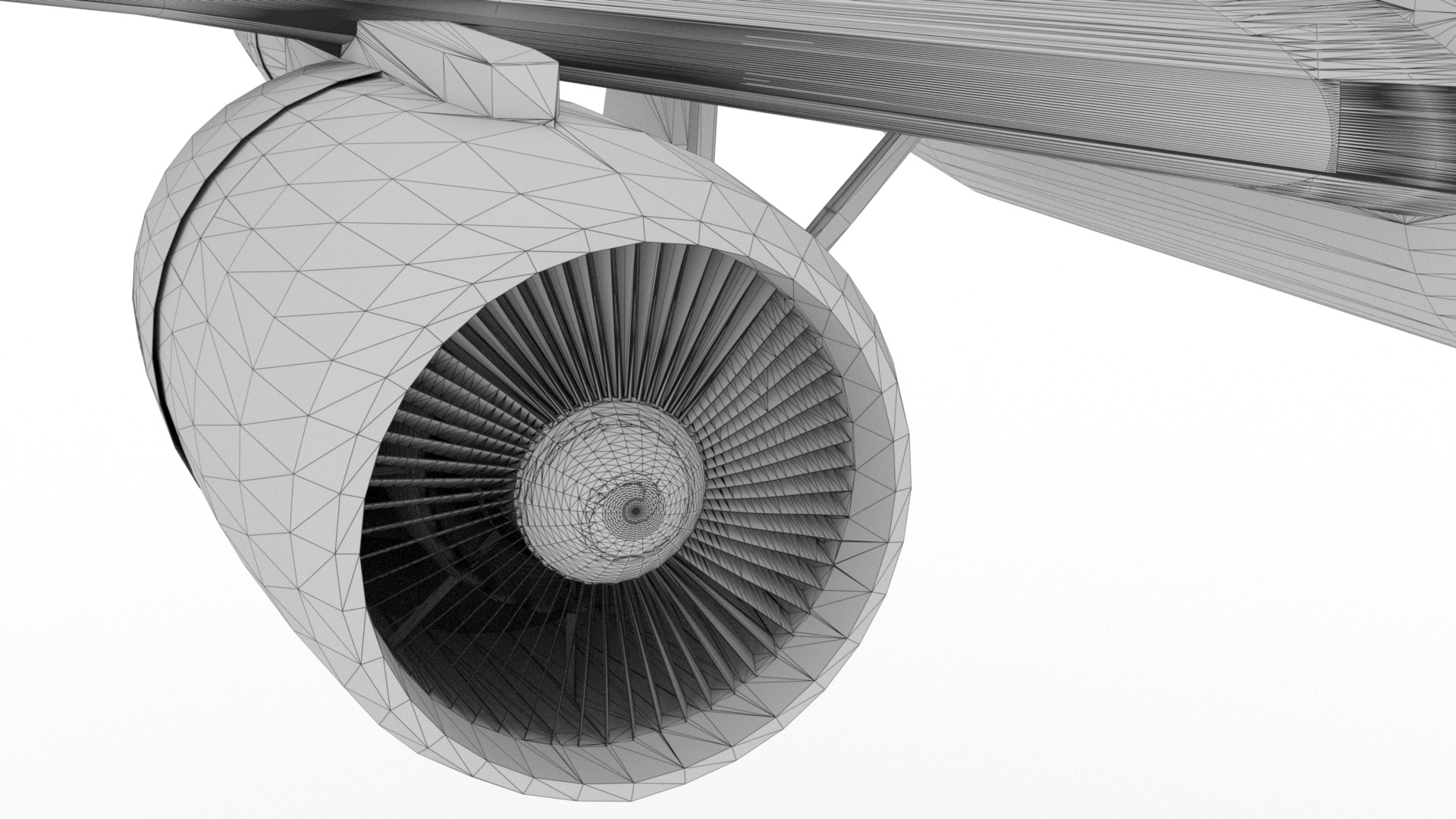}}\hfill
\includegraphics[trim=450 300 500 320,clip,width=0.19\linewidth]{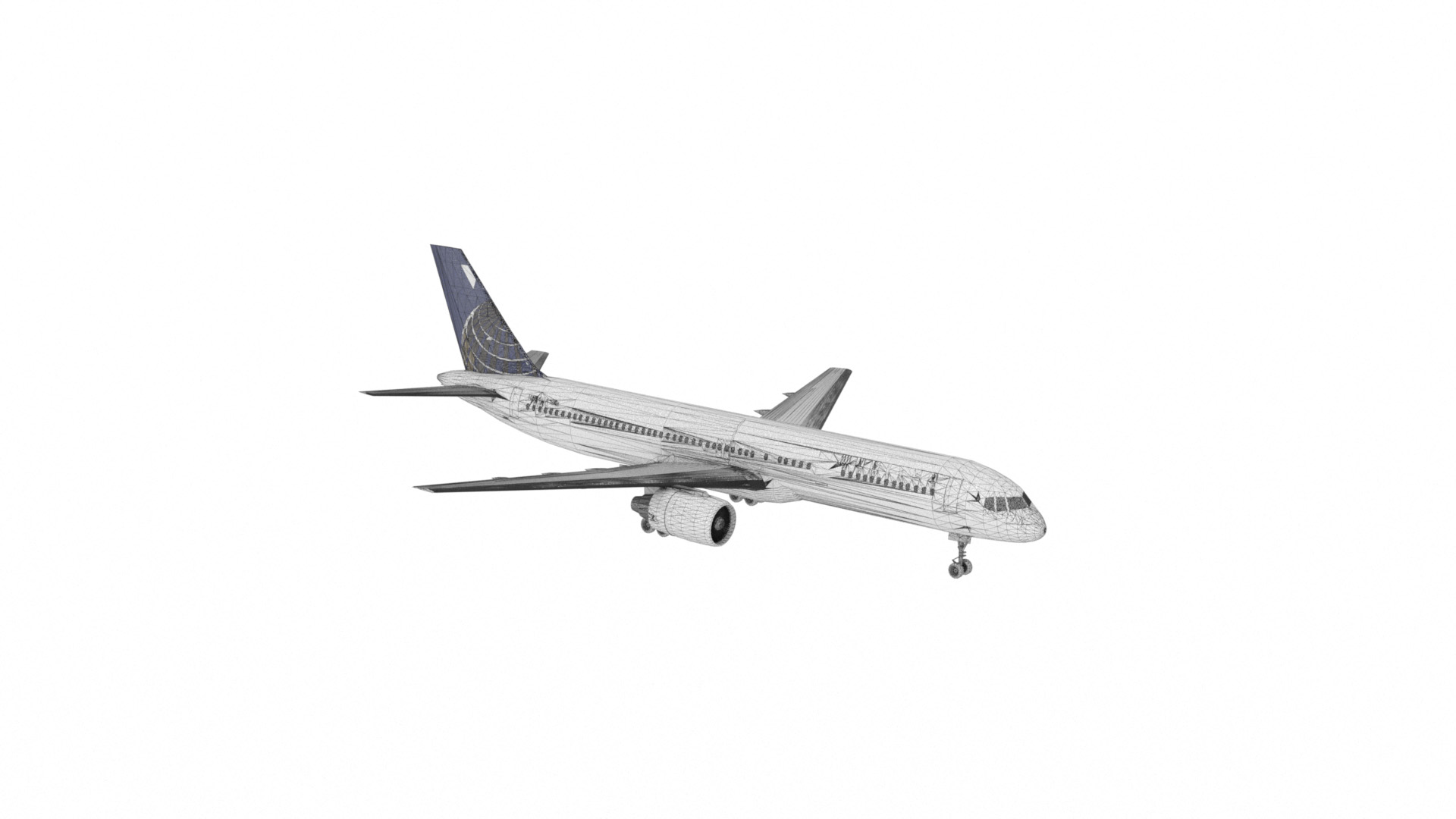}\hfill
\frame{\includegraphics[trim=300 100 600 400,clip,width=0.14\linewidth]{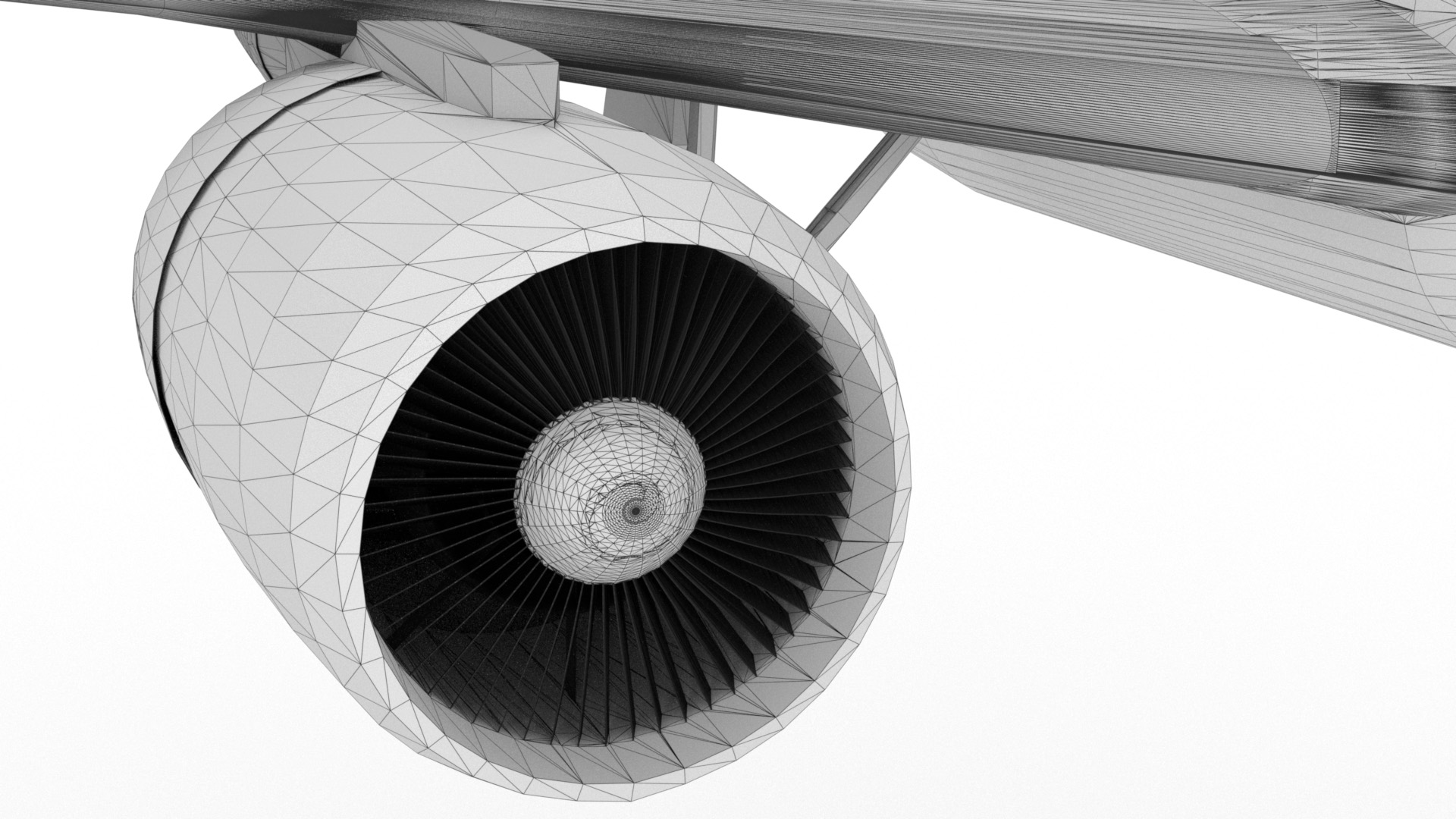}}\\
\begin{minipage}{0.33\linewidth}\centering\footnotesize  VisualRepair + VolumeMesher\end{minipage}\hfill%
\begin{minipage}{0.33\linewidth}\centering\footnotesize  Ours\end{minipage}\hfill%
\begin{minipage}{0.33\linewidth}\centering\footnotesize  Ours with texture\end{minipage}\\
\caption{\textbf{Comparison with existing methods:} We compare mesh repair algorithms on the textured and mis-oriented input model from ShapeNet~\cite{Chang2015shapenet}. None of the existing works can convert the input mesh into watertight manifold mesh while preserving the textures and input details (see the zoom-in view of the engine).}
\label{fig:teaser}
\end{figure*}

Despite the considerable research efforts invested in mesh repair, the challenge of reliably converting a problematic mesh into a watertight manifold mesh, with the input details and UVs preserved as much as possible, remains unsolved. Broadly speaking, existing mesh repair approaches can be categorized into two groups. Local approaches aim at addressing individual defects by analyzing the geometry and topology of a local sub-mesh, as done in~\cite{Hornung2006,Zhao2007,Attene2010,Attene2014}. Local approaches can repair sparse defects while preserving large portions of visual features, but they often lack guarantees and may inadvertently introduce new issues, e.g., self-intersections, during the repair process.
In recent years, researchers have turned their focus to global techniques for superior robustness. On the downside, however, global methods can violate the minimal modification requirement and oftentimes impair the defectless mesh parts due to global conversions and remeshing, as noted in survey~\cite{Attene2013survey}. For example, recent works, VolumeMesher~\cite{Diazzi2021volumemesher} and TetWild~\cite{hu2018tetwild} utilize a BSP tree to partition the ambient space and close gaps and holes by solving a graph cut problem or using winding numbers to filter out interior and exterior volumes. Although these methods can guarantee watertight meshes, their results are sensitive to mis-oriented input meshes. Indeed, a small mis-orientation or nested structures can lead to drastically different output with undesirable modifications to the visual appearance (\autoref{fig:teaser}). In addition, the BSP tree can introduce a massive number of unnecessary faces that are inherited in the output mesh. On a parallel front, ~\cite{Chu2019visualrepair} proposed a surface-based method in which patch orientation and connectivity are globally optimized based on visual guidance, leading to output with minimal modification. Unfortunately, their method splits the inconsistent patches, resulting in many gaps and non-watertight open surfaces.

\begin{table*}
\centering
\caption{Summary of related work. In each column, green indicates the preferred property; red means the method lacks the preferred property; yellow means the method has some, but not all aspects of the preferred property. 
}
\label{tab:compare_methods}
\newcommand{\bad}{\cellcolor{red!20}}
\newcommand{\ok}{\cellcolor{yellow!20}}
\newcommand{\good}{\cellcolor{green!20}}
\newcommand{\non}{\cellcolor{white!20}}
\scalebox{1}{
\begin{tabular}{l|cccccc}
\toprule
\non \textbf{Method}&\non\textbf{Watertightness}&\non\textbf{Face number}&\non\textbf{Preserve shape}&\non\textbf{Preserve UV}&\non\textbf{Memory}&\non\textbf{Speed}\\
\midrule
\non PolyMender (PM)~\cite{Ju2004polymender}             & \good Yes & \good Low  & \ok Medium & \bad No   & \good Low  & \good Fast \\ 
\non TetWild (TW)~\cite{hu2018tetwild}                & \good Yes & \good Low  & \bad Low   & \bad No   & \ok Medium & \bad Slow  \\
\non fTetWild (fTW)~\cite{hu2020ftetwild}               & \good Yes & \good Low  & \bad Low   & \bad No   & \ok Medium & \ok Medium \\
\non AlphaWrapping (AW)~\cite{Portaneri2022AlphaWrapping}   & \good Yes & \ok Medium & \ok Medium & \bad No   & \good Low  & \good Fast \\ 
\non ManifoldPlus (MP)~\cite{huang2020manifoldplus}   & \good Yes & \bad High  & \bad Low  & \bad No  & \ok Medium  & \good Fast  \\ 
\non Takayama et al. (T14)~\cite{takayama2014simple}   &  \bad No & \good Low & \good High  & \good Yes  & \good Low  & \good Fast \\ 
\non VisualRepair (VR)~\cite{Chu2019visualrepair}          & \bad No   & \good Low  & \good High & \good Yes & \good Low  & \bad Slow  \\
\non VolumeMesher (VM)~\cite{Diazzi2021volumemesher}       & \good Yes & \bad High  & \ok Medium & \bad No   & \ok Medium & \good Fast \\ 
\non Ours                                & \good Yes & \good Low  & \good High & \good Yes & \ok Medium & \ok Medium \\
\bottomrule
\end{tabular}
}

\end{table*}

This paper proposes a mesh repair pipeline that combines the merits of all prior local and global algorithms. Our key idea is to guide the global graph cut algorithm using local visual cues. Our method consists of three major steps. In our first (local) step, we propose novel visual measures to quantify the visibility, orientation, and openness of each face. We show that these visual measures can be computed efficiently using GPU ray tracing. Next, we rely on the orientation measure to reorient the faces and use the openness measure to identify and close open surfaces. These local adjustments bootstrap the graph cut, providing well-conditioned initial guess and solution space. Finally, inspired by VolumeMesher~\cite{Diazzi2021volumemesher}, our global step divides the ambient space into polyhedral cells and graph-cuts the interior cells from the exterior, guided by our visual measures. Compared with existing repair algorithms, our key innovations involve:
\begin{itemize}
\item A set of ray-tracing-based visual measures to fix mis-orientations and detect open surfaces, which is further integrated into the graph cut algorithm to produce watertight manifold mesh while preserving visual cues.
\item A constrained simplification post-process to remove unnecessary split faces.
\item A mesh repair algorithm that preserves arbitrary attributes defined on the input mesh, such as UV coordinates.
\end{itemize}
We highlight the effectiveness and robustness of our method on randomly chosen 1000 models from ShapeNet~\cite{Chang2015shapenet} and 400 models from Thingi10K~\cite{Thingi10K}, respectively. Our method outperforms the state-of-the-art in terms of measurements from various aspects such as Hausdorff distance, light field distance (LFD), and PSNR.
\begin{figure*}[ht]
\centering
\includegraphics[width=\linewidth]{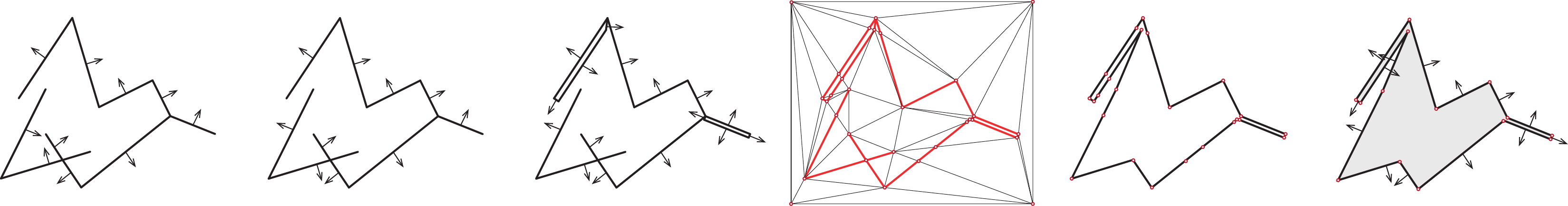}
\vspace{-1.1em}
\begin{flushleft}
{\small\hspace{0.03\linewidth}(a) $M_\text{input}$ \hspace{0.07\linewidth}(b) $M_\text{reoriented}$ \hspace{0.08\linewidth}(c) $M_\text{offset}$ \hspace{0.1\linewidth}(d) $M_\text{partition}$ \hspace{0.07\linewidth}(f) $M_\text{interface}$ \hspace{0.07\linewidth}(e) $M_\text{simplified}$}
\end{flushleft}\vspace{-0.1in}
\caption{\textbf{Our mesh repair pipeline:} Given the input mesh $M_\text{input}$ (a), we first reorient the faces based on the orientation measure to yield $M_\text{reoriented}$ (b). Next, we identify the open faces based on the openness measure and add offsets to close the open faces and arrive at $M_\text{offset}$ (c). Then, we use BSP to partition within the ambient space, getting $M_\text{partition}$ (d), and solve a graph cut problem based on visibility measure to obtain the interface surface $M_\text{interface}$ (f), which is further simplified to remove the redundant faces and vertices (red dot), yielding $M_\text{simplified}$ (e). }
\label{fig:pipeline}
\end{figure*}
\section{Related Work} \label{sec:related}

We summarize the differences between our work and previous papers in~\autoref{tab:compare_methods}. The problem of robustly repairing digital 3D models has been a topic of research for over two decades. For a more in-depth discussion of mesh repair problems and solutions, we refer readers to comprehensive surveys~\cite{Ju2009survey,Attene2013survey}. In general, mesh repair involves remedying the geometric and topological defects from the input mesh, so the output mesh can be used in downstream applications such as mesh processing and simulation. Recently, applications of mesh repair have expanded to 3D printing and learning-related tasks, such as shape analysis and synthesis.

Mesh repairing methods can be broadly classified into two categories: local and global approaches. Local approaches are suitable for input meshes with only sparse defects and remedy the defects by only modifying the mesh structures in a small vicinity. These methods have been widely used to resolve manifold connectivity~\cite{gueziec2001cutting,Attene2009}, close gaps~\cite{Turk1994,Hornung2006}, fill holes~\cite{Podolak2005,Zhao2007}, remove degeneracy~\cite{Attene2010,Campen2010}, and remove self-intersection~\cite{Bischoff2005,Attene2014,attene2018exact}. Since local approaches fix defects only locally, they can preserve as many details as possible. However, most existing local approaches lack guarantees and may introduce new flaws, such as self-intersection.

To address the limitations of local mesh repair approaches, researchers have developed several global methods that leverage volumetric representations to distinguish between interior and exterior volumes, using flood-filling~\cite{Oomes1997,Andujar2002}, line-of-sight information~\cite{Curless1996,Furukawa2007}, distance diffusion~\cite{Guo2006,Masuda2004,Hornung2006}, ray-stabbing~\cite{Nooruddin2003}, parity counting~\cite{spillmann06tetra,Ju2004polymender}, morphology~\cite{Hornung2006,Hetroy2011}, etc. These global methods offer greater robustness in resolving complex defects such as gaps and holes while also ensuring high-quality output. As an example, \cite{Ju2004polymender} proposed PolyMender, a volumetric method based on the octree structure to patch holes for arbitrary input meshes. However, the sharp feature could be aliased due to the voxel-based representation. 

Instead of using regular voxels, recent works~\cite{hu2018tetwild,Diazzi2021volumemesher} partition the space using a binary space partitioning (BSP) tree~\cite{Murali1997} to align the polyhedral cells with the input polygons, so that input-aligned outer shell can be obtained by solving a global segmentation problem. For the input mesh containing open surfaces, Tetwild~\cite{hu2018tetwild} uses winding number~\cite{Jacobson2013} to determine the interior and exterior space, while volume mesher~\cite{Diazzi2021volumemesher} solves a minimum graph cut problem to minimize the total area of the output mesh, so as to close the holes. The same graph cut strategy has been used by \cite{bauchet2020kinetic,labatut2009robust} as well. Unfortunately, these works are based on the assumption that the input faces already have the correct orientations or visibility. But consistent orientations are unavailable in casually collected 3D mesh datasets such as ShapeNet~\cite{Chang2015shapenet}. 
The wrapping technique presented by \cite{Portaneri2022AlphaWrapping,huang2020manifoldplus} provides an alternative approach to repair meshes without relying on face orientation. These methods can produce a strictly enclosing mesh by shifting the faces both inward and outward, but these approaches can corrupt visually sharp features. In contrast, \cite{Chu2019visualrepair} propose a method that utilizes the rasterization pipeline to evaluate the visual significance of each patch and performs a global optimization to ensure consistent orientation and connectivity of the patches. In cases of inconsistent faces, their method splits them to guarantee a manifold output, creating many gaps and holes. We notice that visual measures based on the rasterization pipeline have been adopted in various prior works~\cite{Chu2019visualrepair,wu2022,Gao2022,takayama2014simple}. However, such visual measures cannot account for indirect visibility. This issue is resolved in our novel measures via multi-bounce ray-tracing, which is further integrated into graph cut for visual-preserving segmentation. Using GPU ray-tracing, our measures are also faster to compute than rasterization, leading to improved overall efficacy.
\section{Method} \label{sec:method}
We define a triangle mesh as $M_\bullet\triangleq<\mathcal{V}_\bullet,\mathcal{F}_\bullet>$, where $\mathcal{V}_\bullet$ is a set of vertices and $\mathcal{F}_\bullet$ is a set of triangles connecting vertices. Our mesh repair algorithm requires the input to be a triangle mesh $M_\text{input}$, which is assumed to be visually satisfactory but comes with various geometric and topological defects, including gaps, holes, self-intersection, non-manifold elements, duplicated faces, and inconsistent orientations. Our method generates an output mesh $M_\text{output}$ that is guaranteed to be manifold and watertight, with as few as possible modifications to the visual appearance of the input. Our mesh repair pipeline is demonstrated in \autoref{fig:pipeline}, and we detail each step below.

\subsection{Visual Measures} \label{sec:raytrace}
\begin{figure}[ht]
\centering
\includegraphics[width=0.9\linewidth]{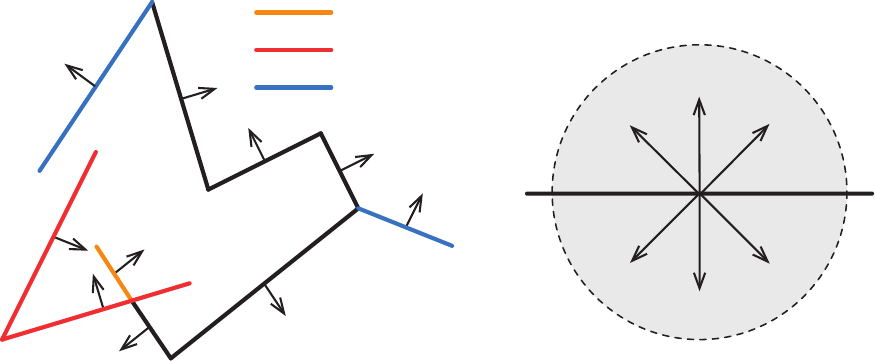}
\put(-133,86){\footnotesize{Invisible}}
\put(-133,75.5){\footnotesize{Misorientated}}
\put(-133,65){\footnotesize{Open surface}}
\put(-25,45){\huge{+}}
\put(-23.5,30){\huge{-}}
\vspace{-2em}
\begin{flushleft}
{\small\hspace{0.37\linewidth}(a)\hspace{0.5\linewidth}(b)}
\end{flushleft}\vspace{-0.1in}
\caption{Given an input mesh (a) with misoriented faces, invisible faces, and open faces, we sample directions (b) on the unit hemisphere for the positive $(+)$ and negative $(-)$ side of the face, and the arrow indicates the face normal.}
\label{fig:visual}
\end{figure}
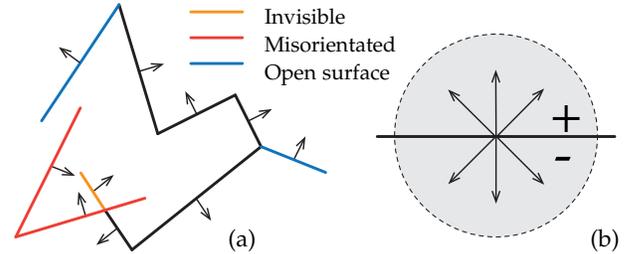

Since our input mesh is visually satisfactory, the visual cues provide strong guidance to our repair algorithm. Therefore, we propose three visual measures to quantify the visibility, orientability, and openness of each face, as illustrated in~\autoref{fig:visual} (a).

We first uniformly sample each face $f_\text{input}^i$, where the number of sample points, $N_\text{s}$, depends on the face area $A(f_\text{input}^i)$:
\begin{align}
 N_\text{s}(f_\text{input}^i)=\max\set{\left\lceil\frac{A(f_\text{input}^i)}{A(M_\text{input})} N_\text{total}\right\rceil, N_{\min}},
\end{align}
where $A(\bullet)$ indicates the surface area of input geometric entity and $N_\text{total}$ is the total sample number over the surface. We set a minimum sample number $N_{\min}$ to avoid under-sampling and we use $\mathcal{S}^i$ to indicate the set of all samples on the face $f_\text{input}^i$, so $|\mathcal{S}^i| = N_\text{s}(f_\text{input}^i)$.

At each sample location, we then use uniformly sampling~\cite{dutre2003global} to generate $N_\text{d}$ directions on the unit hemisphere for positive side $(+)$ and negative side $(-)$ of the face, as shown in \autoref{fig:visual} (b). Unlike conventional ray tracing, which shoots rays from the camera, we shoot rays from the sampled location along the sampled direction. When a ray hits the surface, a random direction over the normal hemisphere is picked as the reflection direction. We consider a ray valid if the ray can hit the bounding box of $M_\text{input}$ within a given number of bounces $N_\text{b}$. At location $\pp\in\mathcal{S}^i$, the total number of valid rays for positive and negative sides of the face are $N^+_\pp$ and $N^-_\pp$, respectively. For example, $\sum_{\pp \in \mathcal{S}_i} (N^+_\pp + N^-_\pp)>0$ means face $f_\text{input}^i$ is visible by at least one of the rays.

Utilizing $N^+_\pp$ and $N^-_\pp$ computed by ray-tracing, we quantify the probability of $f_\text{input}^i$ being visible from outside the mesh using the visibility measure defined as: 
\footnotesize
\begin{align}
\Phi_\text{visibility}(f)=\frac{\max_{\pp\in\mathcal{S}^i} \max\{N^+_\pp, N^-_\pp\}}{N_\text{d}},
\end{align}
\small
and we further classify $f_\text{input}^i$ as being visible if $\Phi_\text{visible}(f_\text{input}^i)>0.5$. We further quantify the probability that $f_\text{input}^i$ is consistently oriented via the orientation measure defined as:
\footnotesize
\begin{align}
\Phi_\text{orientation}(f)=
\begin{cases}
\frac{\sum_{\pp \in \mathcal{S}^i} (N^+_\pp - N^-_\pp)}{\sum_{\pp \in \mathcal{S}^i} (N^+_\pp + N^-_\pp)},& \text{if } \scalemath{0.9}{ \sum_{\pp \in \mathcal{S}^i} (N^+_\pp + N^-_\pp)>0}\\
0,              & \text{otherwise}.
\end{cases}
\end{align}
\normalsize
$\Phi_\text{orientation} = -1$ means there is a large chance that the back side face $f_\text{input}^i$ is much more visible than the front side, and the face should be flipped. Finally, we consider the face an open surface if it has high visibility from both sides, whose probability is quantified by our openness measure:
\footnotesize
\begin{align}
\Phi_\text{openness}(f) = 
\begin{cases}
\max_{\pp\in \tilde{\mathcal{S}}^i} \frac{\min\{N^+_\pp,N^-_\pp\}}{\max\{N^+_\pp,N^-_\pp\}} \frac{N_\pp^+ + N_\pp^-}{2 N_\text{d}},& \text{if } \tilde{\mathcal{S}}^i\neq \varnothing\\
0,                     & \text{otherwise}.
\end{cases}
\end{align}
\normalsize
where $\tilde{\mathcal{S}}^i\subset \mathcal{S}^i$ indicates the visible samples of the face. We have $\Phi_\text{openness}(f_\text{input}^i)\in [0,1]$, and a higher measure indicates $f_\text{input}^i$ is more likely to be an open thin shell.

\subsection{Orientation Adjustment} \label{sub:preprocess}
Guided by $\Phi_\text{orientation}$, our algorithm locally adjusts the misoriented faces as much as possible, which leads to the better performance of the follow-up global graph cut step. To this end, we first remove the duplicated faces sharing the same vertices.
We then group input faces into patches $\mathcal{P}^j = \{f_\text{input}^{ji}\}$ using a flood fill algorithm, such that faces sharing an edge and having consistent orientations will be grouped into the same patch. Next, we calculate a weighted average orientation measure over each patch $\mathcal{P}^j$, which is defined as:
\begin{align}
\Phi_\text{orientation}(\mathcal{P}^j)=
\begin{cases}
\frac{\sum_{f\in \tilde{\mathcal{P}}^j} A(f) \Phi_\text{orientation}(f)}{\sum_{f\in \tilde{\mathcal{P}}^j} A(f)},& \text{if } \tilde{\mathcal{P}}^j \neq \varnothing\\
0,              & \text{otherwise},
\end{cases}
\end{align}
where $\tilde{\mathcal{P}}_j\subset\mathcal{P}^j$ indicates the visible set of faces in the patch $\mathcal{P}^j$. Finally, we flip the patch $\mathcal{P}^j$ if $\Phi_\text{orientation}(\mathcal{P}^j)<0$. Note that for nearly open patches with high visibility measures from both sides, its orientation measure is close to zero, meaning there is no preference for its orientation. The output of this step is denoted as $M_\text{reoriented}.$

\begin{figure}[ht]
\centering
\includegraphics[width=\linewidth]{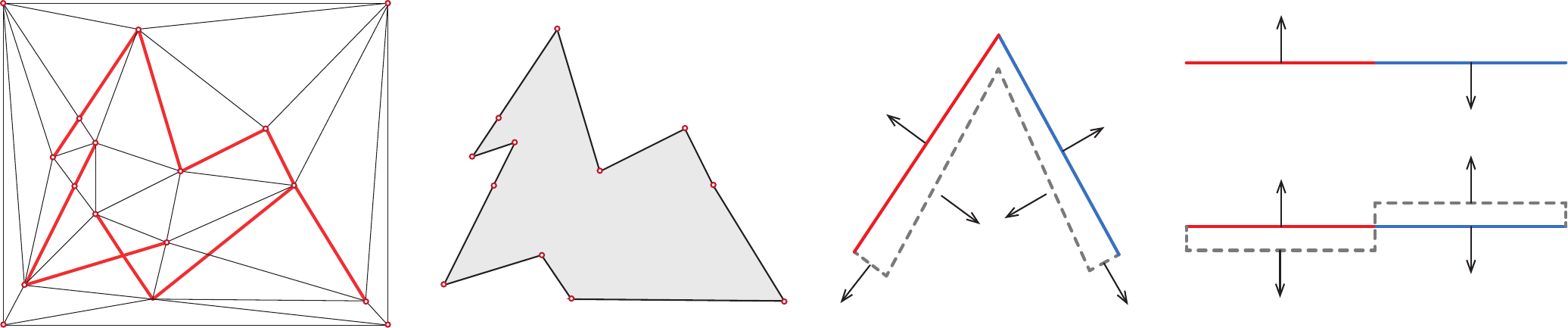}
\vspace{-1.5em}
\begin{flushleft}
{\small\hspace{0.11\linewidth}(a) \hspace{0.22\linewidth}(b) \hspace{0.2\linewidth}(c) \hspace{0.2\linewidth}(d) }
\end{flushleft}\vspace{-0.1in}
\caption{If offset is not applied to the open surface for the example in~\autoref{fig:pipeline}, we end up with $M_\text{partition}$ in (a) and $M_\text{interface}$ in (b), where a large piece of unnecessary volume is introduced, impairing visual appearance. Instead, we offset vertices of the open surface to form a closed thin shell (c), with offset faces denoted as dashed lines. For non-orientable edges (d), we offset vertices multiple times along the normal of each adjacent face (bottom).}
\label{fig:offset}
\end{figure}
\subsection{Offsetting Open Surface}\label{sec:offset}
For the graph cut algorithm to close open surfaces with minimal visual modification, our method offsets the open faces by a small distance. Although graph cut algorithms can find watertight meshes even without such offset, as done in VolumeMesher~\cite{Diazzi2021volumemesher}, it can introduce large unnecessary volumes as demonstrated in \autoref{fig:offset} (ab), impairing the visual appearance. Again, our first step is identifying the open surfaces as guided by $\Phi_\text{openness}$. Unlike the orientation adjustment step, we cannot identify open surfaces in a patchwise manner because a patch may contain inner structures and self-intersections. Instead, we classify each face $f_\text{reoriented}^i$ as an open surface if $\Phi_\text{openess}(f_\text{reoriented}^i)>0.5$. However, offsetting each open face would create too many volumetric cells for the graph cut algorithm, slowing down the overall algorithm. Therefore, after open faces are classified, we group connected, consistent-oriented, open faces into open patches. We offset the vertices on the patch along the negative normal direction with a user-defined distance $d_\text{offset}$ to create thin volumetric shells, as shown in \autoref{fig:offset} (c). The vertex normal is the average normal of adjacent face normals weighted by the face area. Note that, in the case of non-orientable meshes, such as the Mobius strip, grouping neighboring open faces can end up with non-manifold edges with a zero normal vector. In this case, we offset vertices on the non-manifold edge along each adjacent face normal (\autoref{fig:offset} (d)). The output of this step is denoted as $M_\text{offset}$. 


\begin{figure*}[ht!]
\newcommand{\figcap}[1]{\begin{minipage}{0.25\linewidth}\centering#1\end{minipage}}
\includegraphics[width=\linewidth]{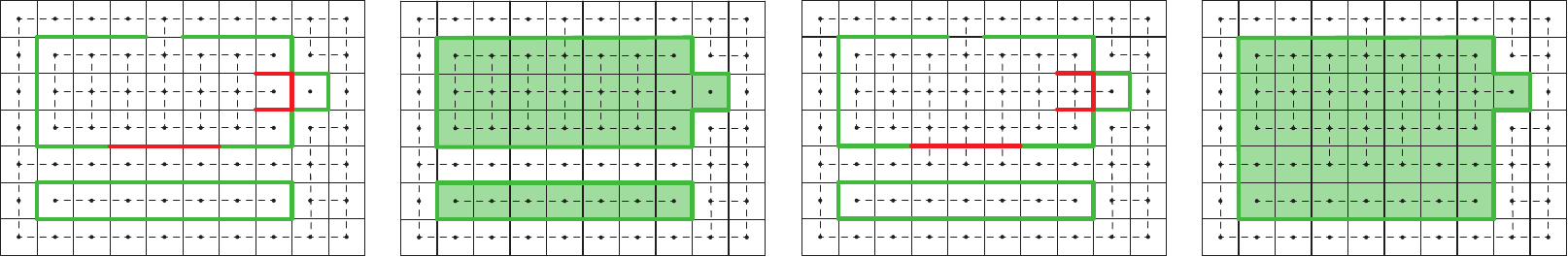}
\figcap{\small (a)}\hfill%
\figcap{\small (b)}\hfill%
\figcap{\small (c)}\hfill%
\figcap{\small (d)}\vspace{-0.05in}
\caption{\textbf{Example of graph cut:} Black dots represent cell nodes, and dashed lines represent the edges between cells. Given the input mesh (a), green and red segments represent visible and invisible faces, respectively, and there is no edge over these visible and invisible faces. After solving the graph cut, the extracted mesh is in green (b). If we keep the edges across invisible faces (c), the graph cut can return an erroneously extracted mesh that closes the entire mesh (d).}
\label{fig:graphcut}
\end{figure*}

\subsection{Space Partition} \label{sub:Partition}
After offsetting open faces, $M_\text{offset}$ may still contain gaps between patches that are not identified as open surfaces. Following the previous work~\cite{Diazzi2021volumemesher}, we next adopt the global step by partitioning the ambient space and solving the graph cut problem to find the interface mesh that closes all gaps.

We initialize the partitioned mesh via a Delaunay tetrahedrization of the vertex set $\mathcal{V}_\text{offset}$. However, such tetrahedrization cannot ensure all the input faces are included in $\mathcal{F}_\text{partition}$. Therefore, we iteratively split the initial partition mesh using two sets of splitting faces, in the same way as constructing the BSP tree. The first set is all the faces in $\mathcal{F}_\text{offset}$. Including all of $\mathcal{F}_\text{offset}$ ensures geometric fidelity, but this is not enough to preserve user-defined surface attributes such as UV coordinates and material IDs. This is because certain edges are shared by two co-planar faces, which are recognized as a single large face by the BSP data structure. Such edges will be erroneously removed from the data structure by the default BSP construction algorithm. If the two neighboring faces have discontinuous surface attributions, we require their shared edge to be included in the BSP data structure. We achieve this by using an arbitrary face passing through the edge to split the partition mesh.

To make the iterative partition process unconditionally robust while preserving origin geometry, we use exact arithmetic during splitting~\cite{Diazzi2021volumemesher} via LPI (Line-Plane Intersection), and TPI (Three-Planes Intersection)~\cite{cherchi2020fast,wang2020exact} for fast exact constructions. This step yields the partitioned mesh $M_\text{partition}$.

\subsubsection{Surface Attribution Preservation}
In order to preserve the surface attributions, e.g., UV coordinates, material IDs, etc., we need to maintain a mapping $\mathcal{M}: \mathcal{F}_\text{input}\to\mathcal{F}_\text{partition}$. Since we use exact arithmetic, $\mathcal{M}(f_\text{partition}^i)$ can be determined by checking whether the barycenter of $f_\text{partition}^i$ lies exactly on some $f_\text{input}^j$, following the strategy used in VolumeMesher~\cite{Diazzi2021volumemesher}. Due to the choice of our splitting surfaces, $\mathcal{M}$ is well-defined, i.e., each $f_\text{partition}^i$ is either contained in some $f_\text{input}^j$ or does not belong to any face of $\mathcal{F}_\text{partition}$ (in which case we let $\mathcal{M}(f_\text{partition}^i)\triangleq\varnothing$).

\subsection{Interface Mesh Extraction} \label{sub:classification}
To perform the global graph cut, we first refine the face orientation (Section~\ref{sec:reorientation}) based on $M_\text{partition}$. Next, we utilize the visibility measure to classify each face in $\mathcal{F}_\text{partition}$ (Section~\ref{sec:classification}). These measures will be used to formulate the objective function in the graph cut (Section~\ref{sec:cellclassify}) to determine the interior/exterior cells, whose interface surface will be the watertight mesh $M_\text{interface}$.

\subsubsection{Face Reorientation} \label{sec:reorientation}
\begin{figure}[ht!]
\centering
\includegraphics[trim=0 50 0 35,clip,width=\linewidth]{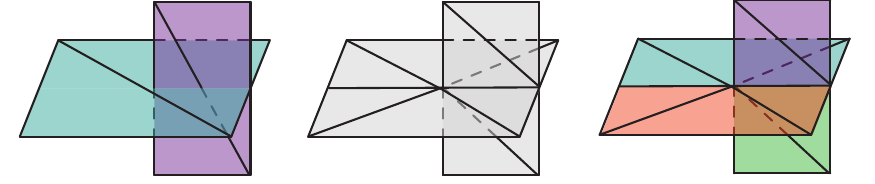}
\vspace{-2em}
\begin{flushleft}
{\small\hspace{0.075\linewidth}(a)\hspace{0.3\linewidth}(b)\hspace{0.3\linewidth}(c)}
\end{flushleft}\vspace{-0.1in}
\caption{Given an input mesh (a) with two intersected patches, our extracted $M_\text{partition}$ (b) contains a non-manifold edge shared by the intersected patches. Starting from isolated faces, we group patches with non-manifold edges labeled by different colors in (c).}
\label{fig:group}
\end{figure}

We use a similar procedure as Section~\ref{sub:preprocess} to reorient $\mathcal{F}_\text{partition}$. Specifically, we use a flood fill strategy to group the faces in $\mathcal{F}_\text{partition}$ into patches, such that no patch contains non-manifold edges, as illustrated in \autoref{fig:group}.
Two patches can be merged if they are co-planar, have consistent orientation, and have no non-manifold edge after merging. Then, we reorient each patch based on orientation measure $\Phi_\text{orientation}(\mathcal{P}^j)$ as defined in Section~\ref{sub:preprocess}.

\subsubsection{Face Classification} \label{sec:classification}
We use the procedure in Section~\ref{sec:raytrace} to compute the visibility measure for each face in $\mathcal{F}_\text{partition}$. As a result, $\mathcal{F}_\text{partition}$ can be classified into three groups: visible faces, invisible faces, and extra faces:
\begin{align*}
\mathcal{F}_\text{partition}^\text{visible}&\triangleq\{f\in\mathcal{F}_\text{parition}\text{ is visible}\land\mathcal{M}(f)\neq\varnothing\}\\
\mathcal{F}_\text{partition}^\text{invisible}&\triangleq\{f\in\mathcal{F}_\text{parition}\text{ is invisible}\land\mathcal{M}(f)\neq\varnothing\}\\
\mathcal{F}_\text{partition}^\text{extra}&\triangleq\mathcal{F}_\text{partition}-\mathcal{F}_\text{partition}^\text{visible}-\mathcal{F}_\text{partition}^\text{invisible}.
\end{align*}
To form the watertight interface surface, our goal is to use as many visible faces and as few extra faces as possible.

\subsubsection{Cell Classification} \label{sec:cellclassify}
We treat each cell in the BSP tree as a node in graph $G$ that can be labeled as either interior or exterior. Each facet of a cell corresponds to an edge in $G$, but no edge is created for mappable faces ($\mathcal{M}(f_\text{partition}^i)\neq\emptyset$), no matter whether the face is visible or invisible as shown in~\autoref{fig:graphcut}.

The faces bordering the interior and exterior cells are guaranteed to form a watertight mesh. We solve for a set of cell labels $l^i$ to produce the interface mesh that maximizes the use of visible faces while minimizing extra faces, which can be formulated as the following minimal cut problem:
\begin{align} \label{eq:graphcut}
E(\mathcal{L})=\sum_{l^i} D(l^i) + \sum_{e^{ij}} S(l^i,l^j),
\end{align}
where $l^i \in \{I,E\}$ indicates the $i$th cell $c^i$ to be either \underline{I}nterior or \underline{E}xterior. In~\autoref{eq:graphcut}, the first data cost is formulated as:
\footnotesize
\begin{align}
D(l^i)=
\begin{cases}
\sum_{f\subset c^i \land f\in\mathcal{F}_\text{partition}^\text{visible} \land f\text{ has inward normal}}A(f) & \text{if }l^i=I\\
\sum_{f\subset c^i \land f\in\mathcal{F}_\text{partition}^\text{visible} \land f\text{ has outward normal}}A(f) & \text{if }l^i=E
\end{cases}.
\end{align}
\normalsize
In other words, $D(l^i)$ penalizes incorrectly oriented faces. If a cell $c^i$ is chosen to be interior, then its visible faces should have normals facing outward. Similarly, an exterior cell $c^i$ should have visible faces facing inward. The second edge cost simply penalizes the use of any extra faces, defined as:
\footnotesize
\begin{align}
S(l^i,l^j) =
\begin{cases}
A(f), & \text{if } f\subset c^i\cap c^j \land f\in\mathcal{F}_\text{partition}^\text{extra} \land l^i\neq l^j \\
0,    & \text{otherwise}
\end{cases}.
\end{align}
\normalsize
It is well-known that, as long as the regular condition~\cite{kolmogorov2004energy}:
\begin{align*}
S(I,I)+S(E,E)\leq S(I,E)+S(E,I),
\end{align*}
holds, the problem of binary graph cut has a polynomial complexity algorithm. It is trivial to see the regular condition holds in our case as $S(I,I)=S(E,E)=0$ and $S(I,E)=S(E,I)\geq 0$. The output of this step is denoted as $M_\text{interface}$.

\subsection{Constrained Simplification}
Although our global step guarantees a watertight output, it could also incur many redundant, small facets. We could remove them using conventional mesh simplifiers, e.g., QEM-based mesh reduction~\cite{Garland1997QEM}, but these methods would lead to inverted faces or incur expensive computation to check for inverted faces at each simplification step. Instead, we introduce a constrained mesh simplification to reduce $M_\text{interface}$. We first detect the geometric and UV patch boundaries (Section~\ref{sec:boundarydetect}) and then re-triangulate each patch (Section~\ref{sec:triangulation}) to significantly reduce the face number, while complying with detected boundaries.

\subsubsection{Boundary Detection} \label{sec:boundarydetect}
There are two types of boundaries we want to preserve during the simplification. The first type is the geometric boundaries. We use a flood fill strategy to group co-planar faces with consistent orientation from $\mathcal{F}_\text{interface}$ and extract the boundary of each group as the geometry boundary. Additionally, we traverse all edges in $M_\text{input}$ and find all UV patch boundaries. Notice that although these UV boundaries are preserved during partition, these UV boundaries may be split into segments, and only parts of them are in $\mathcal{M}_\text{interface}$ after mesh extraction. We use the intersection between the origin UV boundaries and the geometric boundaries of each group to find these segments, which uses the rational number to ensure accuracy. 

\subsubsection{Constrained Triangulation} \label{sec:triangulation}
Our constrained triangulation complies with the geometry and texture boundaries found in the previous step. We first use edge-collapse to remove any vertex whose degree is two or adjacent to co-linear edges and check for face intersection before collapse operation to prevent any self-intersection. In particular, we check if the resulting faces intersect with all other faces within the extended bounding box of $P_i$ with extended length $l_\text{extended}$ before each ear-cut operation. Then, we use the constrained ear-cut triangulation~\cite{held2001fist} that obeys geometry and texture boundaries. After triangulation, we get the simplified mesh $M_\text{simplified}$ with much fewer faces and vertices. It should be noted that both edge collapse and ear-cut triangulation techniques introduce no new vertices, operating solely within the interior of edges or faces. Consequently, these processes do not introduce any self-intersection.

\subsection{Topological Correction}
It is worth noting that the extracted mesh from Section~\ref{sec:classification} is a watertight combinatorial 3-manifold with boundary~\cite{Diazzi2021volumemesher}, meaning that it may contain non-manifold edges and vertices. In this case, we split these non-manifold edges and vertices to recover manifoldness. Since our edge-collapse and triangulation always produce edges with an even number of adjacent triangles, our method is guaranteed to output watertight and manifold meshes~\cite{rossignac1999matchmaker,Attene2009}. The output mesh is denoted as $M_\text{simplified}$.

\subsection{Recovering Surface Attributes}
There are three types of faces in $\mathcal{F}_\text{simplified}$: inherited faces from $F_\text{output}$, offset faces due to Section~\ref{sec:offset}, and extra faces defined in Section~\ref{sub:classification}. We recover inherited faces' attributes from $M_\text{input}$ using barycentric interpolation. The offset faces' attributes are copied from their original faces. In our experiment, most faces can be traced back to their original faces. However, for those extra faces created for closing holes and gaps, we do not have prior knowledge. To assign surface attributes, we perform a flood fill and iteratively set the attributes of extra faces by averaging from their one-ring neighboring vertices.

\begin{figure}[b!]
\newcommand{\figcap}[1]{\begin{minipage}{0.49\linewidth}\centering#1\end{minipage}}
\includegraphics[trim=570 50 650 300,clip,width=0.30\linewidth]{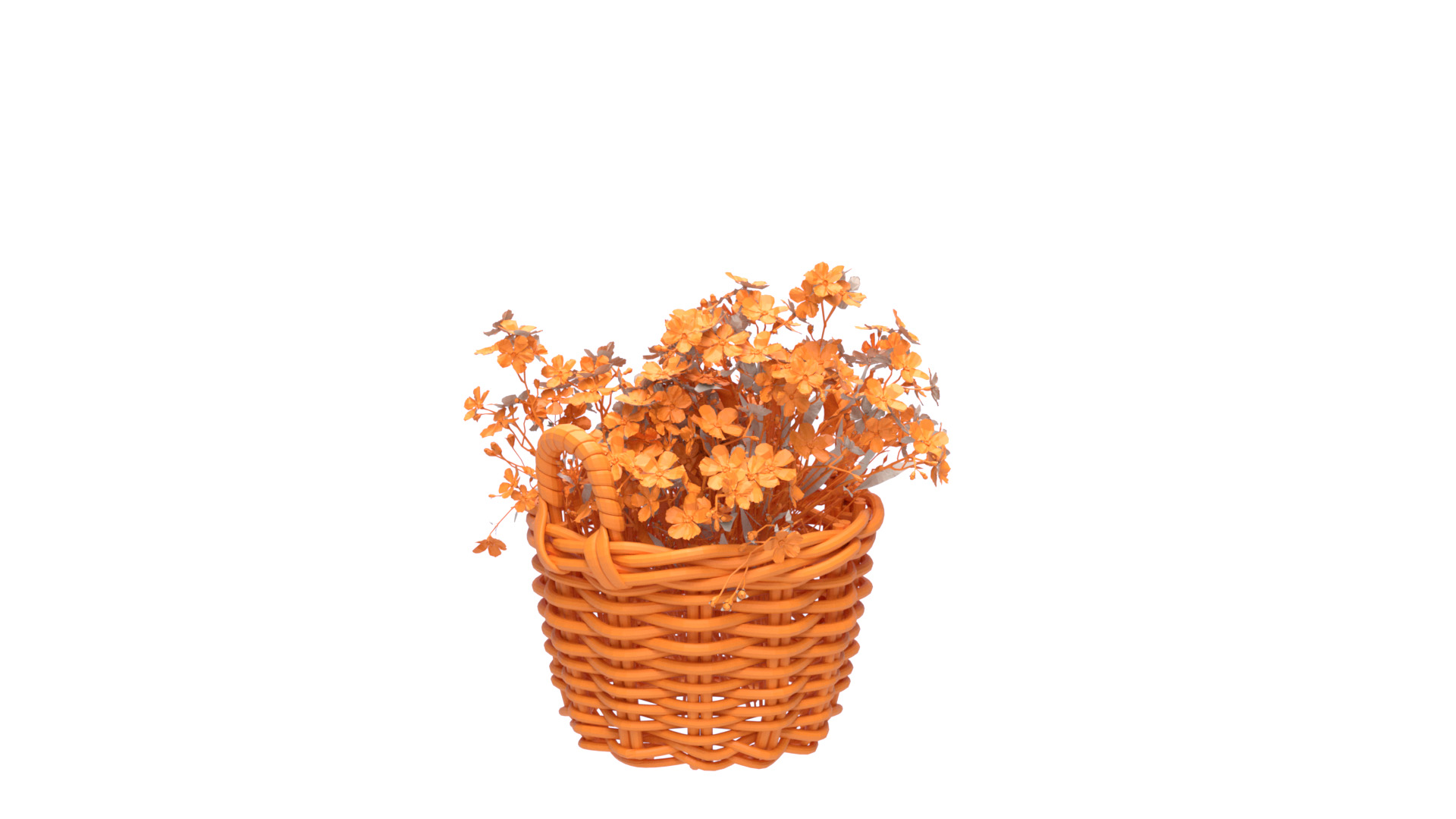}\hfill
\frame{\includegraphics[trim=500 50 650 250,clip,width=0.19\linewidth]{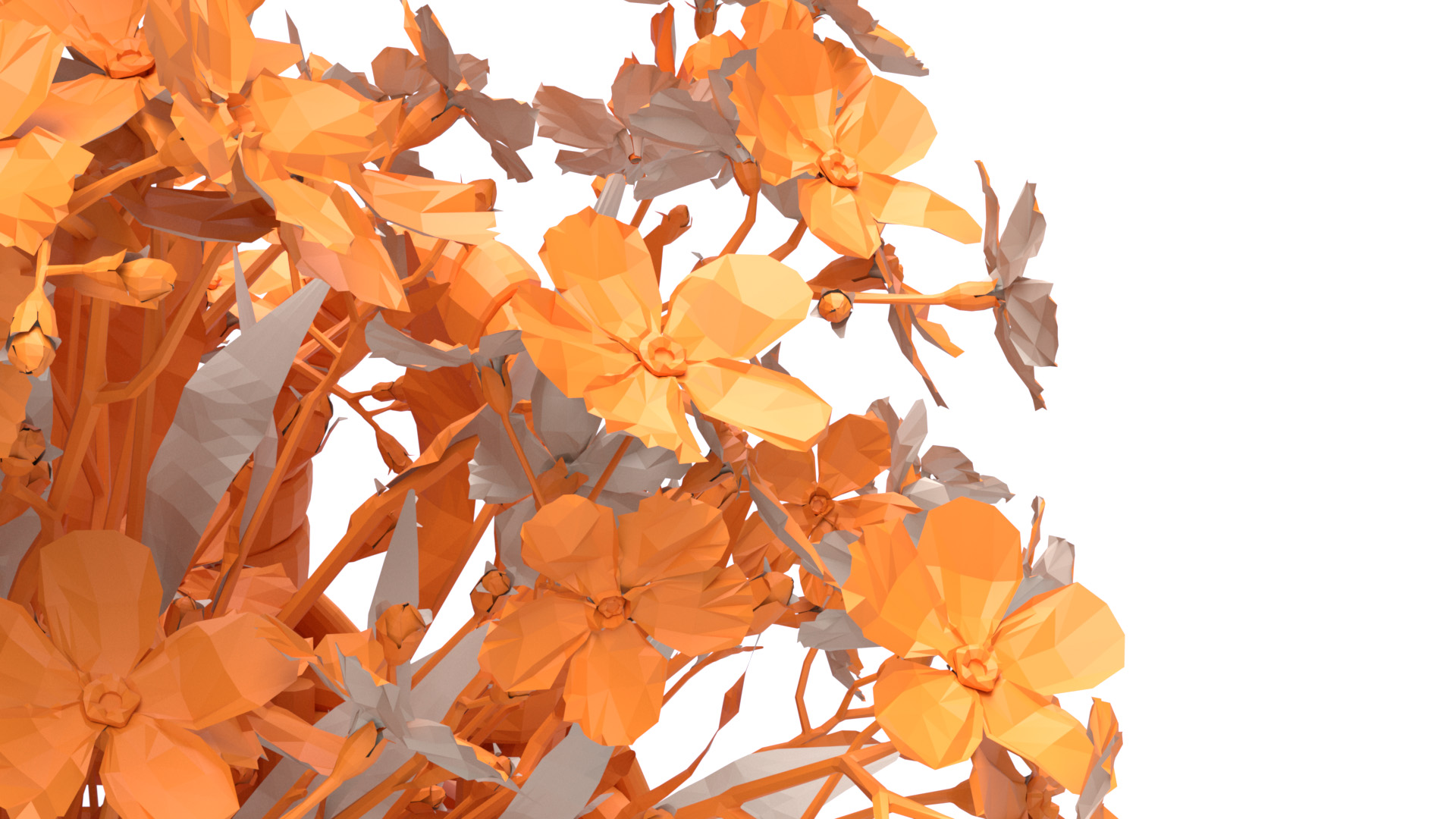}}\hfill
\includegraphics[trim=570 50 650 300,clip,width=0.30\linewidth]{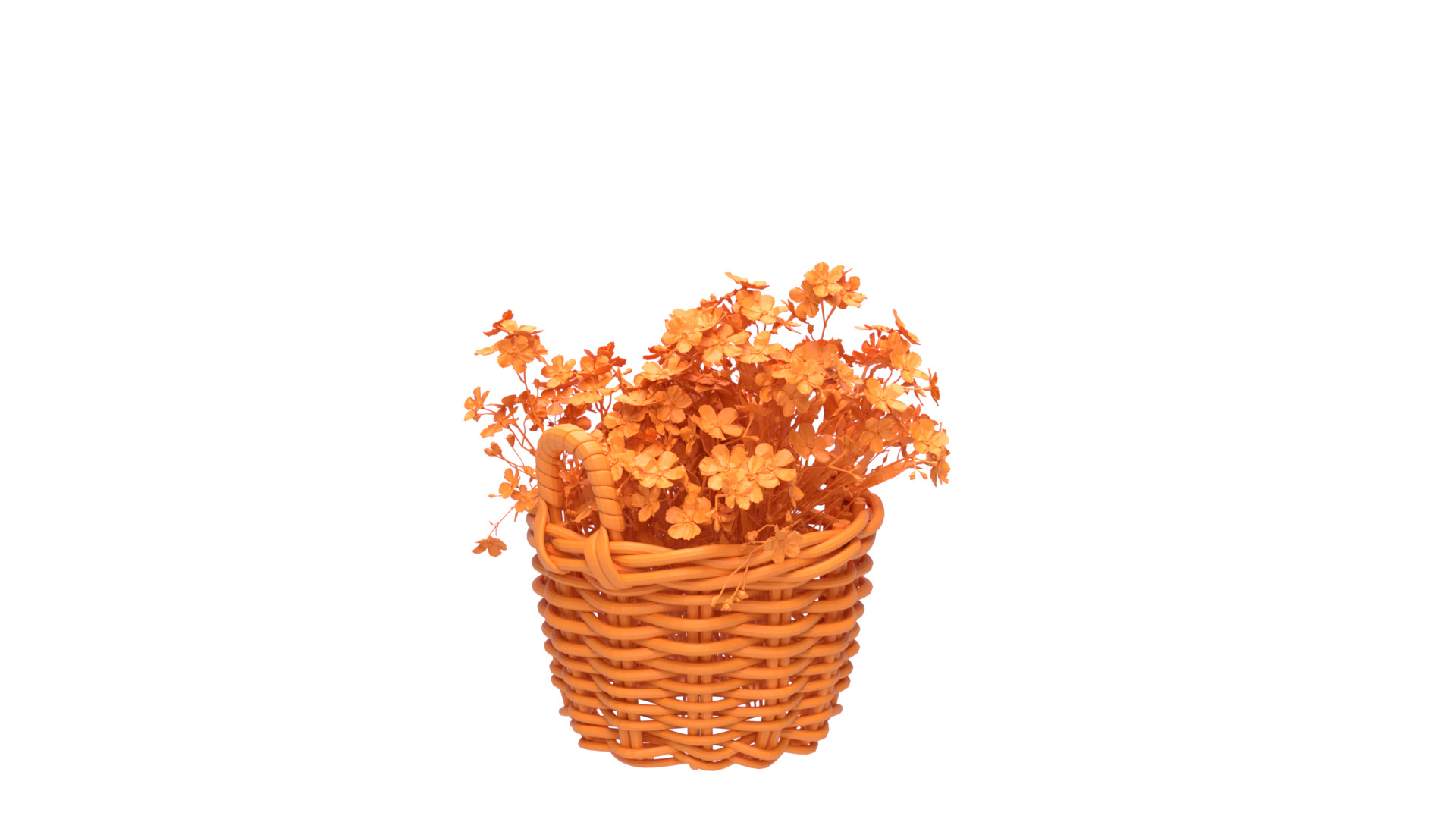}\hfill
\frame{\includegraphics[trim=500 50 650 250,clip,width=0.19\linewidth]{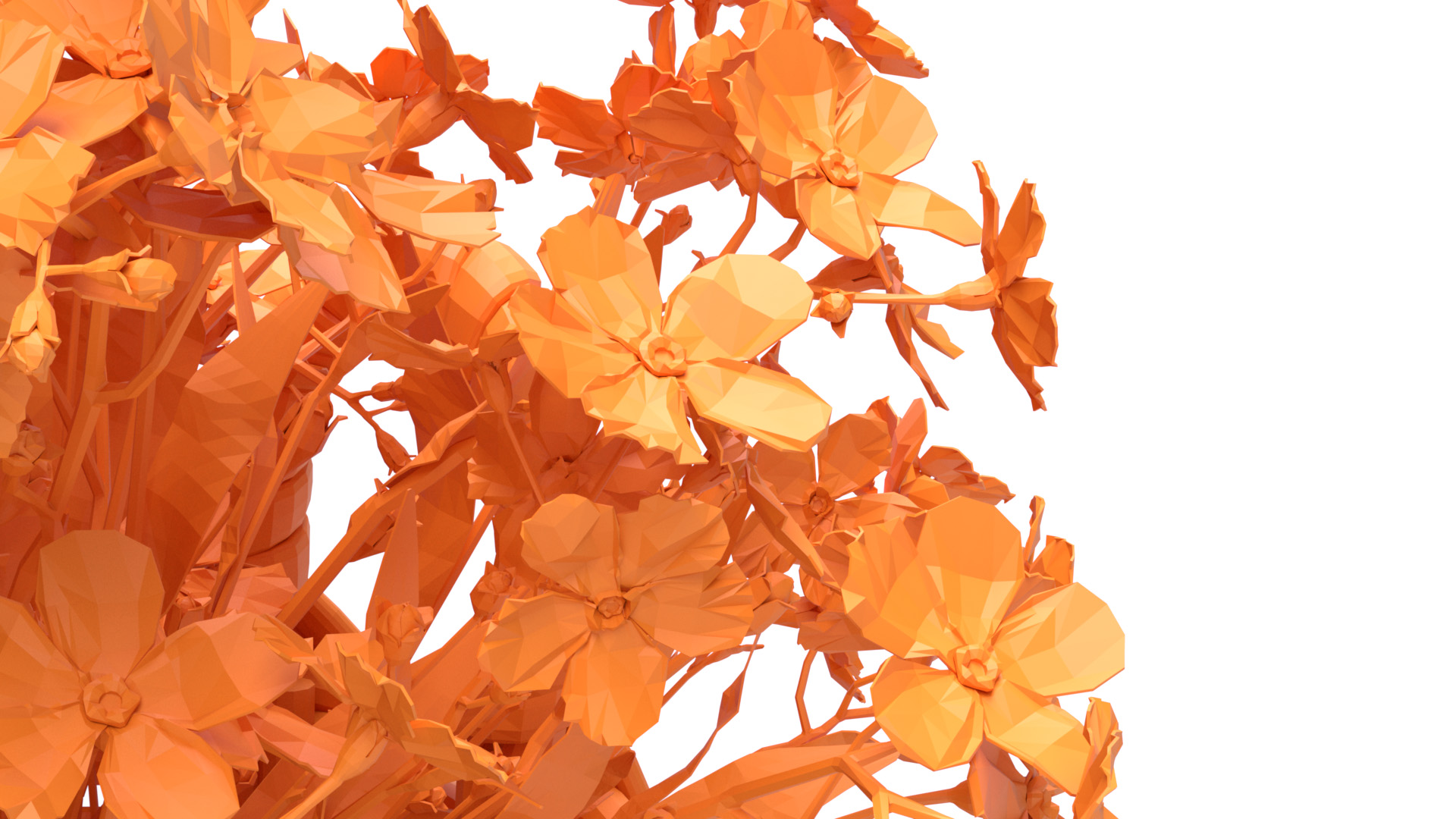}}\\
\figcap{\small Input }\hfill%
\figcap{\small Ours with $d_\text{offset} = {D}/{20000}$ }\\
\figcap{\small -- }\hfill%
\figcap{\small $\text{HD} = 6.9e-3, \text{LFD} = 86$ } \\
\includegraphics[trim=570 50 650 300,clip,width=0.30\linewidth]{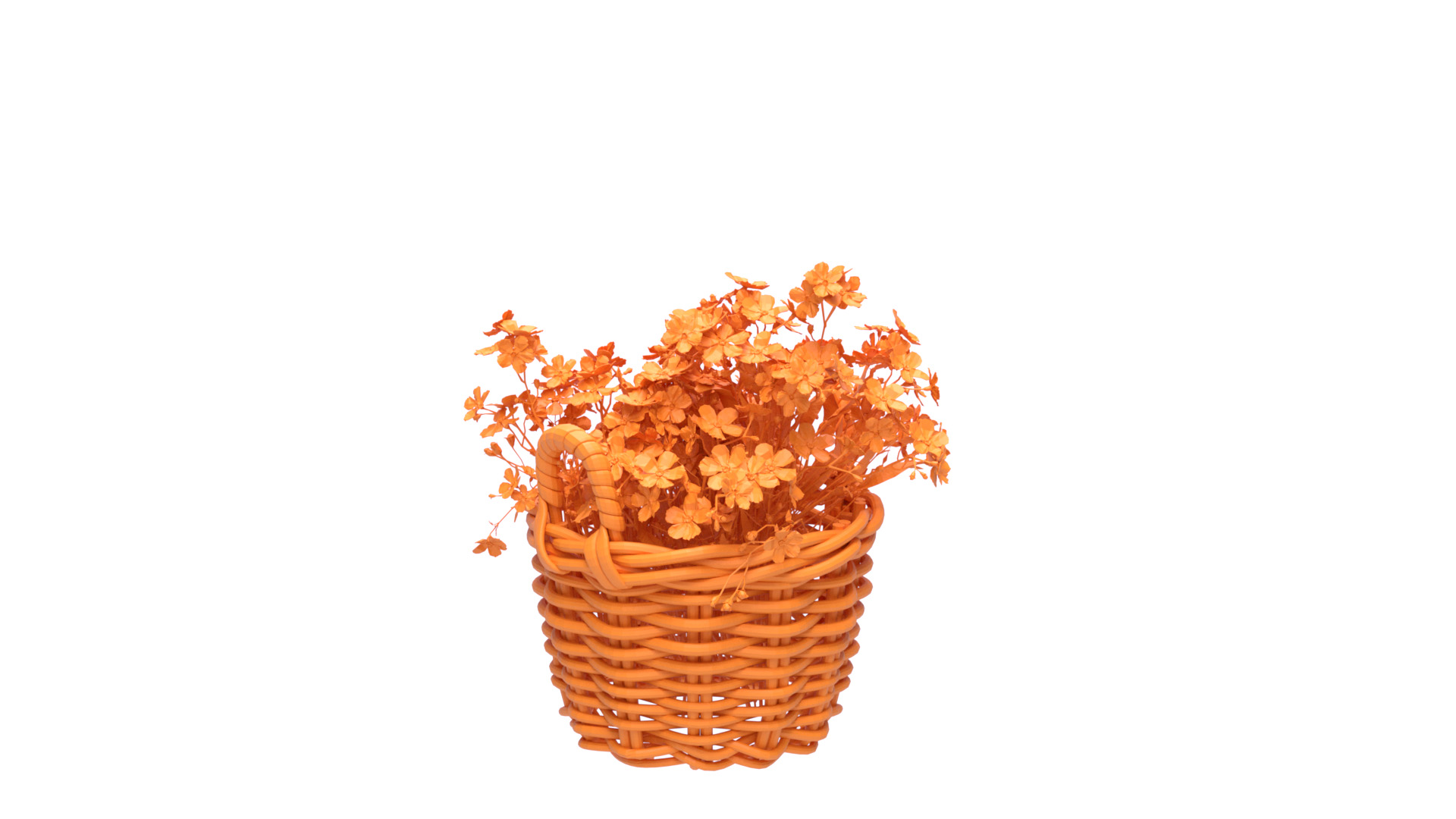}\hfill
\frame{\includegraphics[trim=500 50 650 250,clip,width=0.19\linewidth]{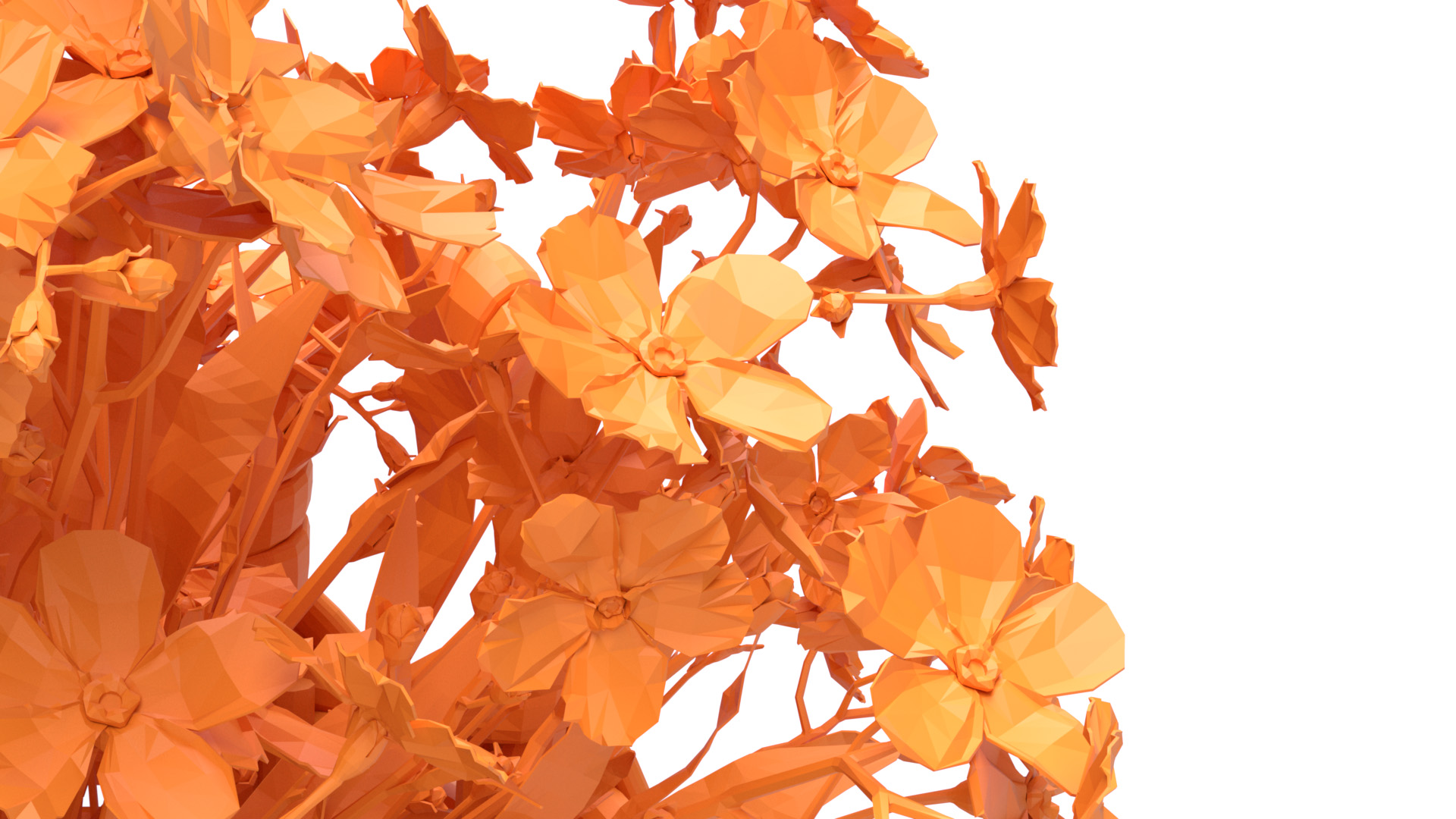}}\hfill
\includegraphics[trim=570 50 650 300,clip,width=0.30\linewidth]{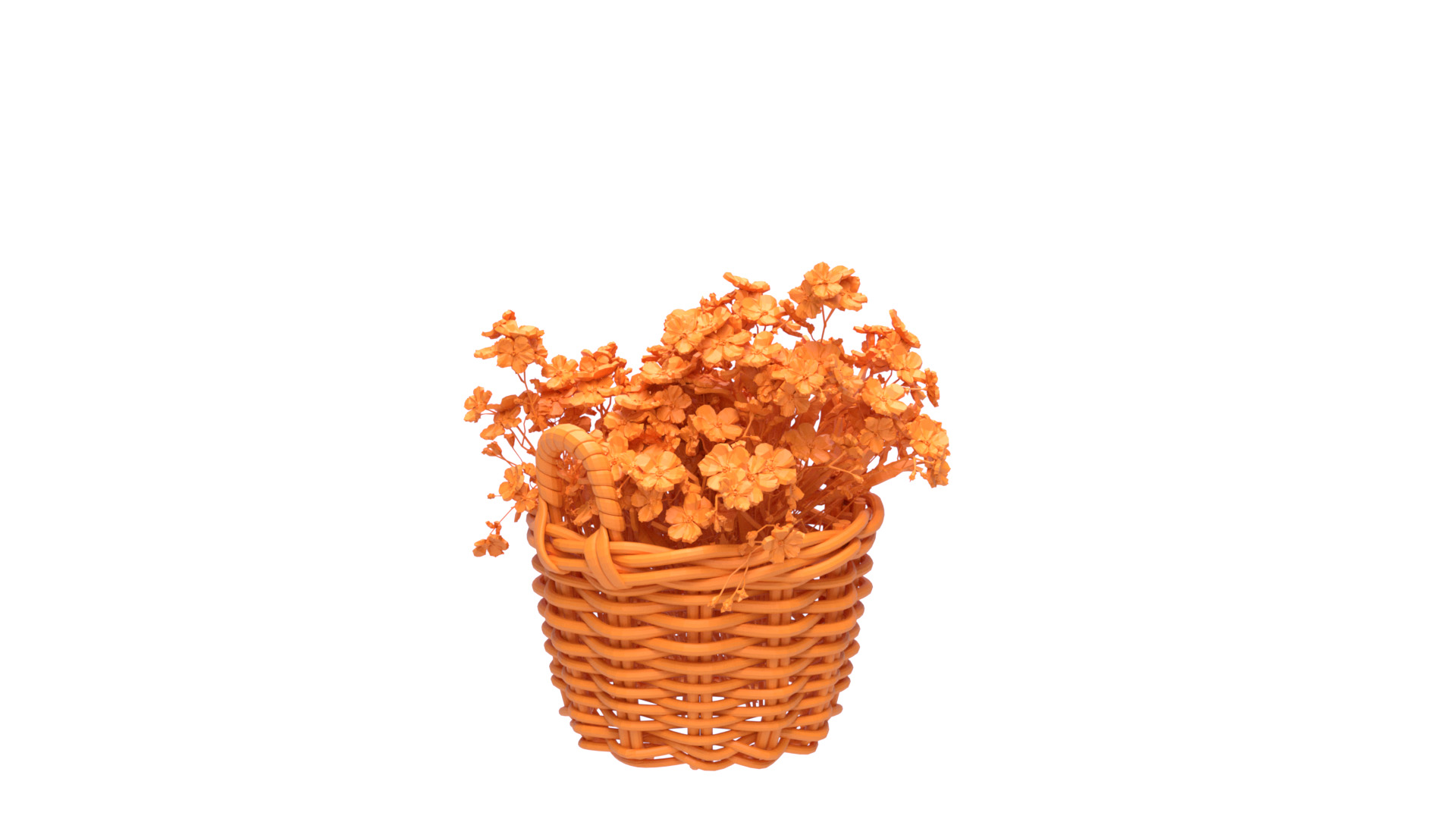}\hfill
\frame{\includegraphics[trim=500 50 650 250,clip,width=0.19\linewidth]{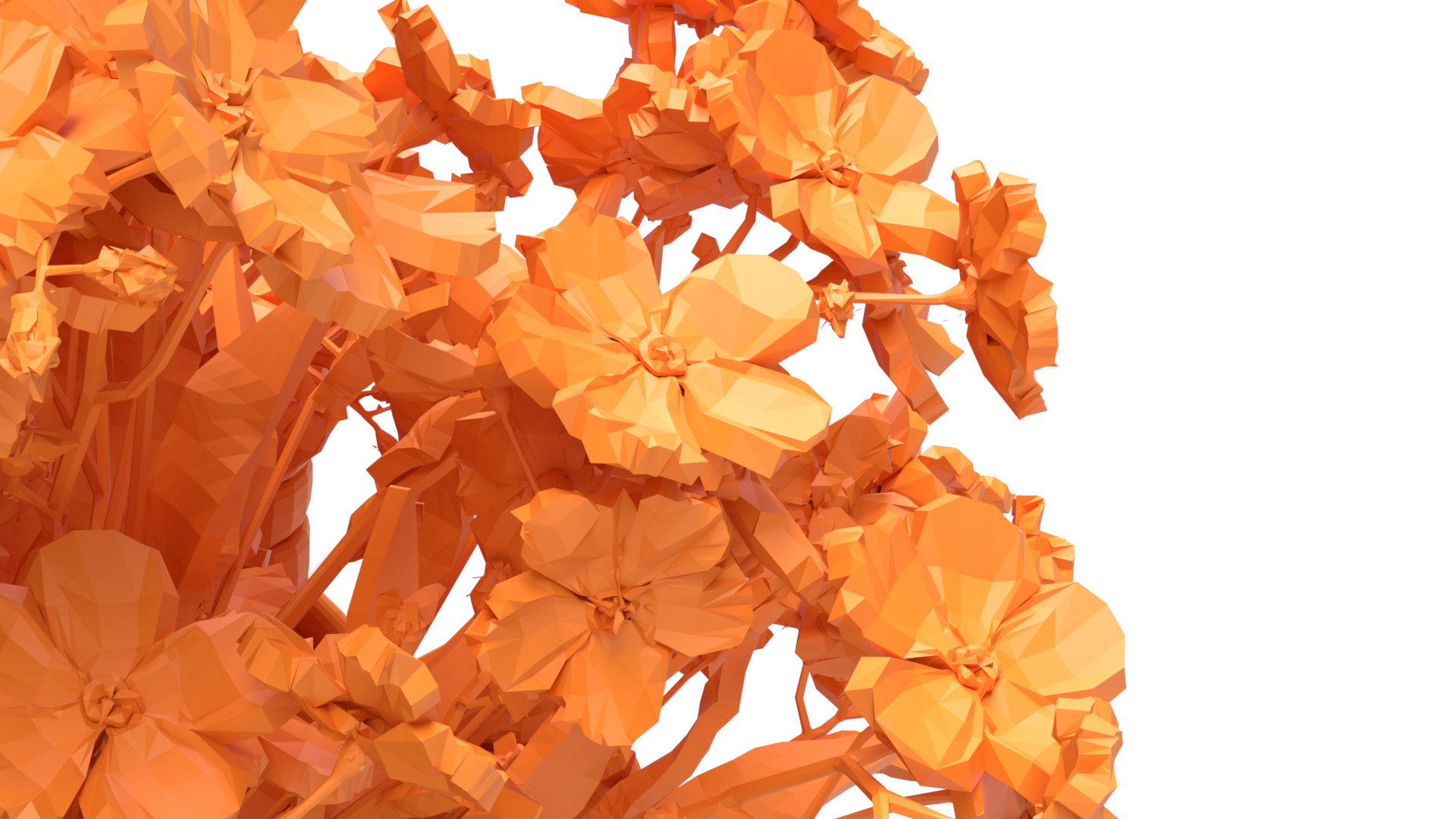}}\\
\figcap{\small Ours with $d_\text{offset} = {D}/{2000}$ } \hfill%
\figcap{\small Ours with $d_\text{offset} = {D}/{200}$ }\\
\figcap{\small $\text{HD} = 6.8e-3, \text{LFD} = 86 $ } \hfill%
\figcap{\small $\text{HD} = 7.1e-3, \text{LFD} = 520 $ }\vspace{-0.0in}
\caption{\textbf{Study on $d_\text{offset}$:} Given an input plant model containing numerous open faces, our method can reliably convert it into watertight manifold meshes, enabling control over Hausdorff distance (HD) and light field distance (LFD) by adjusting $d_\text{offset}$. $D$ refers to the diagonal length of the model's bounding box. The front side is orange, while the back side is white.}
\label{fig:ablation_off}
\end{figure}

\begin{figure*}[ht]
\newcommand{\figcap}[1]{\begin{minipage}{0.16\linewidth}\centering#1\end{minipage}}
\includegraphics[trim=300 300 400 50,clip,width=0.16\linewidth]{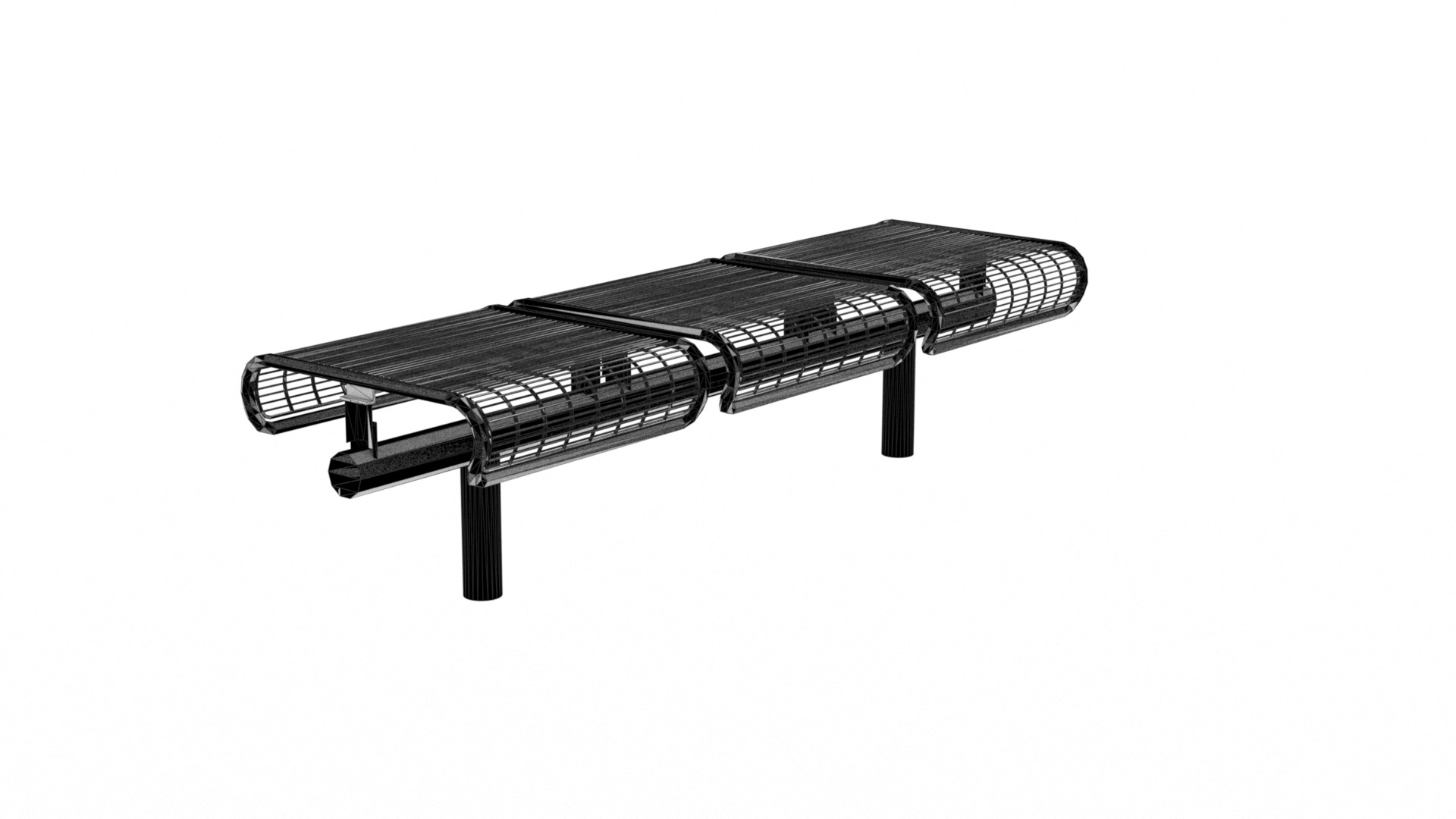}\hfill
\includegraphics[trim=300 300 400 50,clip,width=0.16\linewidth]{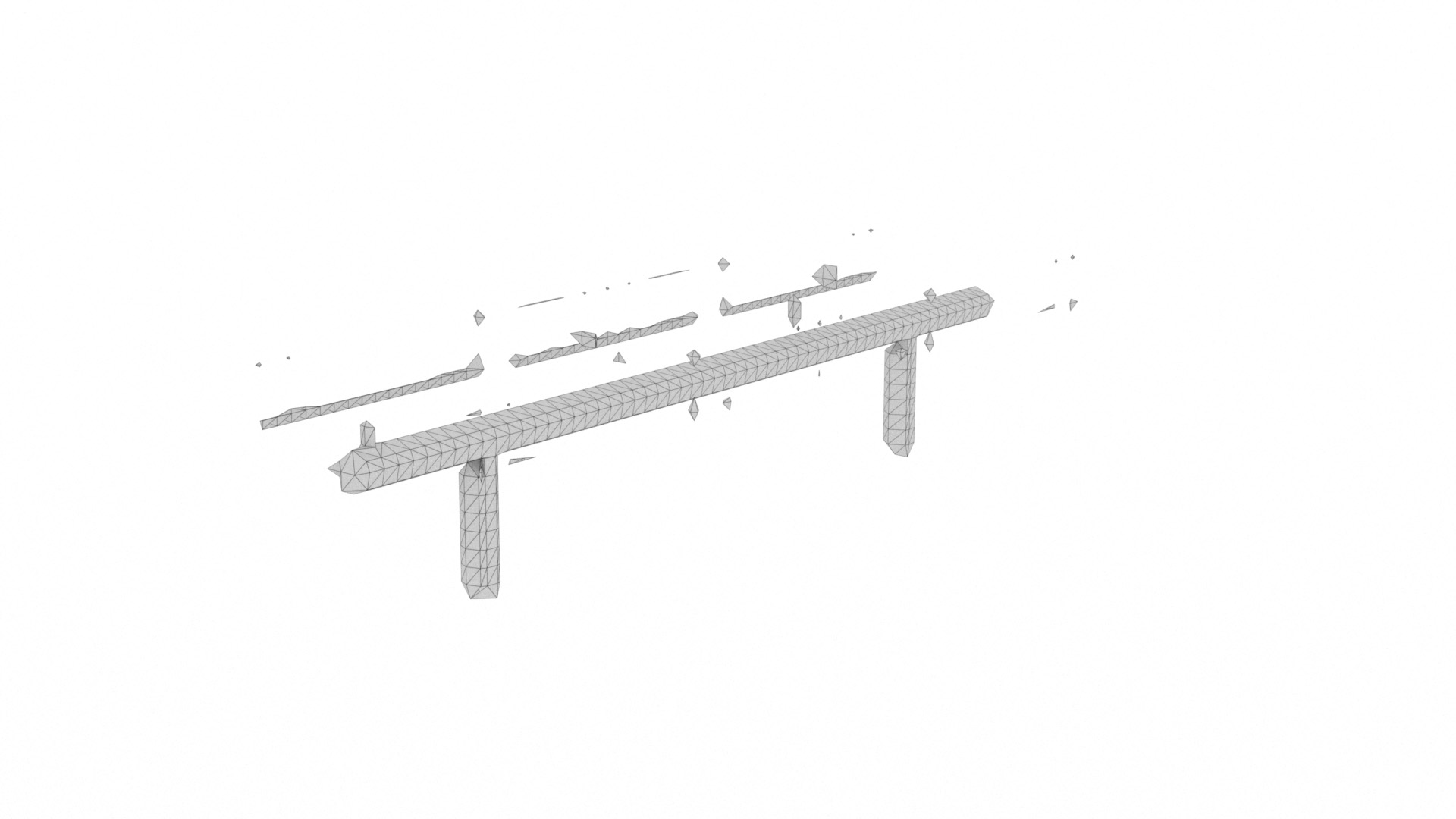}\hfill
\includegraphics[trim=300 300 400 50,clip,width=0.16\linewidth]{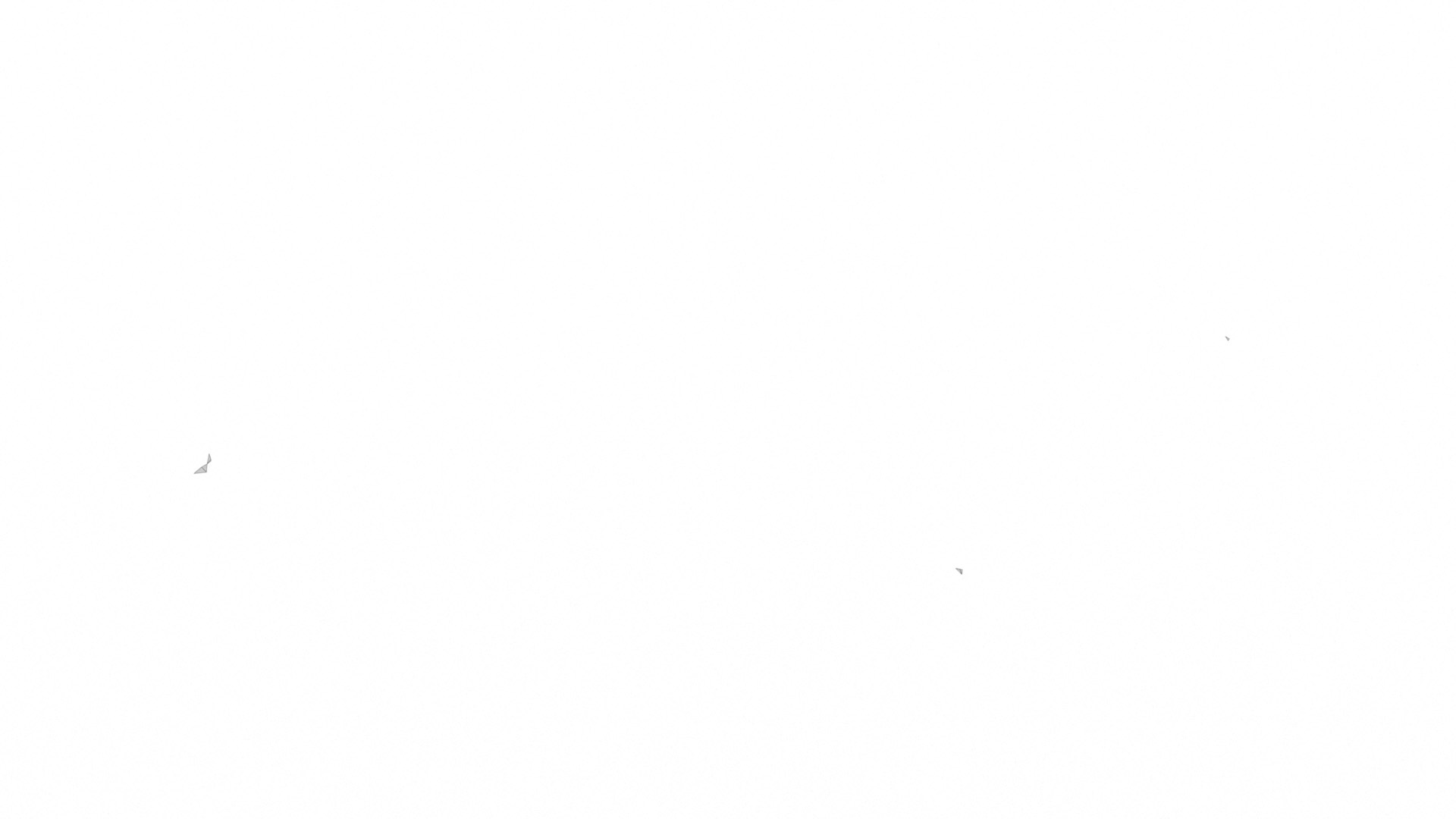}\hfill
\includegraphics[trim=300 300 400 50,clip,width=0.16\linewidth]{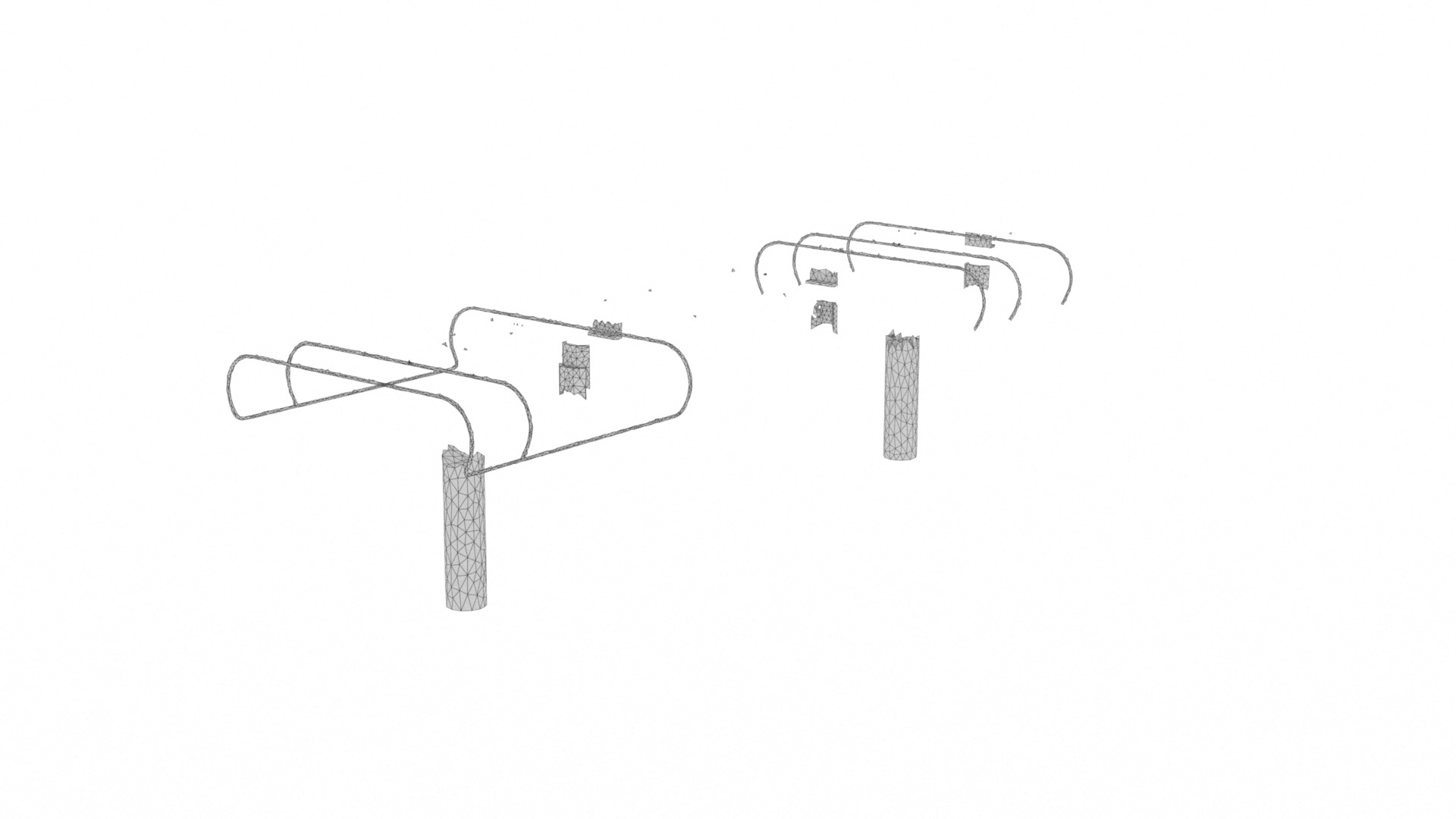}\hfill
\includegraphics[trim=300 300 400 50,clip,width=0.16\linewidth]{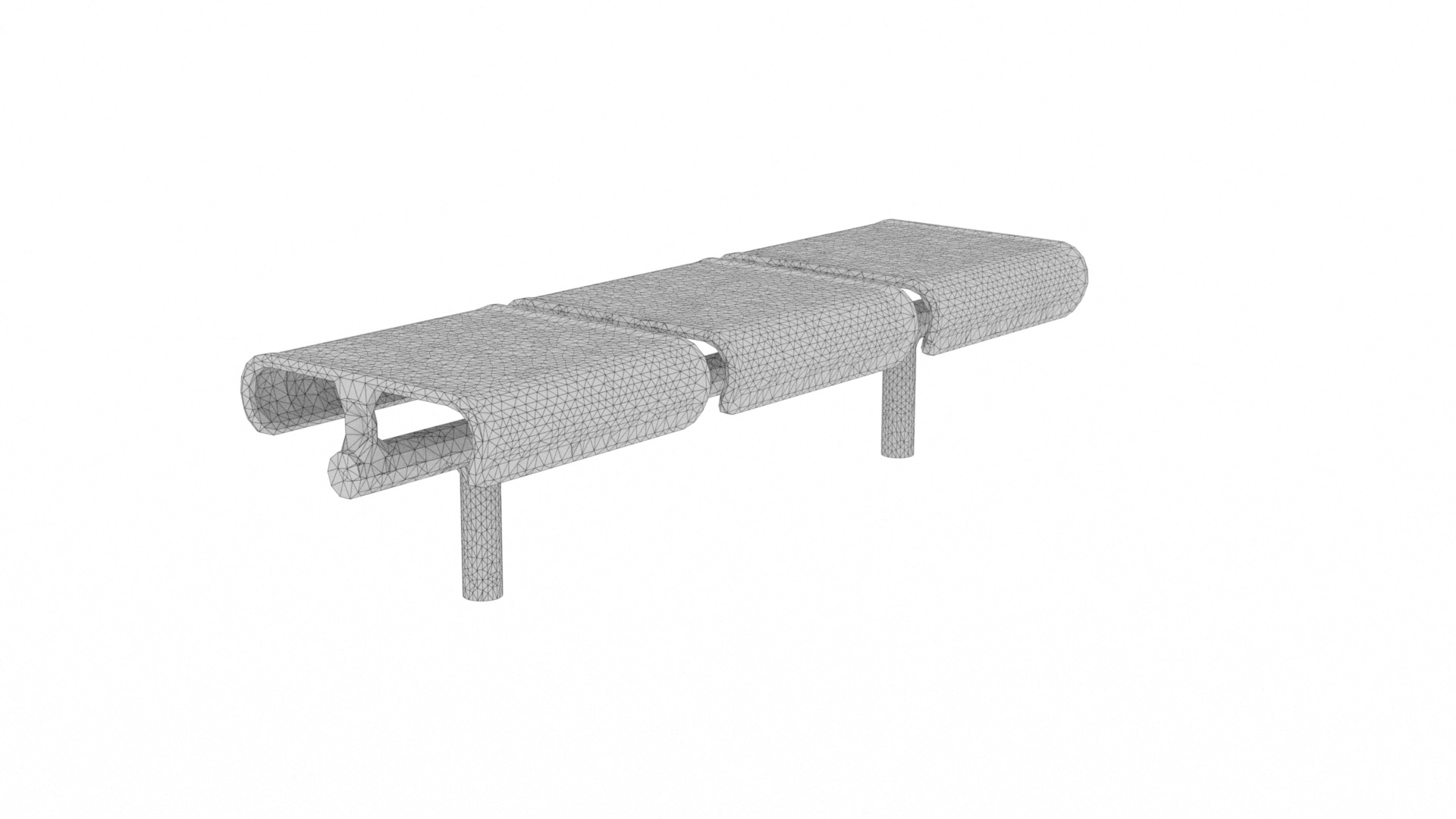}\hfill
\includegraphics[trim=300 300 400 50,clip,width=0.16\linewidth]{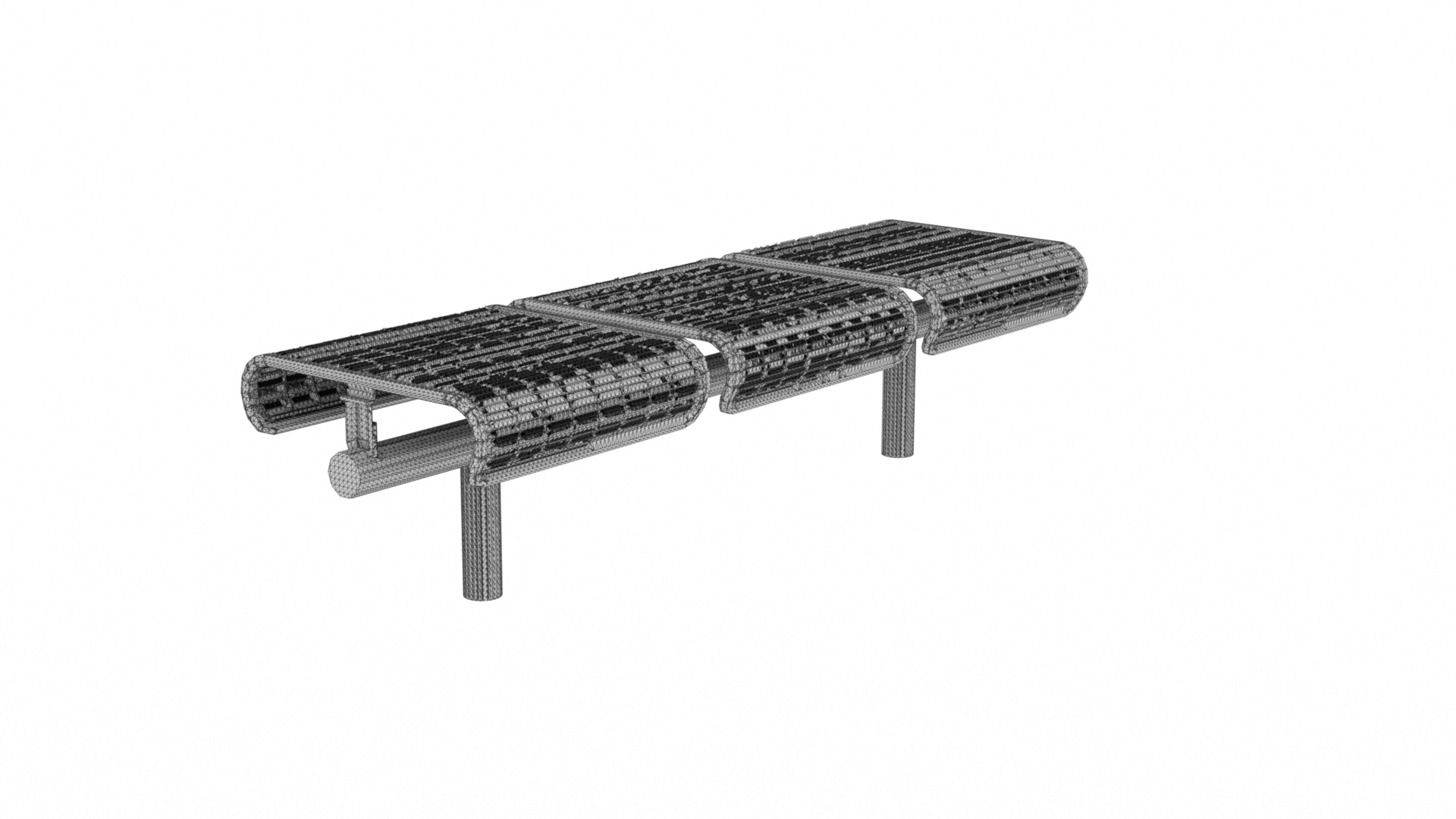}\\
\figcap{\tiny Chair} \hfill%
\figcap{\tiny PM (2K,0.10,16882,25.0)} \hfill%
\figcap{\tiny TW (0K,0.50,60968,23.5)} \hfill%
\figcap{\tiny fTW (8K,0.12,15868,23.8)} \hfill%
\figcap{\tiny AW (24K,0.01,2860,26.8)} \hfill%
\figcap{\tiny MP (457K,0.01,2868,18.1)} \\
\includegraphics[trim=300 300 400 50,clip,width=0.16\linewidth]{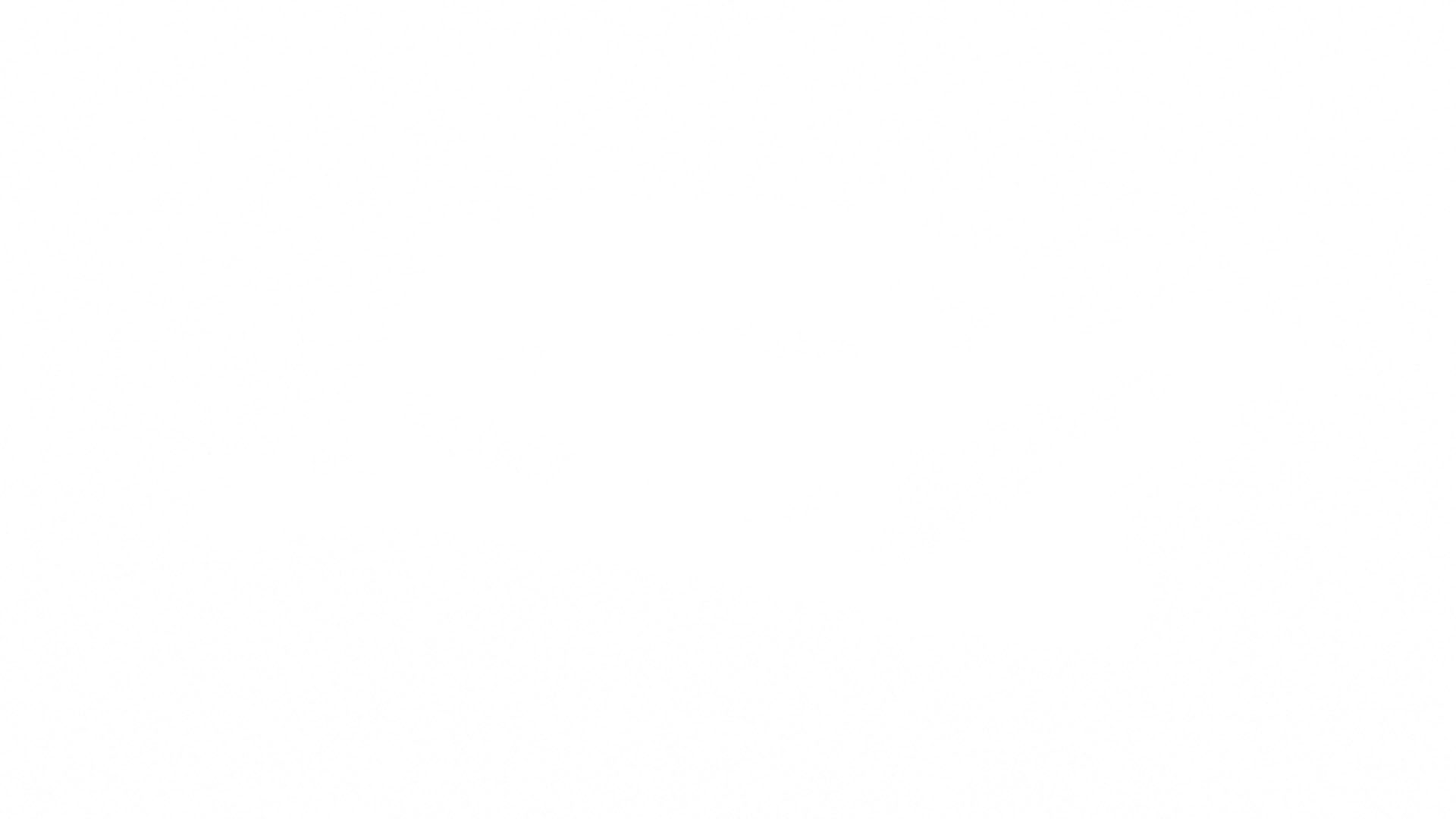}\hfill
\includegraphics[trim=300 300 400 50,clip,width=0.16\linewidth]{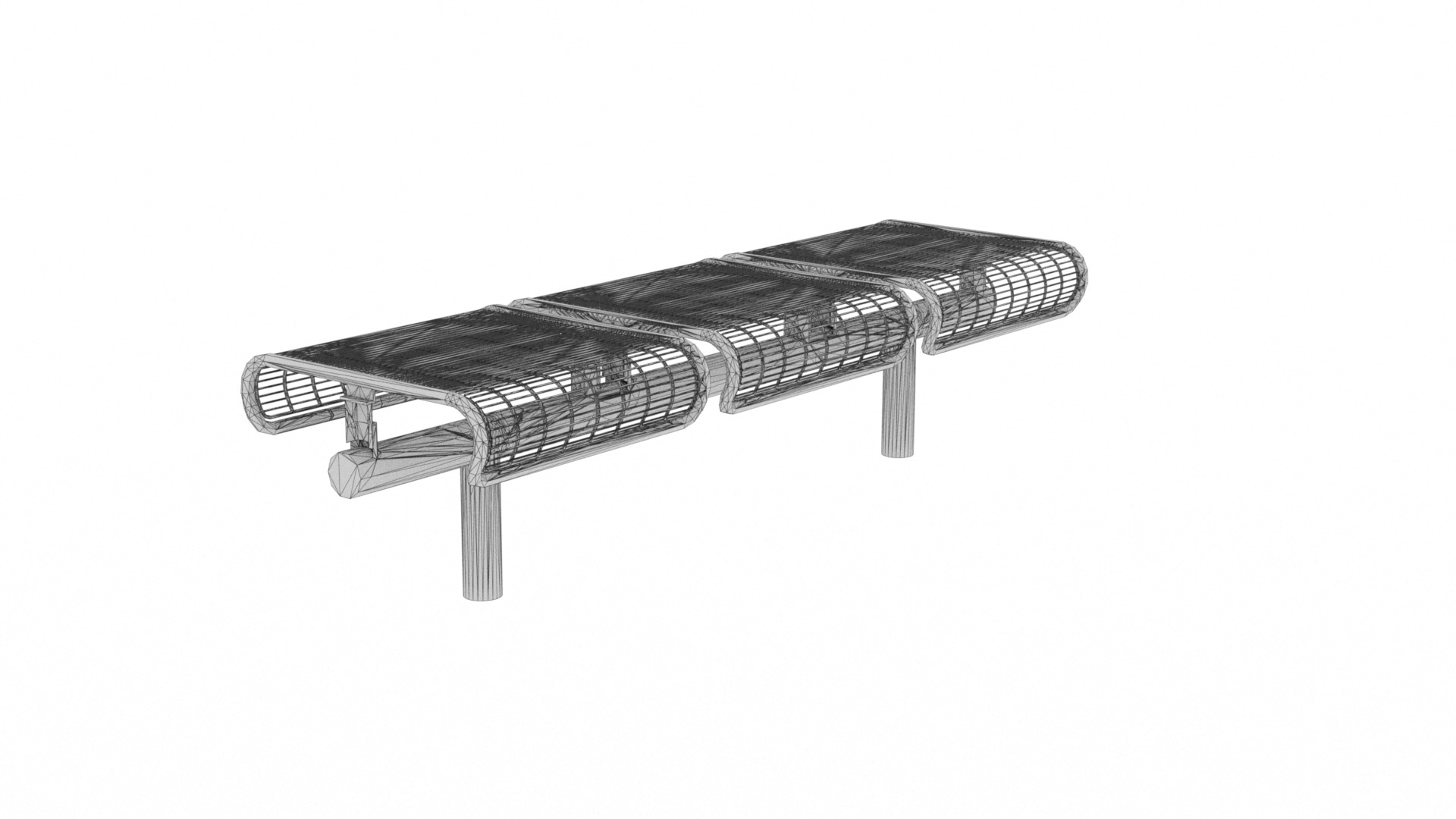}\hfill
\includegraphics[trim=300 300 400 50,clip,width=0.16\linewidth]{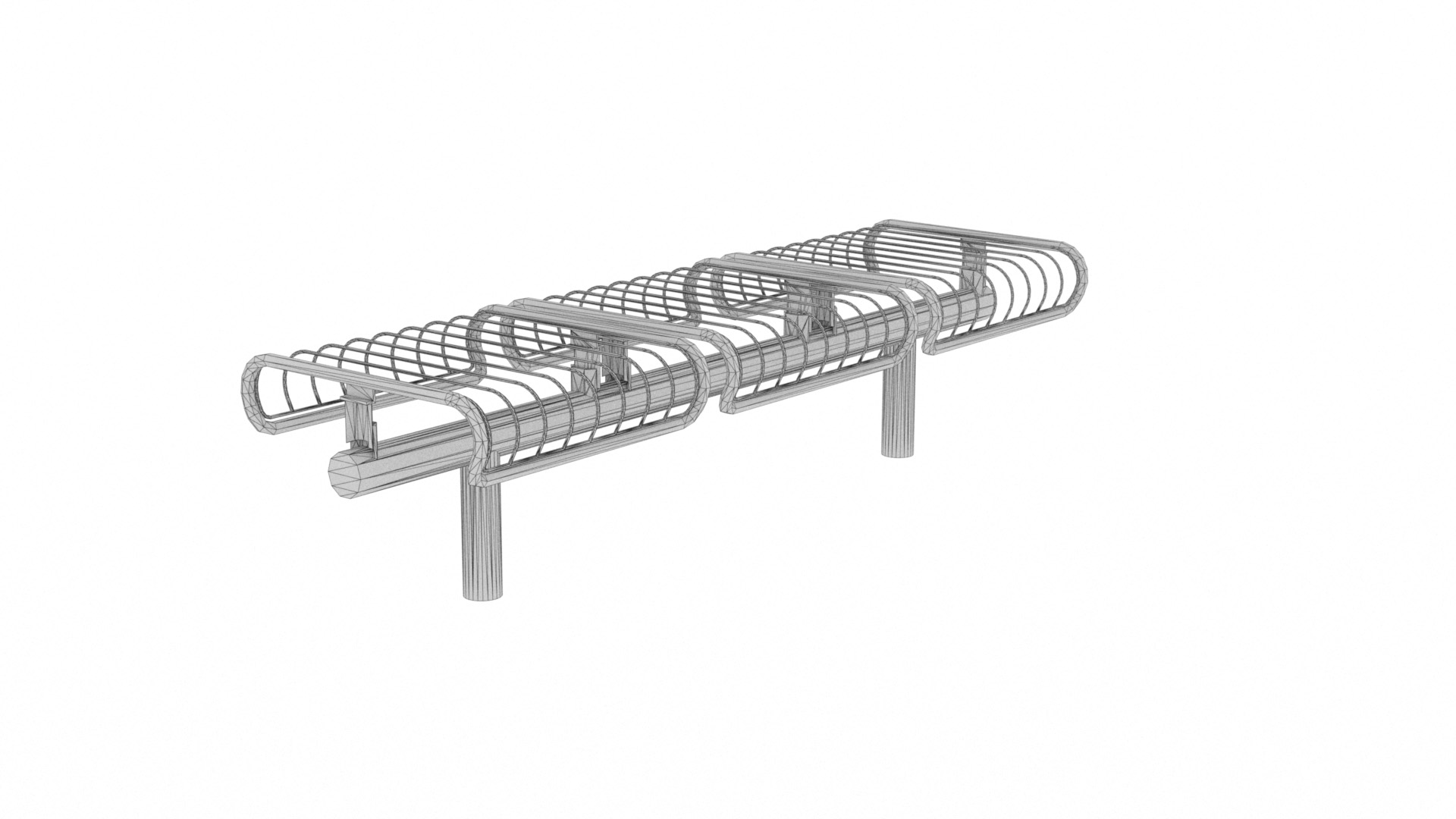}\hfill
\includegraphics[trim=300 300 400 50,clip,width=0.16\linewidth]{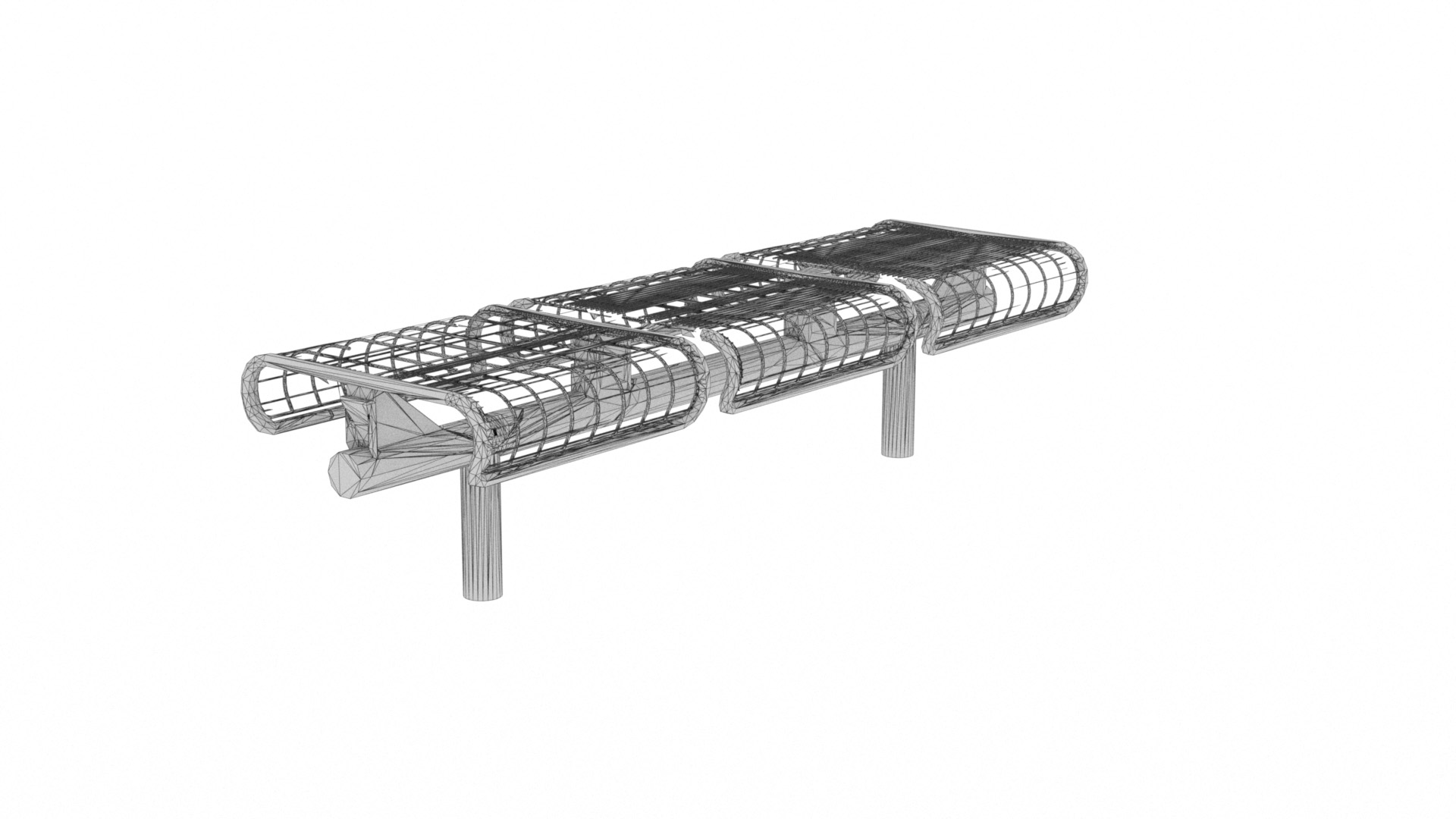}\hfill
\includegraphics[trim=300 300 400 50,clip,width=0.16\linewidth]{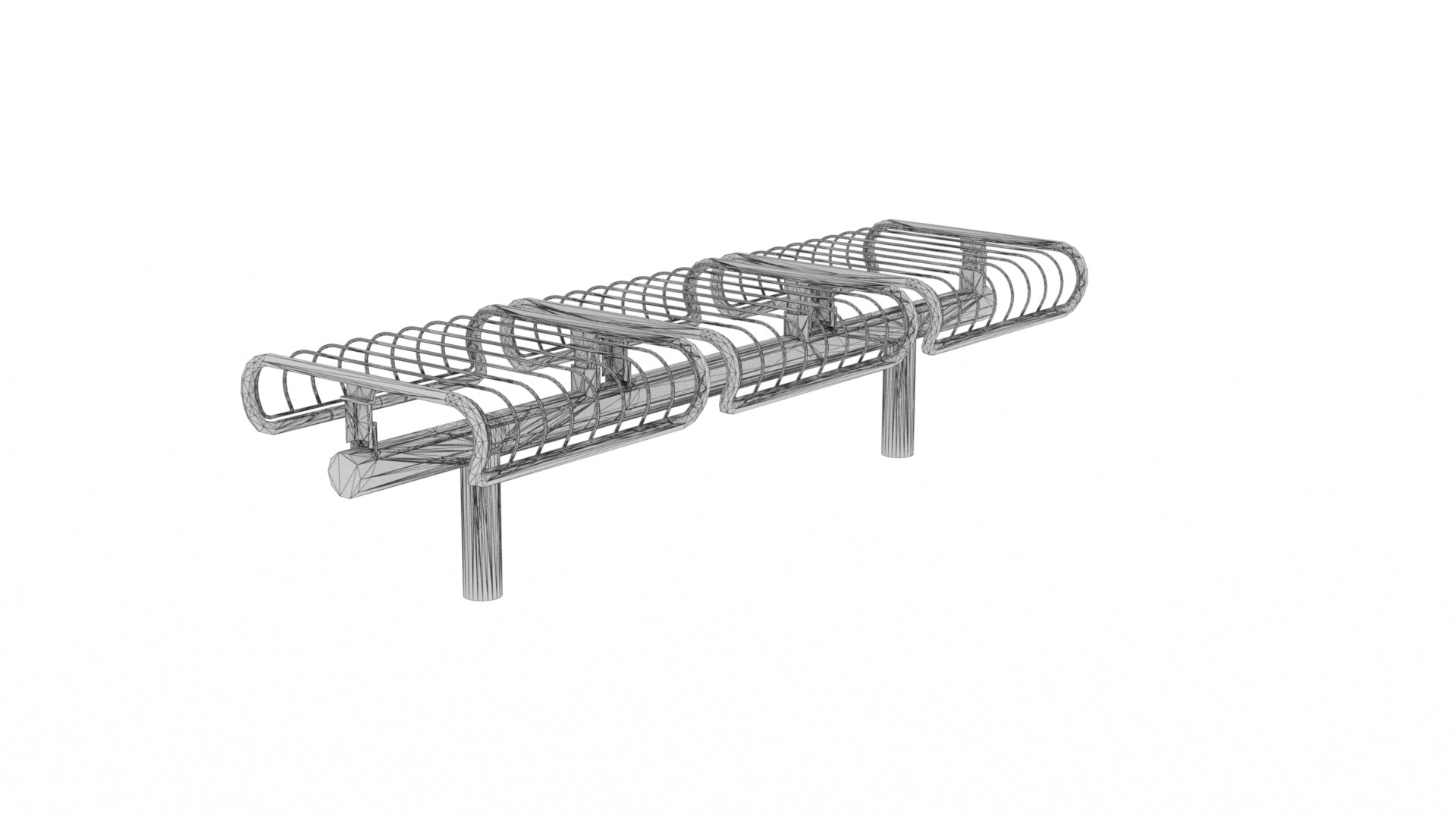}\hfill
\includegraphics[trim=300 300 400 50,clip,width=0.16\linewidth]{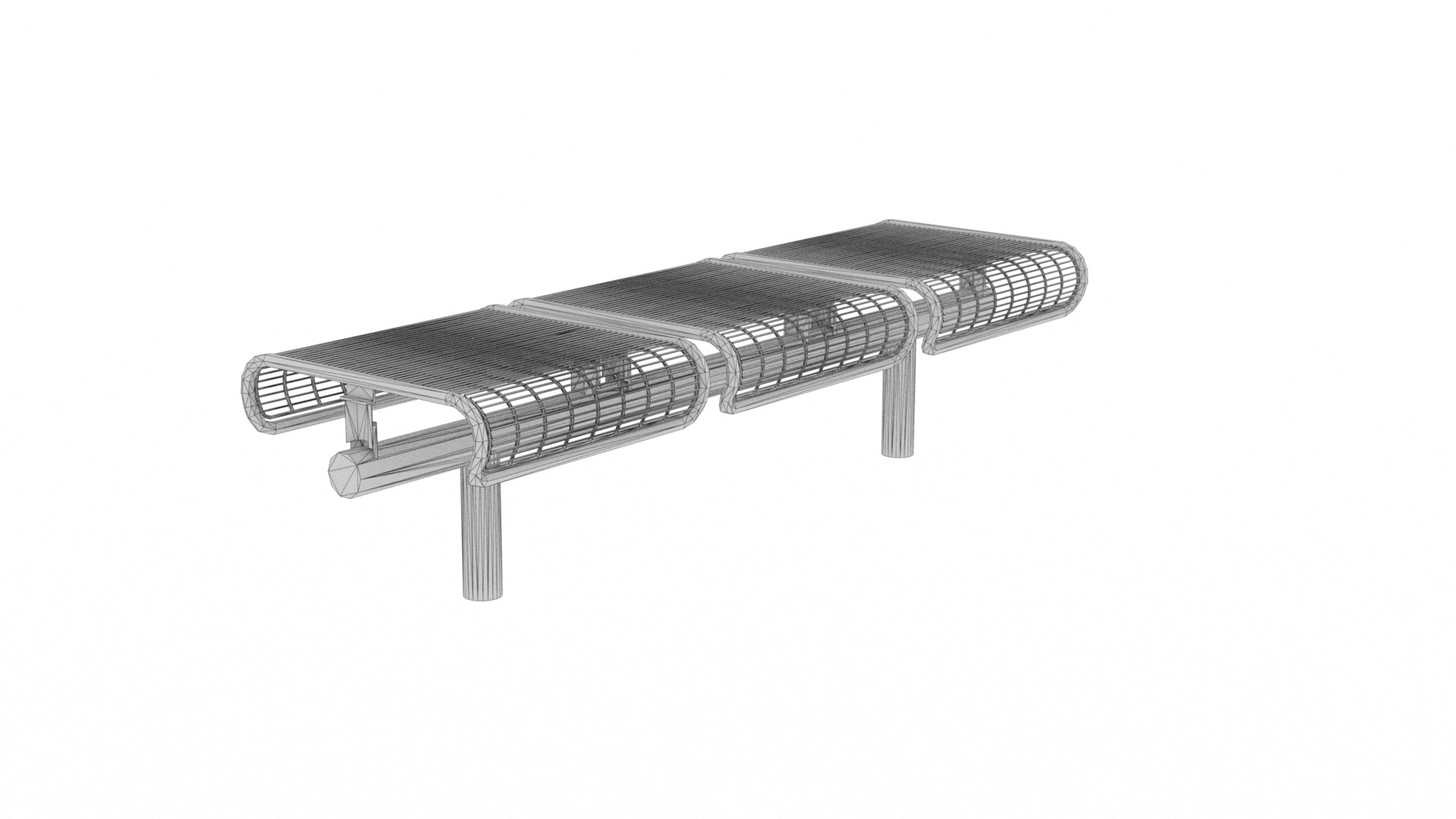}\\\vspace{0.25in}
\figcap{\tiny ~~} \hfill%
\figcap{\tiny T14+VM (168K,0.01,56,48.4)} \hfill%
\figcap{\tiny VR (14K,0.01,3484,30.8)} \hfill%
\figcap{\tiny VM (125K,0.03,2368,31.8)} \hfill%
\figcap{\tiny VR+VM (66K,0.01,3492,30.8)} \hfill%
\figcap{\tiny Ours (20K,0.01,6,51.4)} \\
\includegraphics[trim=400 80 400 50,clip,width=0.16\linewidth]{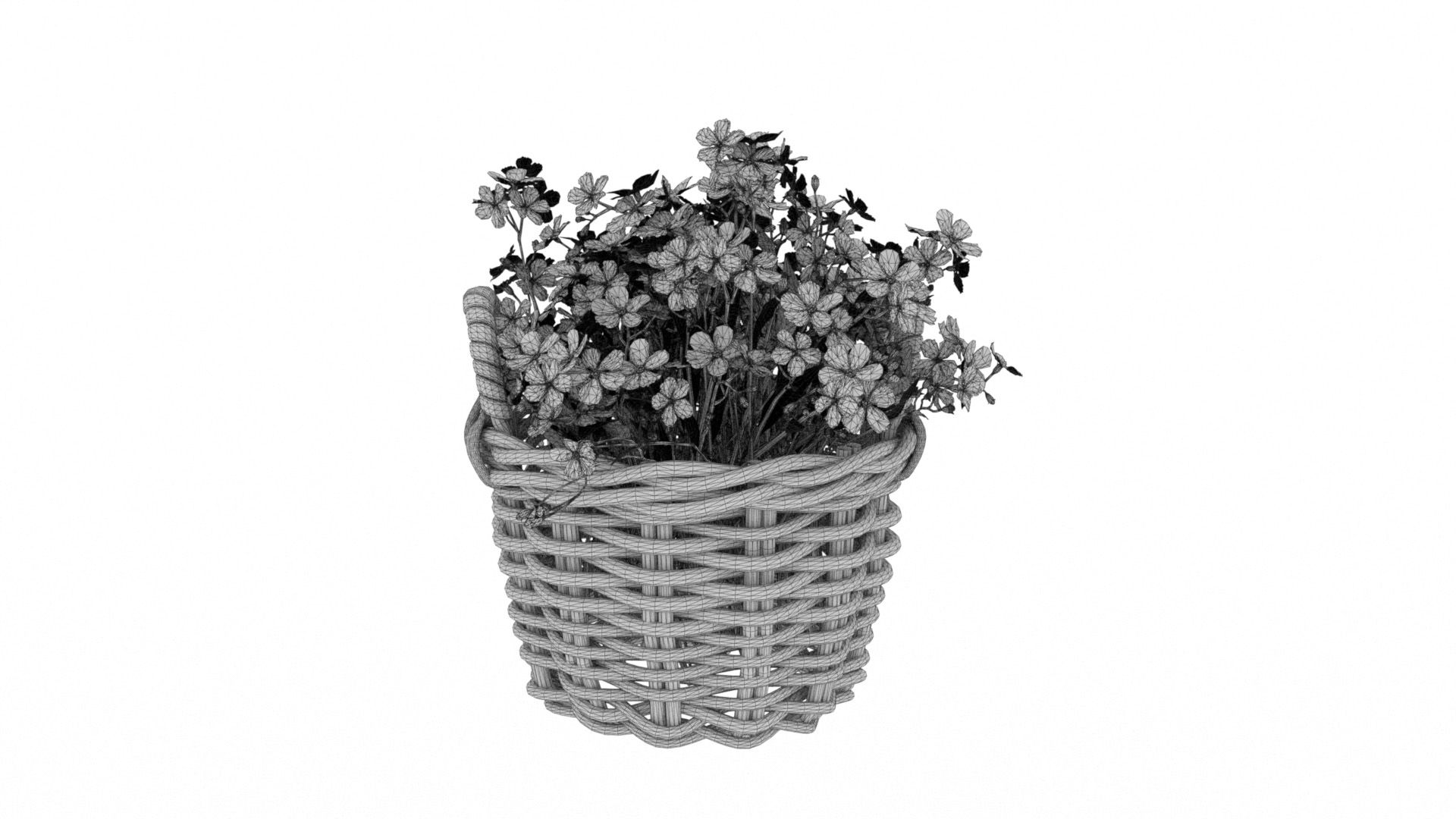}\hfill
\includegraphics[trim=400 80 400 50,clip,width=0.16\linewidth]{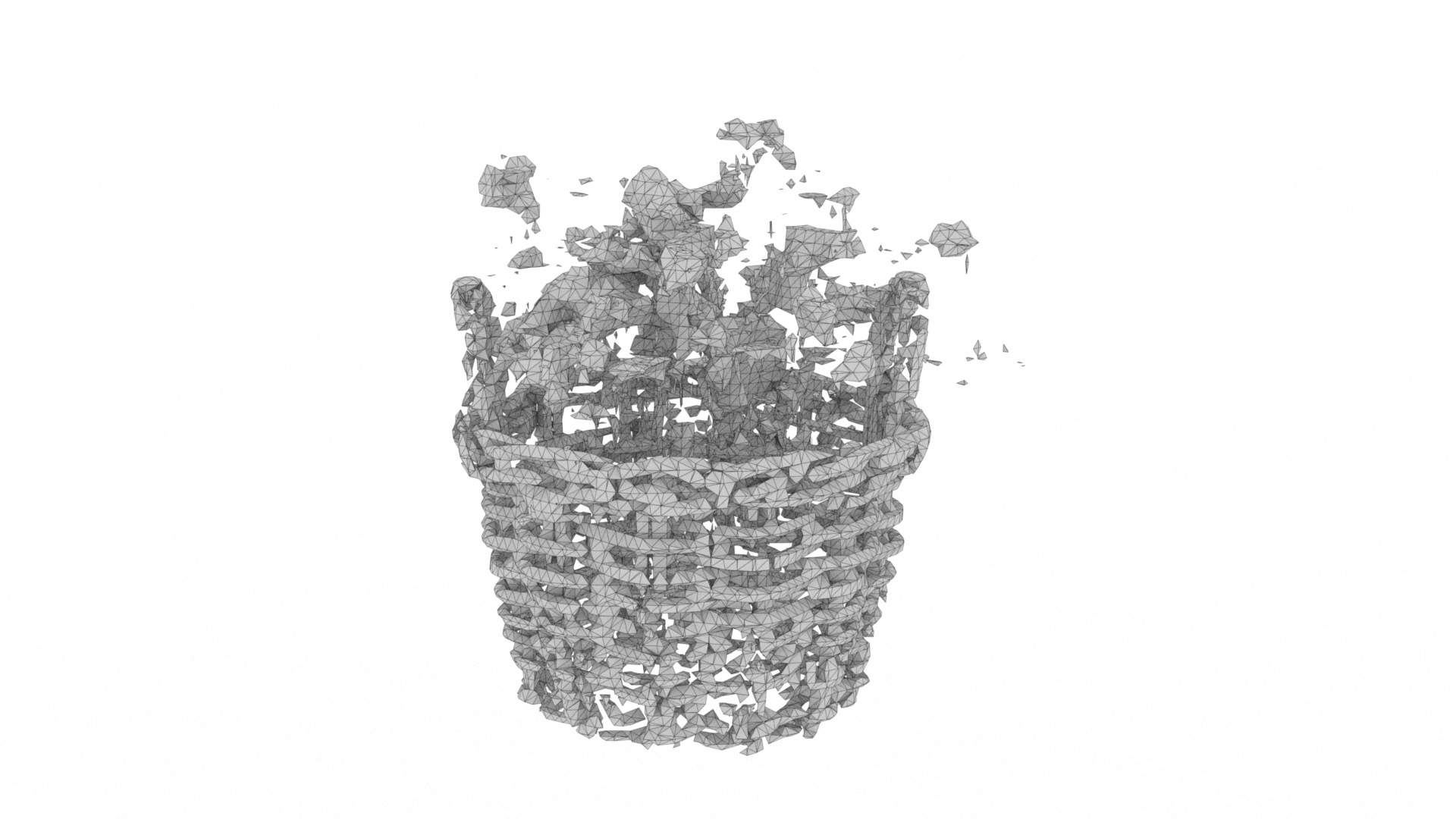}\hfill
\includegraphics[trim=400 80 400 50,clip,width=0.16\linewidth]{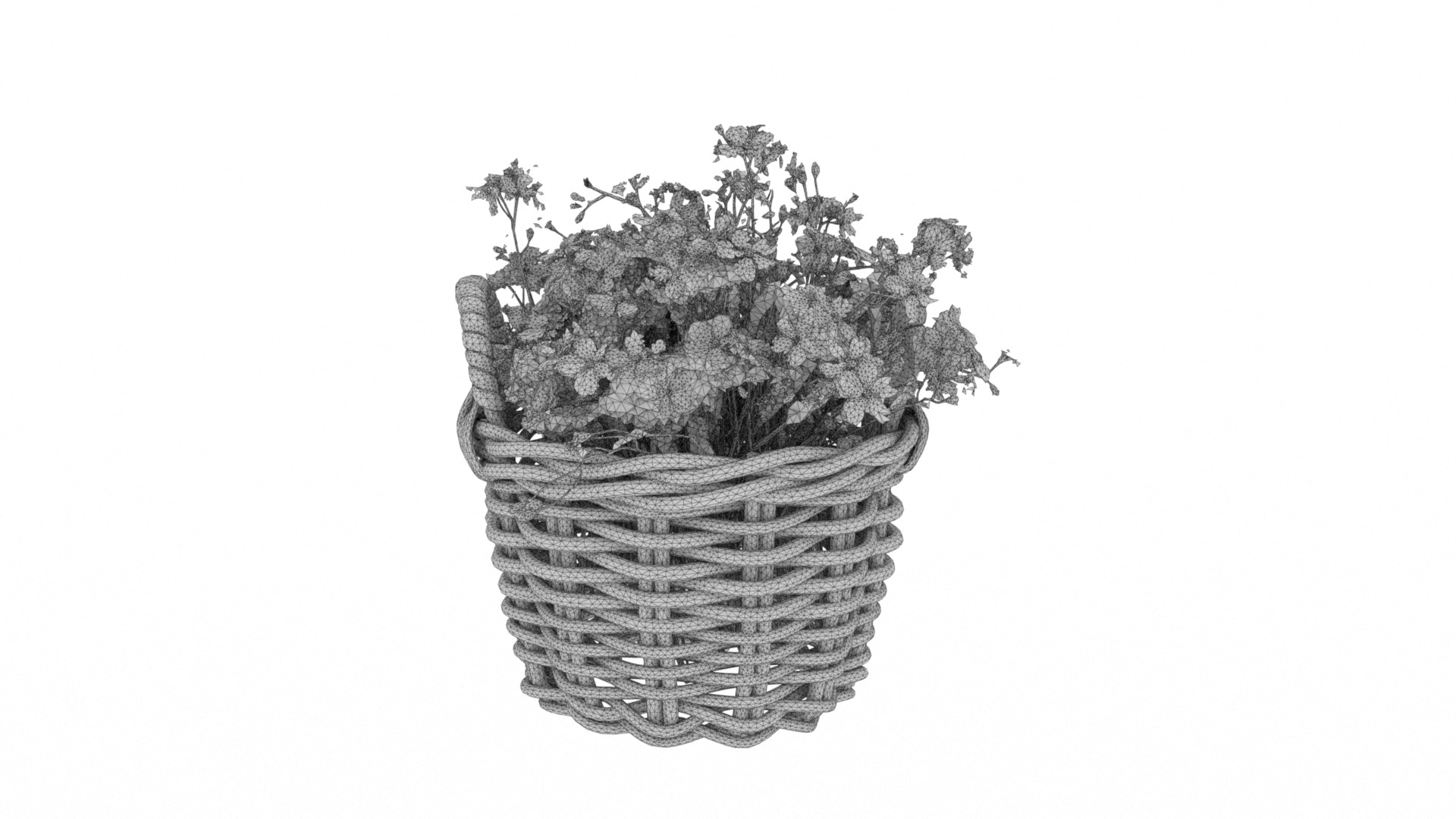}\hfill
\includegraphics[trim=400 80 400 50,clip,width=0.16\linewidth]{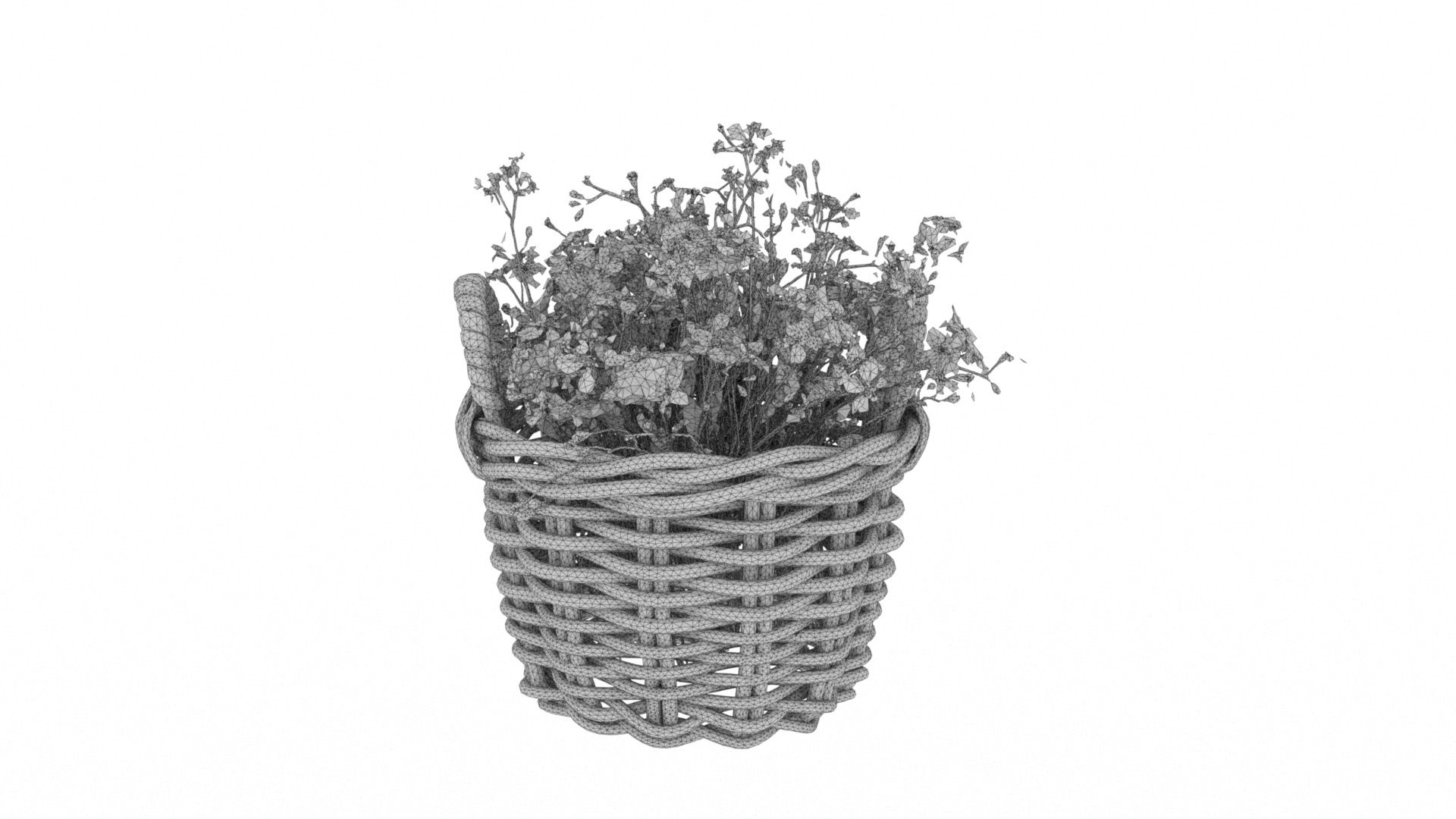}\hfill
\includegraphics[trim=400 80 400 50,clip,width=0.16\linewidth]{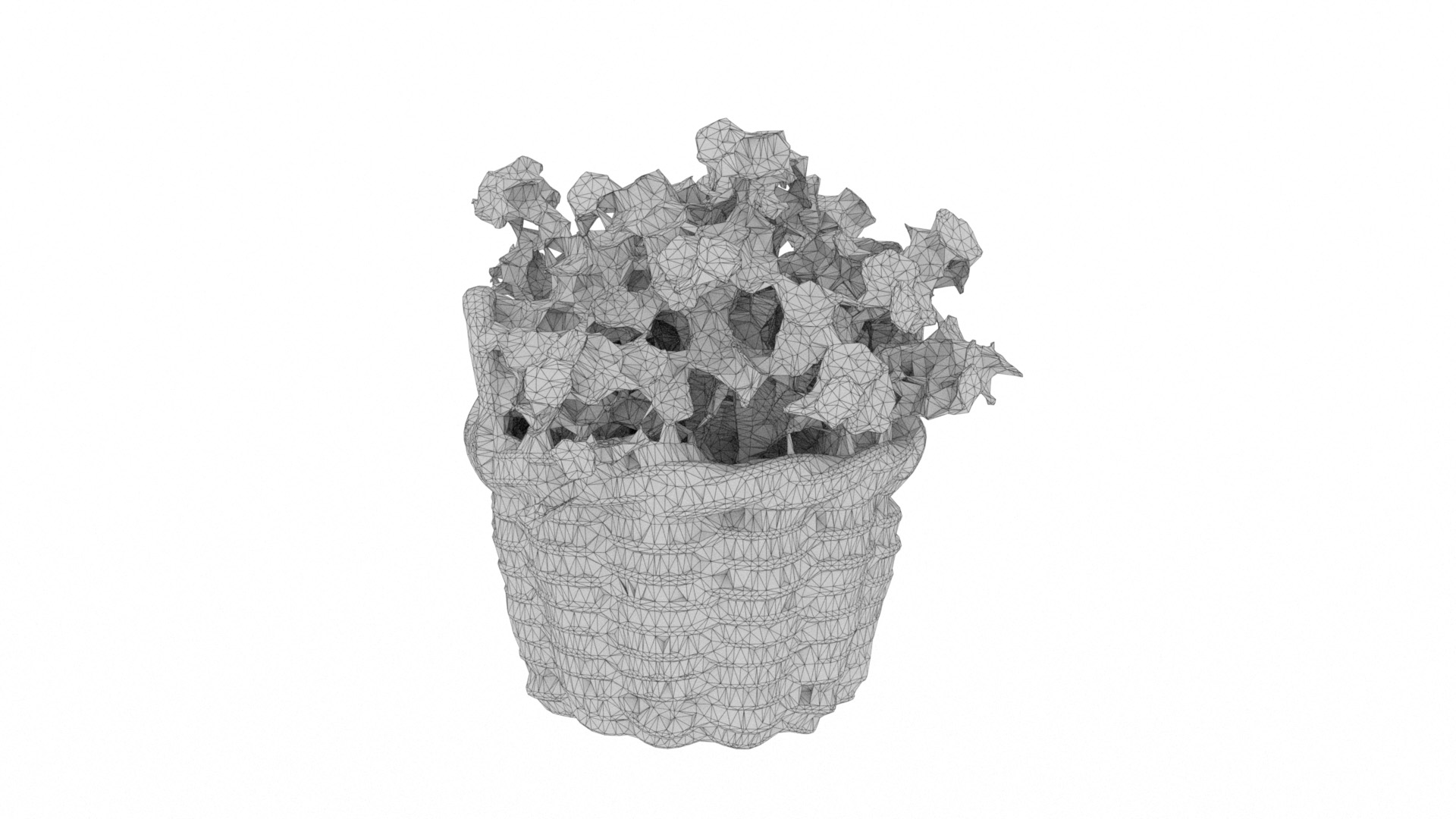}\hfill
\includegraphics[trim=400 80 400 50,clip,width=0.16\linewidth]{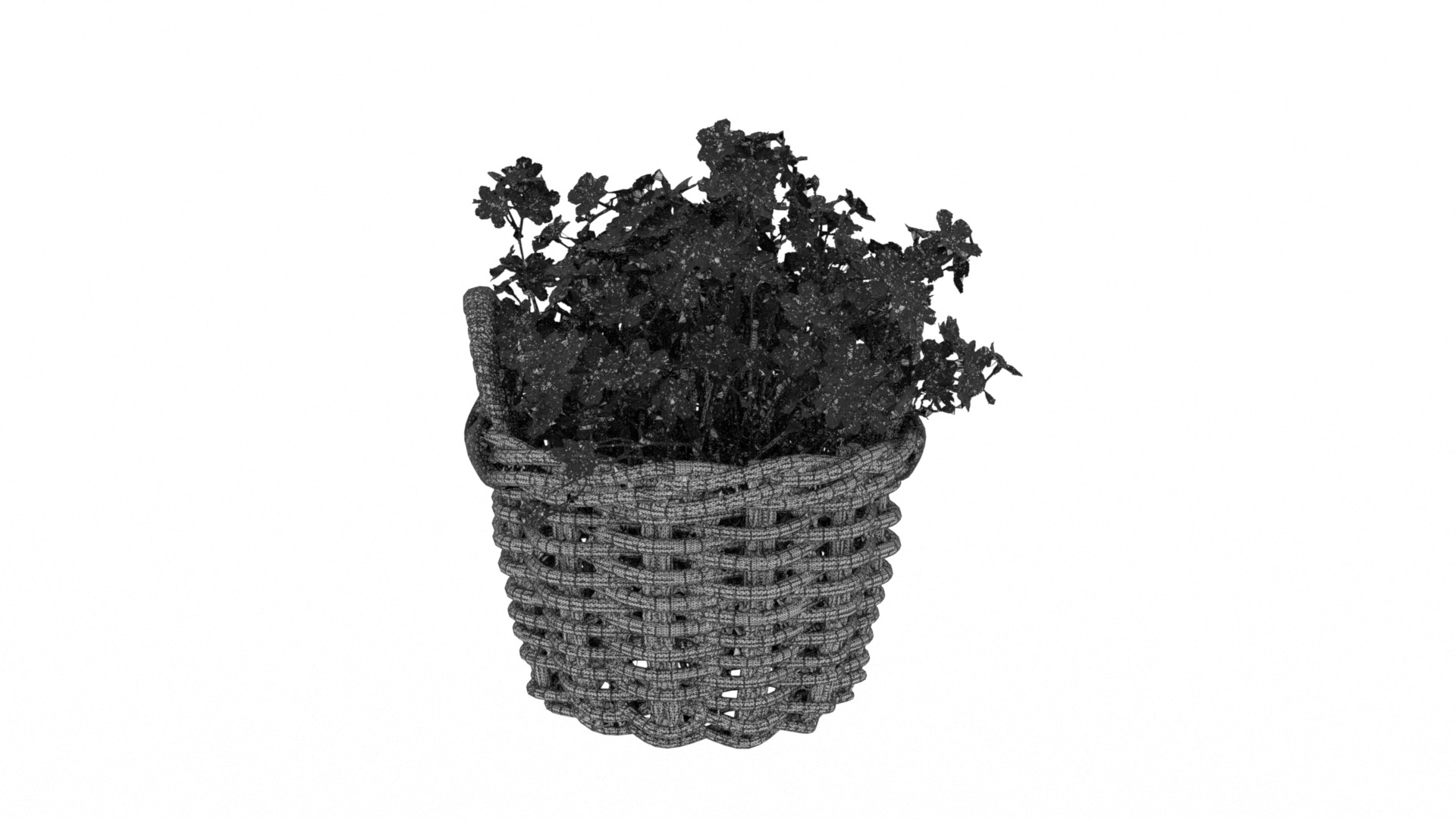}\\
\figcap{\tiny Flowers} \hfill%
\figcap{\tiny PM (44K,0.05,2534,23.7)} \hfill%
\figcap{\tiny TW (409K,0.03,1068,24.5)} \hfill%
\figcap{\tiny fTW (312K,0.03,1662,24.3)} \hfill%
\figcap{\tiny AW (99K,0.02,912,25.7)} \hfill%
\figcap{\tiny MP (5486K,0.01,168,21.8)} \\
\includegraphics[trim=400 80 400 50,clip,width=0.16\linewidth]{images/white.jpg}\hfill
\includegraphics[trim=400 80 400 50,clip,width=0.16\linewidth]{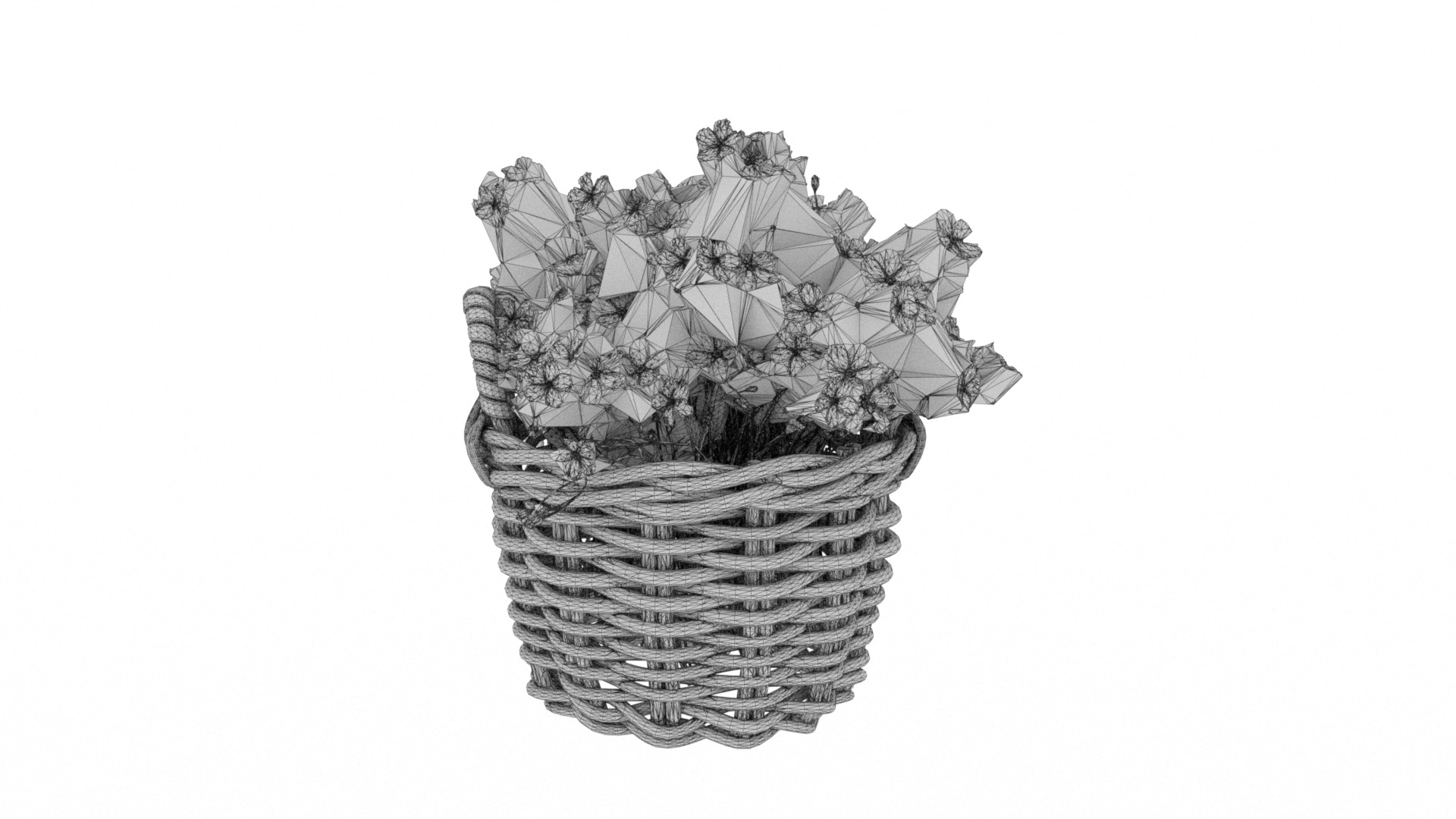}\hfill
\includegraphics[trim=400 80 400 50,clip,width=0.16\linewidth]{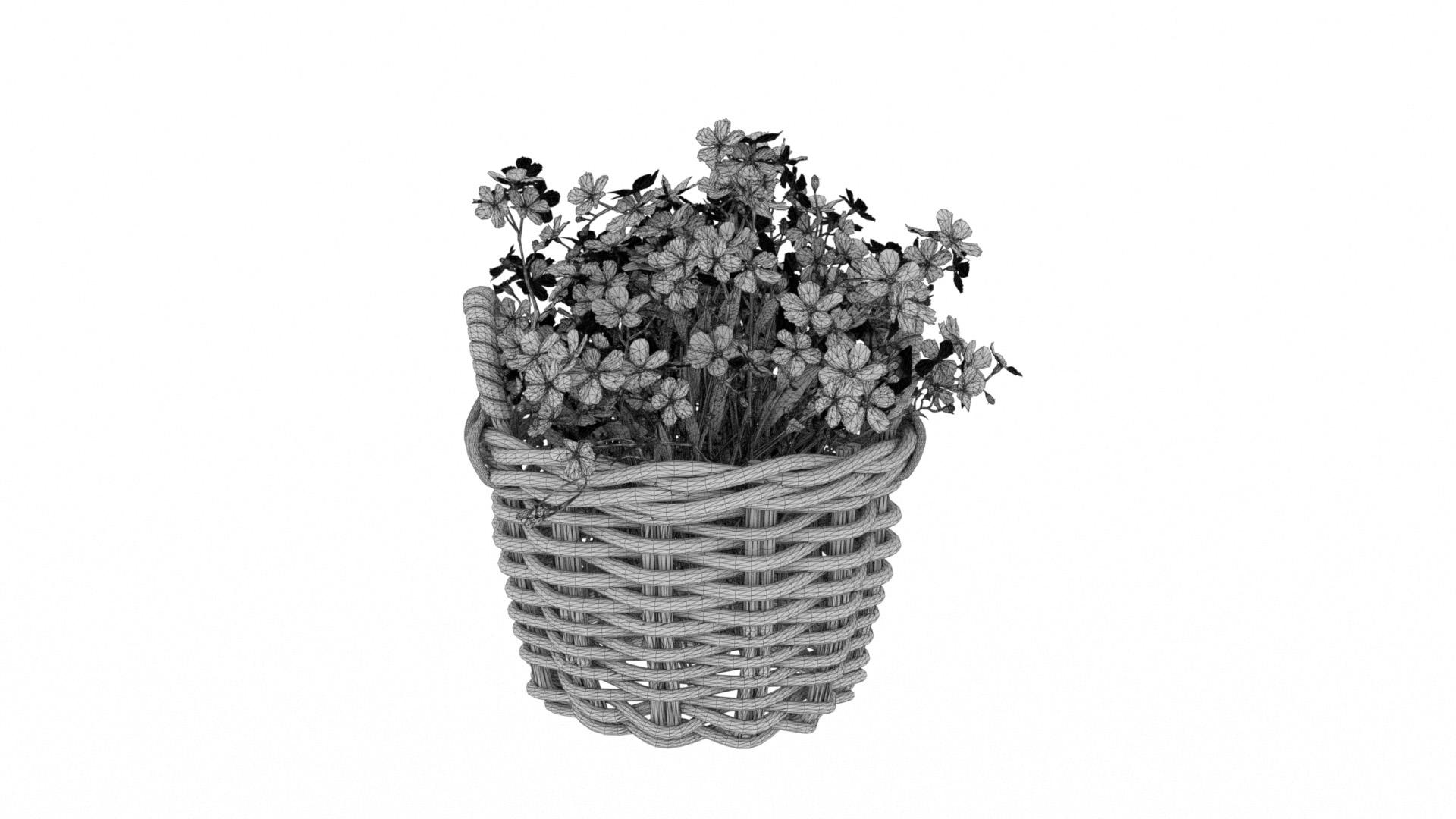}\hfill
\includegraphics[trim=400 80 400 50,clip,width=0.16\linewidth]{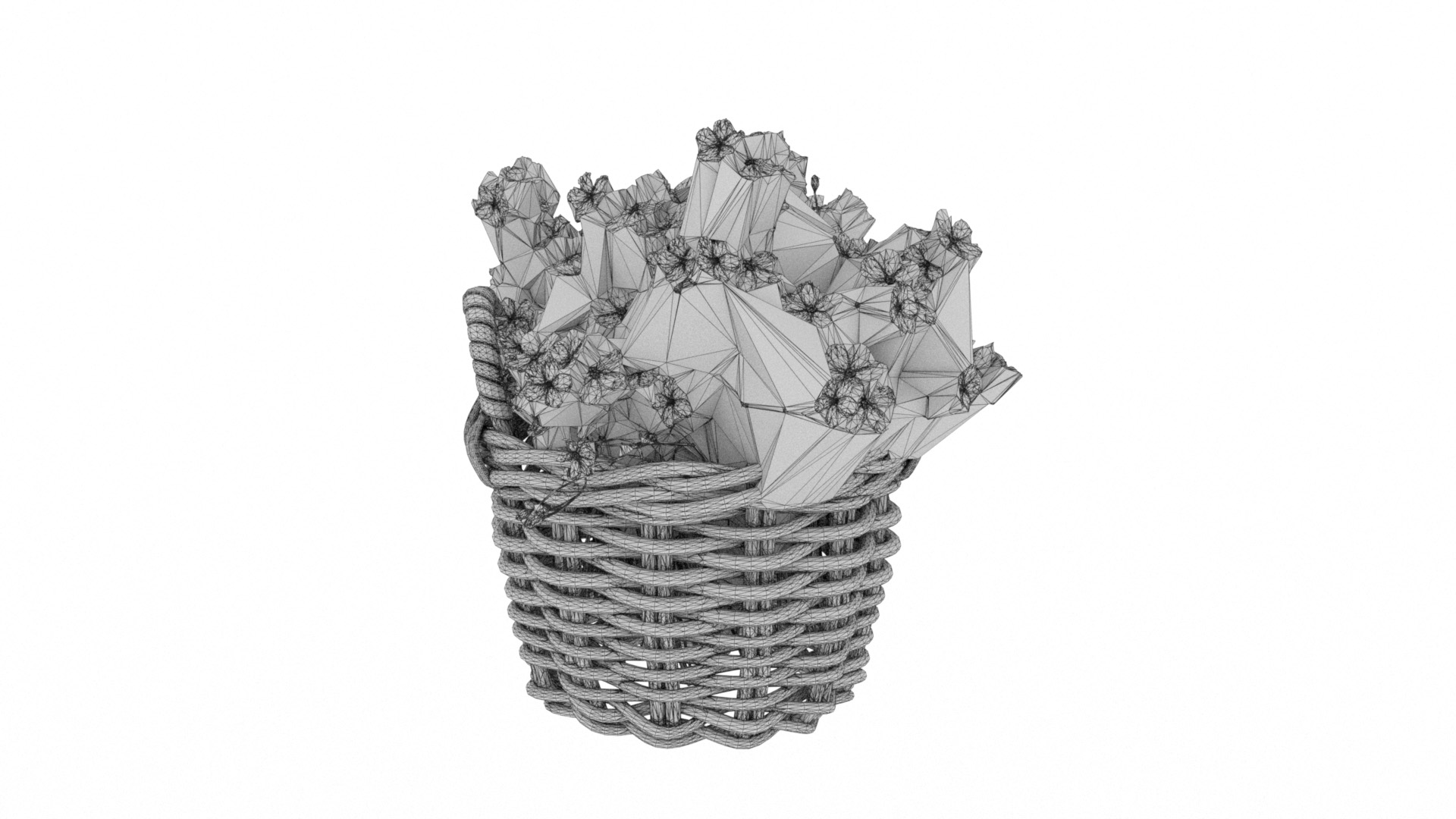}\hfill
\includegraphics[trim=400 80 400 50,clip,width=0.16\linewidth]{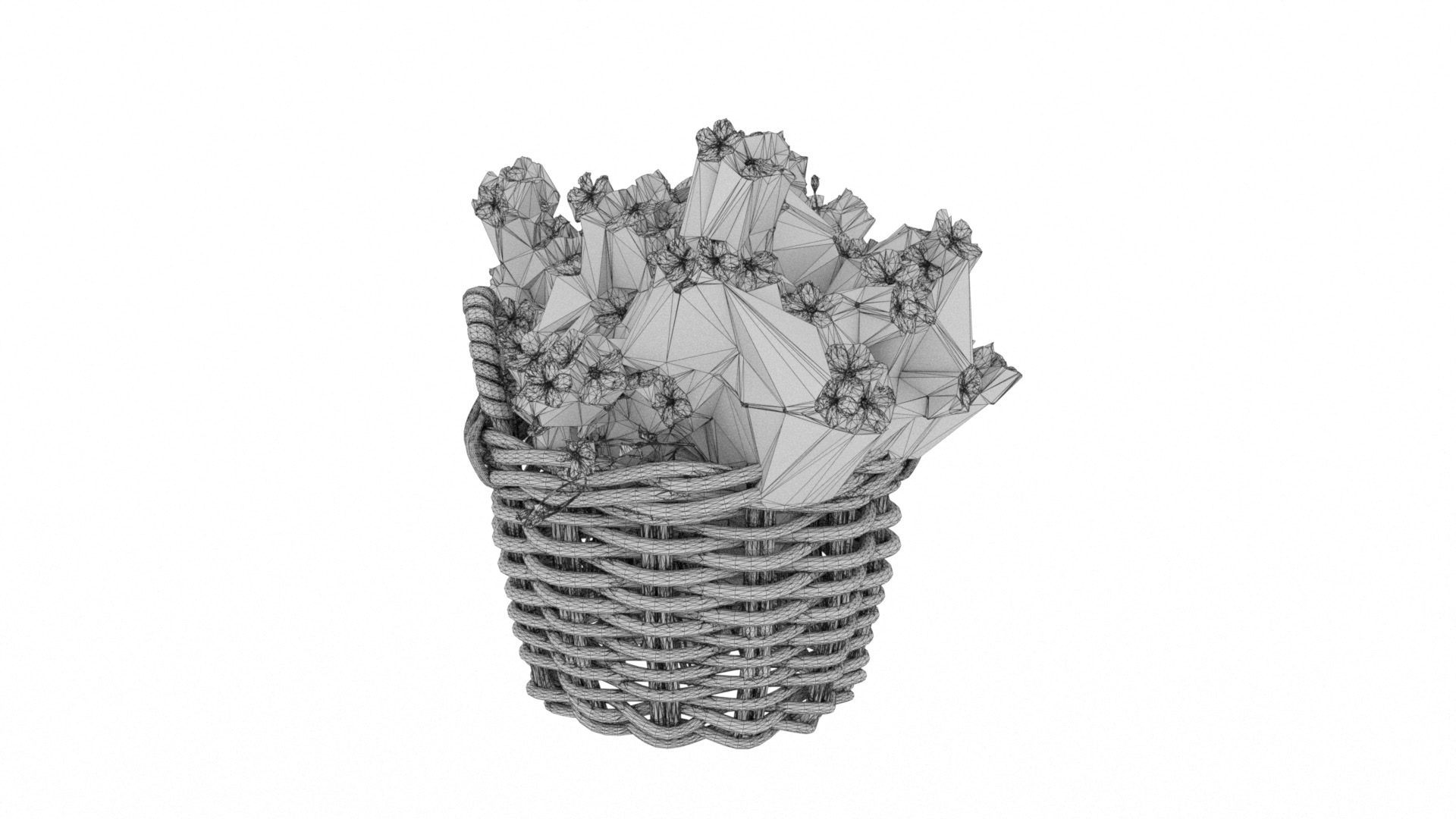}\hfill
\includegraphics[trim=400 80 400 50,clip,width=0.16\linewidth]{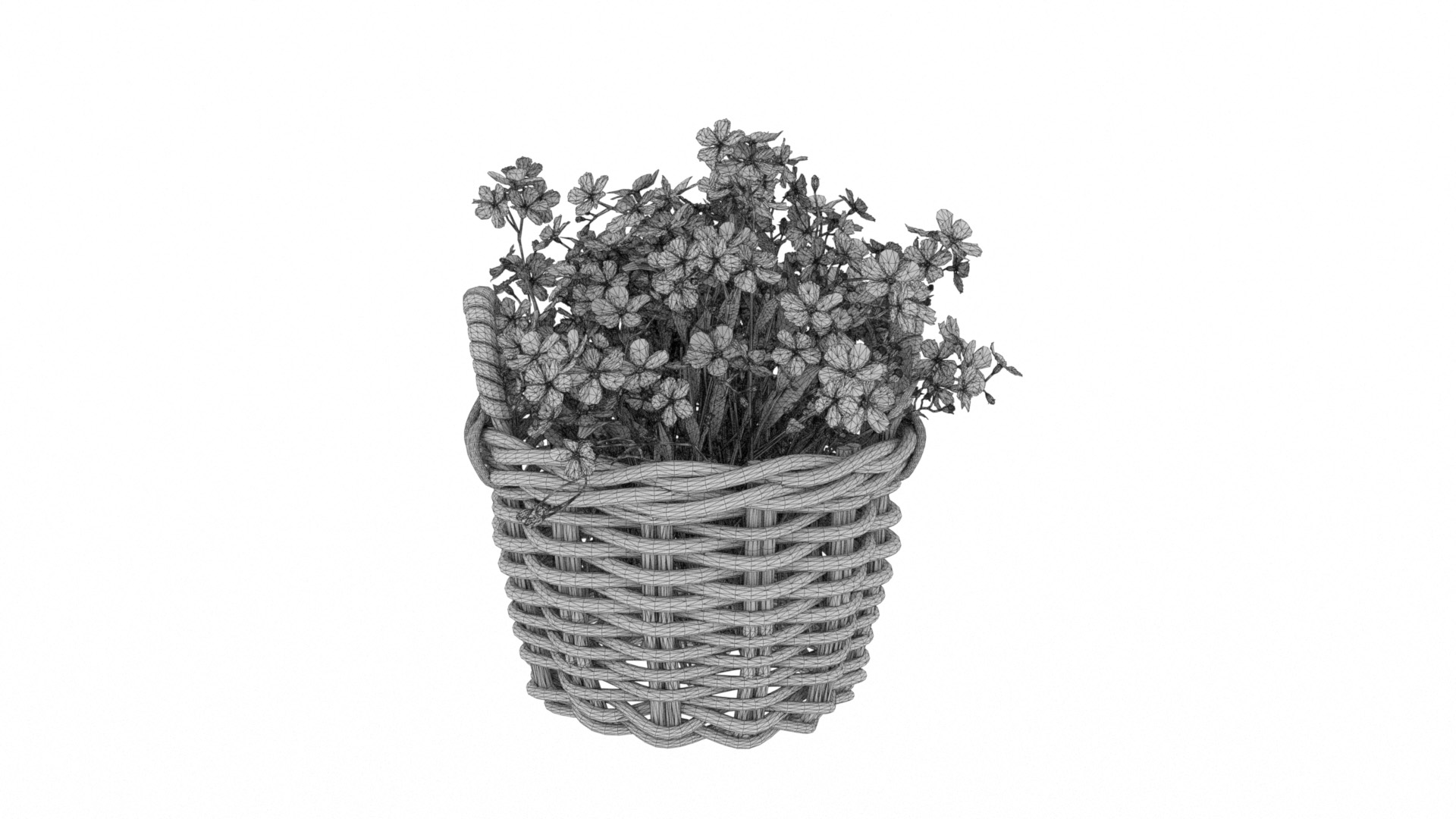}\\\vspace{0.25in}
\figcap{\tiny ~~} \hfill%
\figcap{\tiny T14+VM (1675K,0.04,1328,28.3)} \hfill%
\figcap{\tiny VR (640K,0.01,0,22.7)} \hfill%
\figcap{\tiny VM (1316K,0.04,1482,27.4)} \hfill%
\figcap{\tiny VR+VM (1794K,0.04,1492,27.4)} \hfill%
\figcap{\tiny Ours (1483K,0.01,12,45.3)} \\
\includegraphics[trim=200 100 200 250,clip,width=0.16\linewidth]{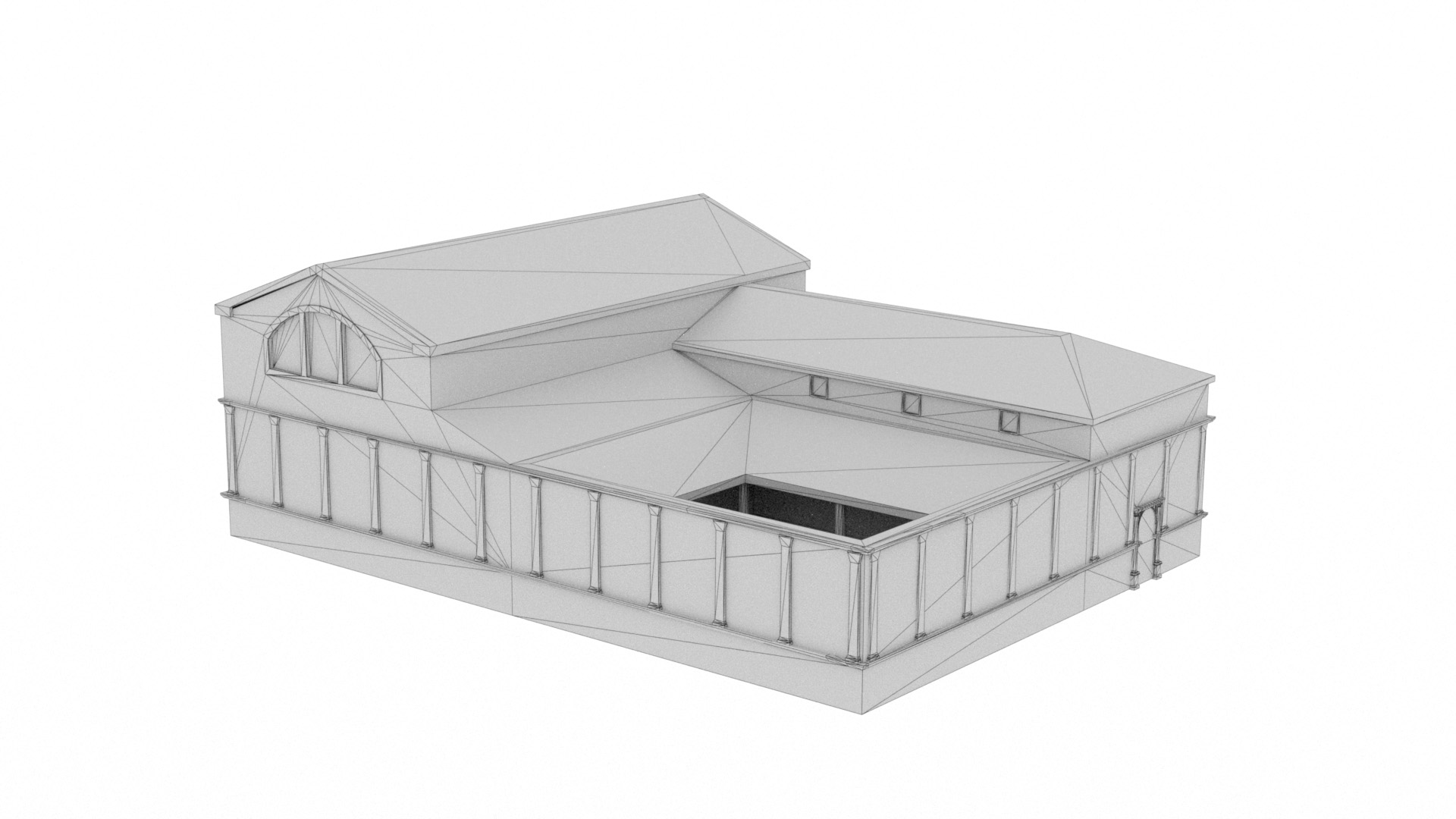}\hfill
\includegraphics[trim=200 100 200 250,clip,width=0.16\linewidth]{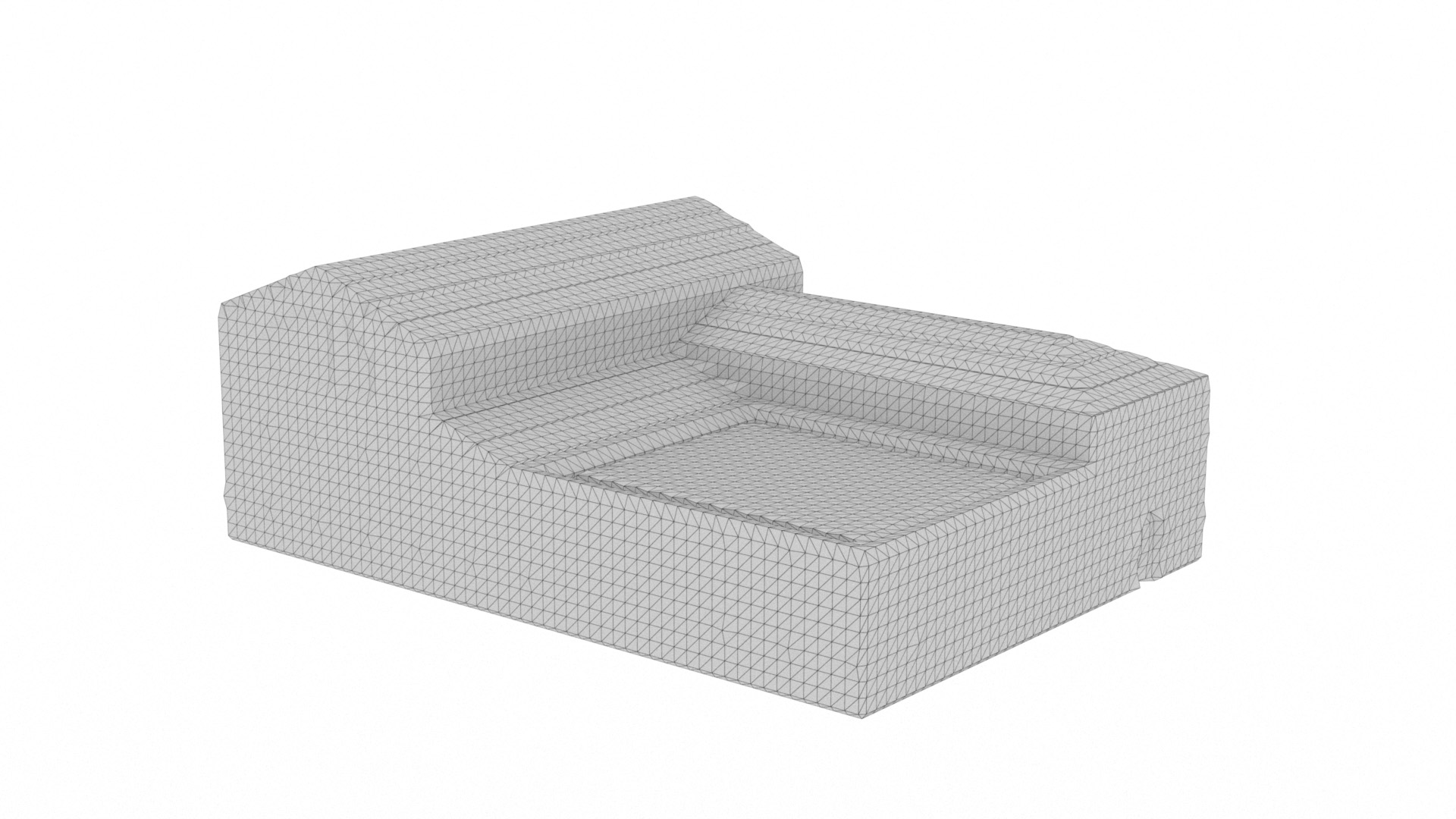}\hfill
\includegraphics[trim=200 100 200 250,clip,width=0.16\linewidth]{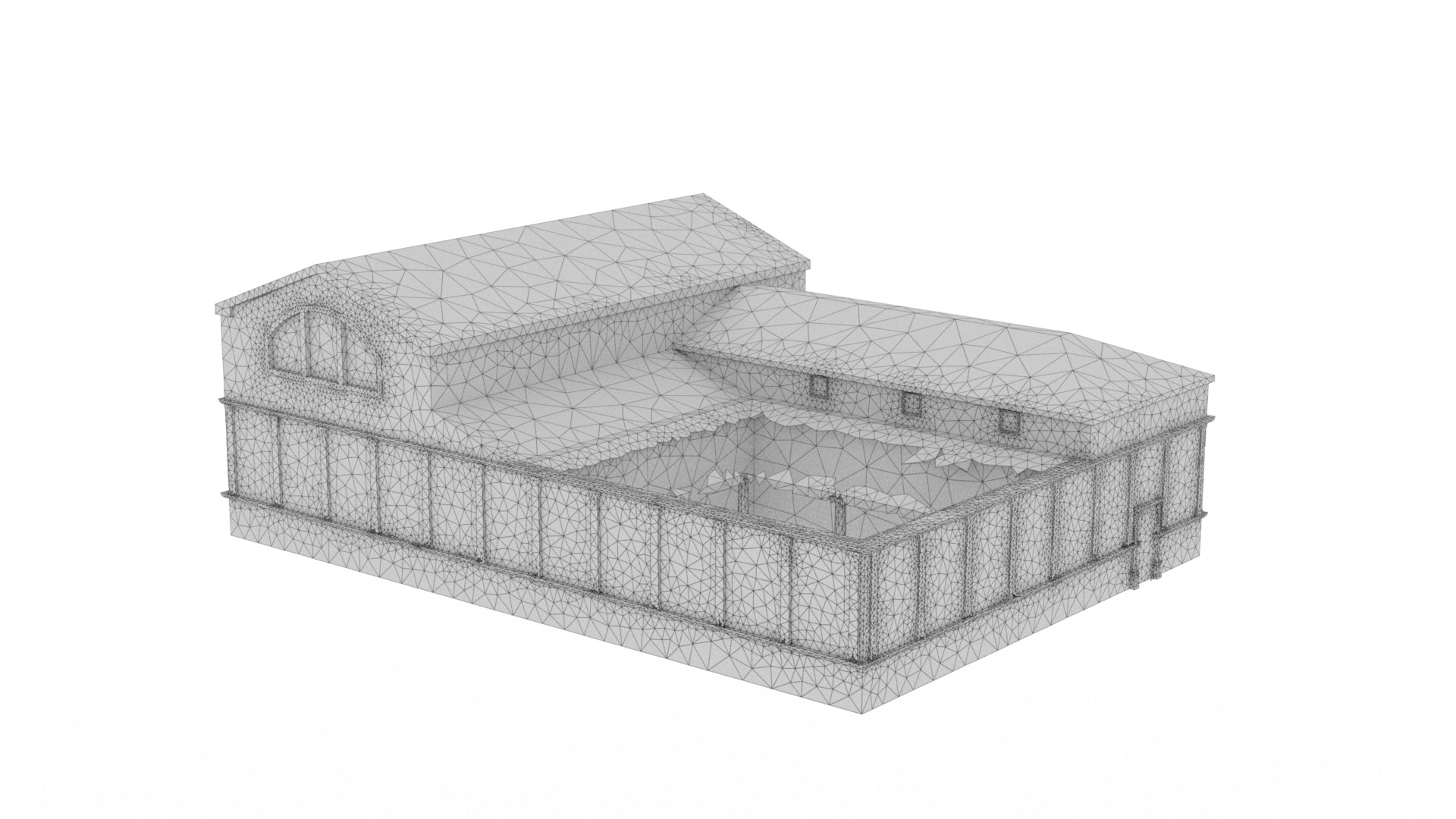}\hfill
\includegraphics[trim=200 100 200 250,clip,width=0.16\linewidth]{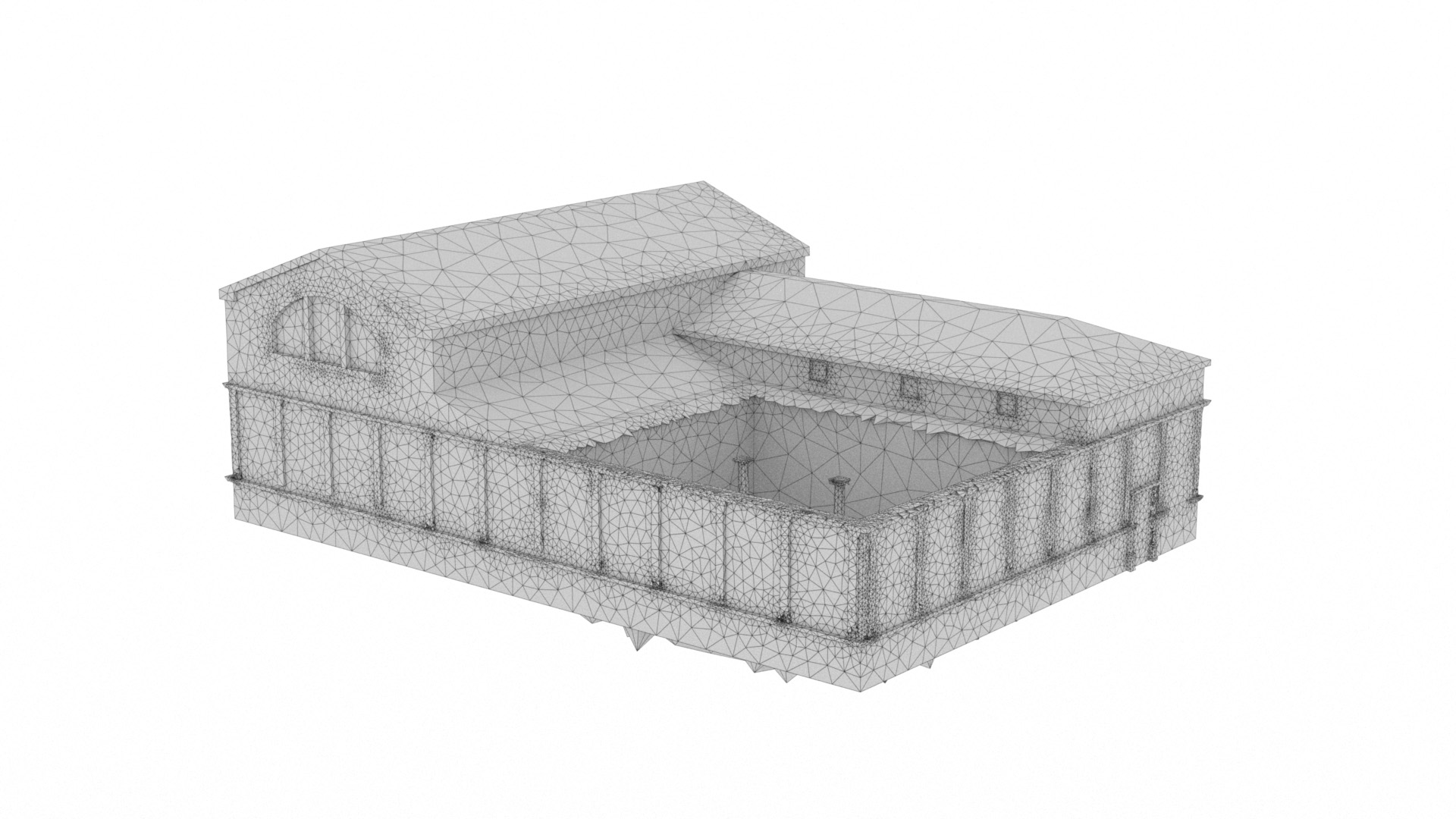}\hfill
\includegraphics[trim=200 100 200 250,clip,width=0.16\linewidth]{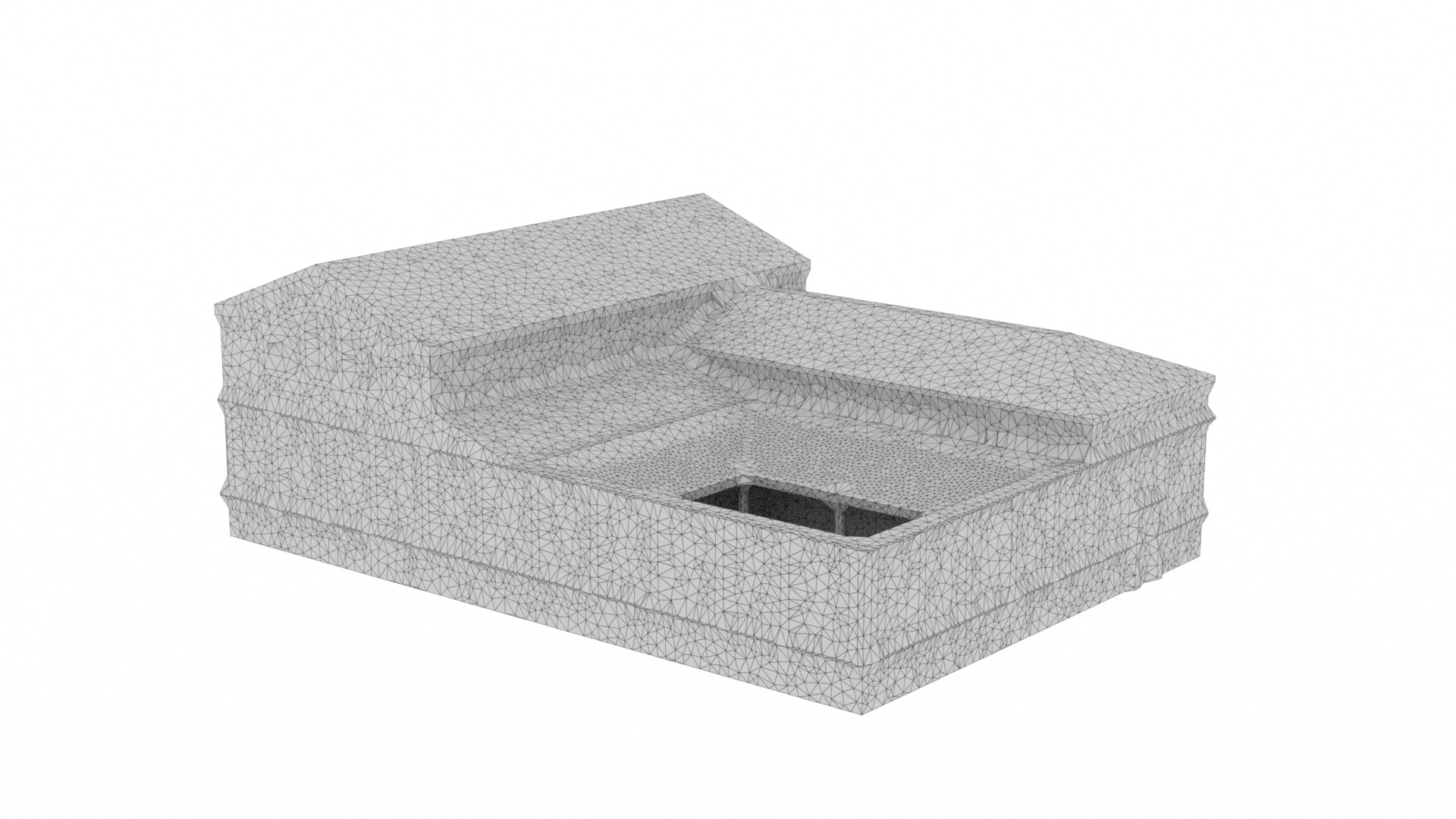}\hfill
\includegraphics[trim=200 100 200 250,clip,width=0.16\linewidth]{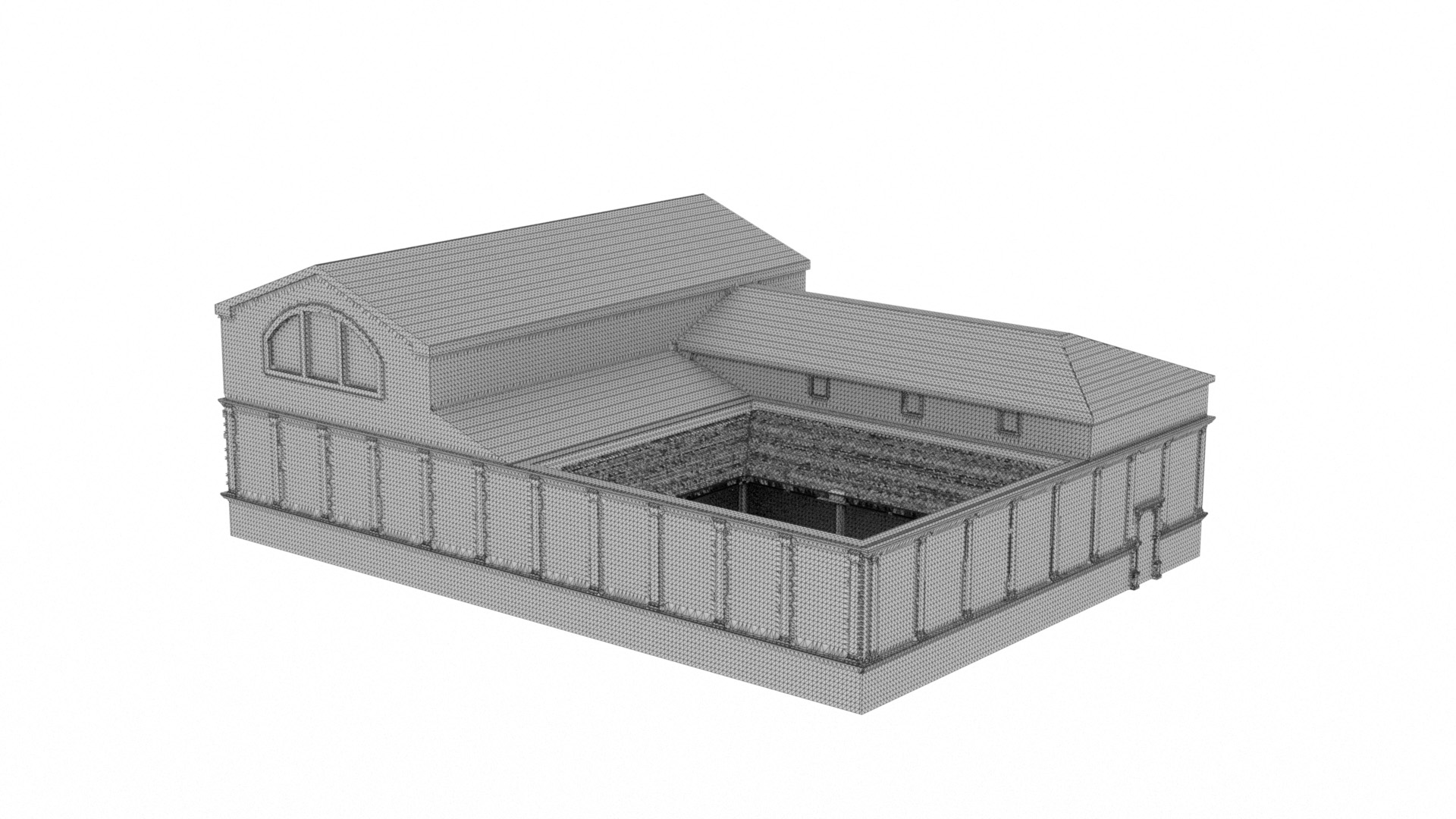}\\
\figcap{\tiny Roman} \hfill%
\figcap{\tiny PM (18K,0.09,450,25.8)} \hfill%
\figcap{\tiny TW (46K,0.02,954,28.1)} \hfill%
\figcap{\tiny fTW (46K,0.02,632,23.9)} \hfill%
\figcap{\tiny AW (53K,0.02,42,33.9)} \hfill%
\figcap{\tiny MP (312K,0.02,10,21.0)} \\
\includegraphics[trim=200 100 200 100,clip,width=0.16\linewidth]{images/white.jpg}\hfill
\includegraphics[trim=200 100 200 100,clip,width=0.16\linewidth]{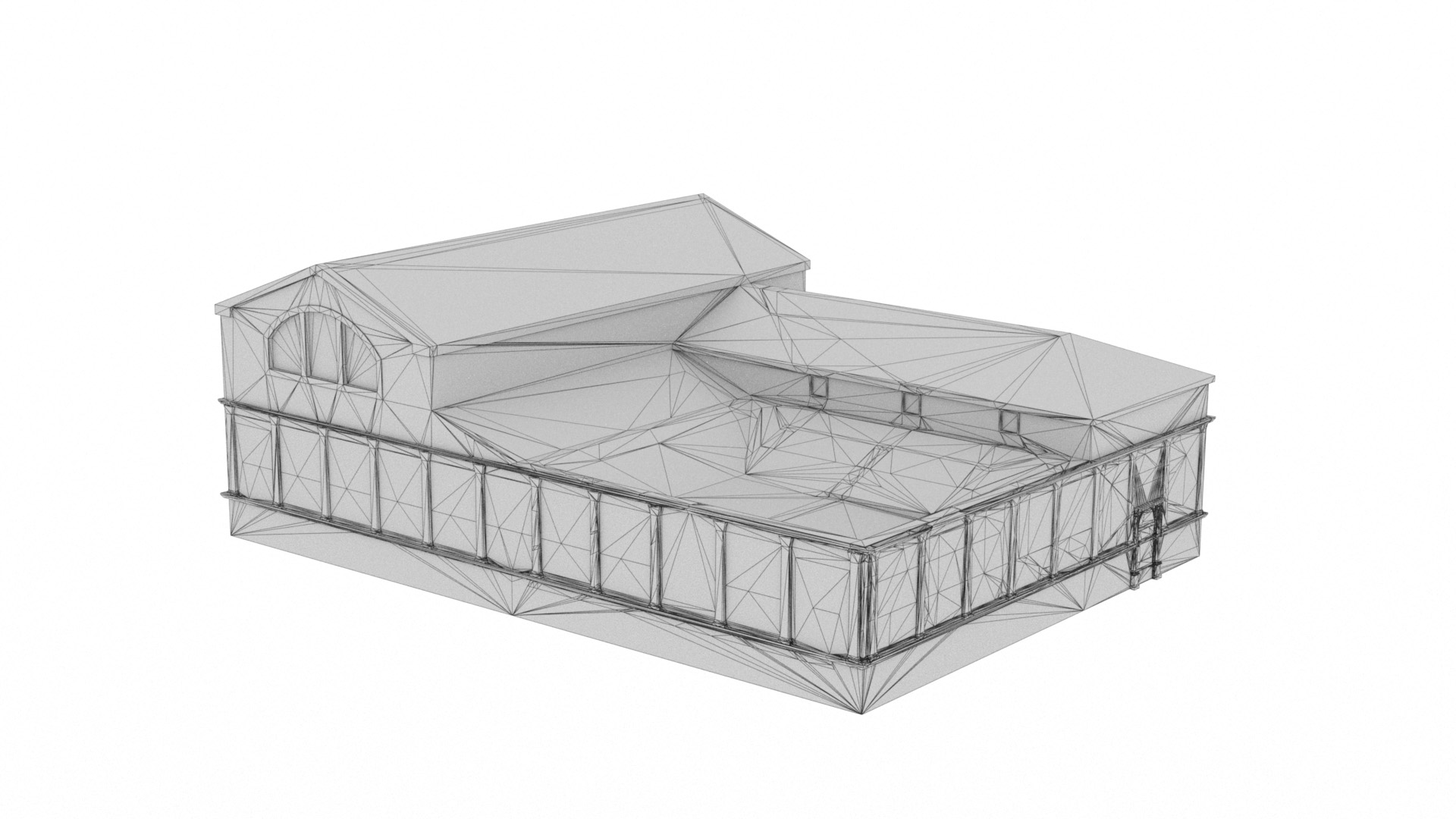}\hfill
\includegraphics[trim=200 100 200 100,clip,width=0.16\linewidth]{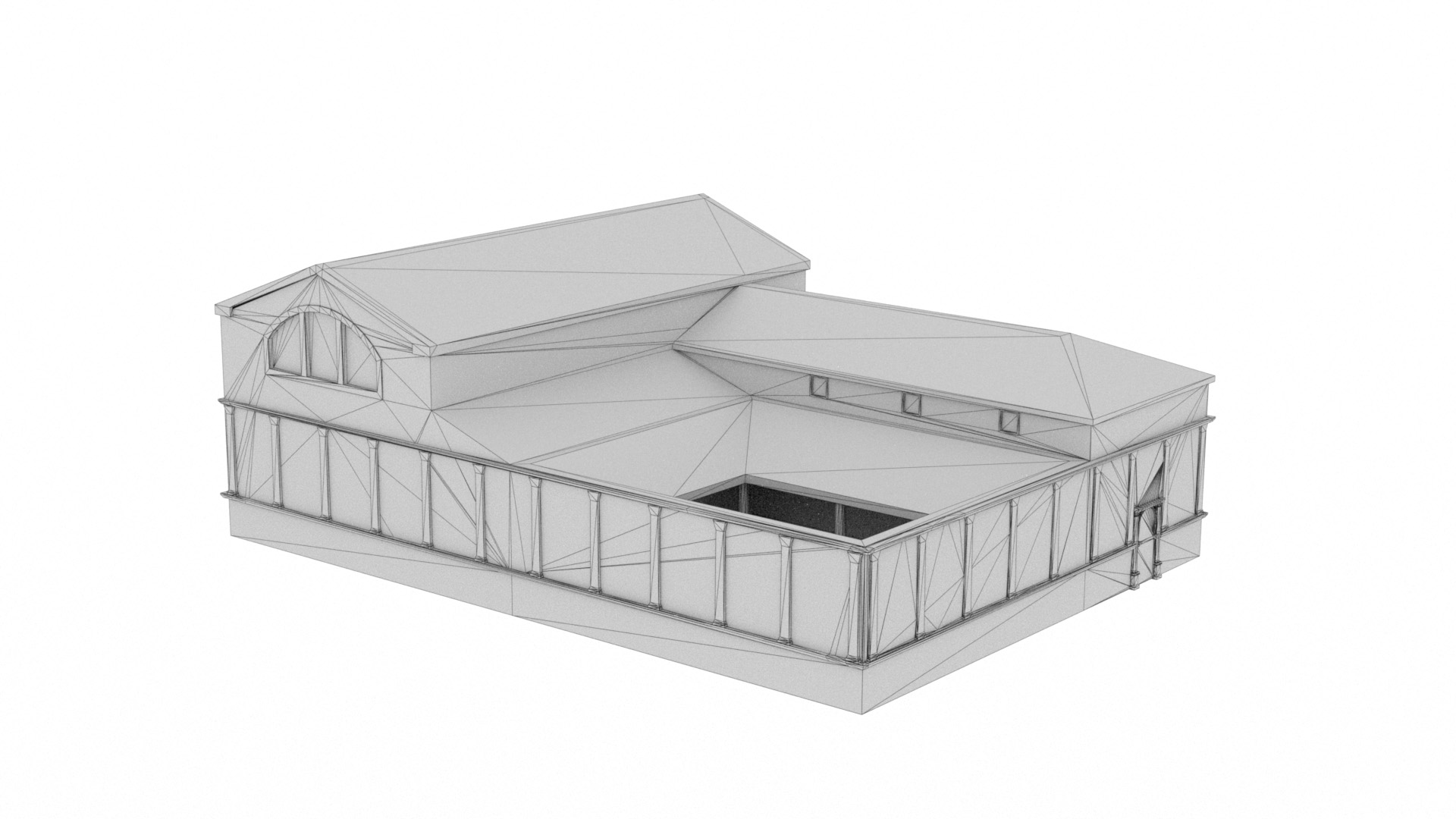}\hfill
\includegraphics[trim=200 100 200 100,clip,width=0.16\linewidth]{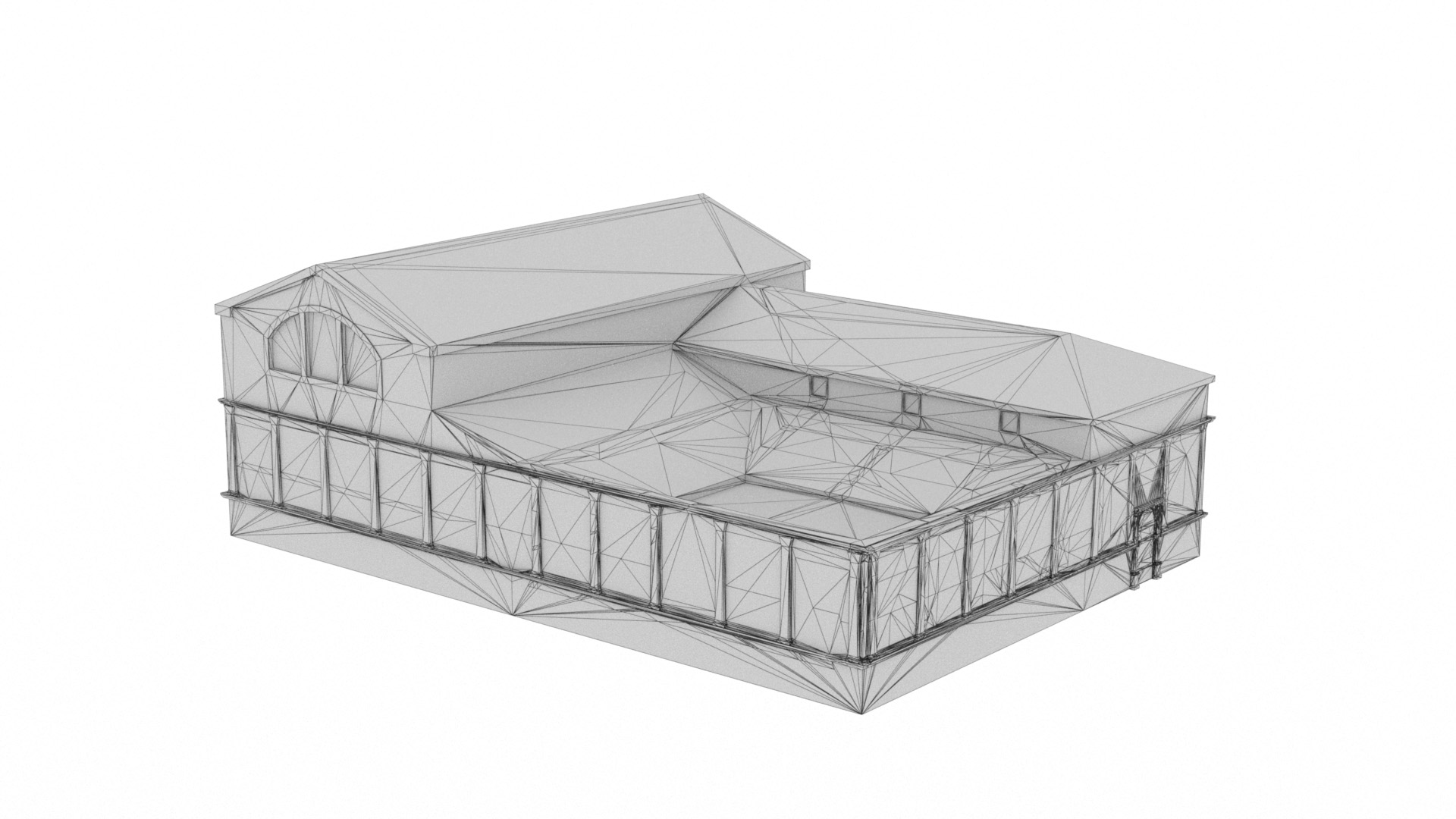}\hfill
\includegraphics[trim=200 100 200 100,clip,width=0.16\linewidth]{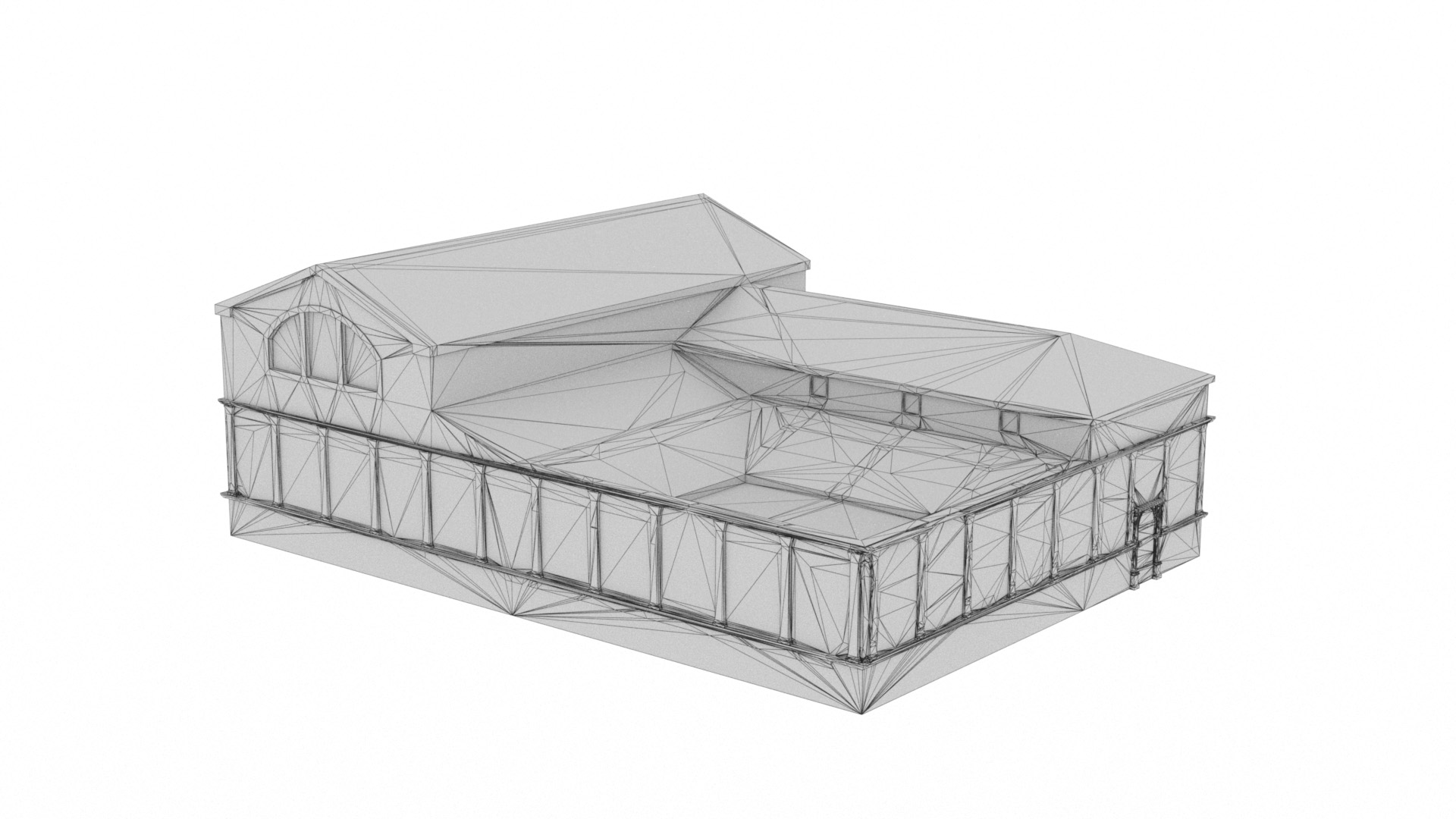}\hfill
\includegraphics[trim=200 100 200 100,clip,width=0.16\linewidth]{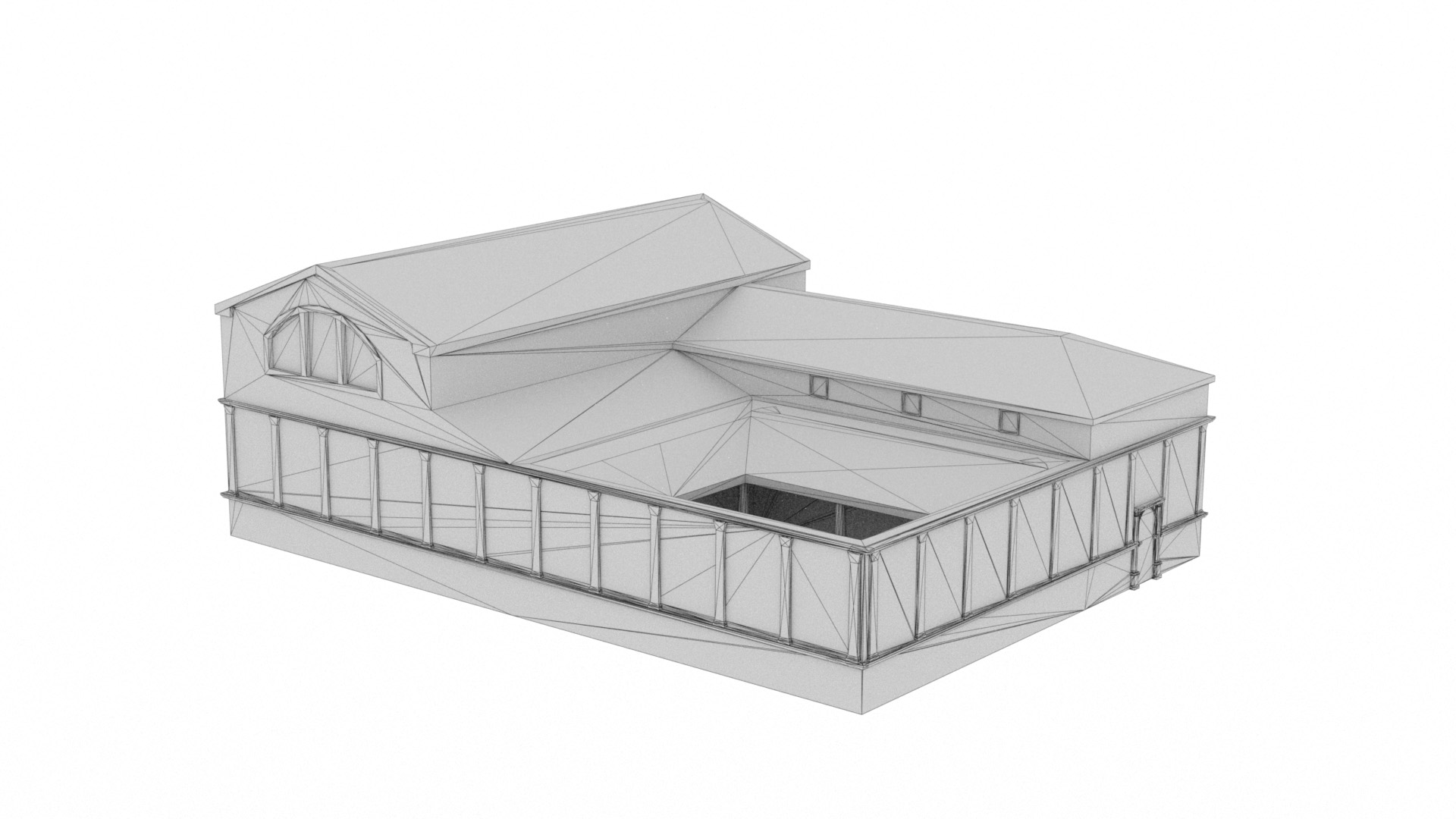}\\
\figcap{\tiny ~~} \hfill%
\figcap{\tiny T14+VM (31K,0.05,4,40.5)} \hfill%
\figcap{\tiny VR (12K,0.003,0,31.9)} \hfill%
\figcap{\tiny VM (32K,0.05,2,38.8)} \hfill%
\figcap{\tiny VR+VM (35K,0.05,4,38.9)} \hfill%
\figcap{\tiny Ours (14K,0.01,0,49.6)} 
\caption{\textbf{Results:} We choose ``Chair'', ``Flowers'', and ``Roman'' to demonstrate the issues of existing methods and our superiority over them. $(\bullet,\bullet,\bullet,\bullet)$ indicates face number, Hausdorff distance, LFD, and PSNR. }
\label{fig:results}
\end{figure*}

\section{Results} \label{sec:result}
We implement our framework in C++ with CGAL, libigl, and Eigen. We use Optix~\cite{Parker2010optix} to compute visual metrics via ray tracing on GPU and a fast approximate energy minimization solver~\cite{boykov2001fast} to solve the graph cut. We did all experiments on a computer with an AMD Ryzen Threadripper 3970X 32-Core Processor at 3.69 GHz and 256 GB RAM. We use $N_\text{total}=2e7$, which is sufficiently large for over-sampling all the models in our input dataset, and $N_{\min}=5$ for good coverage of mesh surface, and sample $N_{d}=5$ directions on the unit hemisphere for both sides of $f_\text{input}^i$. Each ray has a maximum bounce number $N_\text{b} = 10$. For the ear-cut triangulation used in constrained simplification, an extension distance is required and set to $l_\text{extended}=D/1000$, with $D$ being the diagonal length of the model's bounding box. 

\textbf{Offset Distance.}
The offset distance $d_\text{offset}$ controls the thickness of the thin shells created for open faces. We illustrate the impact of the offset distance $d_\text{offset}$ using a flower model with different  $d_\text{offset}$. \autoref{fig:ablation_off} demonstrates our method can robustly convert the input model with hundreds of open faces to a watertight mesh, where the Hausdorff distance between output and input meshes is controlled by $d_\text{offset}$. We use $d_\text{offset} = D/20000$ for the rest of the examples. On average, the introduction of intersected faces by offset faces amounts to approximately 280 pairs across the 1400 testing models (with an average of 33,829 faces) in \autoref{fig:results}. In the majority of cases, due to the presence of intersecting open faces in the original model, our offset faces will inevitably intersect with these existing intersections.

\begin{table*}[ht!]
\centering
\caption{\label{tab:statistic} Statistics of 1000 models from ShapeNet and 400 models of Thingi10K, including the percentage of results that are finished within 1 hour, watertight and manifold, respectively, and average face number, HD, LFD, PSNR, peak memory, and time usage. Smaller values are more desired for all listed numerical metrics except PSNR. $^*$ indicates PNSR of our result with UVs, while all other methods do not have UVs. Note that Models in Thingi10K do not have UVs.}
\scalebox{1}
 	{
\begin{tabular}{|l|ccccccccc|}
\hline
\multicolumn{10}{|c|}{ShapeNet} \\ \hline
 &Finishes($<$ 1h) &Watertight & Manifold & Face \#  & HD & LFD & PSNR & Mem. & Time (s) \\ \hline
Input       &--  & \hspace{0.5em}82\% & \hspace{1em}2\%& \hspace{0.5em}32K   & --    & --    & --  & -- & --   \\ 
PM   &100\% & 100\% & 100\% & \hspace{0.5em}14K  & 0.06 & $2.2e3$ & 29.0 & \hspace{1.0em}7 Mb & \hspace{1.em}0.4 \\ 
TW      & 99.5\%& 100\% & \hspace{0.5em}23\% & \hspace{1em}4K  & 0.18   & $3.5e4$ & 22.2 & 943 Mb & 259.4  \\ 
fTW     & 99.7\%& 100\% & 100\% & \hspace{1em}10K & 0.18 & $1.3e4$ & 26.3 & 714 Mb &114.8  \\ 
AW &100\%& 100\% & 100\% & \hspace{0.5em}34K& 0.03 & $2.3e2$ & 35.3 & \hspace{0.5em}96 Mb& \hspace{1.em}{6.0} \\
VR & 97.9\%& \hspace{0.5em}18\% & 100\% & \hspace{0.5em}22K & 0.02 & $1.0e1$ & 48.1 & 111 Mb & 176.7 \\ 
VM & 100\%& 100\% & \hspace{0.5em}16\% & \hspace{0.5em}95K & 0.05 & $1.4e3$ & 32.8 & 313 Mb & \hspace{1.em}{3.6}   \\
VR+VM  &   97.9\%    & 100\%  & \hspace{0.5em}37\%  & 180K   & 0.03 & $2.0e2$ & 49.4 & 381 Mb & 186.8 \\
MP     & 100\% & 100\% & 100\% &325K    & 0.02 &$8.0e1$ & 36.9 & 217MB & \hspace{1em}3.8 \\ 
T14+VM   &  100\%   & 100\% & 31\% &  117K  & 0.03 &$2.4e2$ & 46.6 &299MB  &\hspace{0.5em}11.8  \\ 
Ours     &100\%     & 100\% & 100\% & \hspace{0.5em}29K& 0.02  & $1.0e0$ &  58.9 ($58.8^*$) & 722 Mb  & \hspace{0.5em}{16.9} \\ \hline
\multicolumn{10}{|c|}{Thingi10K} \\ \hline
Input  & --       &5\% &97\% &  40K & --    & --    & --  & -- & --   \\ 
PM    &   100\%   &100\% &100\% &  17K  & 0.02   & $9.5e2$    & 31.7  & 8MB & 0.9   \\ 
TW     &     100\%   &100\% &87\% &  13K  & 0.006    & $3.3e2$    & 47.9  & 213MB & 77.7   \\ 
fTW    &   100\%     &100\% &100\% &  10K  & 0.006    & $3.3e2$    & 46.3  & 143MB & 24.0  \\ 
AW     & 100\%  &100\% & 100\%& 27K   & 0.014    & $1.1e2$    & 37.2  & 88MB & 4.4   \\ 
VR    &  90\%  &66\% &92\% & 34K  & 0.0001    & $5.0e$-2    & 67.9  & 149MB & 406.2   \\ 
VM     &  100\%   &100\% &81\% & 128K   & 0.006    & $1.9e2$    & 64.1  & 258MB & 4.4   \\ 
VR+VM   &  90\%     & 100\%&84\% &  119K  & 0.003    & $7.7e1$    & 65.2  & 259MB & 411.4   \\ 
MP     &   100\% &100\% &100\% & 416K  & 0.006    & $3.8e1$    & 50.8  & 274MB & 7.8   \\ 
T14+VM    &  98\%      &100\% &75\% & 110K   & 0.004    & $9.0e0$ & 64.1 & 237MB & 24.9   \\ 
Ours   &  100\%   &100\% &100\% &   36K & 0.001    & $5.5e0$    & 66.7  & 755MB & 26.7   \\  \hline
\end{tabular}
}
\end{table*}

\textbf{Maximum Bounce Number.} The maximum bounce number $N_b$ governs the ray-tracing process's ability to navigate complex structures. When $N_b=0$, no bounces are allowed, potentially misguiding the representation of hard-to-see structures. Conversely, when $N_b=10$, the exploration is thorough, facilitated by multiple reflections. We conducted tests with varying max bounce numbers using the 500 models from ShapeNet~\cite{Chang2015shapenet}. For $N_b=$0, 1, 2, 3, 5, and 10, we obtained HD/LFD/PSNR values of $0.03/3.3/57.1$, $0.02/2.0/58.3$, $0.02/0.9/59.1$, $0.02/0.7/59.5$, and $0.02/0.7/60.2$, respectively. Subsequently, we employed $N_b=10$ for all subsequent experiments in this paper.

\begin{figure*}[ht]
\newcommand{\figcap}[1]{\begin{minipage}{0.15\linewidth}\centering#1\end{minipage}}
\includegraphics[trim=800 80 800 0,clip,width=0.09\linewidth]{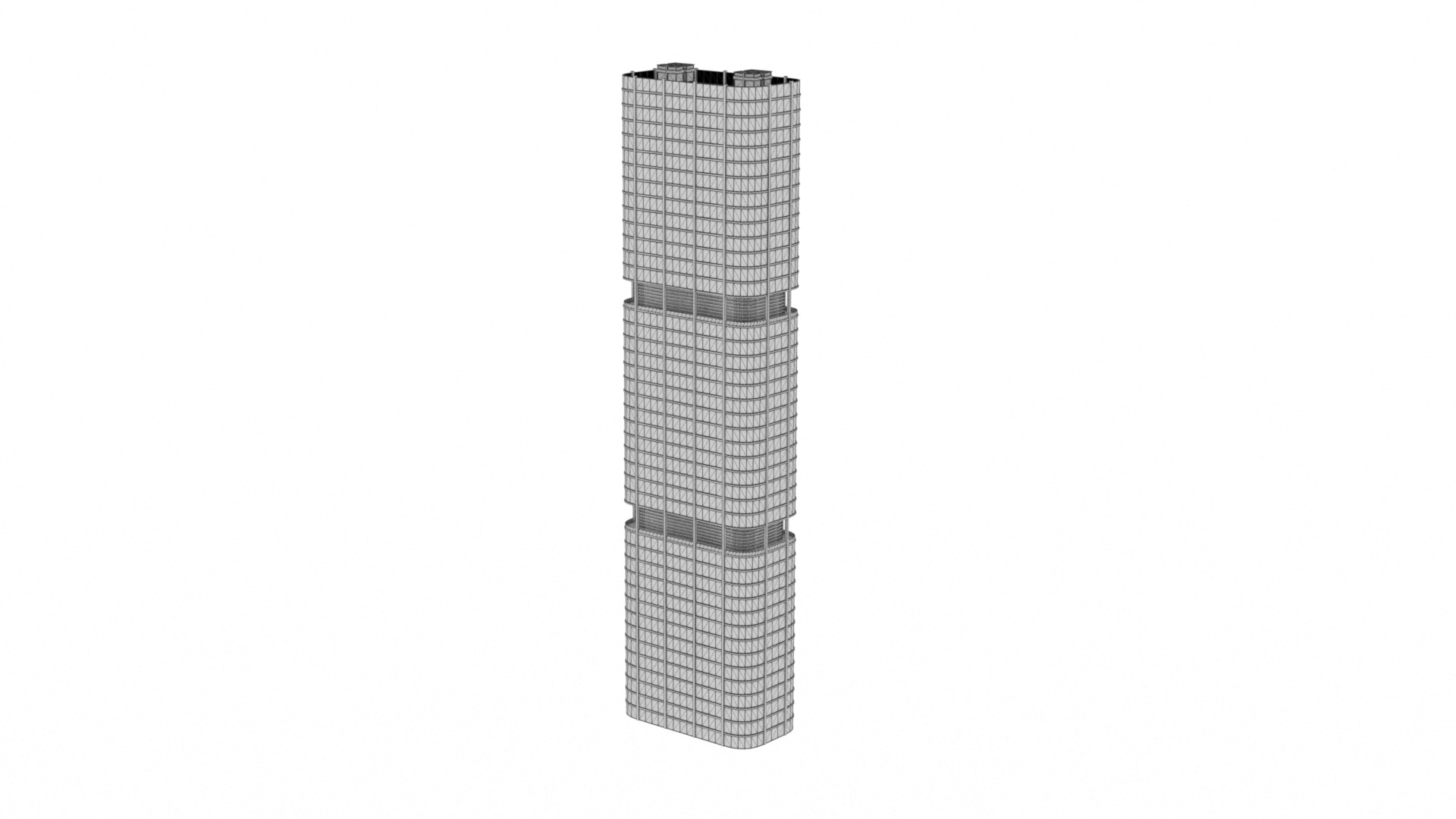}\hfill
\includegraphics[trim=800 80 800 0,clip,width=0.09\linewidth]{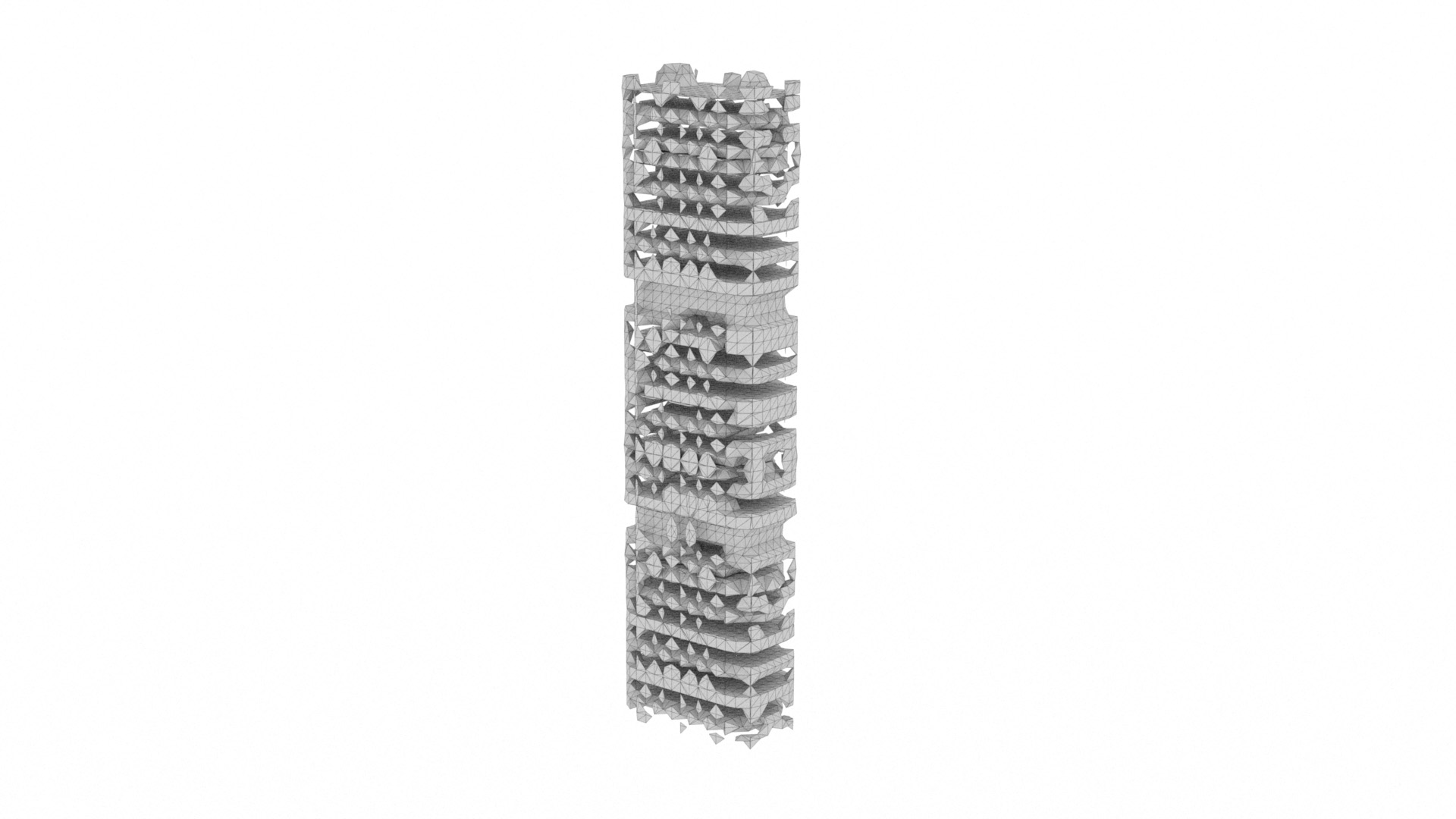}\hfill
\includegraphics[trim=800 80 800 0,clip,width=0.09\linewidth]{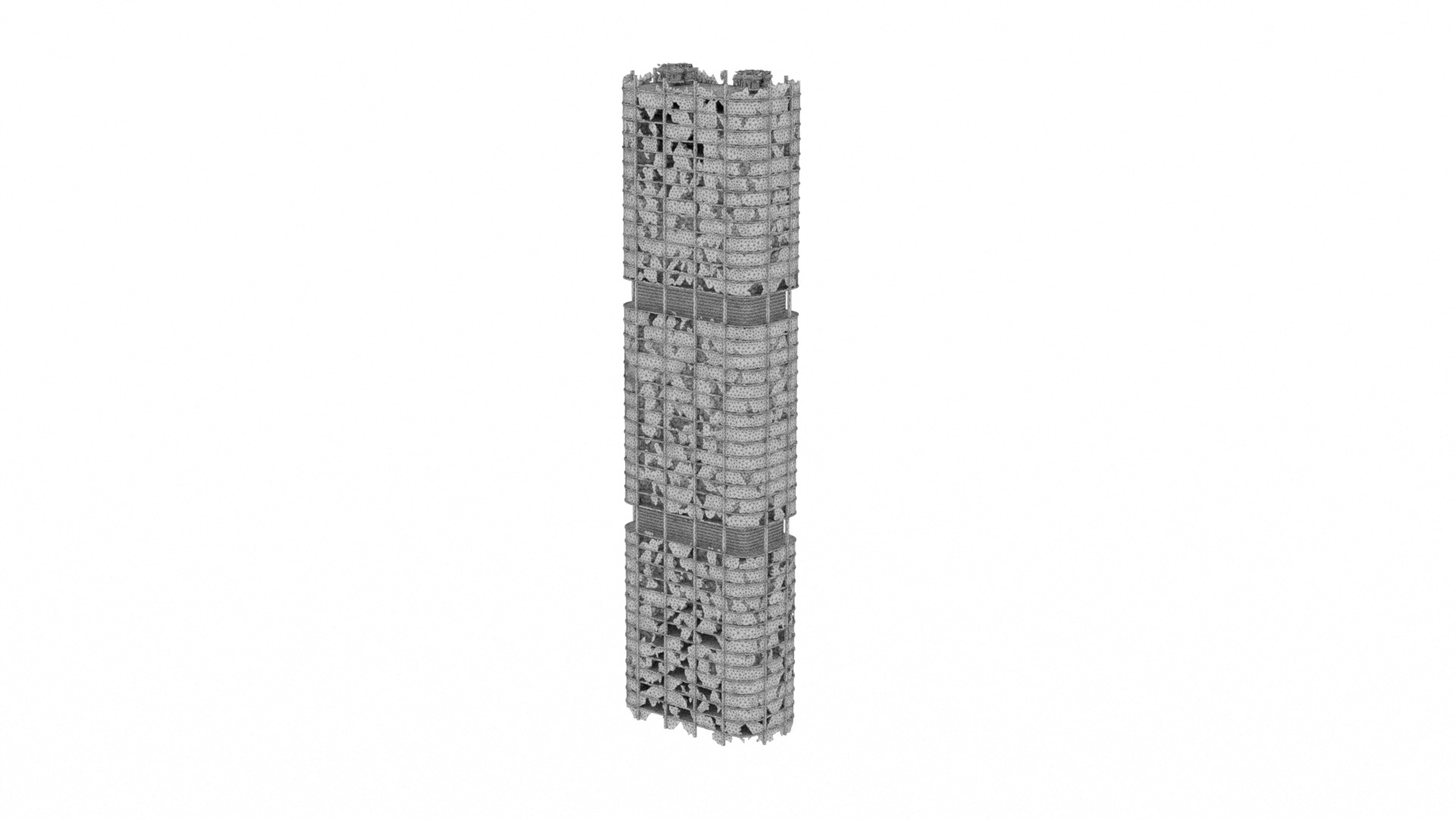}\hfill
\includegraphics[trim=800 80 800 0,clip,width=0.09\linewidth]{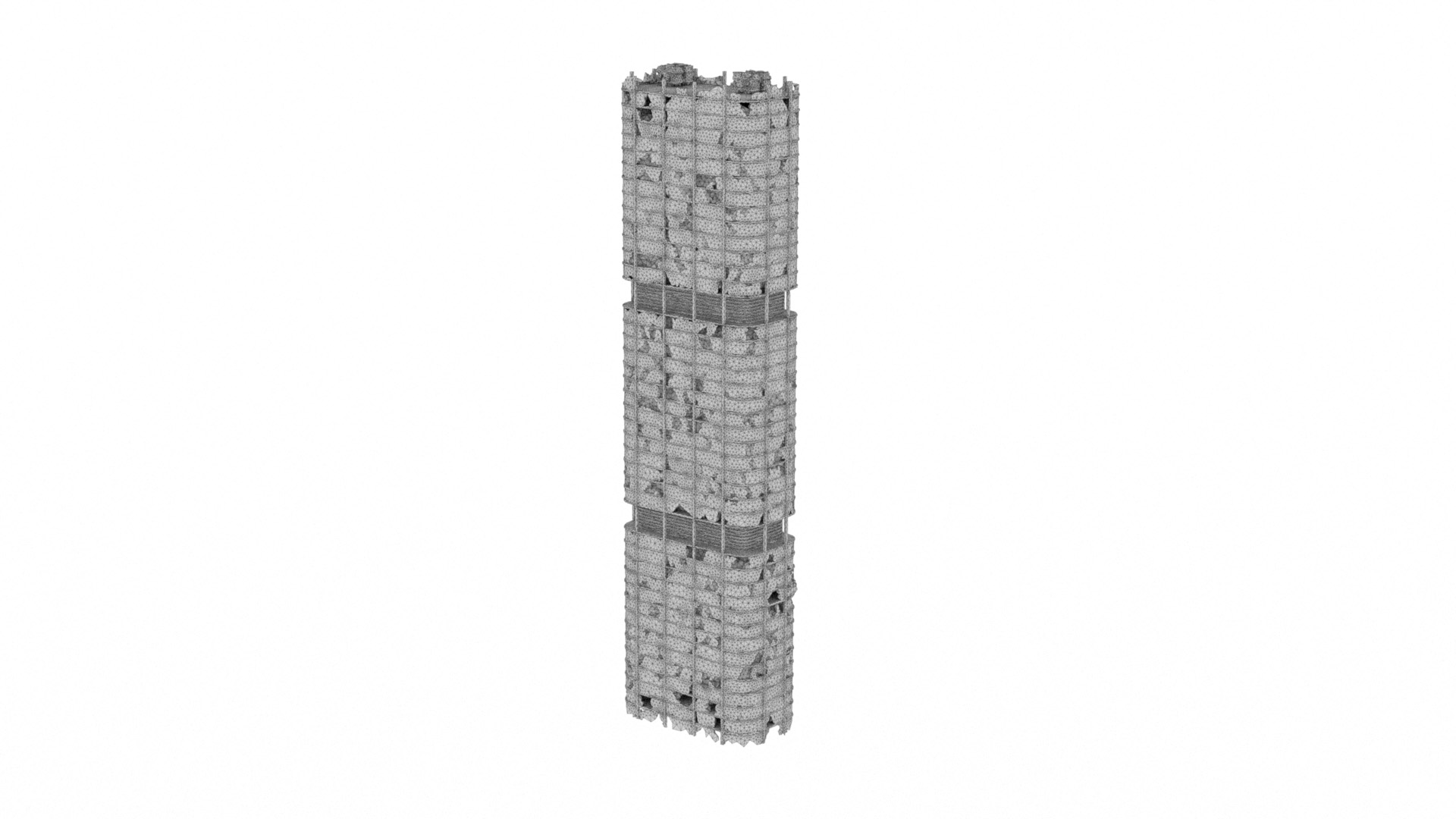}\hfill
\includegraphics[trim=800 80 800 0,clip,width=0.09\linewidth]{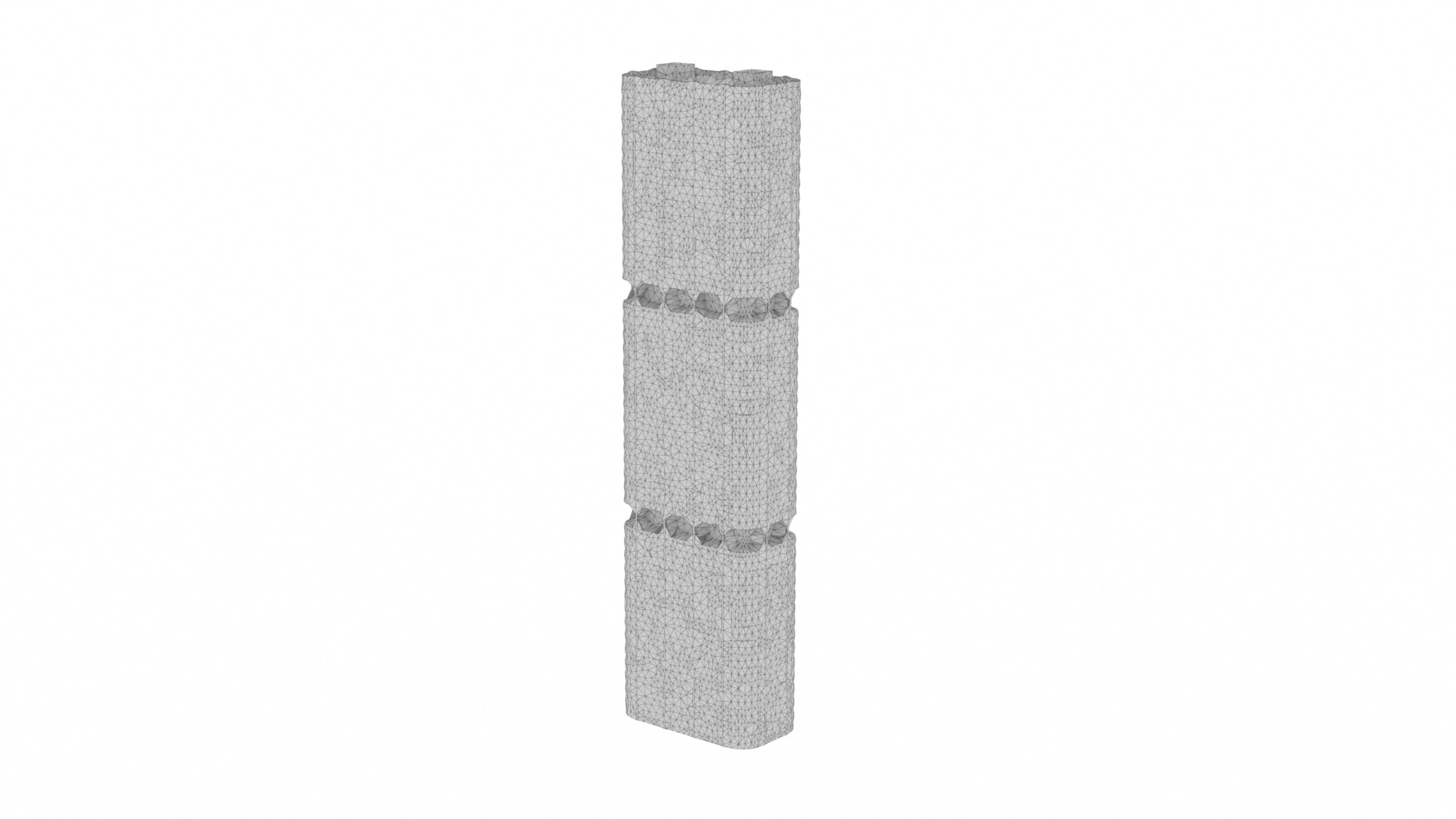}\hfill
\includegraphics[trim=800 80 800 0,clip,width=0.09\linewidth]{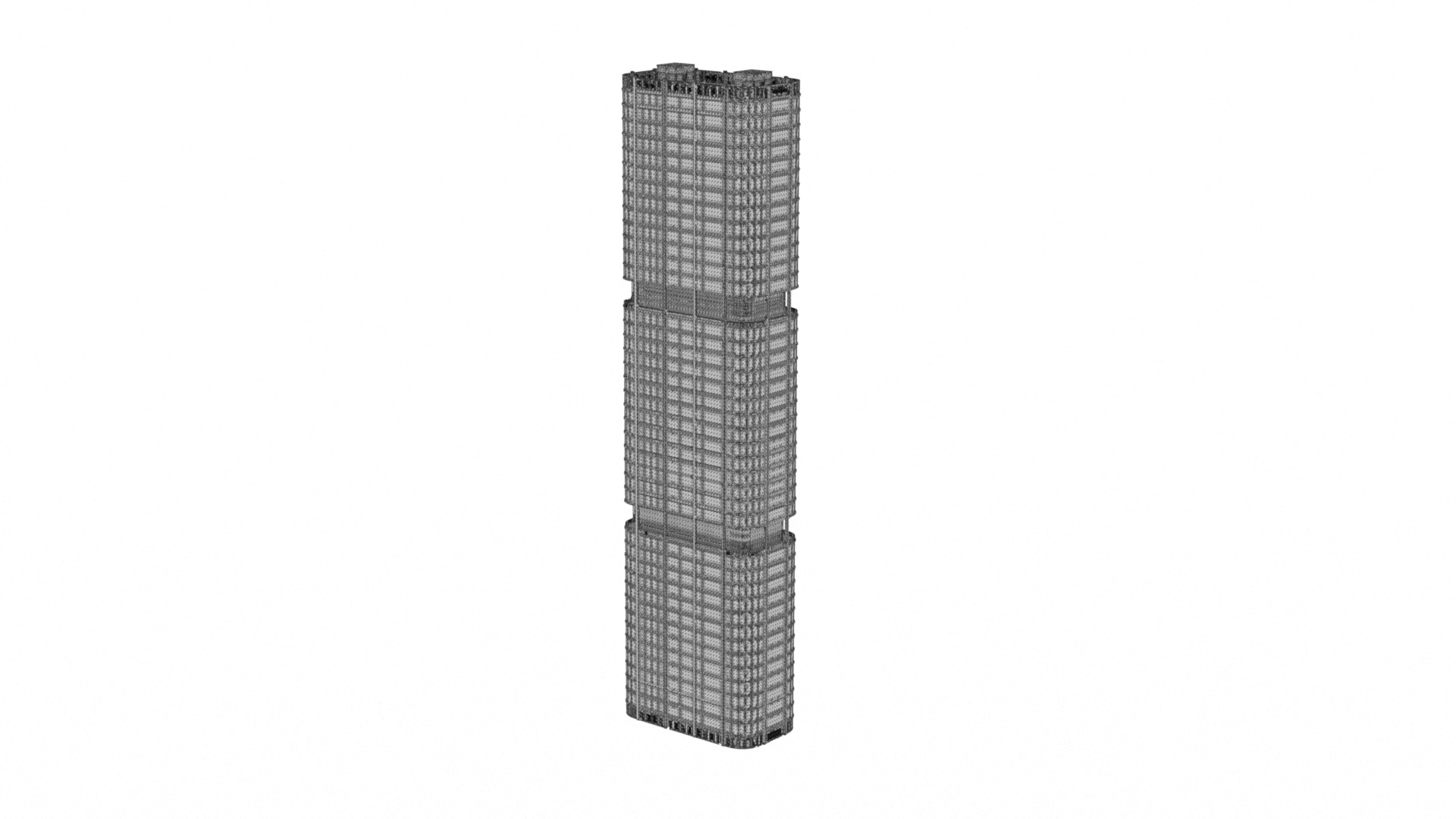}\\
\figcap{\tiny Skyscraper} \hfill%
\figcap{\tiny PM (19K,0.02,684,23.5)} \hfill%
\figcap{\tiny TW (249K,0.02,360,28.4)} \hfill%
\figcap{\tiny fTW (187K,0.02,730,28.8)} \hfill%
\figcap{\tiny AW (23K,0.02,56,31.6)} \hfill%
\figcap{\tiny MP (431K,0.02,34,30.4)} \\
\includegraphics[trim=800 80 800 0,clip,width=0.09\linewidth]{images/white.jpg}\hfill
\includegraphics[trim=800 80 800 0,clip,width=0.09\linewidth]{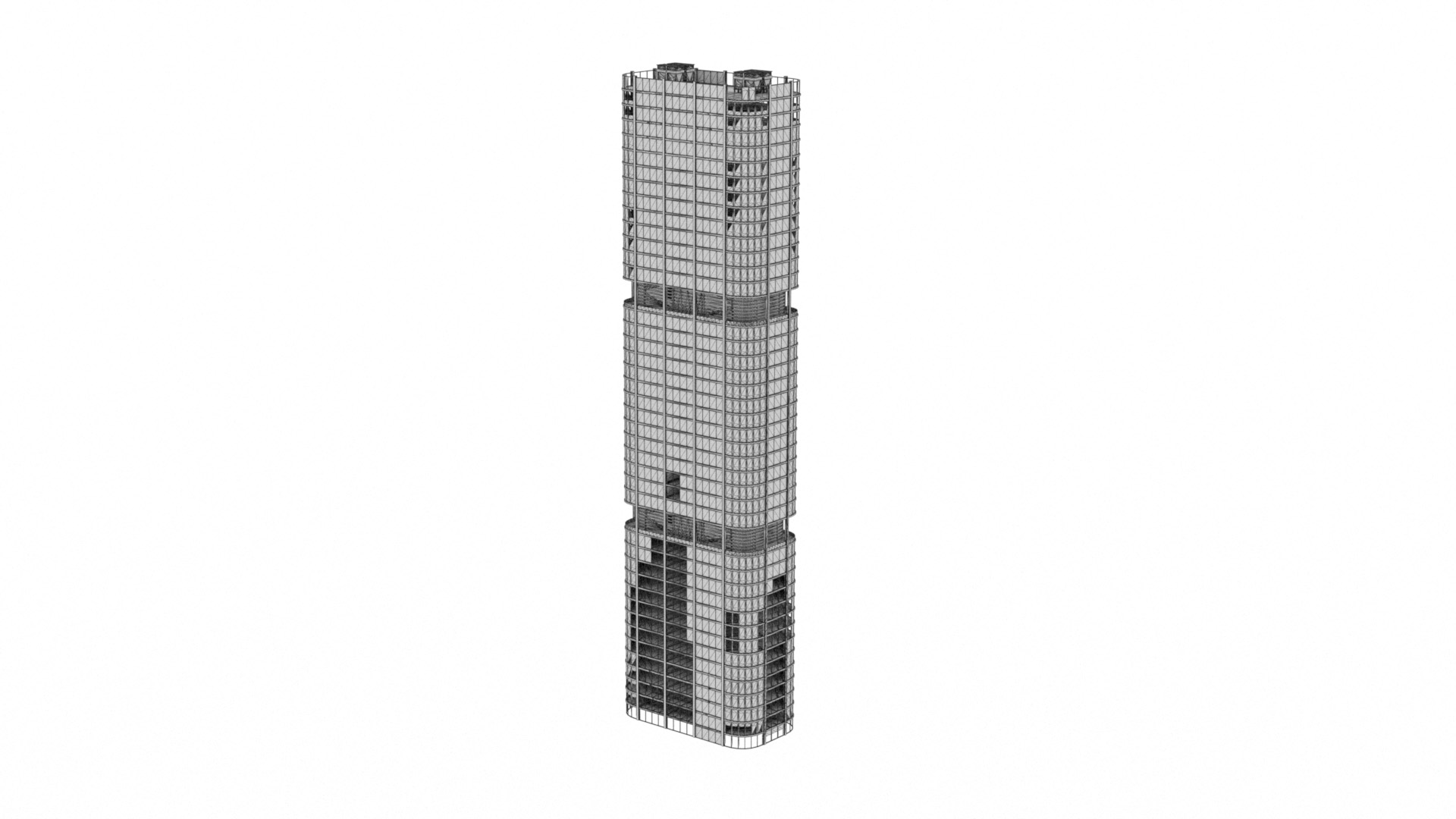}\hfill
\includegraphics[trim=800 80 800 0,clip,width=0.09\linewidth]{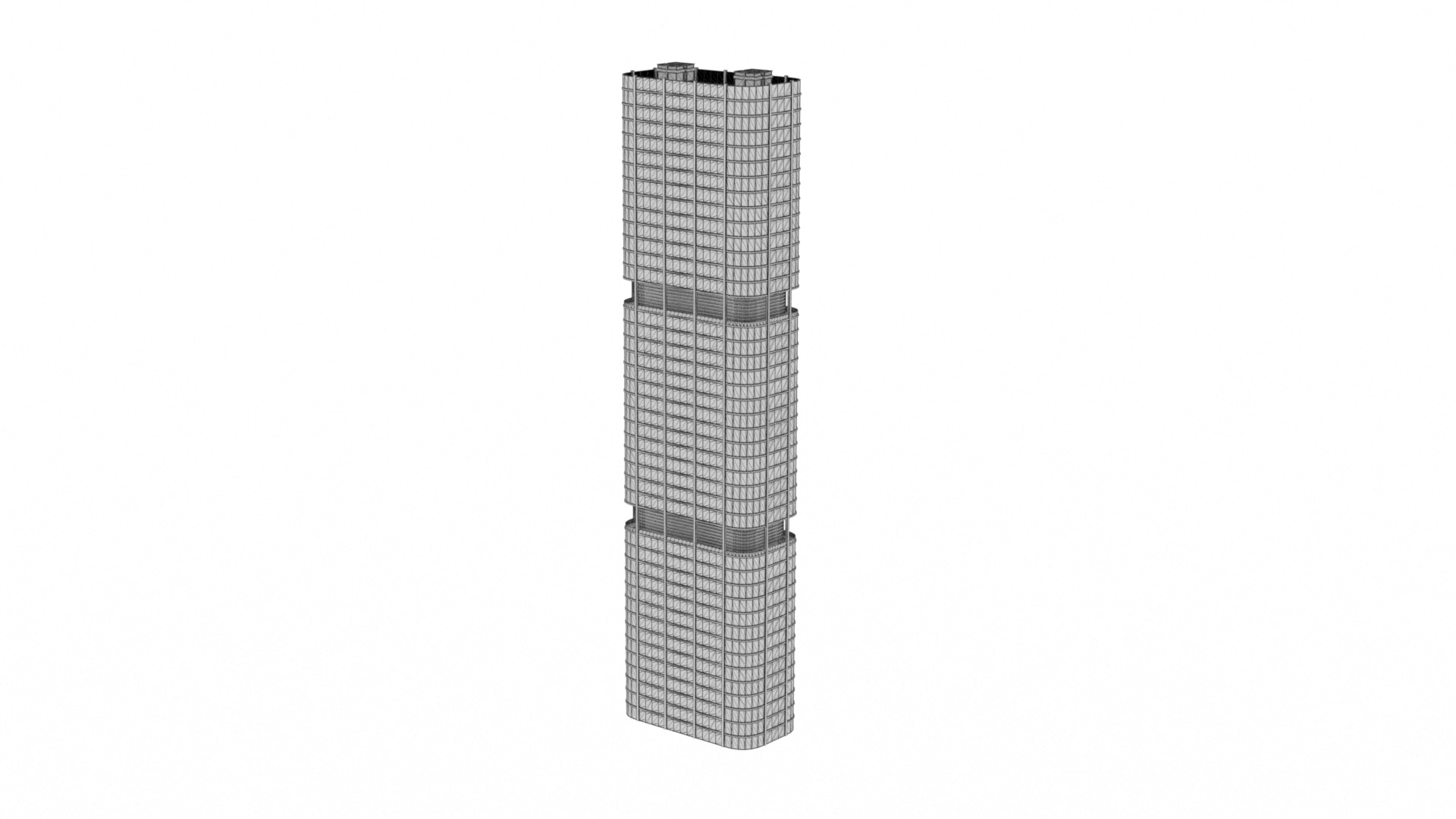}\hfill
\includegraphics[trim=800 80 800 0,clip,width=0.09\linewidth]{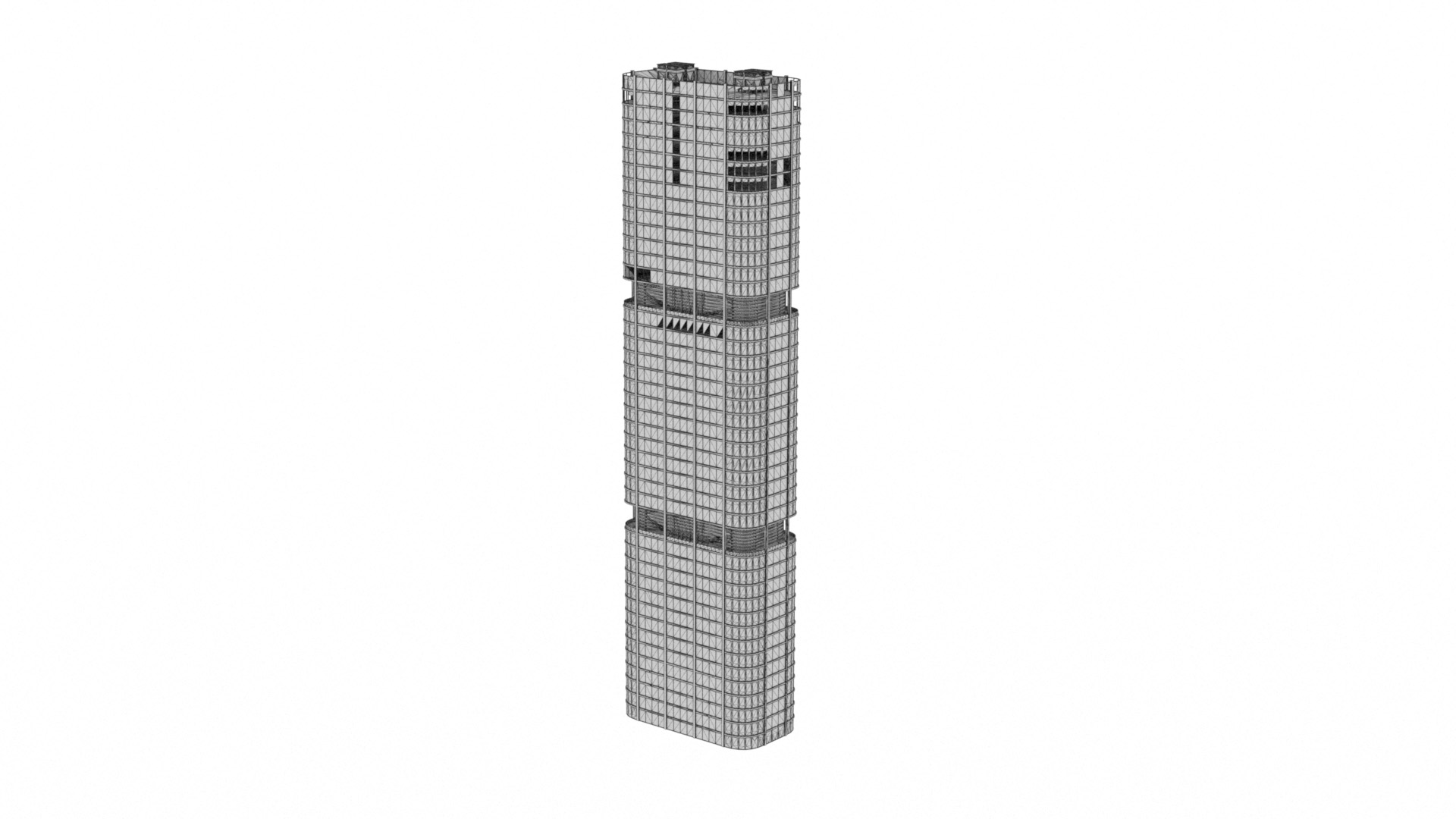}\hfill
\includegraphics[trim=800 80 800 0,clip,width=0.09\linewidth]{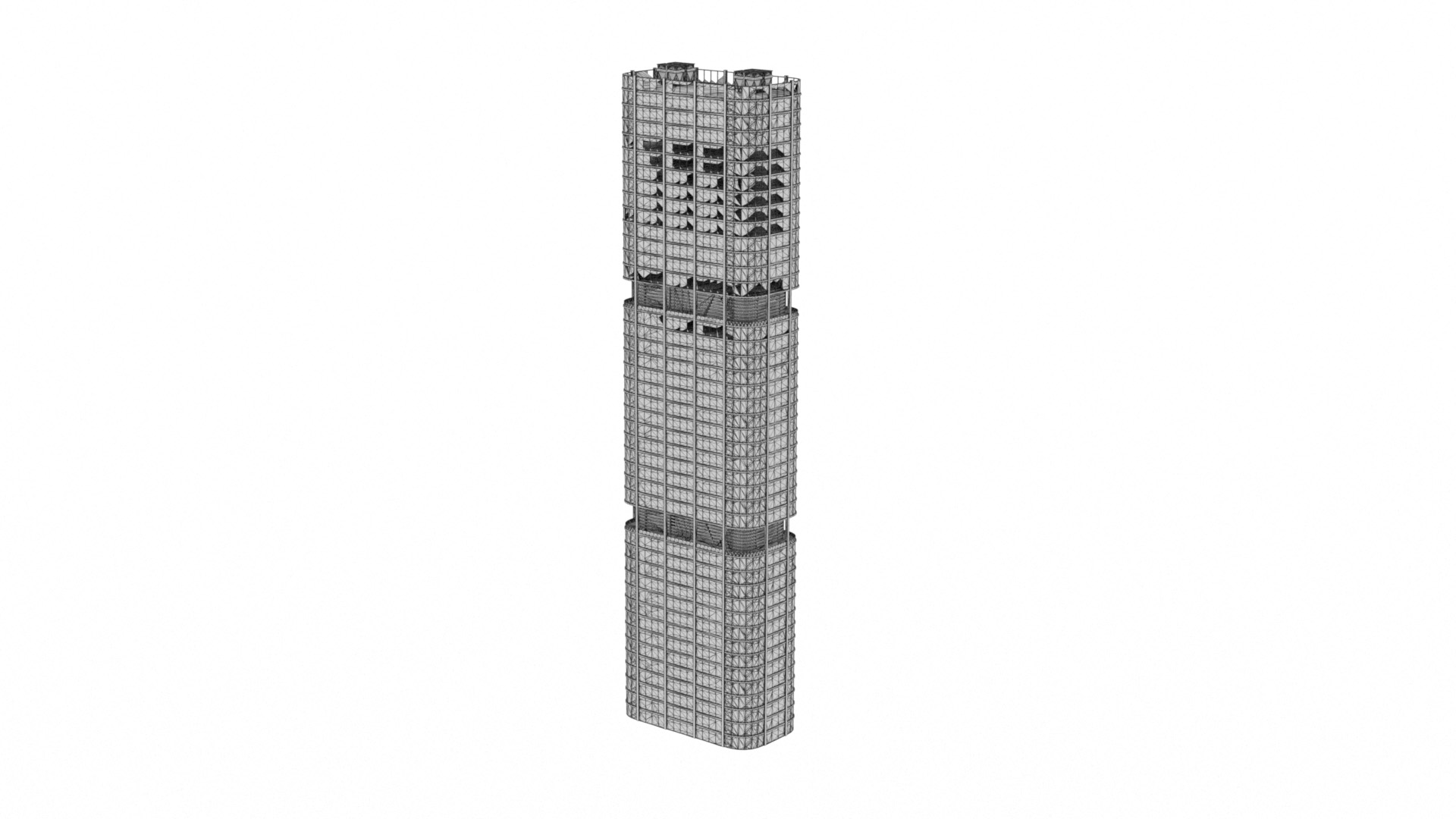}\hfill
\includegraphics[trim=800 80 800 0,clip,width=0.09\linewidth]{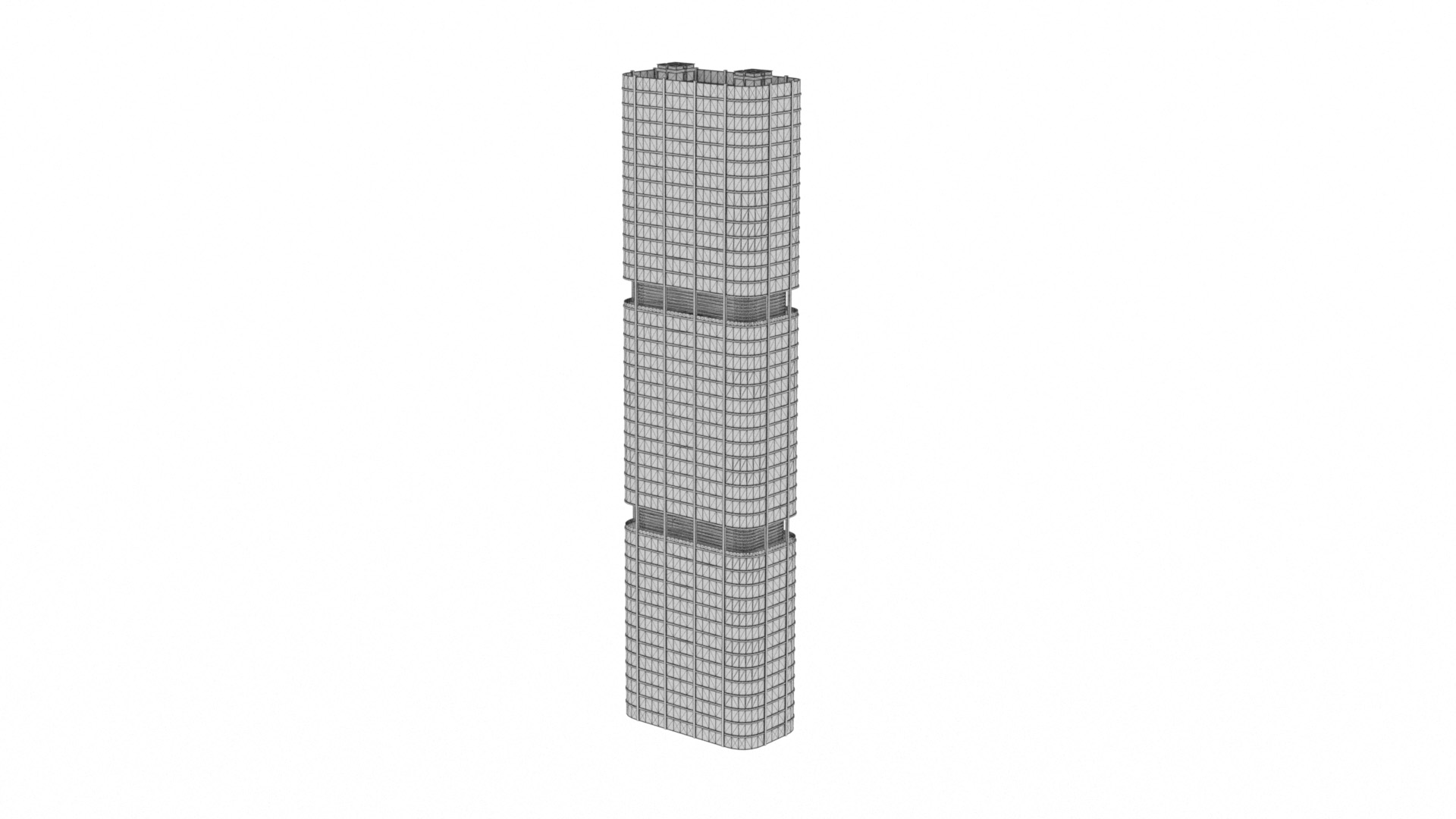}\\
\figcap{\tiny ~~} \hfill%
\figcap{\tiny T14+VM (1364K,0.02,374,28.8)} \hfill%
\figcap{\tiny VR (304K,0.02,0,30.5)} \hfill%
\figcap{\tiny VM (1734K,0.01,154,32.5)} \hfill%
\figcap{\tiny VR+VM (1936K,0.01,250,31.0)} \hfill%
\figcap{\tiny Ours (634K,0.02,0,36.1)} 
\caption{\textbf{Skyscraper:} We choose ``Skyscraper'' to demonstrate the issues of existing methods and our superiority over them. $(\bullet,\bullet,\bullet,\bullet)$ indicates face number, Hausdorff distance, LFD, and PSNR. }
\label{fig:building}
\end{figure*}
\begin{figure}[t!]
\centering
\rotatebox{90}{\small \hspace*{.1\linewidth}Ours \hspace*{.1\linewidth}Input }
\includegraphics[trim=0 0 0 0,clip,width=0.95\linewidth]{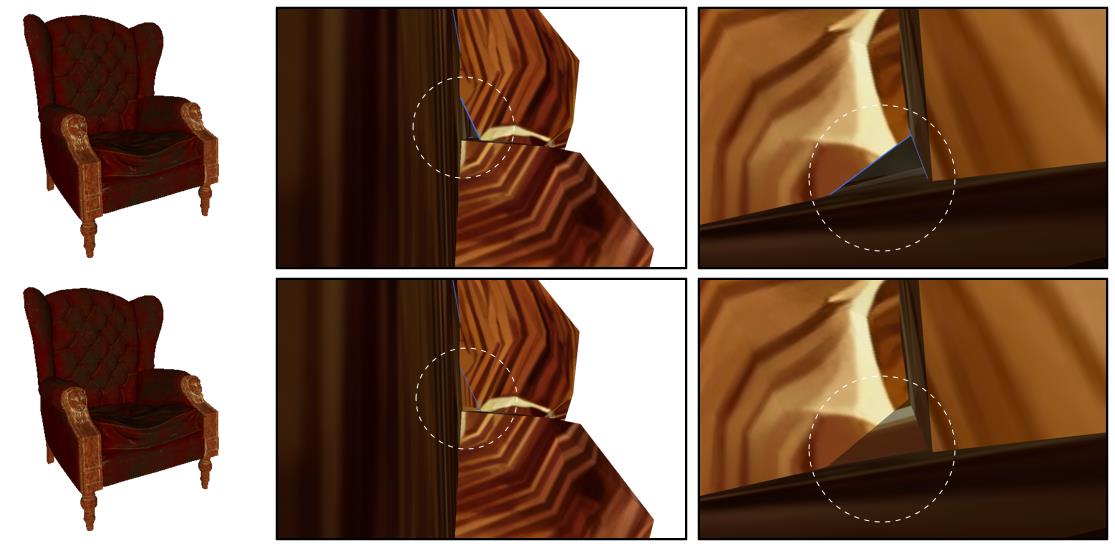}\\
\caption{\textbf{Example of UV recovering:} Top, Input textured model with holes; bottom, Our repaired model, with holes filled and UV, recovered.}
\label{fig:chair}
\end{figure}

\textbf{Compared to Existing Methods.} We compare our method with several state-of-the-art mesh repair methods, including PolyMender (PM)~\cite{Ju2004polymender}, TetWild (TW)~\cite{hu2018tetwild}, fTetWild (fTW)~\cite{hu2020ftetwild}, VisualRepair (VR)~\cite{Chu2019visualrepair}, VolumeMesher (VM)~\cite{Diazzi2021volumemesher}, combining of VisualRepair and VolumeMesher (VR+VM), combining of Takayama et al.~\cite{takayama2014simple} and VolumeMesher (T14+VM), and AlphaWrapping (AW)~\cite{Portaneri2022AlphaWrapping} (we use hyper-parameters $\alpha=D/100$, $\delta=D/3000$ for AW as suggested in their work) on randomly chosen 1000 models from ShapeNet~\cite{Chang2015shapenet} and 400 models from Thingi10K~\cite{Thingi10K}. All results can be found in \url{https://github.com/VisualGuidedMeshRepair/dataset}. The resulting statistics are collected in~\autoref{tab:statistic}, and we pick five examples (``Chair'' and ``Table'' are from ShapeNet) to demonstrate the issues of the existing methods in~\autoref{fig:results},~\autoref{fig:building}, and~\autoref{fig:chair}. Note that for the ``Chair'' model, TW's result loses a large portion of the shape due to the mis-oriented input faces.

\textbf{Watertightness, Manifoldness, and Number of Faces. }
Unlike VR~\cite{Chu2019visualrepair}, our proposed method guarantees a watertight and manifold output, which only subtracts the input mesh without filling gaps and holes. Other approaches, such as TW and VM, only ensure a combinatorial 3-manifold with boundary output that may contain non-manifold edges. Moreover, VM's output is unnecessarily complex due to BSP partitioning, resulting in tripling the number of faces in $M_\text{input}$. In contrast, our constrained simplification step significantly reduces the face count to approximately the same level as $M_\text{input}$.

\setlength{\columnsep}{5pt}
\begin{figure}[t!]
\scalebox{0.9}{
\includegraphics[trim=60 60 10 40,clip,width=\linewidth]{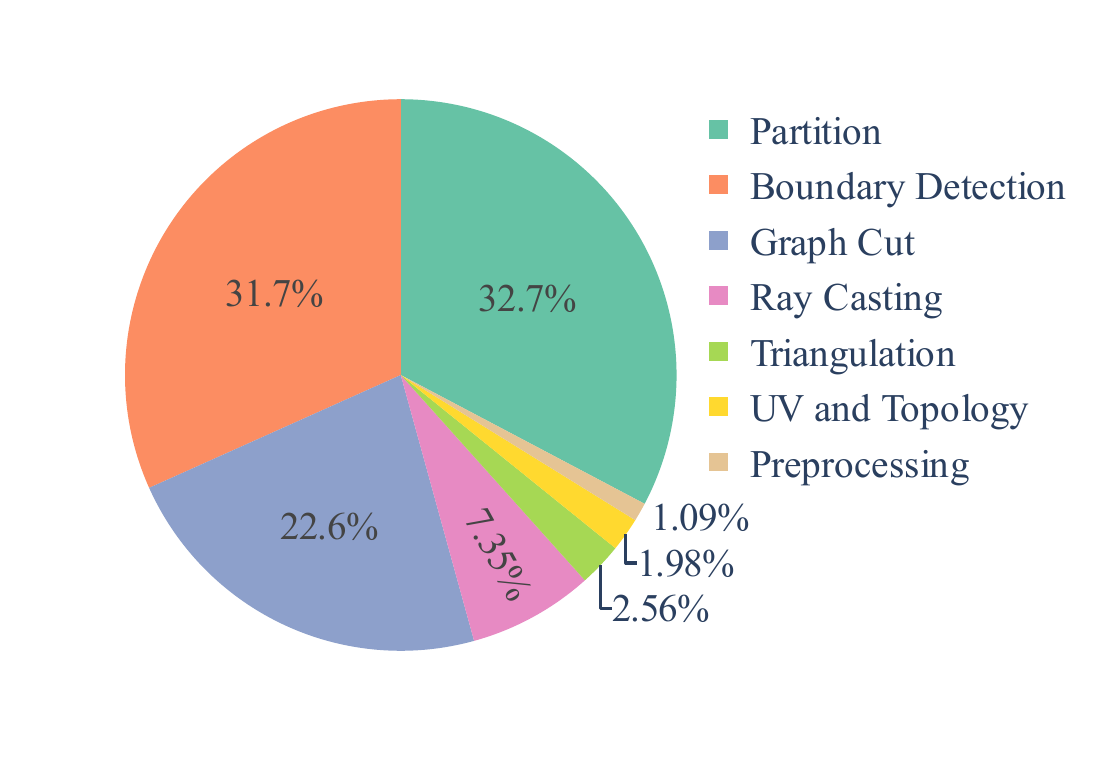}}
\captionsetup{justification=centering}
\caption{{Time breakdown}}
\label{fig:timepie}
\end{figure}  

\begin{figure*}[ht]
\newcommand{\figcap}[1]{\begin{minipage}{0.245\linewidth}\centering#1\end{minipage}}
\includegraphics[trim=350 200 350 200,clip,width=0.245\linewidth]{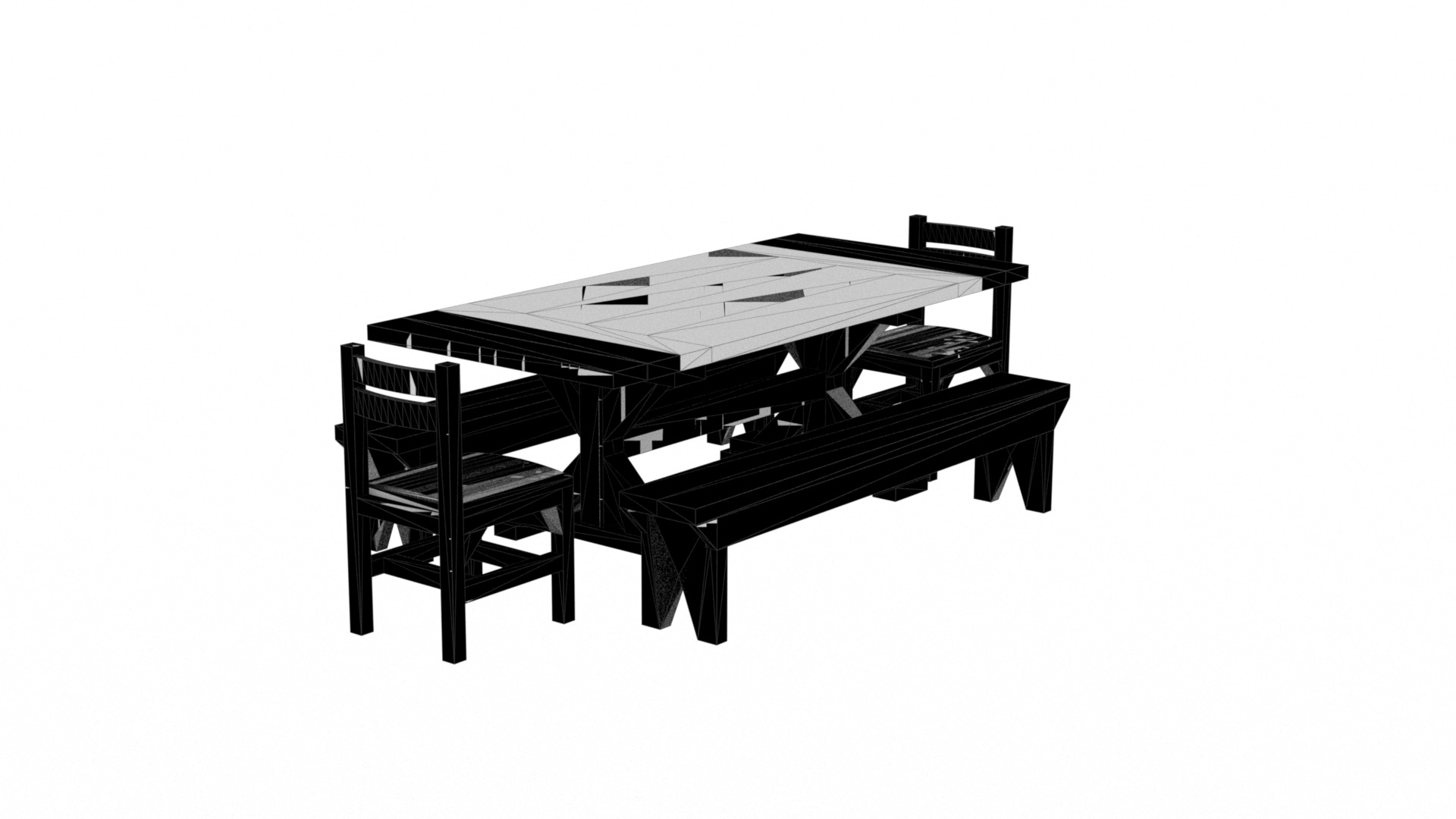}\hfill
\includegraphics[trim=350 200 350 200,clip,width=0.245\linewidth]{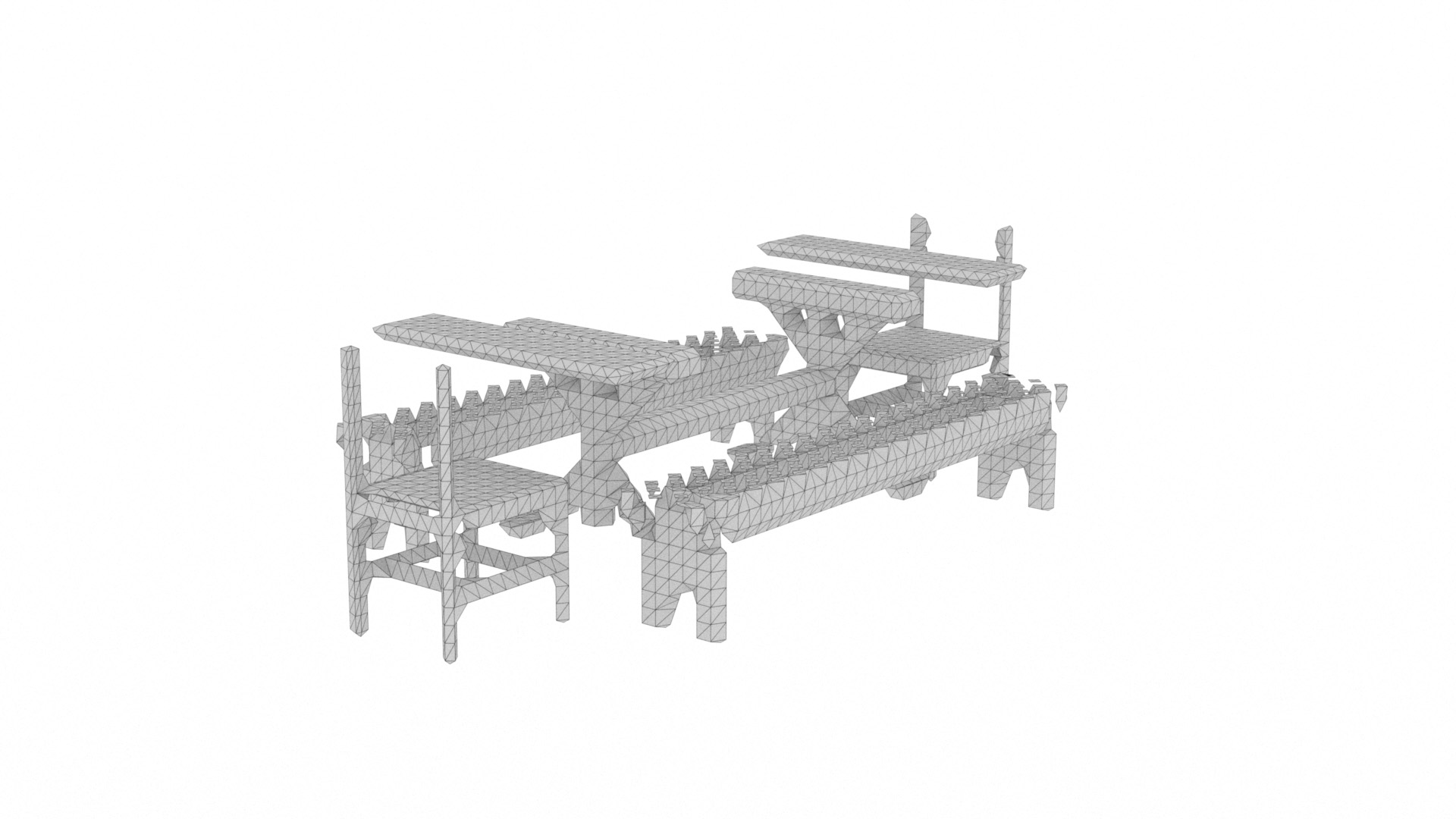}\hfill
\includegraphics[trim=350 200 350 200,clip,width=0.245\linewidth]{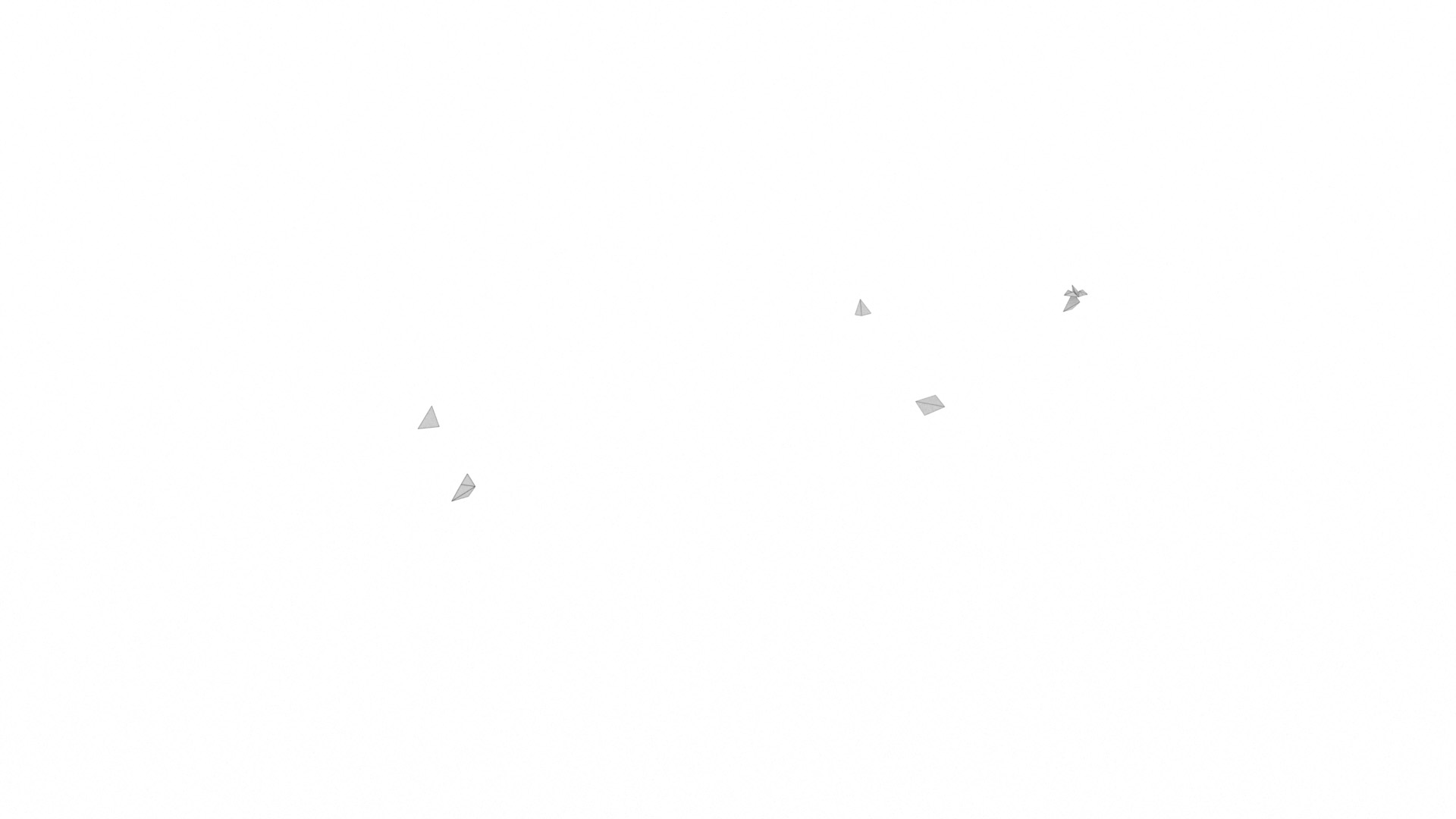}\hfill
\includegraphics[trim=350 200 350 200,clip,width=0.245\linewidth]{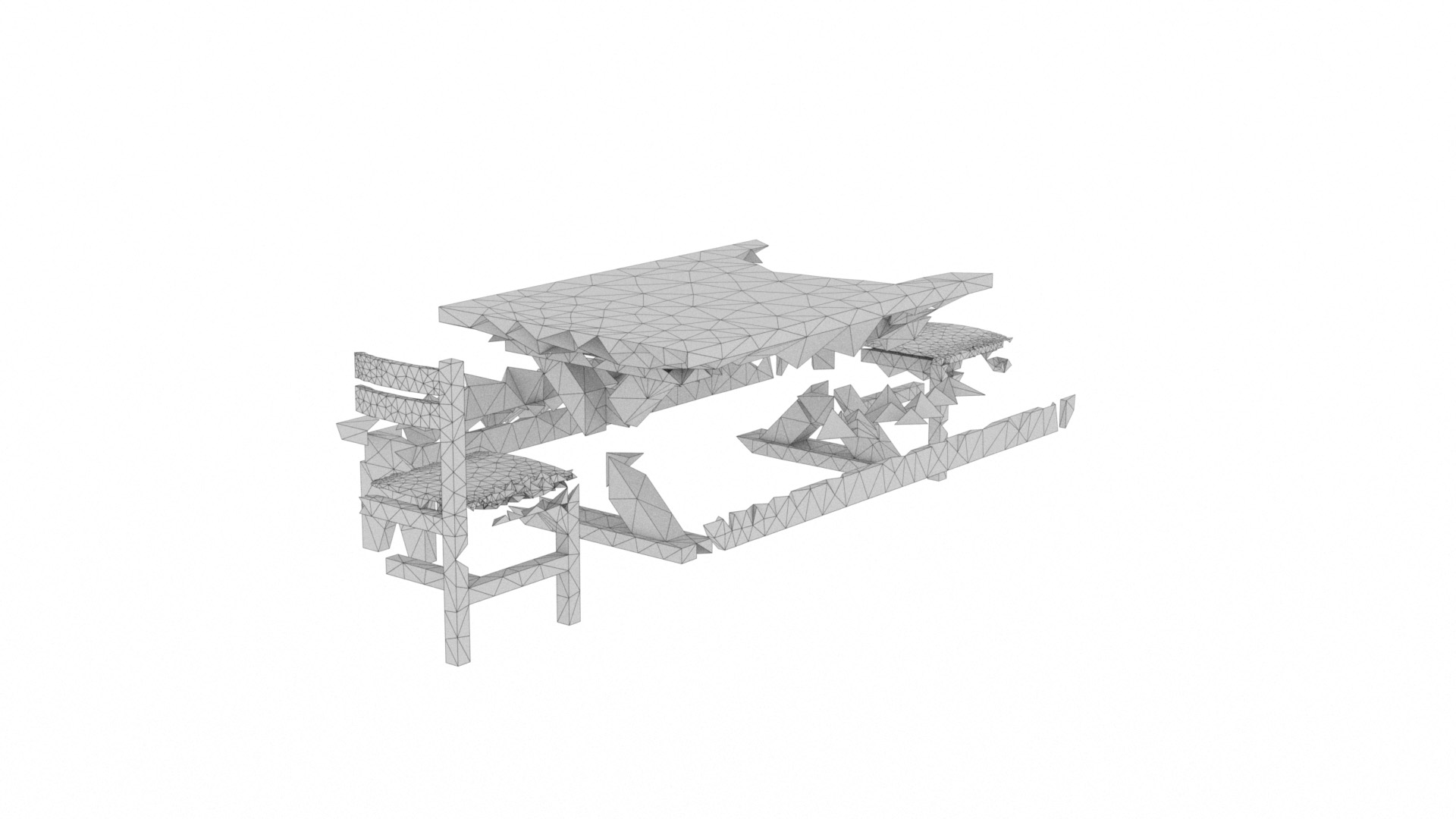}\\
\figcap{\footnotesize Input} \hfill%
\figcap{\footnotesize PolyMender  (9K,0.05,4158,22.3)} \hfill%
\figcap{\footnotesize TetWild (18,0.29,51862,19.9)} \hfill%
\figcap{\footnotesize fTetWild (7K,0.12,5074,22.3)} \\

\includegraphics[trim=350 200 350 200,clip,width=0.245\linewidth]{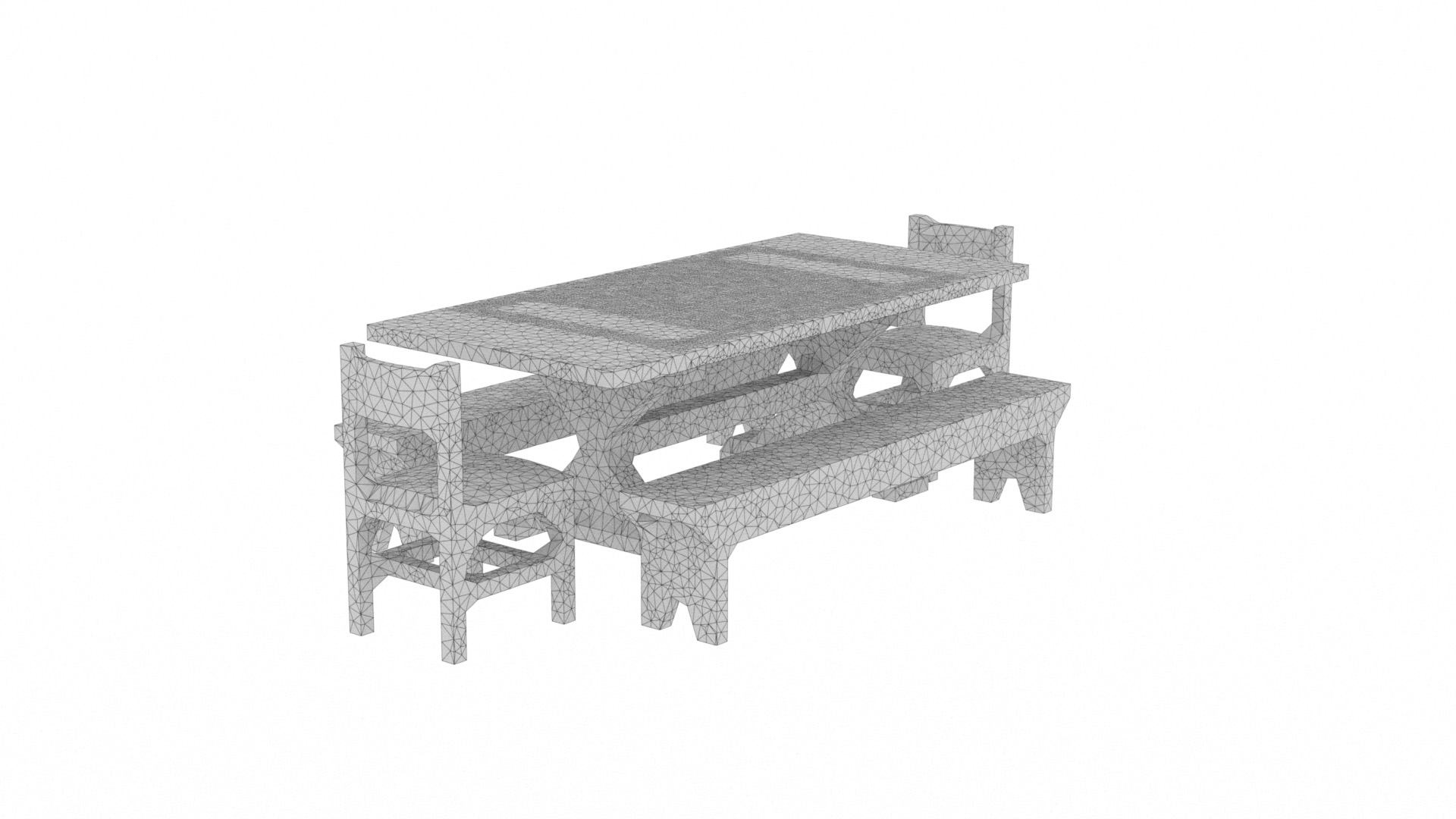}\hfill
\includegraphics[trim=350 200 350 200,clip,width=0.245\linewidth]{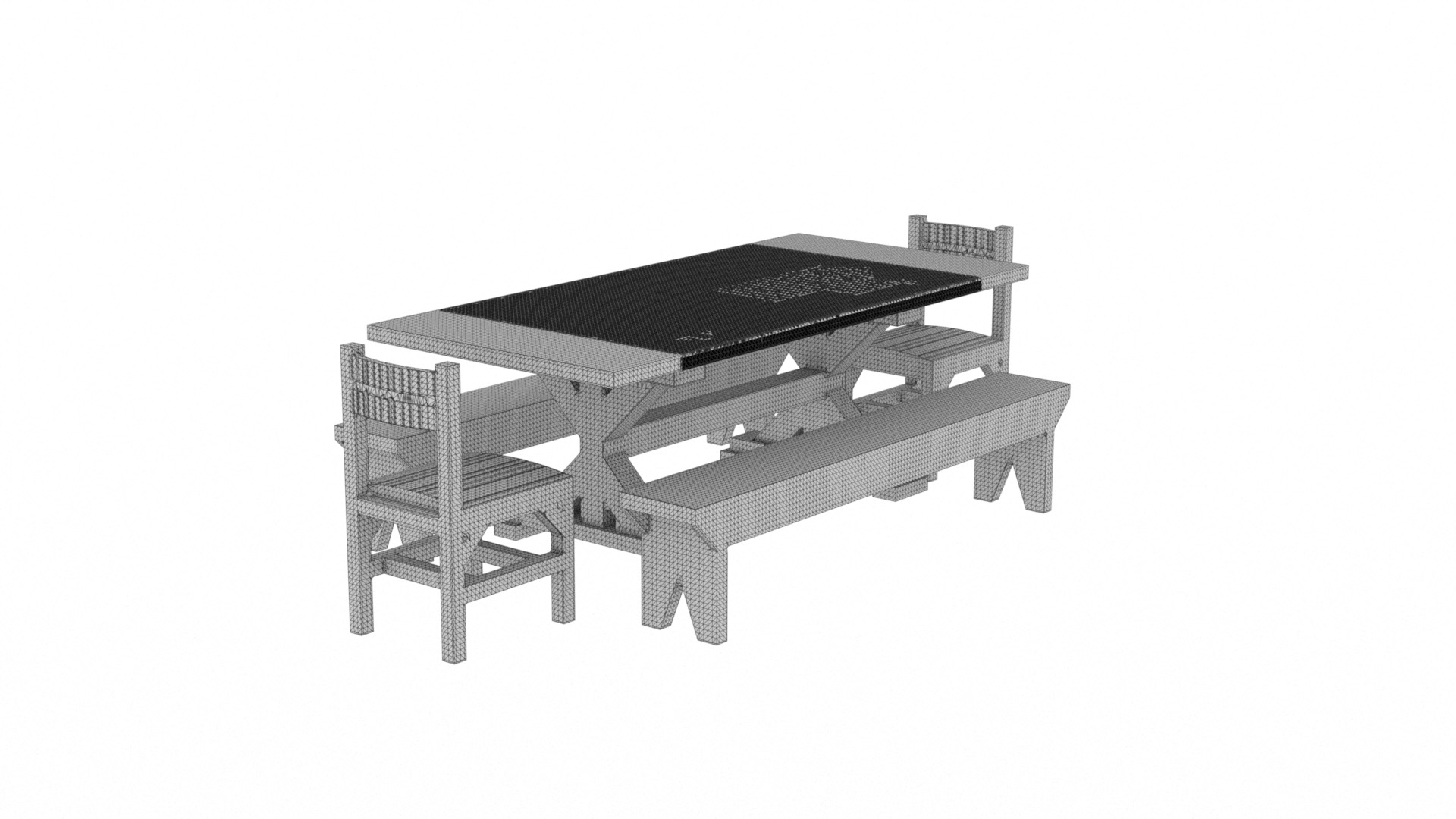}\hfill
\includegraphics[trim=350 200 350 200,clip,width=0.245\linewidth]{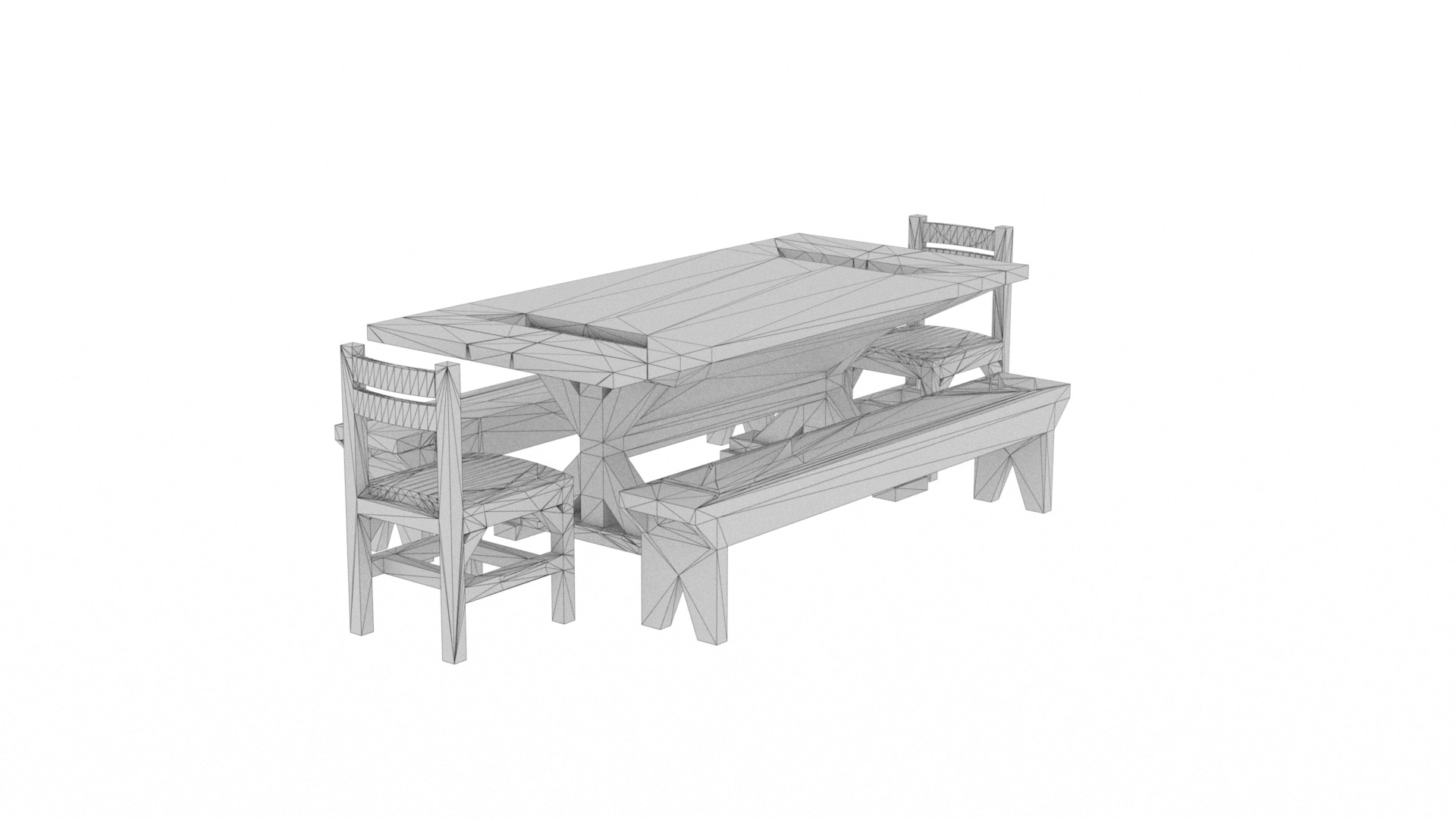}\hfill
\includegraphics[trim=350 200 350 250,clip,width=0.245\linewidth]{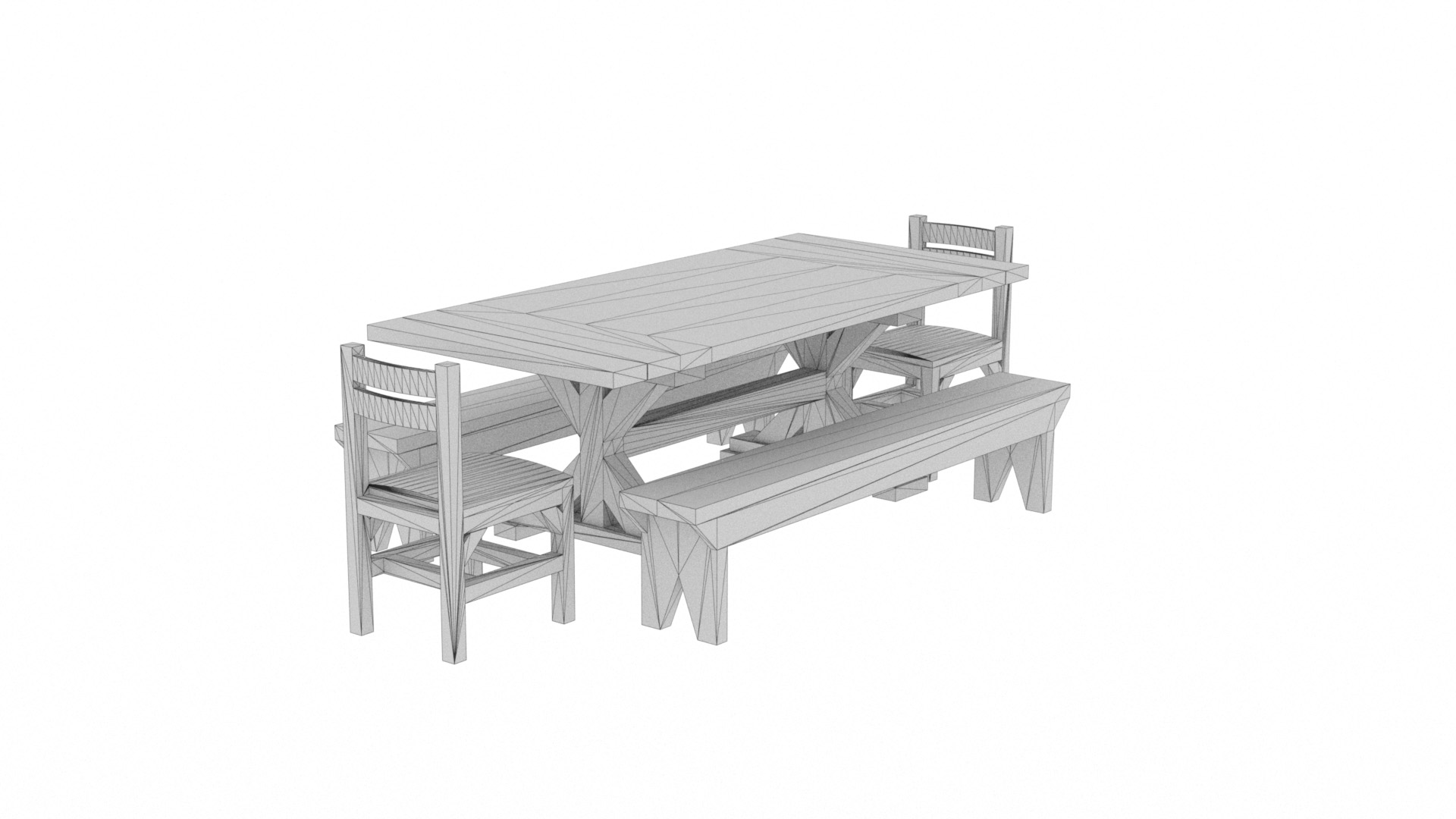}\\
\figcap{\footnotesize AlphaWrapping (40K,0.01,164,31.8)}\hfill
\figcap{\footnotesize ManifoldPlus (158K,0.003,160,22.9)}\hfill
\figcap{\footnotesize T14+VM (6K,0.03,394,31.6)}\hfill
\figcap{\footnotesize VisualRepair (2K,0.001,0,23.3)}\\

\includegraphics[trim=350 200 350 250,clip,width=0.245\linewidth]{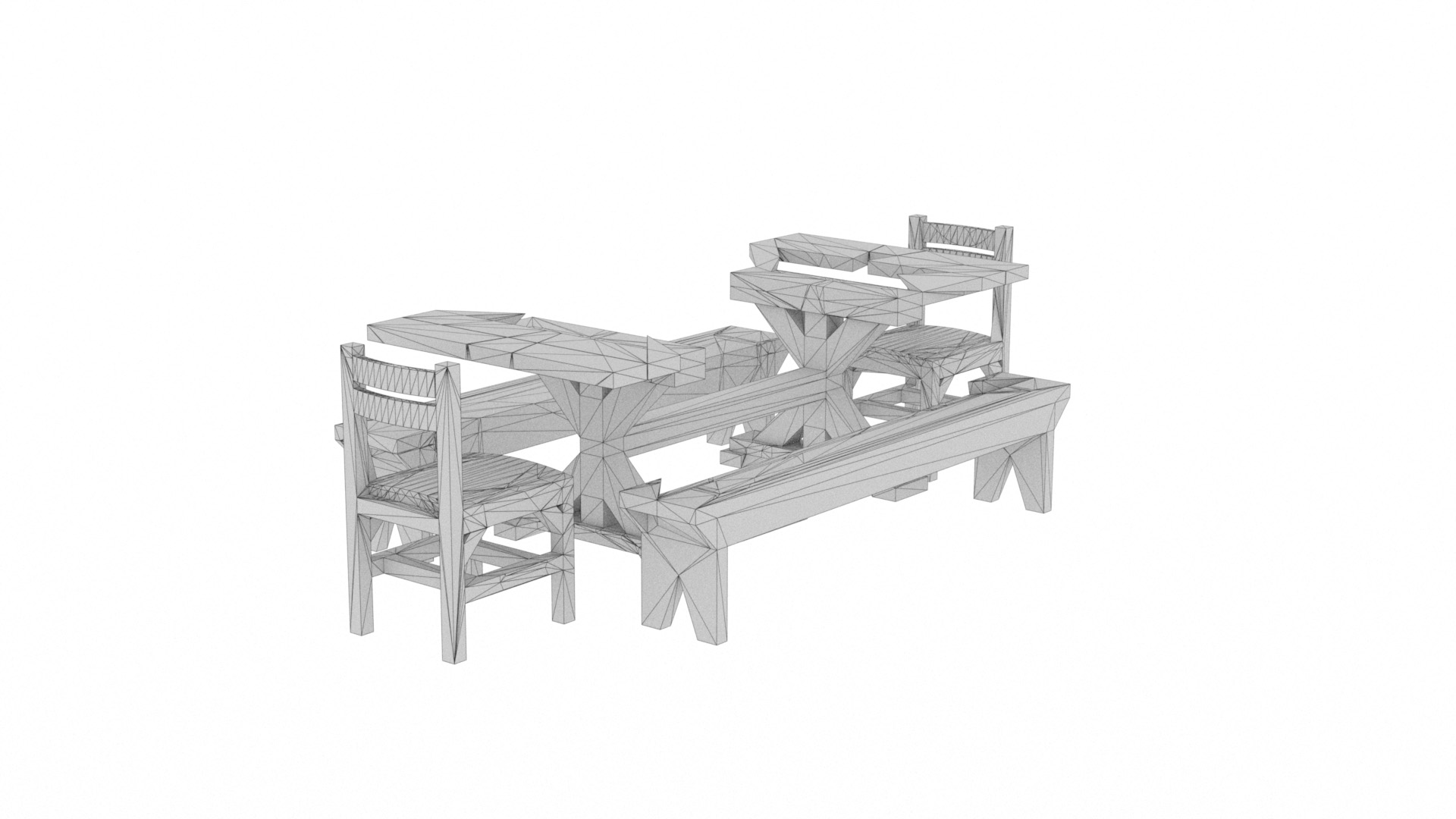}\hfill
\includegraphics[trim=350 200 350 250,clip,width=0.245\linewidth]{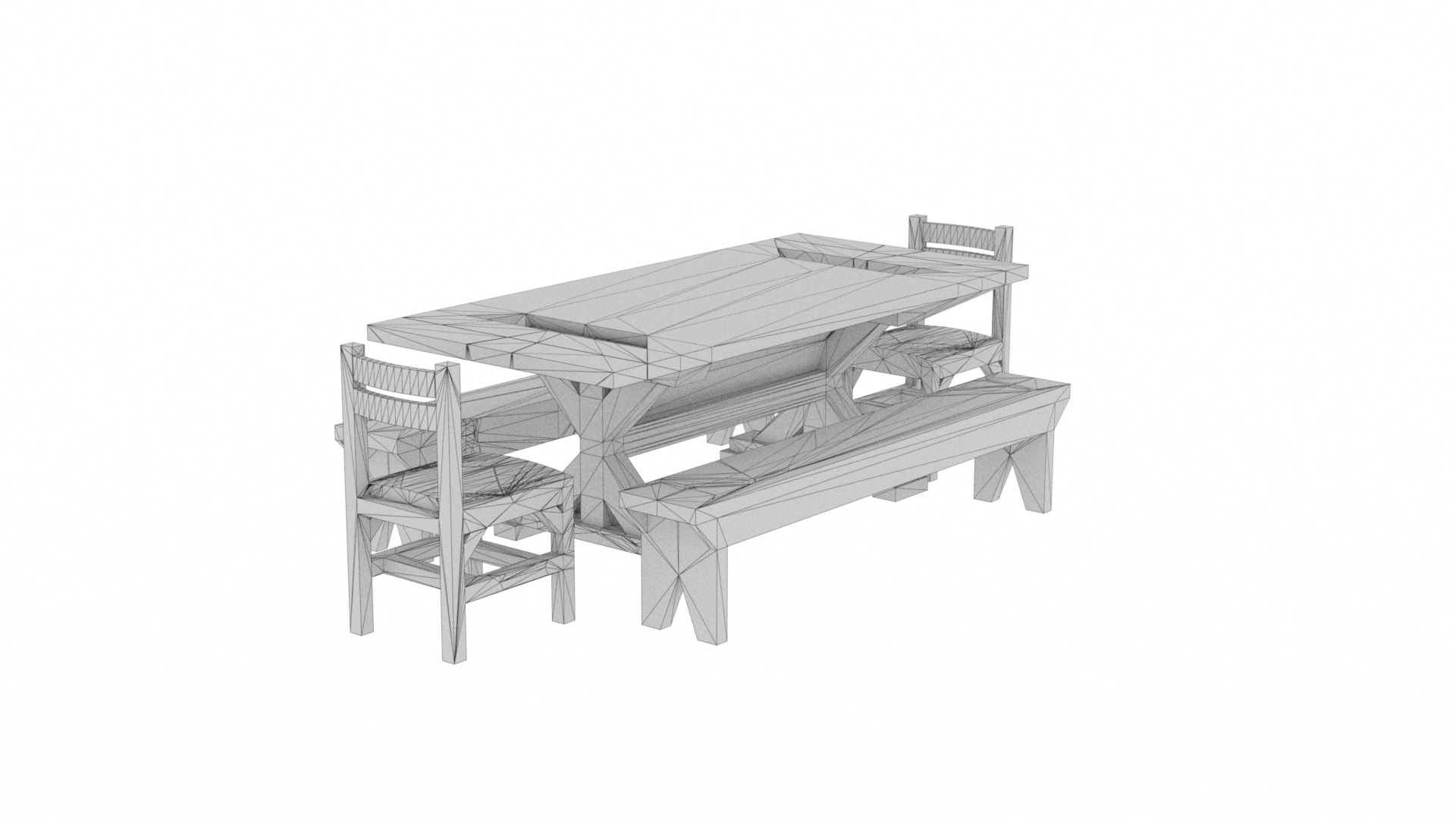}\hfill
\includegraphics[trim=350 200 350 250,clip,width=0.245\linewidth]{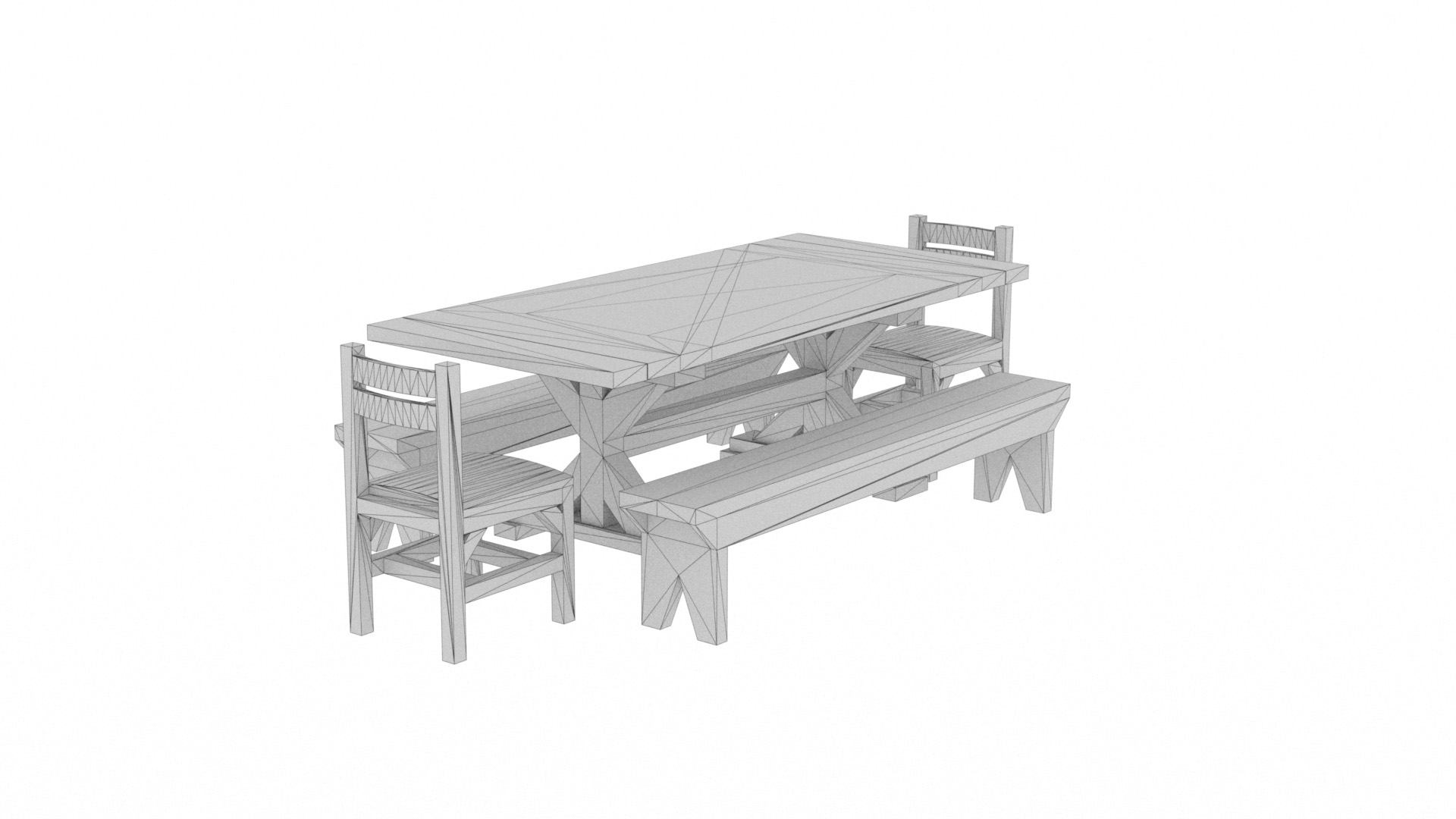}\hfill
\includegraphics[trim=500 200 700 450,clip,width=0.245\linewidth]{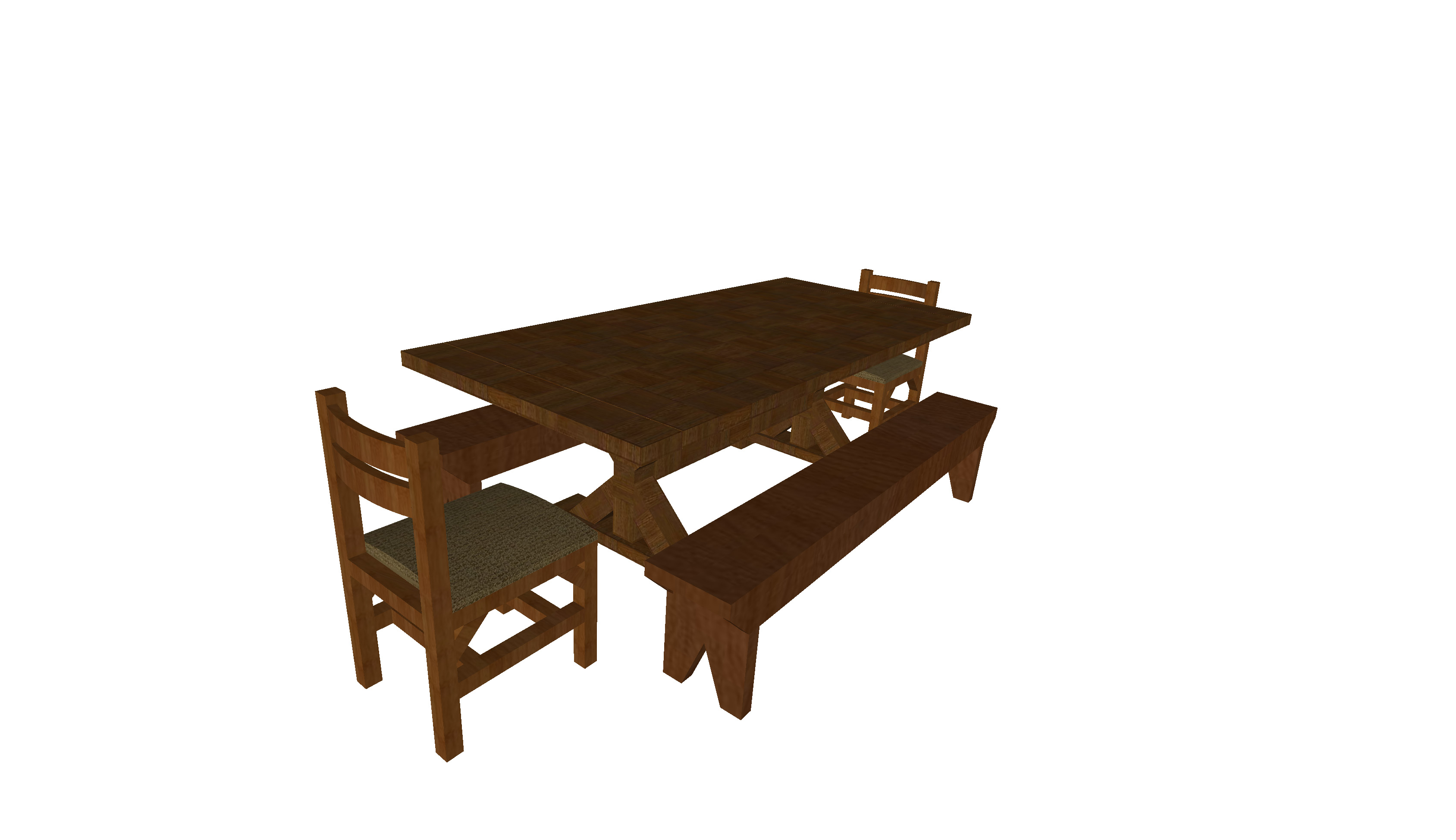}\\
\figcap{\footnotesize VolumeMesher (6K,0.04,2484,26.4)} \hfill%
\figcap{\footnotesize VR+VM (8K,0.03,400,32.6)} \hfill%
\figcap{\footnotesize Ours (2K,0.001,0,56.5) } \hfill%
\figcap{\footnotesize Ours with texture (2K,0.001,0,51.5) } \vspace{-0.in}
\caption{\textbf{Table:} We choose ``Table'' to demonstrate the issues of existing methods and our superiority over them. $(\bullet,\bullet,\bullet,\bullet)$ indicates face number, Hausdorff distance, LFD, and PSNR. }
\label{fig:table}
\end{figure*}

\begin{figure*}[ht]
\centering
\newcommand{\figcap}[1]{\begin{minipage}{0.105\linewidth}\centering#1\end{minipage}}
\includegraphics[trim=450 150 650 200,clip,width=0.105\linewidth]{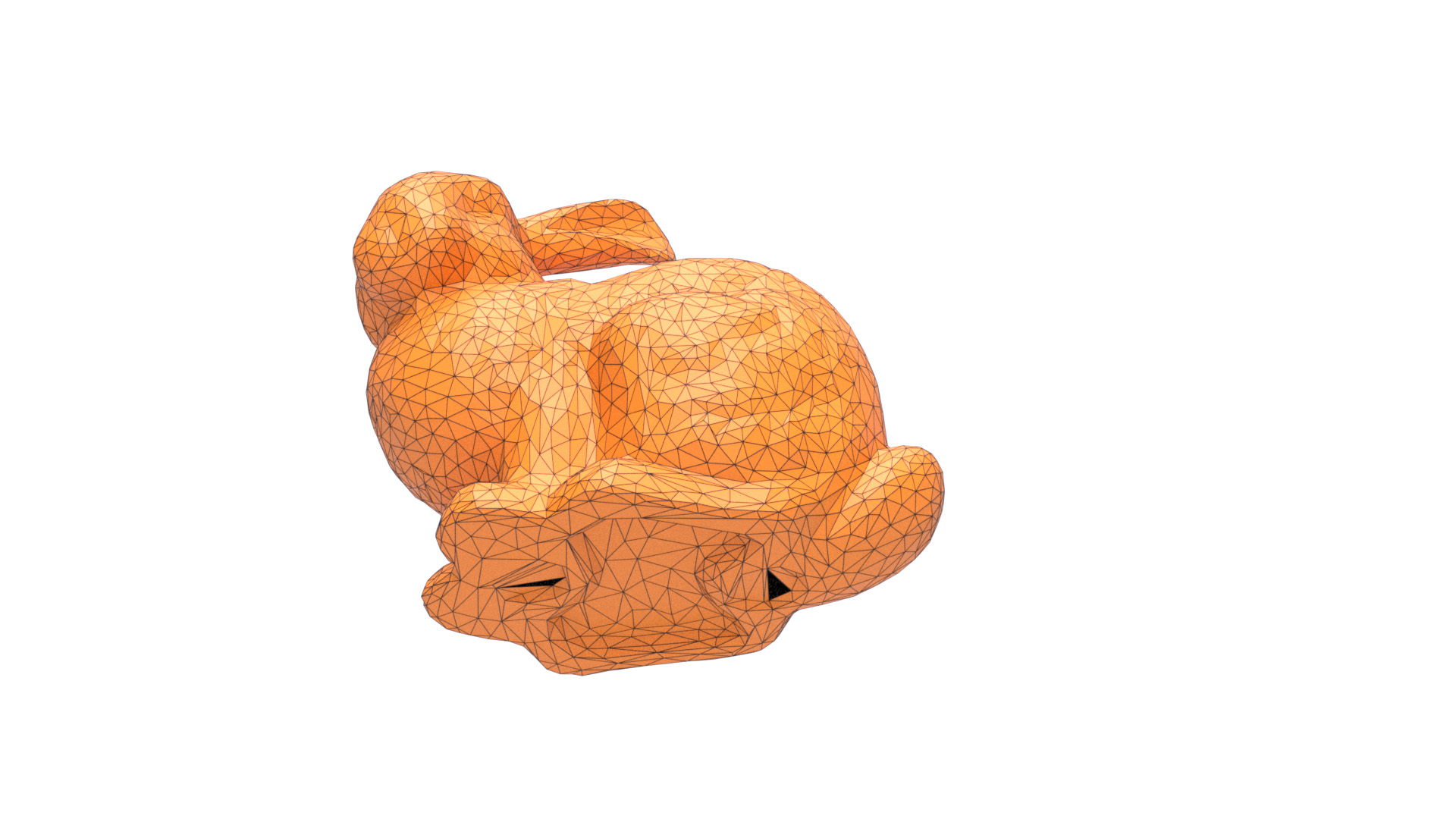}\hfill
\includegraphics[trim=450 150 650 200,clip,width=0.105\linewidth]{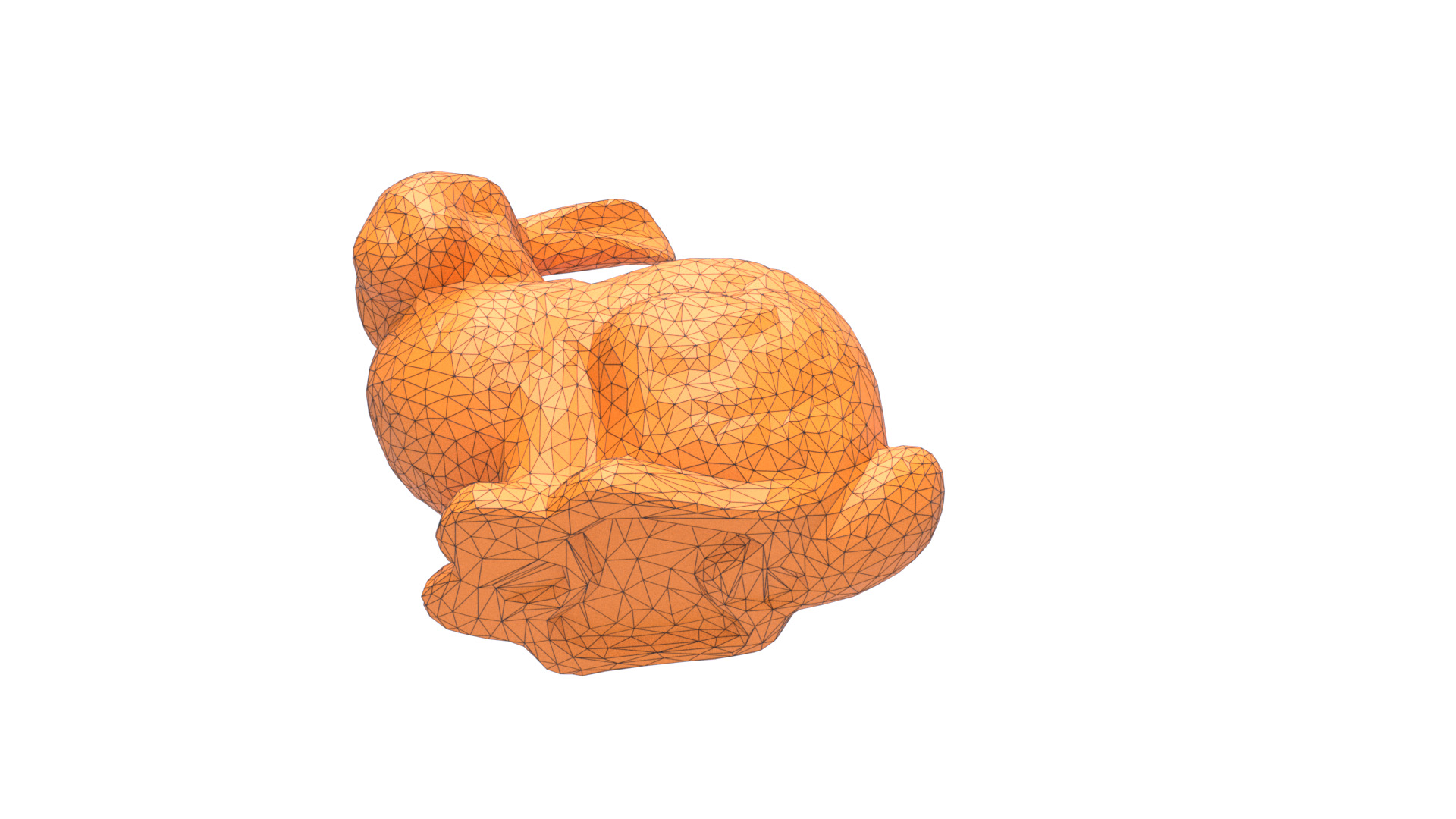}\hfill
\includegraphics[trim=450 150 650 200,clip,width=0.105\linewidth]{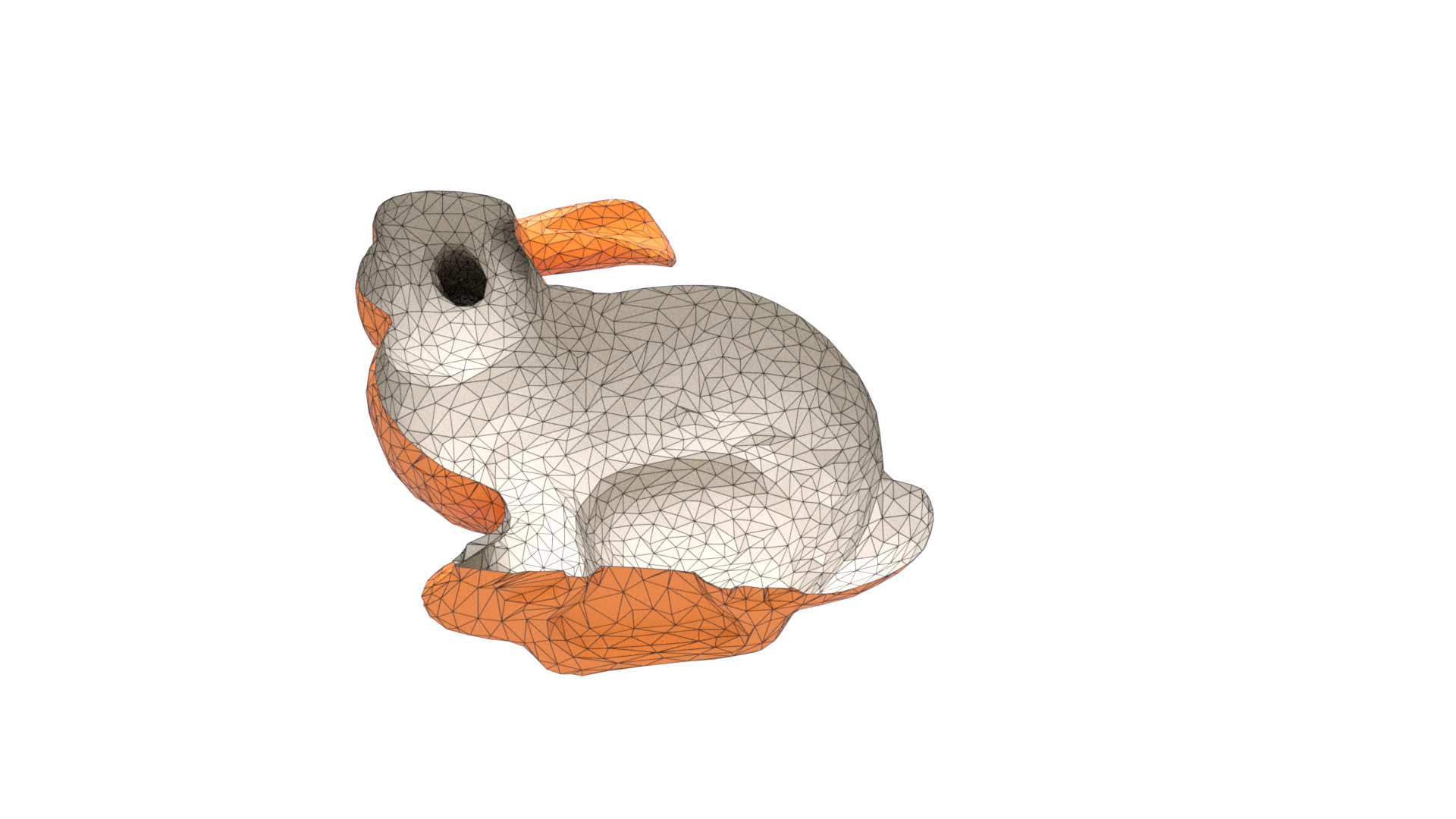} \rulesep \hfill
\includegraphics[trim=450 150 650 200,clip,width=0.105\linewidth]{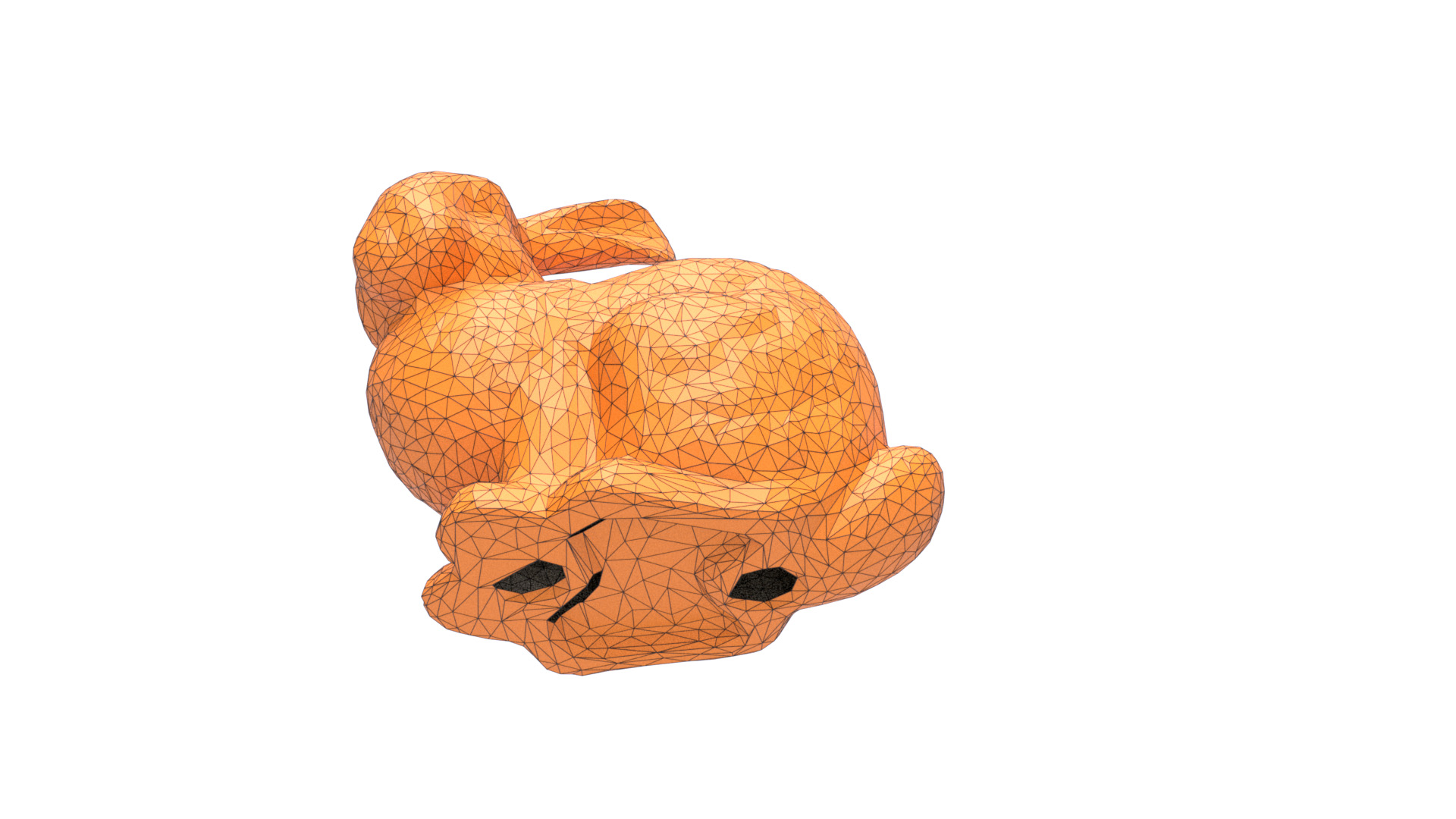}\hfill
\includegraphics[trim=450 150 650 200,clip,width=0.105\linewidth]{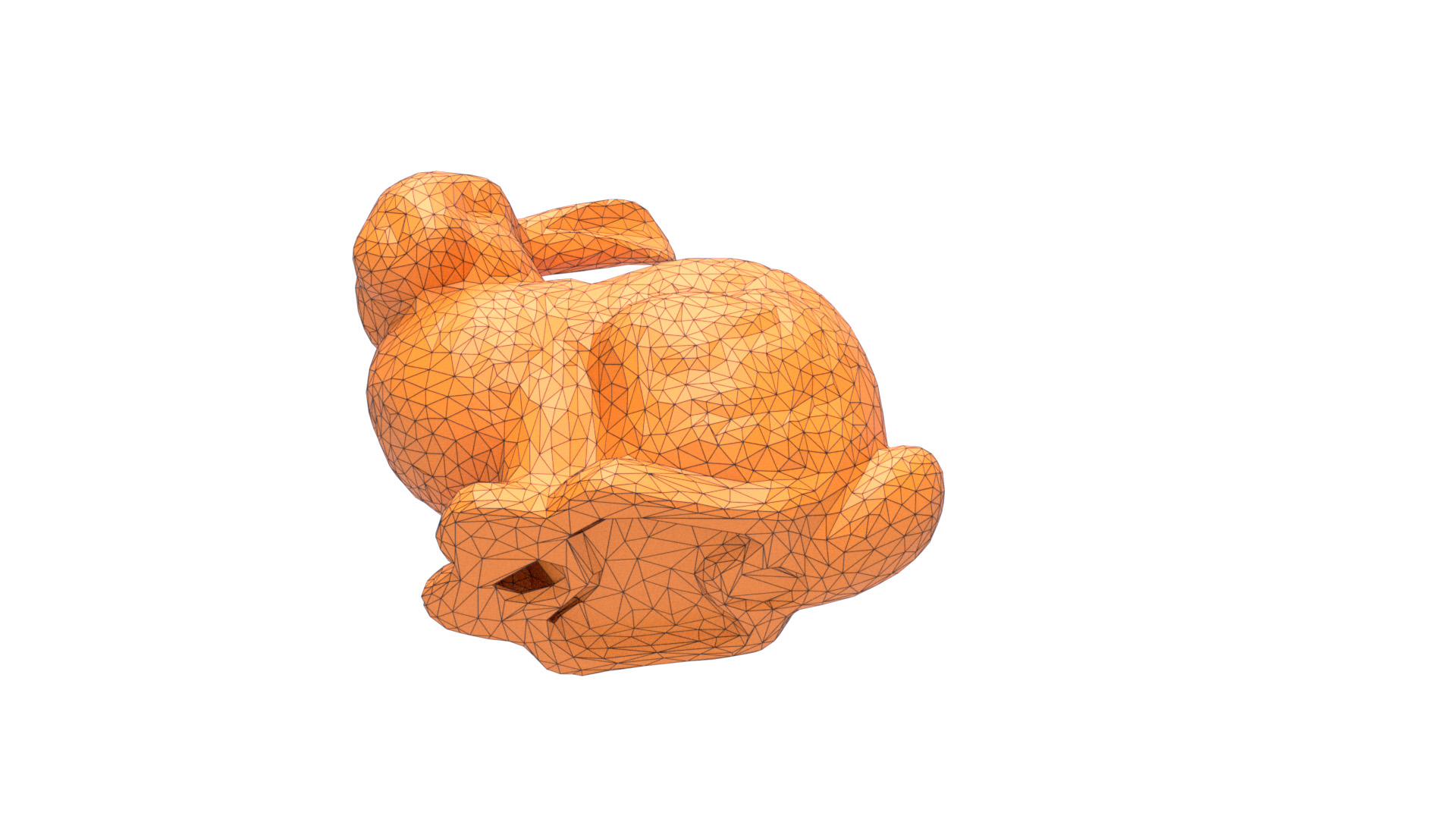}\hfill
\includegraphics[trim=450 150 650 200,clip,width=0.105\linewidth]{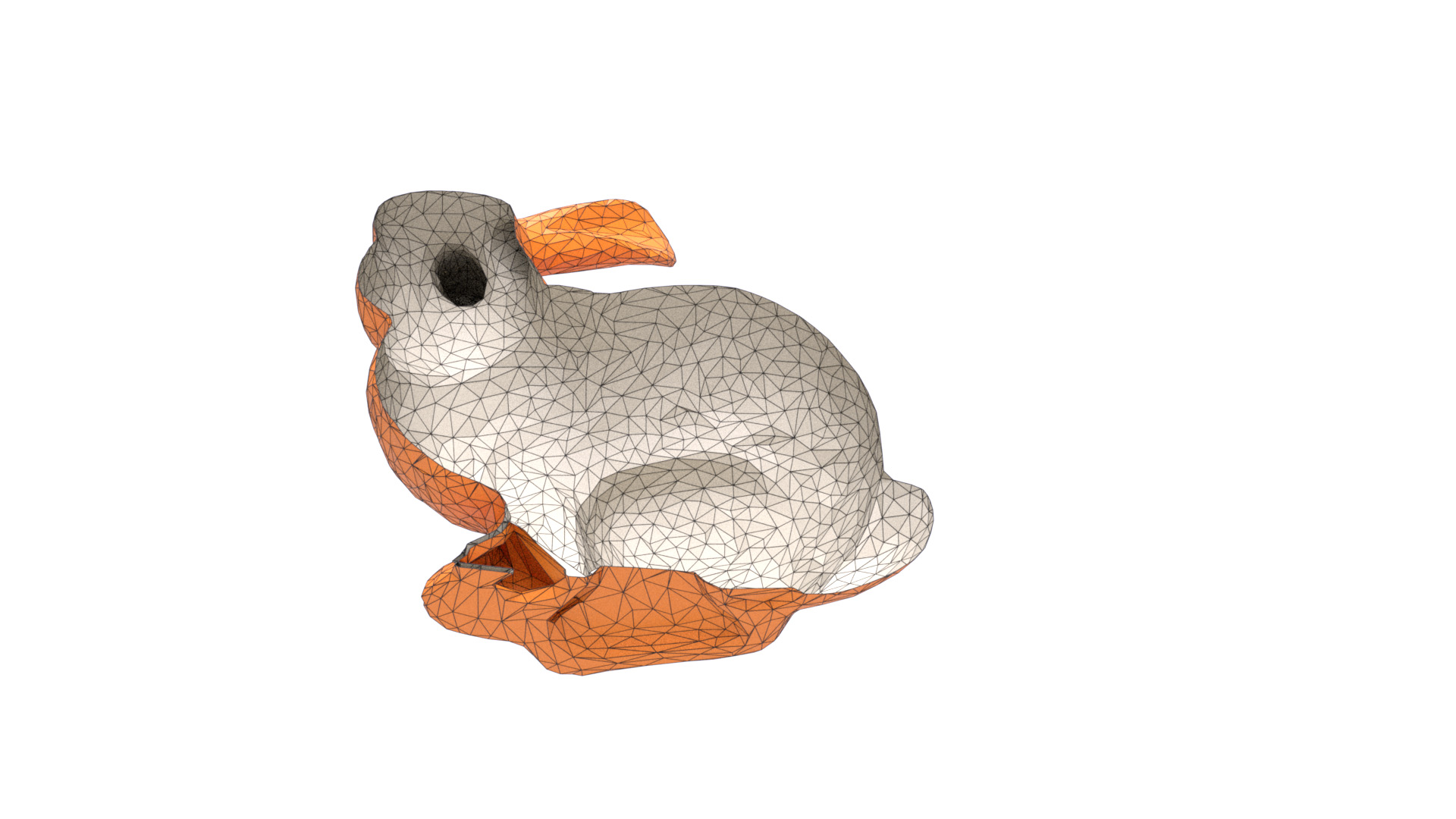} \rulesep \hfill
\includegraphics[trim=450 150 650 200,clip,width=0.105\linewidth]{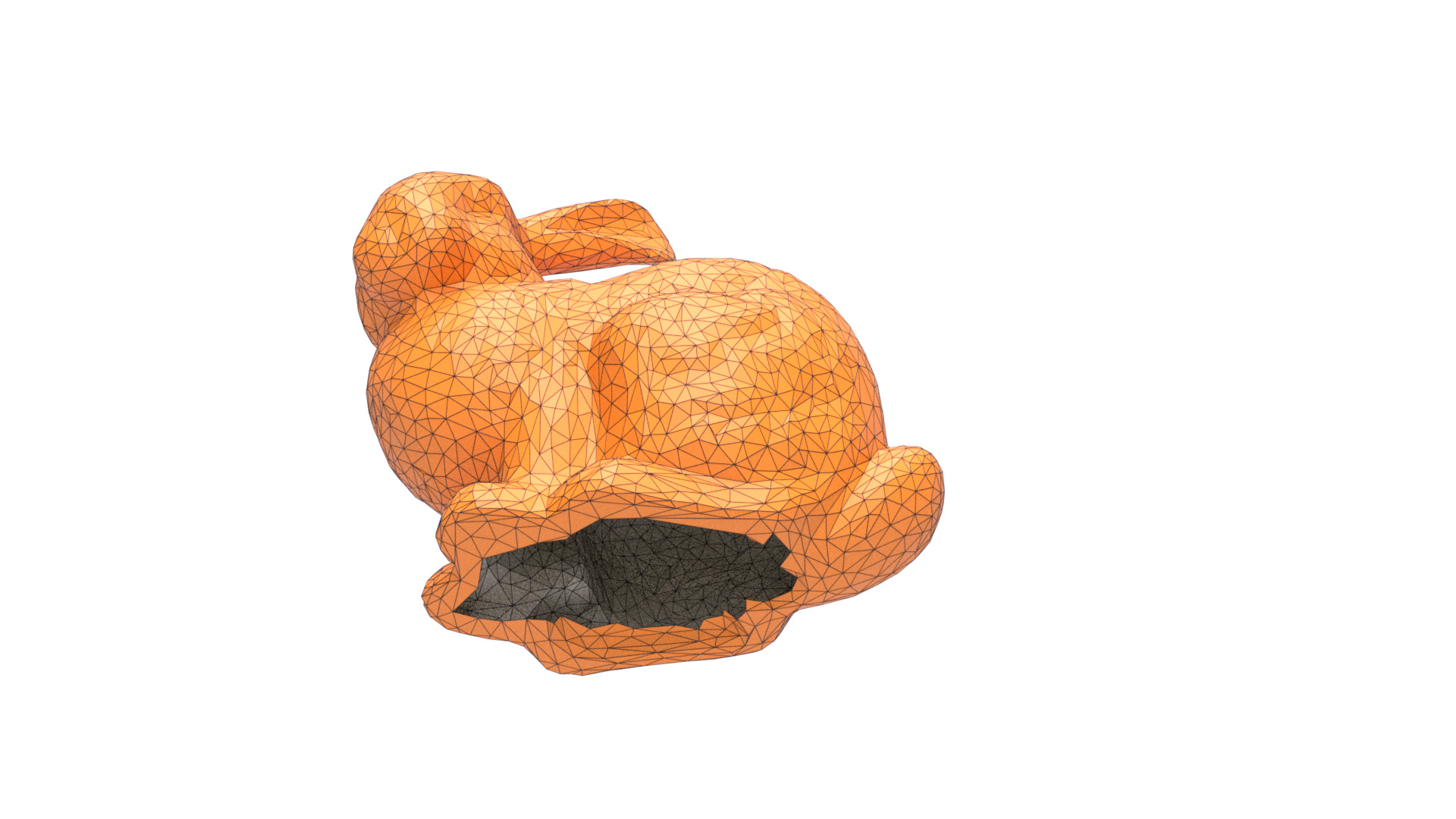}\hfill
\includegraphics[trim=450 150 650 200,clip,width=0.105\linewidth]{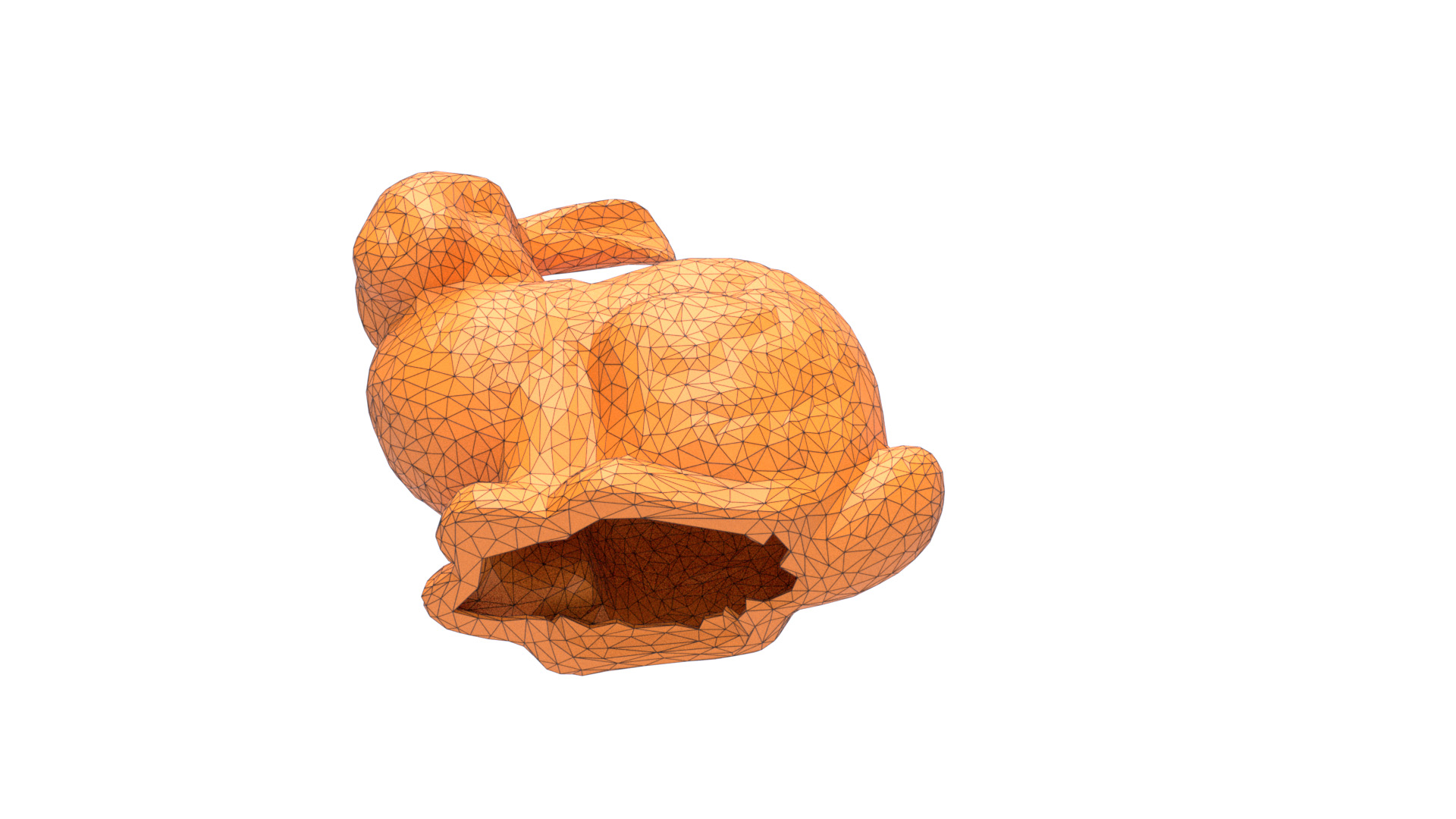}\hfill
\includegraphics[trim=450 150 650 200,clip,width=0.105\linewidth]{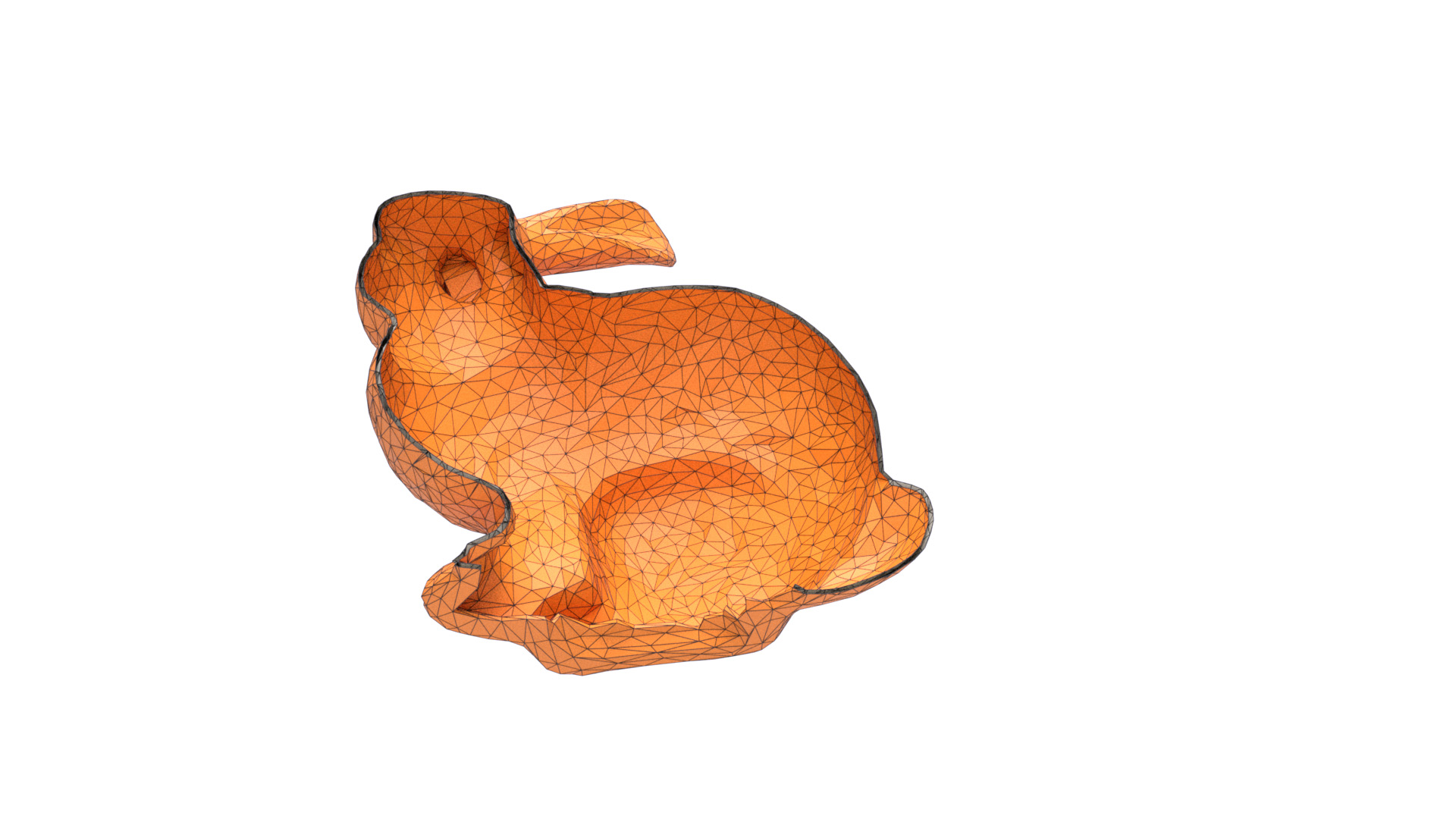}\\\vspace{-0.2em}
\figcap{\small Input }\hfill%
\figcap{\small Ours } \hfill%
\figcap{\small Cut view } \hfill
\figcap{\small Input }\hfill%
\figcap{\small Ours } \hfill%
\figcap{\small Cut view } \hfill
\figcap{\small Input }\hfill%
\figcap{\small Ours } \hfill%
\figcap{\small Cut view }\\\vspace{-0.2em}
\caption{\textbf{Study on hole size:} From left to right, as the hole size in the "Bunny" model increases from small to large, more faces are progressively identified as open, leading to their filling by our method, ultimately forming a thin shell. The front side is orange, while the back side is white.}
\label{fig:hole}
\end{figure*}

\textbf{HD, LFD, and PSNR. }
We utilize three qualitative metrics, Hausdorff distance (HD), light-field distance (LFD), and peak signal-to-noise ratio (PSNR), to evaluate the output quality in comparison to the input. Notably, we render the input mesh with double face rendering as the reference and render others' and our results with single face colored in black. Due to misorientation of the input from ShapeNet, methods that rely on input orientations to determine the interior and exterior, such as PM, TW, fTW, and VM, cannot produce a result even close to the input. Additionally, ``Flowers'' in \autoref{fig:results} demonstrate that VM might close the surface with lots of unnecessary volumes for the open faces. VM also fails to close the court due to the open surface under the eave in ``Roman'', and VM always solves for the smallest surface area, which might close the concave structure. ``Skyscraper'' shows VM, TW, and fTW have difficulty handling the model with inner structures. On the other hand, AlphaWrapping, which does not require consistent orientation, can suffer from the blurring of input geometric details and sharp features if the surface is offset by a large distance (see ``Skyscraper''). PM fails to preserve sharp features due to limited octree resolution (see ``Chair''). For ShapeNet, although VR scores well in HD and LFD, its optimization cannot ensure correct patch orientation across the entire mesh, resulting in a low PSNR score, for which one reason is that visual measures in PM are based on rasterization, which may not capture small or occluded faces accurately, such as the frames on the chair (\autoref{fig:results}) and engine on the plane (\autoref{fig:teaser}). 

For the models in Thingi10K, which does not have complex interior structure, only VR can get better HD/LFD/PSNR than ours; there are 40 out of 400 models that VR cannot produce the results within 1 hour due to the high input face number. 
For a more fair comparison, we manually tune the parameters of fTW, PM, and AW to match the same face number with ours for the ``Flower'' example. Unfortunately, fTW is out of 64G memory. More importantly, fTW cannot handle open faces or nested structures as it uses the winding number to determine the interior/exterior. Given the same face number, HD/LFD/PSNR of PM (0.025/1492/20.1) and AW (0.01/200/25.8) are still worse than ours (0.01/12/45.3). We further tune AW and manage to produce a similar HD and LFD as ours. However, even with 10x more faces, AW (0.01/56/30.4) is still unable to achieve a similar quality.
In summary, our method outperforms all listed state-of-the-art techniques. In addition, unlike other methods, our method can close the hole and propagate UV for the newly added faces from neighboring faces, as shown in~\autoref{fig:chair}.

\textbf{Memory and Time Usage. } Due to offset faces and extra cuts for boundary edges, ours need more memory and computational time than VM. Fortunately, our visual evaluation step based on ray tracing is more efficient than VR, in which reading back results from the rasterization pipeline causes the CPU/GPU to get stuck and easily takes hours for large models. We also plot the time breakdown of our pipeline in~\autoref{fig:timepie}, where our ray-tracing step only takes $7\%$ of the computation time. Boundary detection occupies almost one-third of the computation time due to using rational numbers. Our processing time does not solely depend on mesh size. The complexity of the input, the number of intersection faces, and the size of the output all have impacts.

\textbf{Compared to VR+VM and T14+VM.} As previously stated, VM is a volumetric approach that can generate a mesh without holes, yet it relies on correct input orientation. In contrast, VR and T14 can adjust orientation but is unable to fill gaps. Therefore, we use VR and T14 to orientate the faces and use VM to mend the topology, denoted as VR+VM and T14+VM. As anticipated, these approaches perform better in PSNR than any other existing technique that guarantees watertightness. Nevertheless, the resulting mesh contains numerous faces due to the additional face division caused by VR. Furthermore, our method achieves around 10 PSNR advantages over the VR+VM and T14+VM approaches, for which one reason is that our method incorporates visual guidance throughout the entire repair process and optimization procedure. On the other hand, VR+VM and T14+VM only apply the visual metric to the surface optimization step, i.e., the graph cut stage, thus has not fully exploited the critical visual cues. For instance, in ``Chair'' example, since the rasterization cannot capture tiny faces, VR and T14 cannot correct all face orientations, which leads to incorrect output from VM. The same issue can be observed in \autoref{fig:teaser} as well. For ``Table'' in ~\autoref{fig:results}, VM, VR+VM, and T14+VM all discard part of the table because the graph cut in VM solves for the minimal surface area, which is not always the case.

\textbf{Filling Holes.} Due to the stochastic nature of ray-tracing, our method can be sensitive to the sample density in visual measure computation. Such sensitivity is most obvious when a surface is open by a small hole, so we have performed a study on the hole size. In~\autoref{fig:hole}, we observe that a small hole size leads to the closure of the hole, as there are insufficient rays shot from the interior of the bunny to identify any face as an open surface. As the hole size increases, more faces are progressively identified as open, leading to their filling by our method, ultimately forming a thin shell. 

\begin{figure}[t!]
\newcommand{\figcap}[1]{\begin{minipage}{0.49\linewidth}\centering#1\end{minipage}}
\hfill
\includegraphics[trim=450 10 320 100,clip,width=0.49\linewidth]{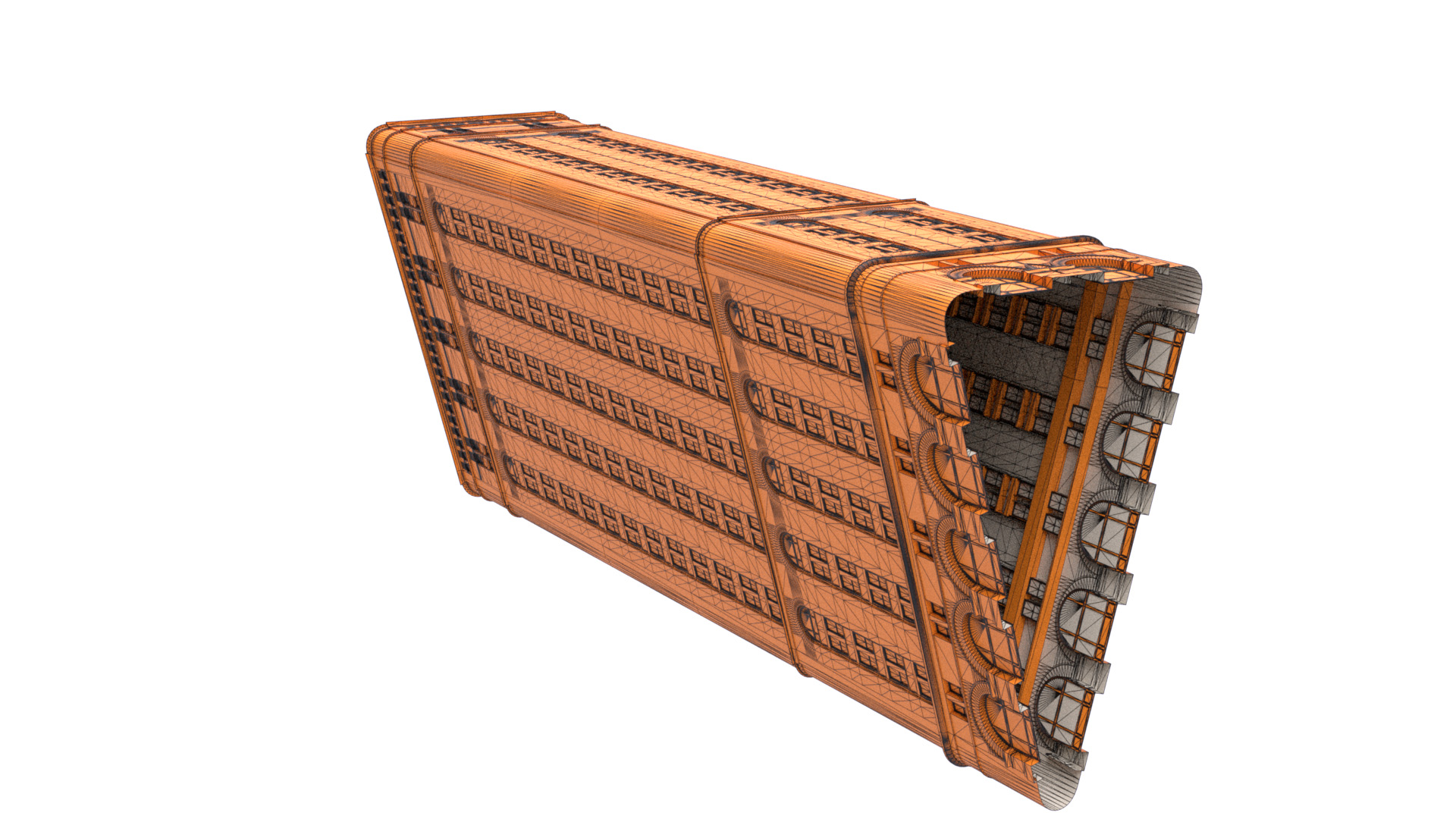}\hfill
\includegraphics[trim=450 10 320 100,clip,width=0.49\linewidth]{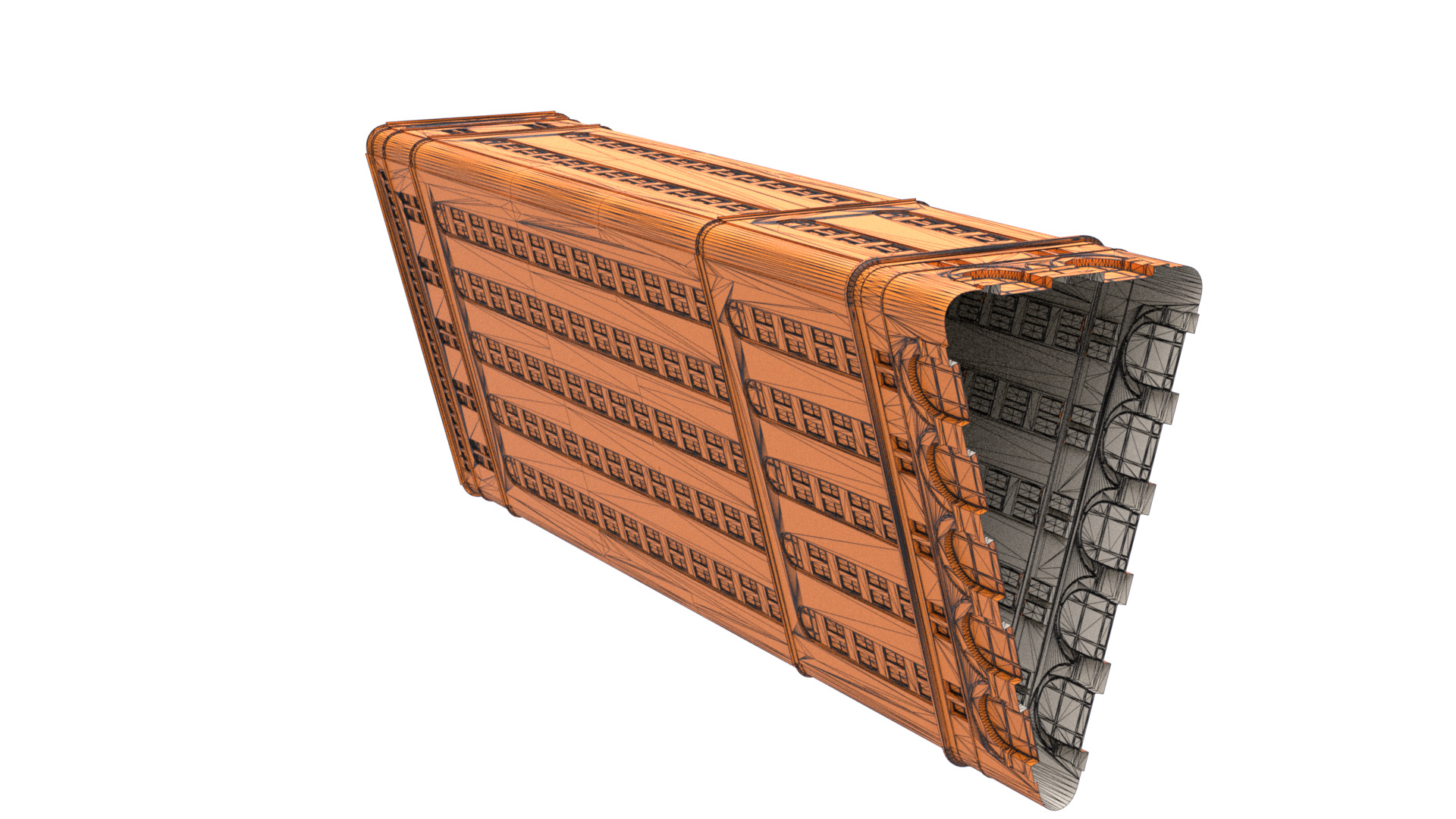}\hfill\\
\vspace{-2em}
\begin{flushleft}{\small \hspace{6.5em} Input \hspace{12.6em} Ours}\end{flushleft}
\captionsetup{justification=centering}\vspace{-0.1in}
\caption{{Example of preserving hole}}
\label{fig:boundary}
\end{figure}

\textbf{Preserving Holes. } Although our method closes the hole and converts the large open surface into a thin shell by default, our system allows users to specify the open boundary to be preserved during the repair (\autoref{fig:boundary}). Particularly, our system first closes the hole and marks all corresponding faces. After repair, all marked faces are removed to recover the boundary.

\begin{figure}[ht]
\newcommand{\figcap}[1]{\begin{minipage}{0.105\linewidth}\centering#1\end{minipage}}
    \centering
    \rotatebox{90}{\small \hspace*{.1\linewidth} Full visual guidance }\hfill
    \includegraphics[trim=250 0 200 100,clip,width=0.47\linewidth]{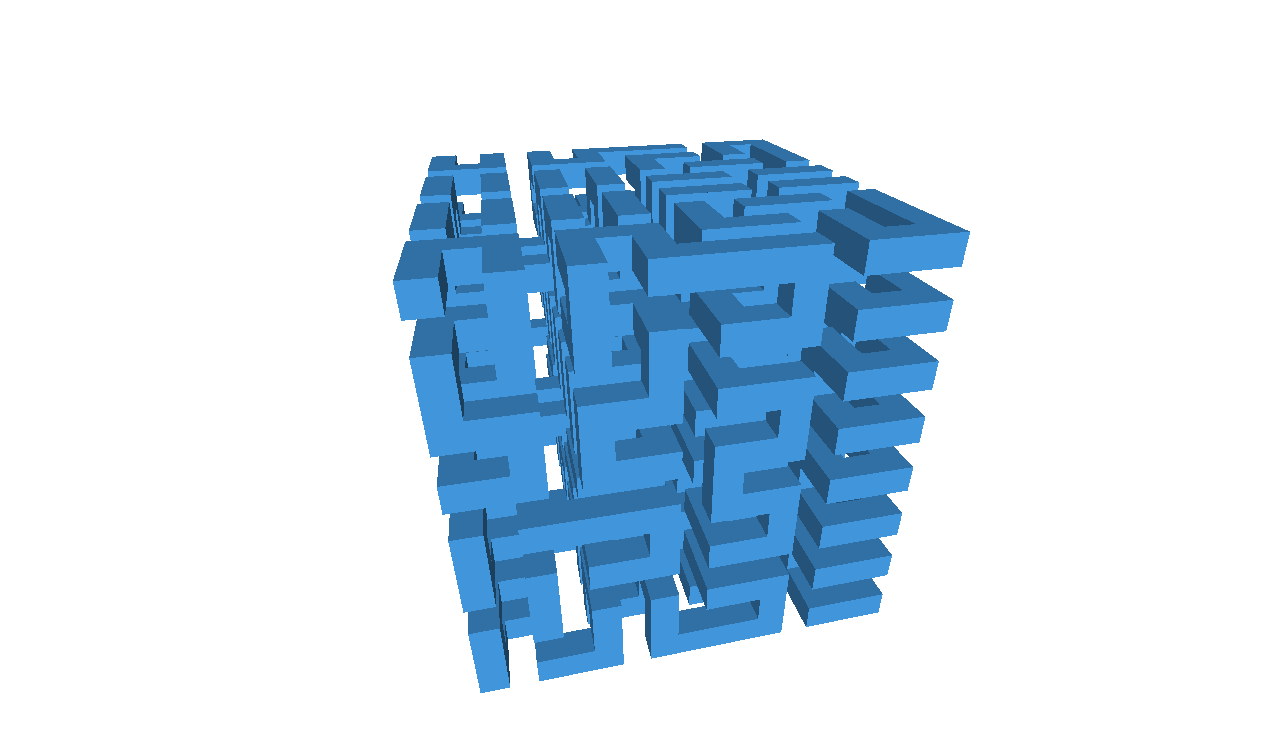}\hfill
    \includegraphics[trim=250 0 200 100,clip,width=0.47\linewidth]{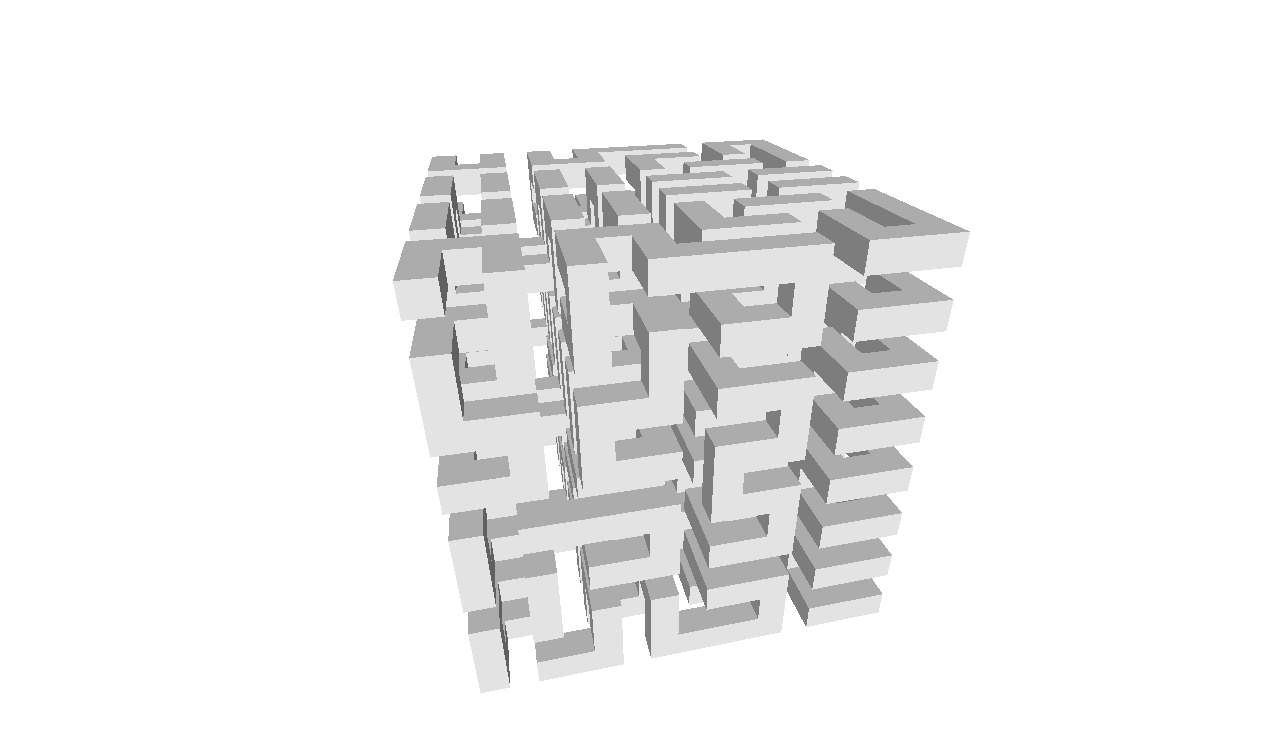}\hfill
    \rotatebox{90}{\small  \hspace*{.05\linewidth}  Partial visual guidance }\hfill
    \includegraphics[trim=250 0 200 100,clip,width=0.47\linewidth]{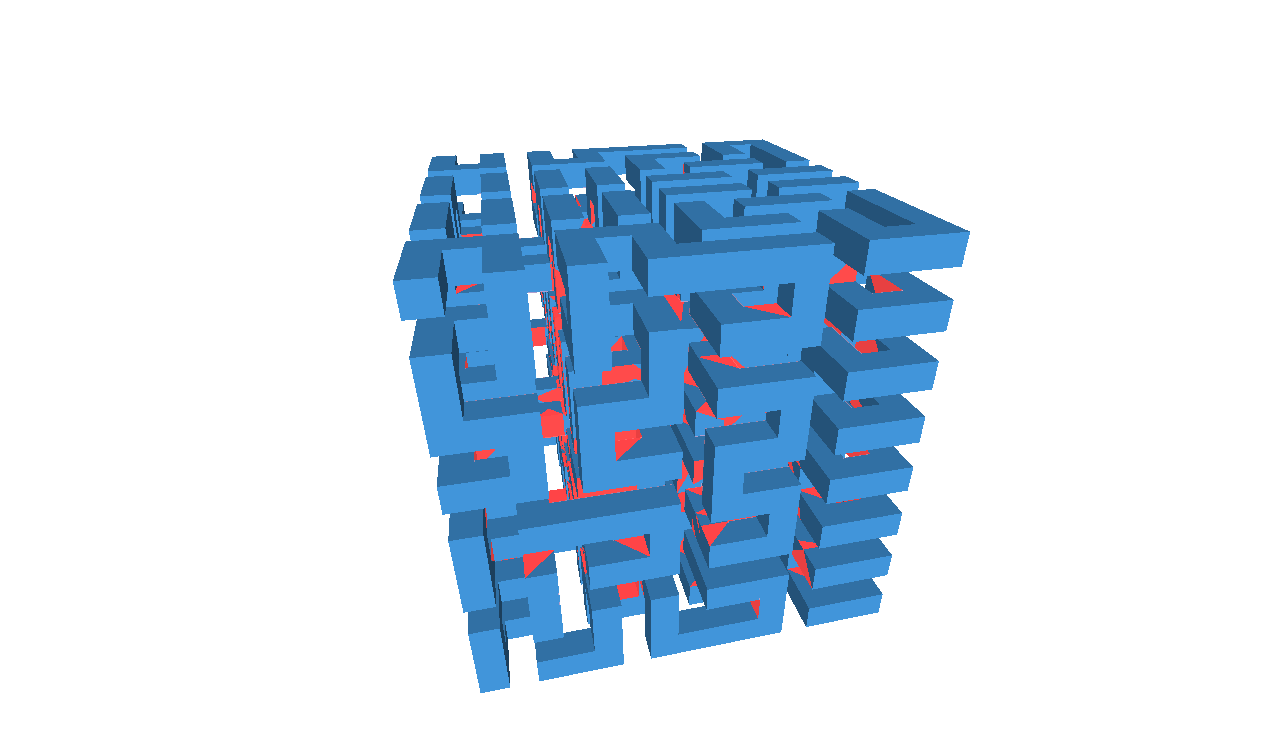}\hfill
    \includegraphics[trim=250 0 200 100,clip,width=0.47\linewidth]{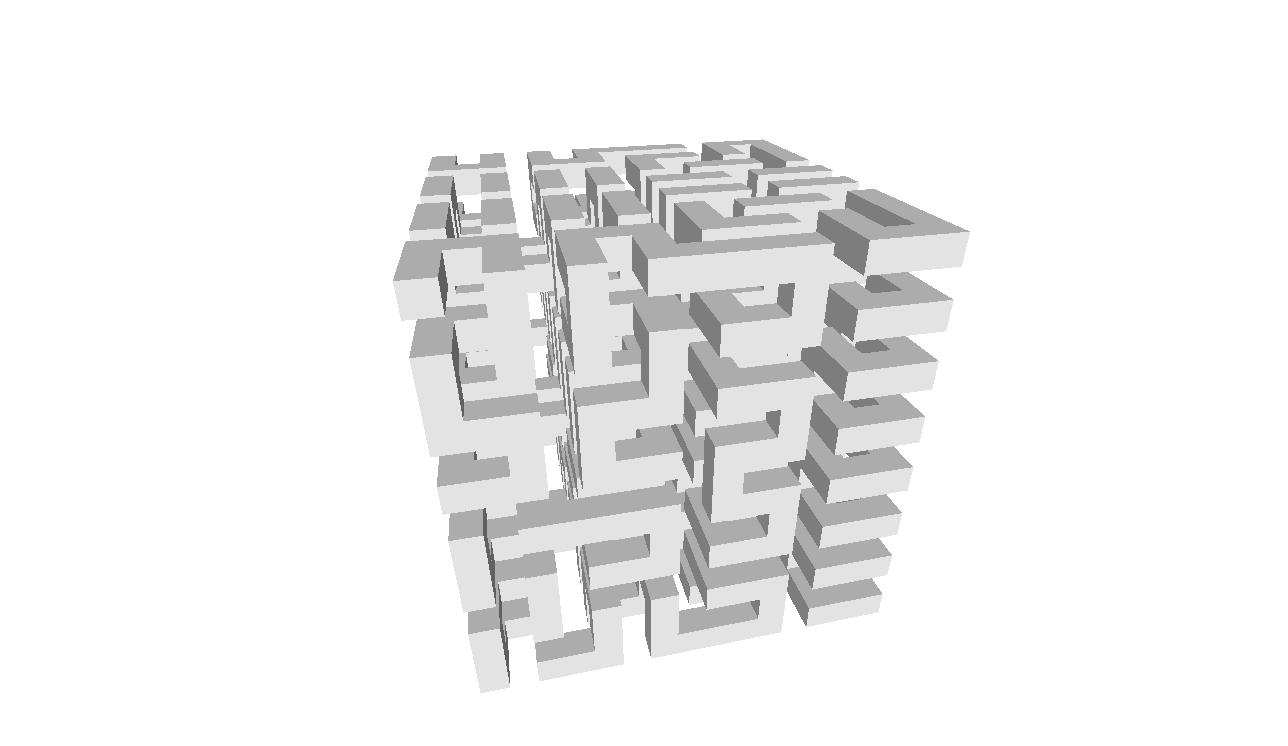}\\\vspace{-0.1in}
    \begin{minipage}{0.49\linewidth}\centering Visibility\end{minipage}\hfill
    \begin{minipage}{0.49\linewidth}\centering Results\end{minipage}\\\vspace{-0.1in}
    \caption{\textbf{Example of dealing with hard-to-see structures:}  The Hilbert Cube model from Thingi10K exhibits intricate structures. The visibility (left) and the outcomes (right) of our method under complete visual guidance (top) and partial visual guidance (bottom) are shown.  \textbf{Blue} indicates visible faces, while \textbf{Red} represents faces that is not visible by the ray with zero bounces. Remarkably, in both scenarios, our approach achieves zero HD. }
    \label{fig:cube}
\end{figure}

\textbf{Hard-to-see Structures.} The design of ray casting and graph cut makes our method robustly handle complex structures. We tested our method on a Hilbert Cube from Thingi10K using default parameters $N_b=10$ and $N_\text{total}=2\times10^7$ to obtain the complete visual guidance and $N_b=0$ and $N_\text{total}=2\times10^3$ to obtain only partial visual guidance, as is shown in \autoref{fig:cube}. Under the default parameters, the visibility results demonstrate that multiple reflections enable a comprehensive exploration of hard-to-see structures. Even with partial visual guidance, the edge breaking for invisible faces during the graph cut process allows our method to respect input geometry. In both cases, the output meshes are with zero HD.

\textbf{Ablation Study.} We scrutinize and analyze the efficacy of pivotal components of our method, including visual guidance, graph cut, and constrained simplification, using the "Chair" model illustrated in \autoref{fig:results}. As shown in \autoref{tab:ablation}, without visual guidance, all faces are considered visible and are not reoriented. This leads to subpar performance in terms of HD, LFD, and PSNR due to the presence of misoriented or open faces. Without graph cut entails retaining steps prior to the graph cut, as the subsequent steps rely on its outcomes. Consequently, the outcome lacks watertightness and manifold properties, resulting in the model being unrepaired. Lastly, after simplification, the face count reduces from 169K to 20K.

\begin{table}[ht]
\caption{\label{tab:ablation} {Ablation study on each component of our method} }
\centering
\scalebox{0.8}
 	{
\begin{tabular}{|l|cccccc|}
\hline
&Watertight & Manifold & Face \#  & HD & LFD & PSNR  \\ \hline
Input        &Yes  &No & 27K  & --  & --    & --    \\ 
No Visual guidance  &Yes  &Yes & 17K  & 0.12      & 7956    & 27.3     \\ 
No graph cut   & No &No & 13K  & 0    & 0    & 56.7    \\ 
No simplification & Yes &Yes &  169K & 0.01    & 6    & 51.4     \\ 
Ours     &Yes&Yes & 20K &  0.01  &  6   &    51.4  \\  \hline
\end{tabular}
}
\end{table}

\begin{figure}[ht]
\newcommand{\figcap}[1]{\begin{minipage}{0.24\linewidth}\centering#1\end{minipage}}
\rotatebox{90}{\small \hspace*{.0\linewidth}Input}\hfill%
\includegraphics[trim=150 350 150 350,clip,width=0.24\linewidth]{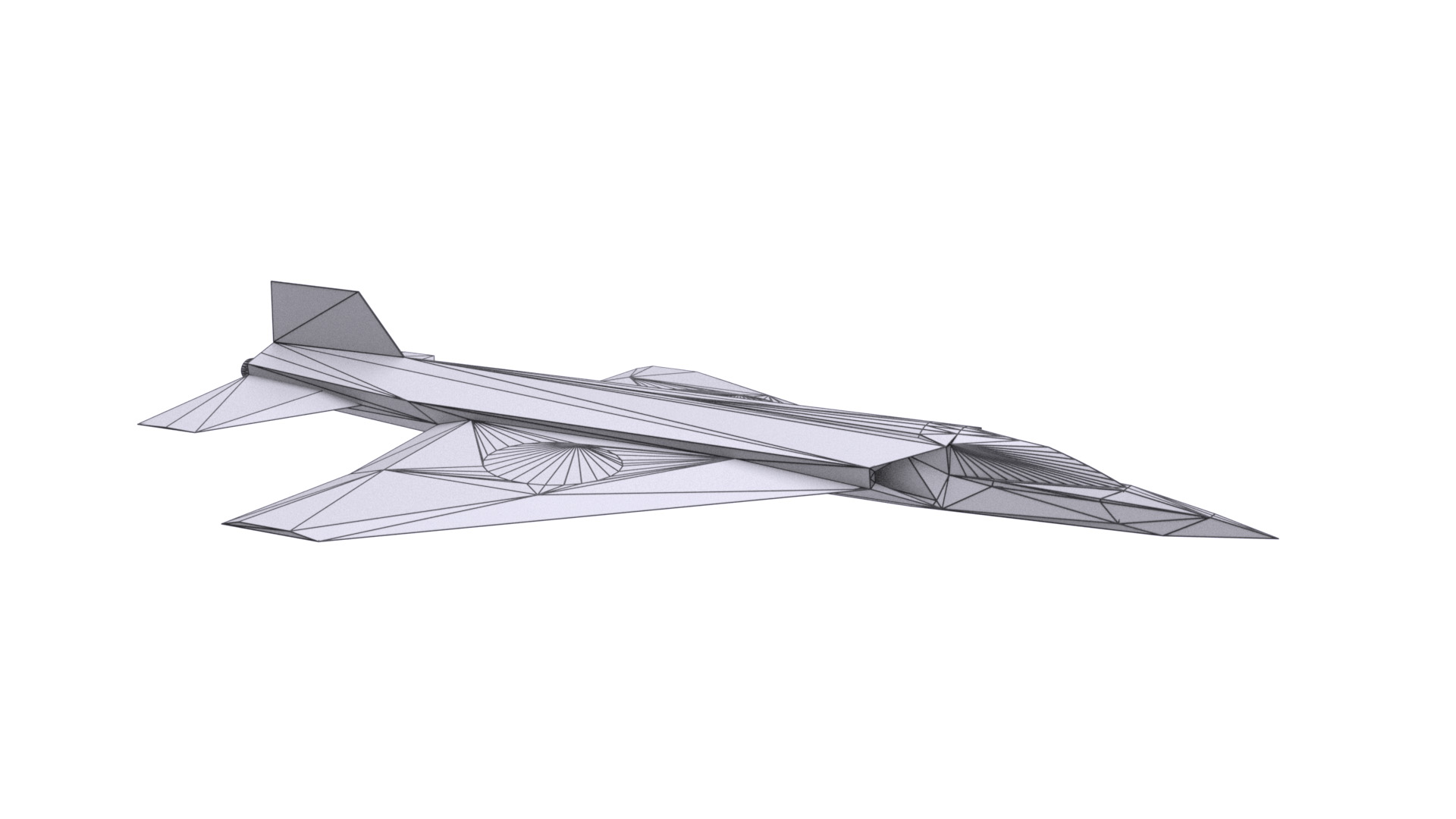}\hfill
\includegraphics[trim=150 350 150 350,clip,width=0.24\linewidth]{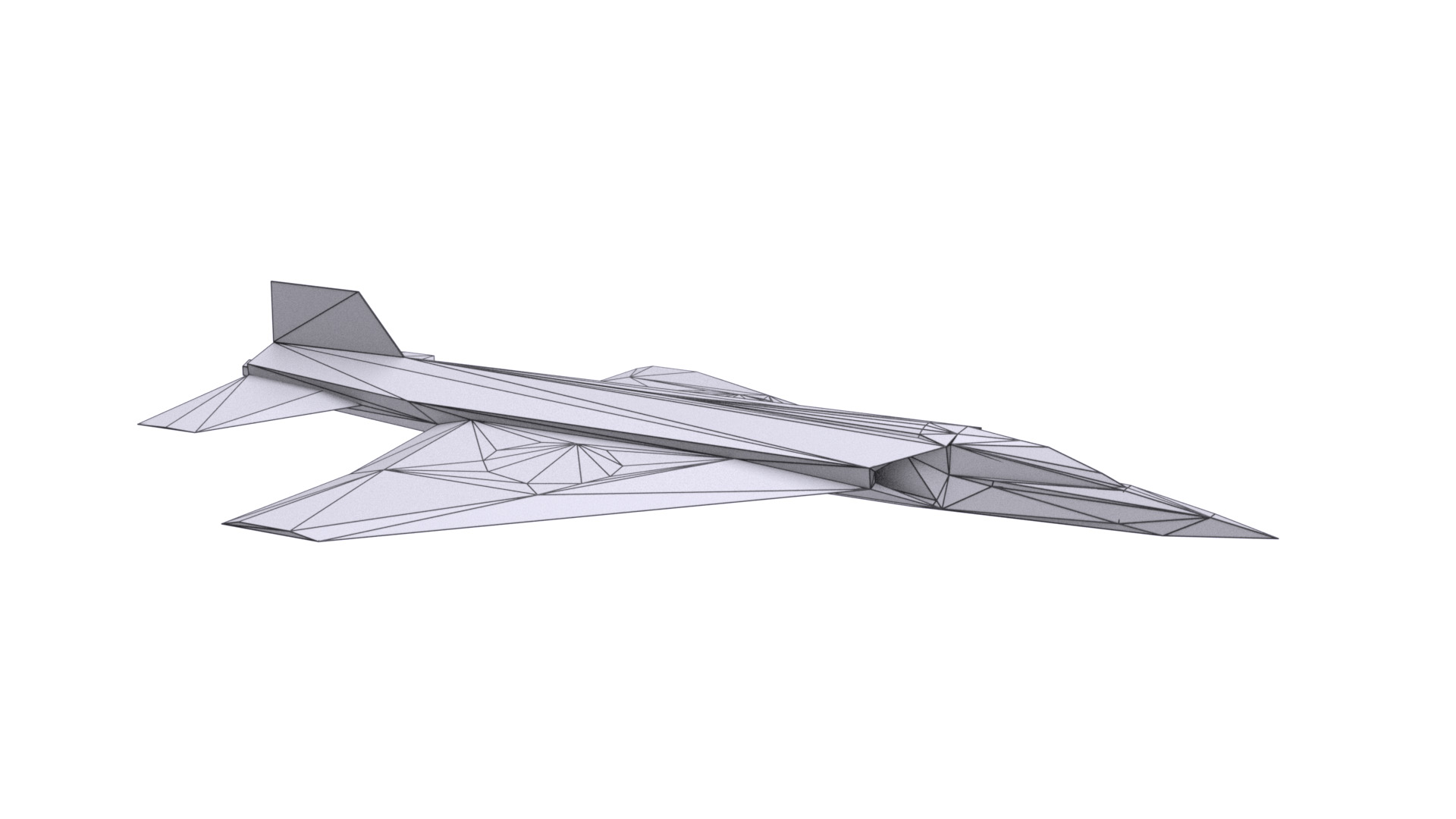}\hfill
\includegraphics[trim=150 350 150 350,clip,width=0.24\linewidth]{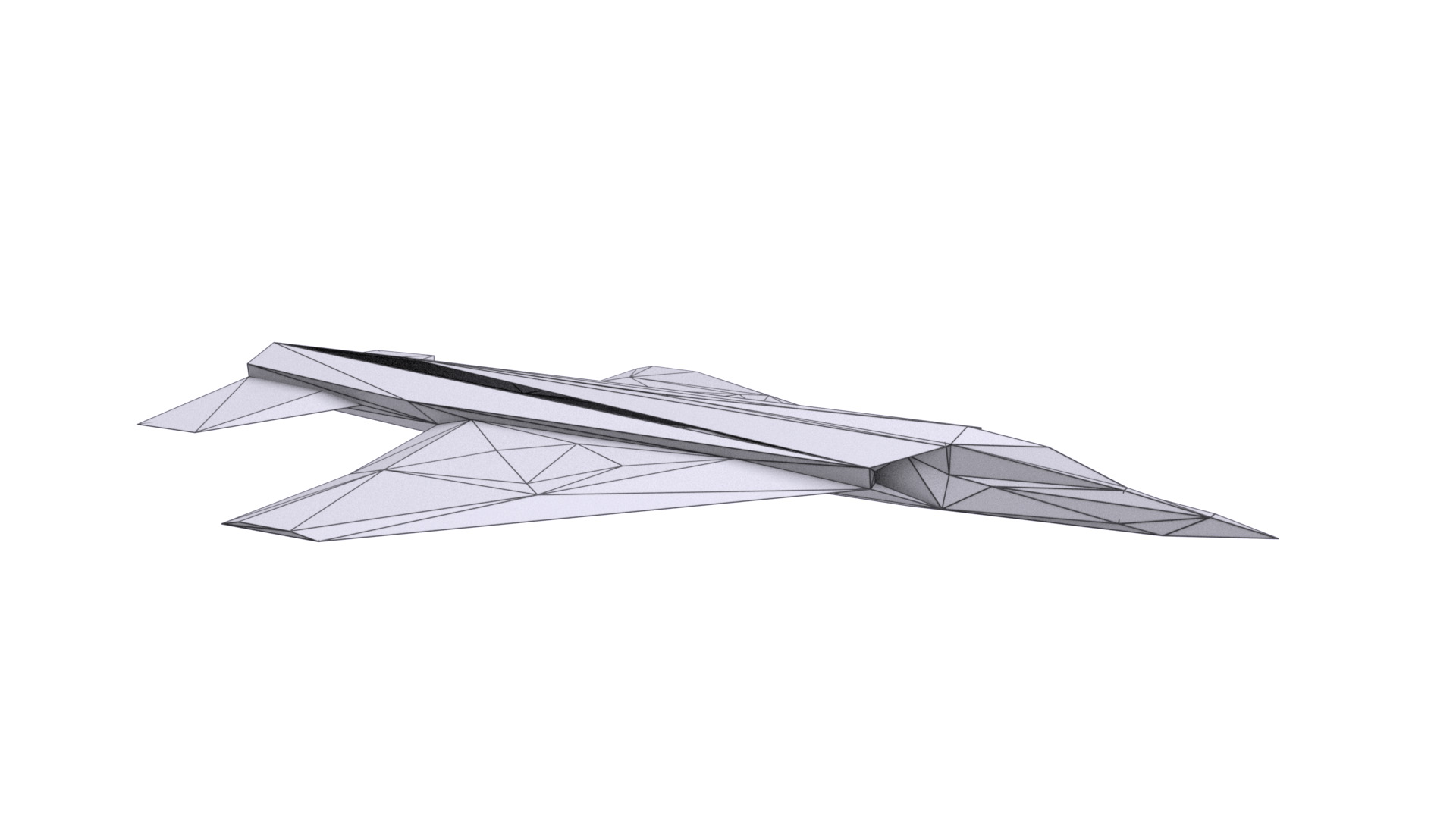}\hfill
\includegraphics[trim=150 350 150 350,clip,width=0.24\linewidth]{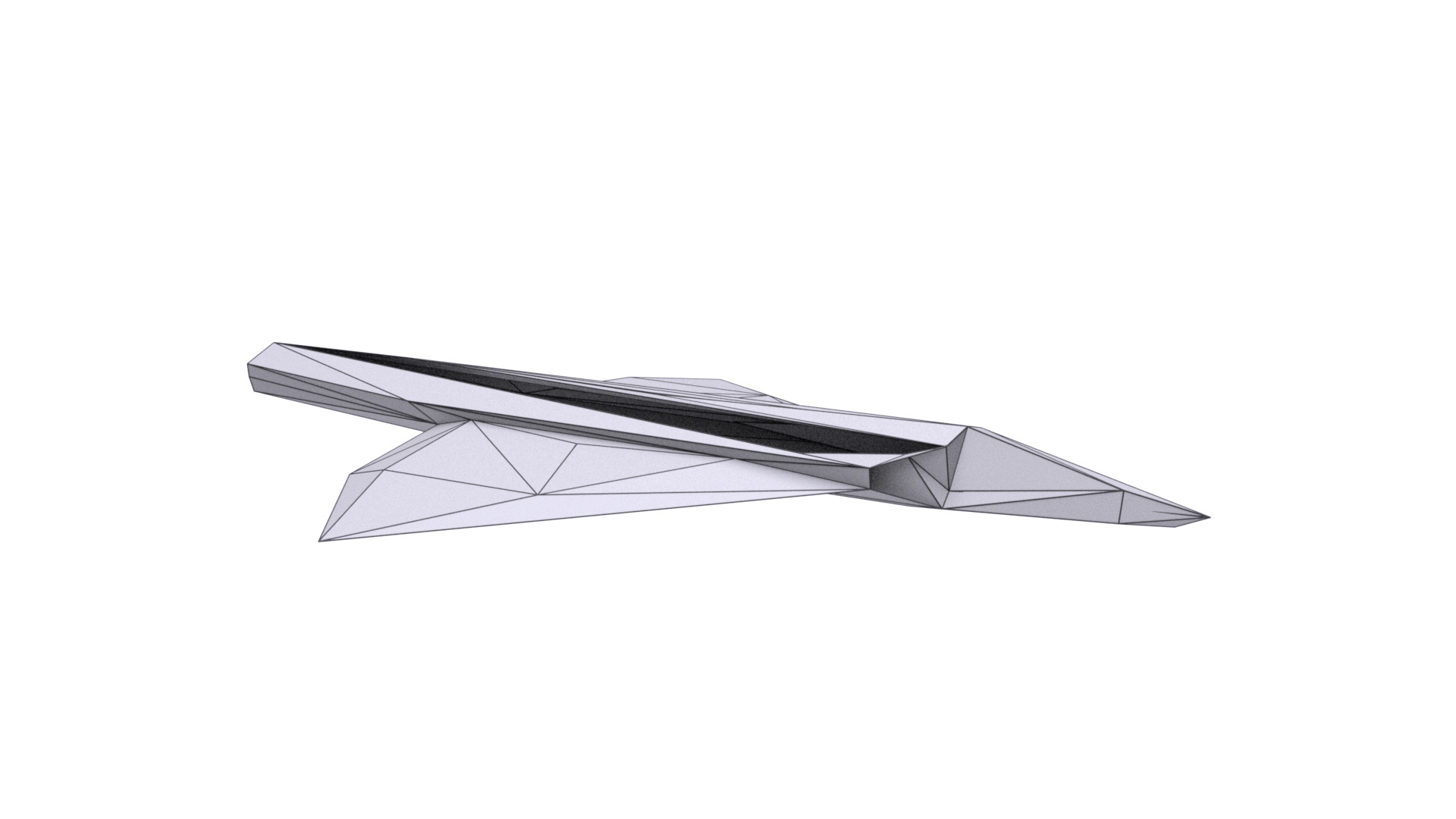}\\
\figcap{\small 2786} \hfill%
\figcap{\small 1000 }\hfill%
\figcap{\small 500 } \hfill%
\figcap{\small 250 } \\
\rotatebox{90}{\small \hspace*{.0\linewidth}Ours}\hfill%
\includegraphics[trim=150 350 150 250,clip,width=0.24\linewidth]{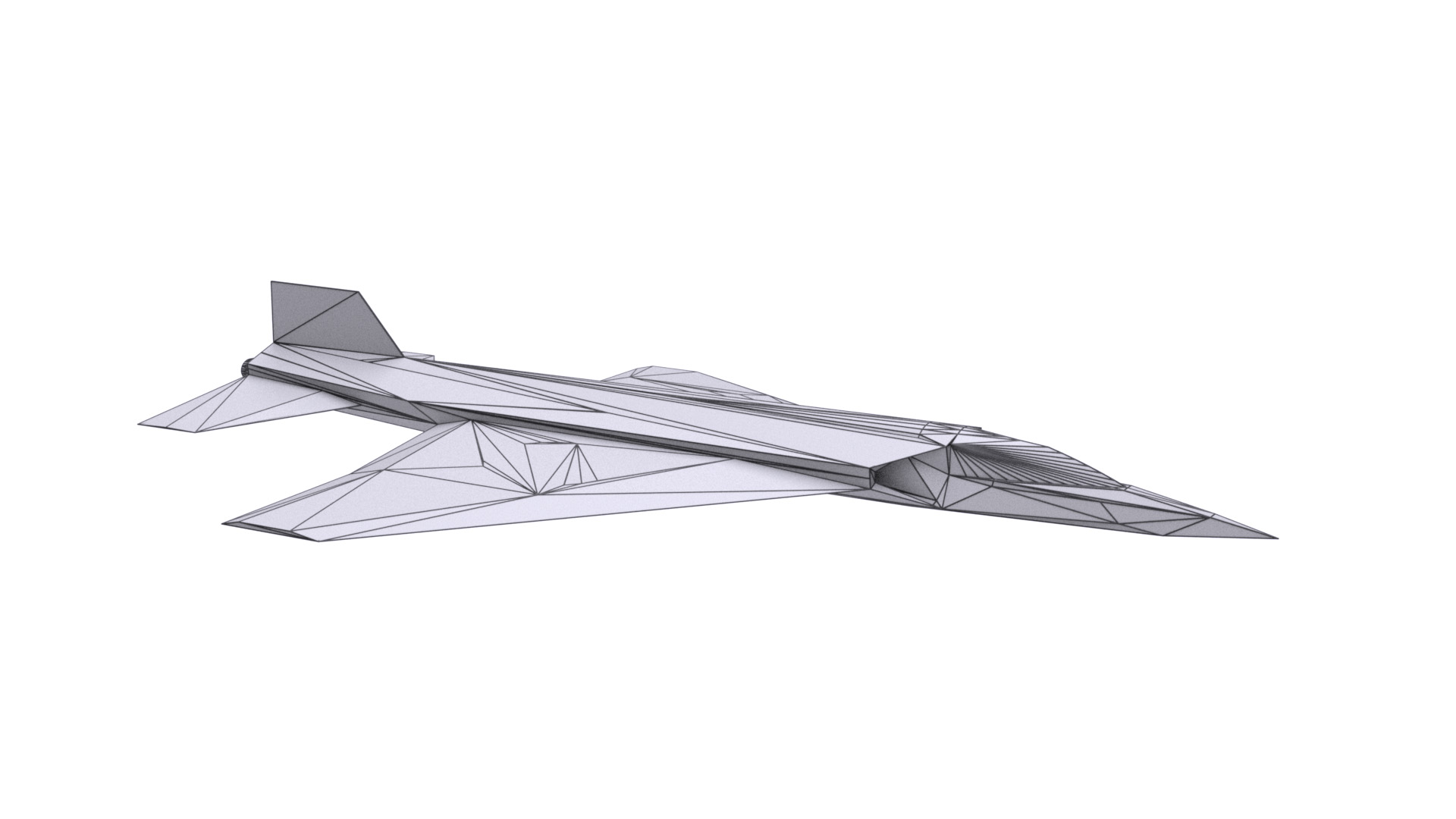}\hfill
\includegraphics[trim=150 350 150 250,clip,width=0.24\linewidth]{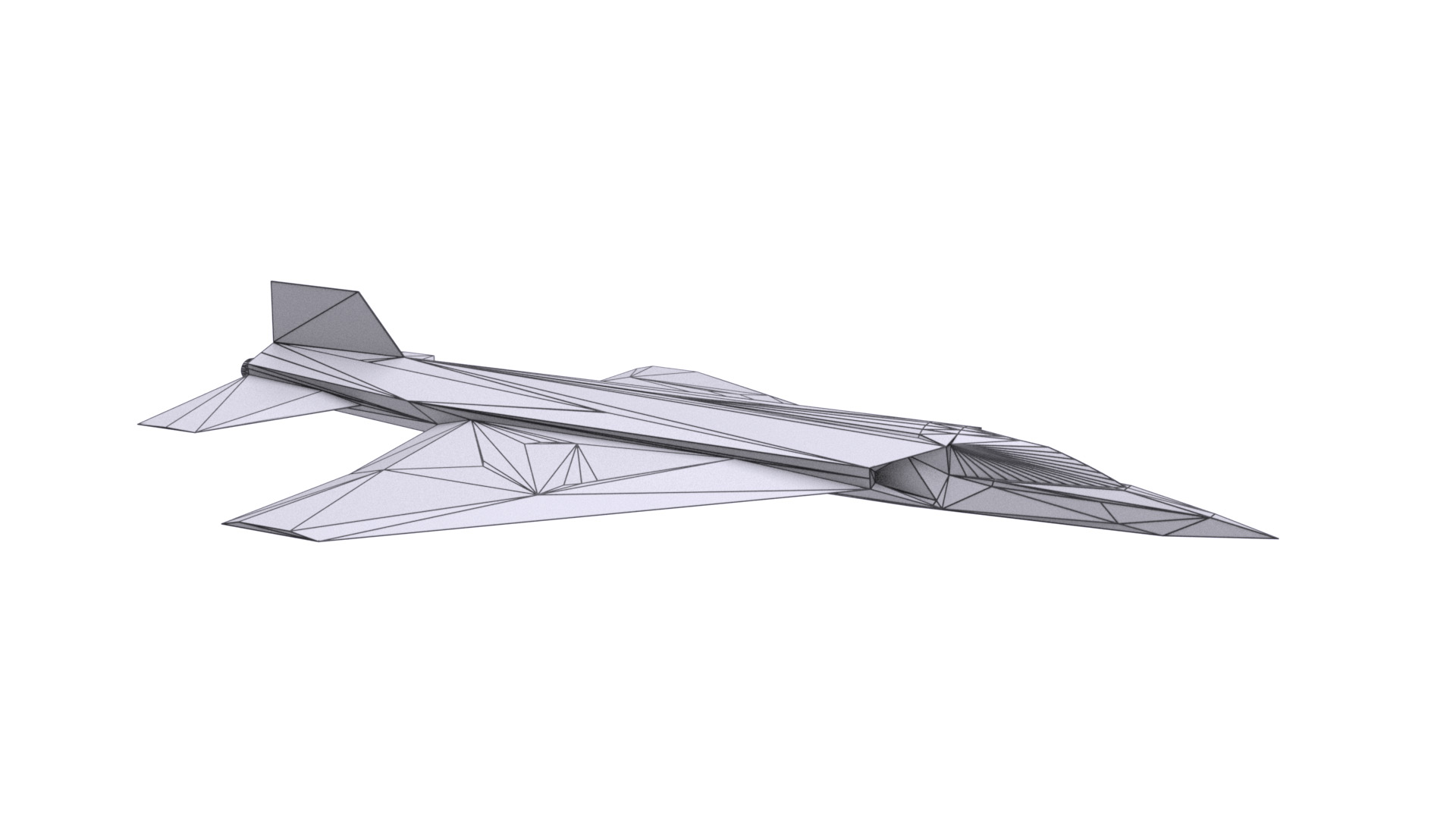}\hfill
\includegraphics[trim=150 350 150 250,clip,width=0.24\linewidth]{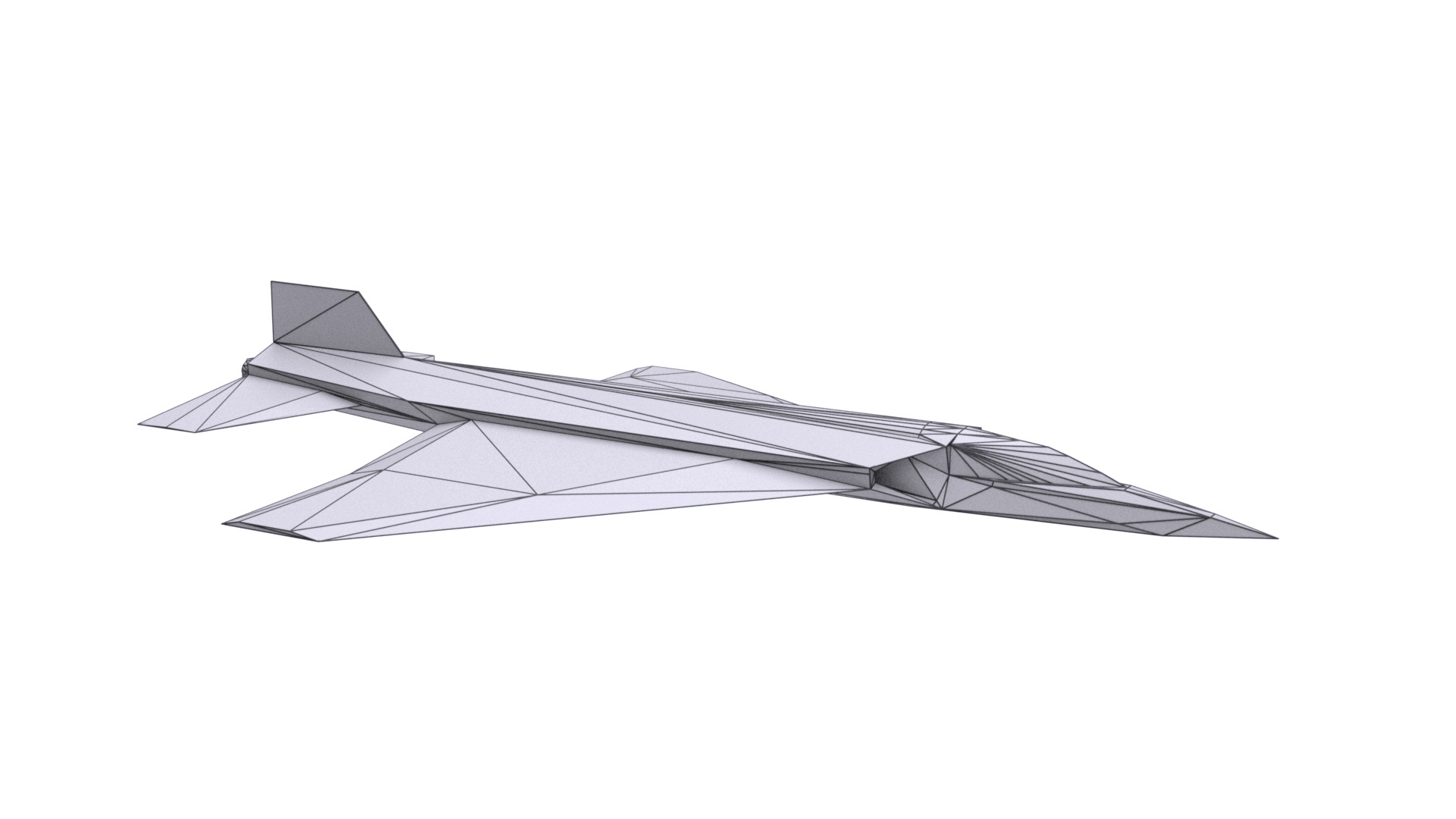}\hfill
\includegraphics[trim=150 350 150 250,clip,width=0.24\linewidth]{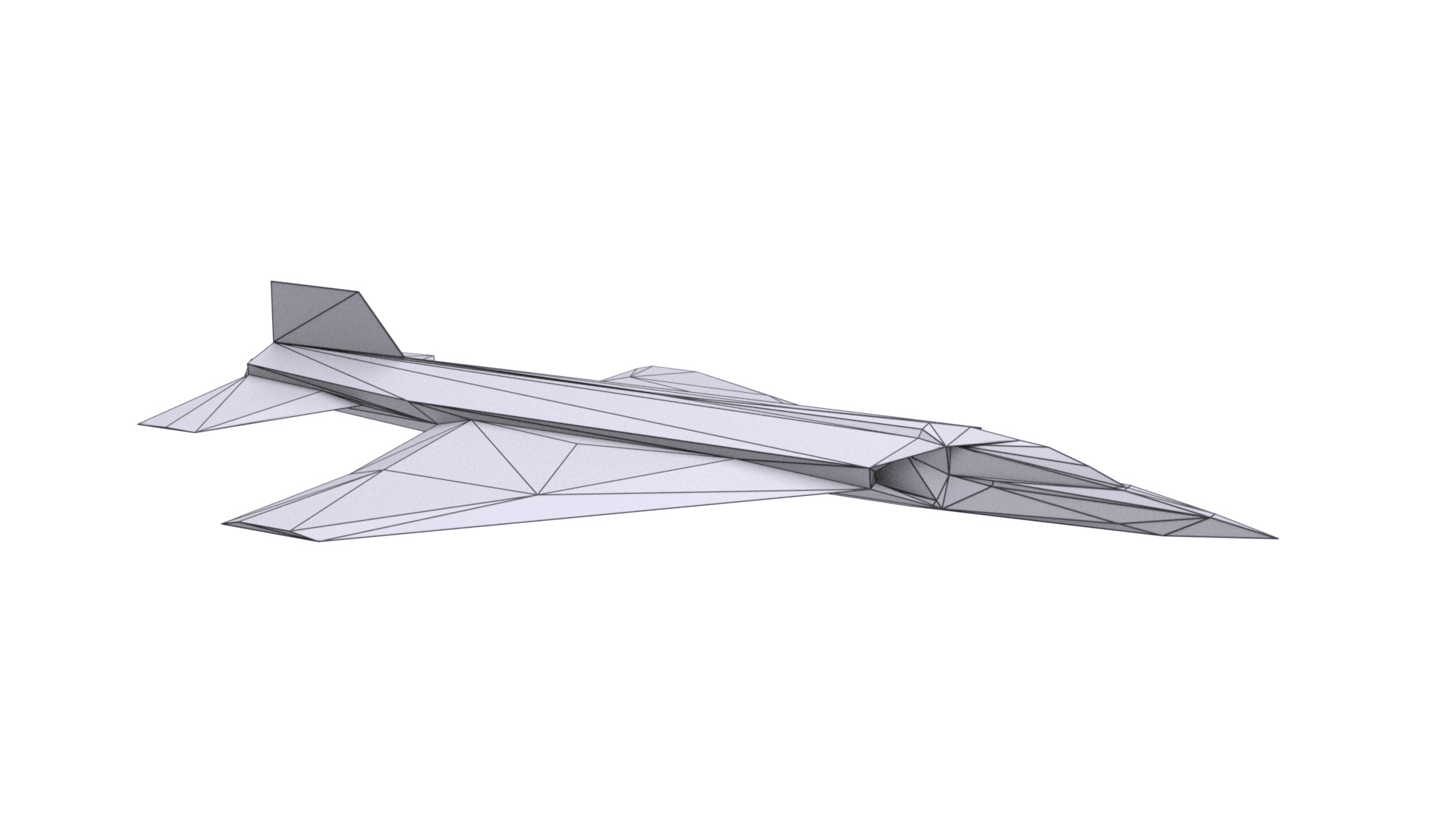}\\
\figcap{\small 2020} \hfill%
\figcap{\small 1000 }\hfill%
\figcap{\small 500 } \hfill%
\figcap{\small 250 } \\
\vspace{-.15in}
\caption{\textbf{Example application of mesh simplification:} The input airplane model is shown without (top) and with (bottom) using our mesh repair before making quadratic error metric (QEM) simplification~\cite{Garland1997QEM}. The model repaired by our method can better preserve its original shape after simplification.}
\label{fig:QEM}
\end{figure}

\begin{figure}[ht]
\begin{flushleft}{\small \hspace{2.5em} $M_0$ \hspace{2.0em} $M_1$ \hspace{0.7em} $M_0$ not $M_1$ \hspace{1.3em} $M_0$ \hspace{2.0em} $M_1$ \hspace{0.7em} $M_0$ not $M_1$}\end{flushleft}
\includegraphics[trim=700 100 700 50,clip,height=0.4\linewidth]{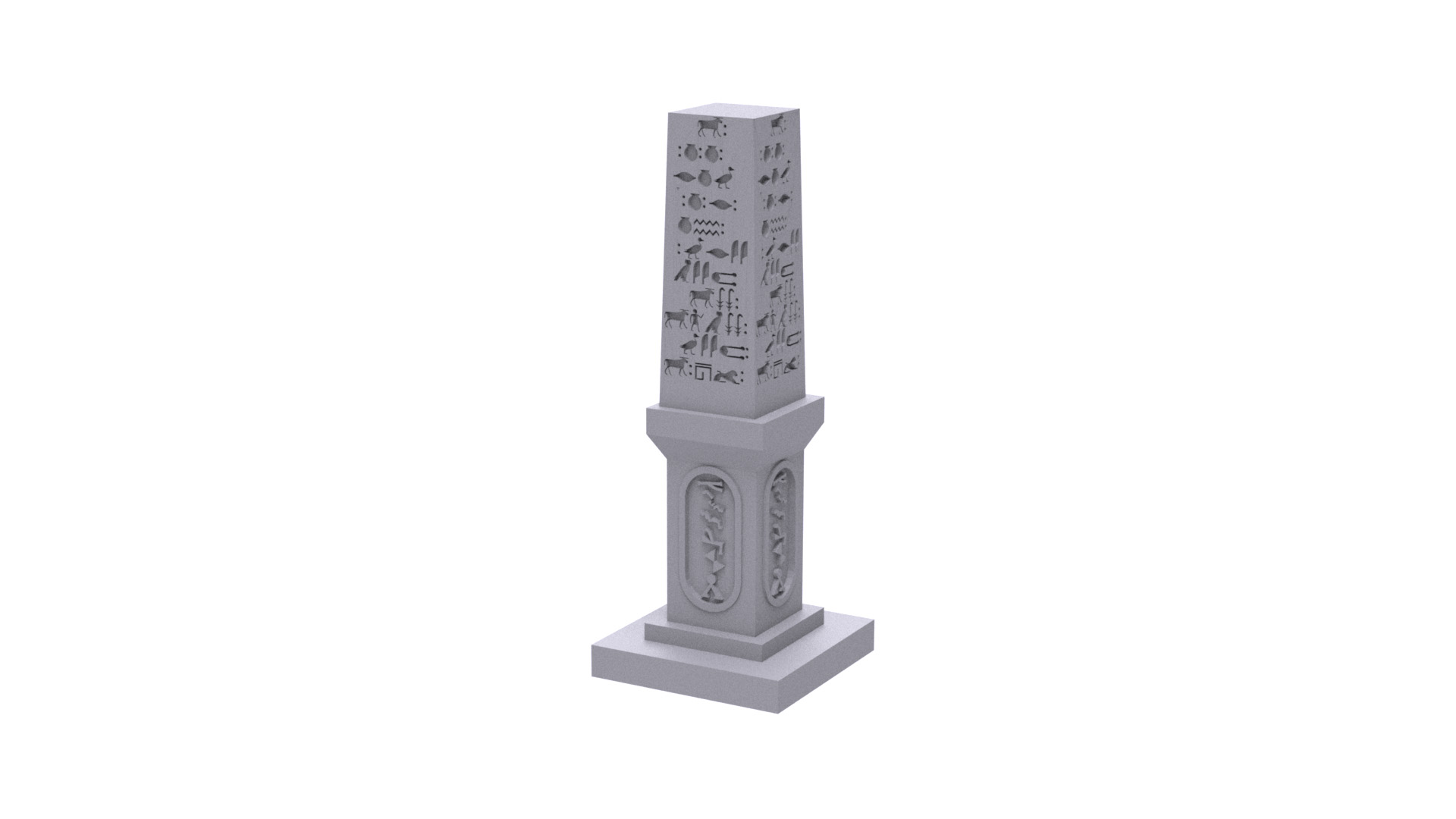}\hfill
\includegraphics[trim=900 100 900 50,clip,height=0.4\linewidth]{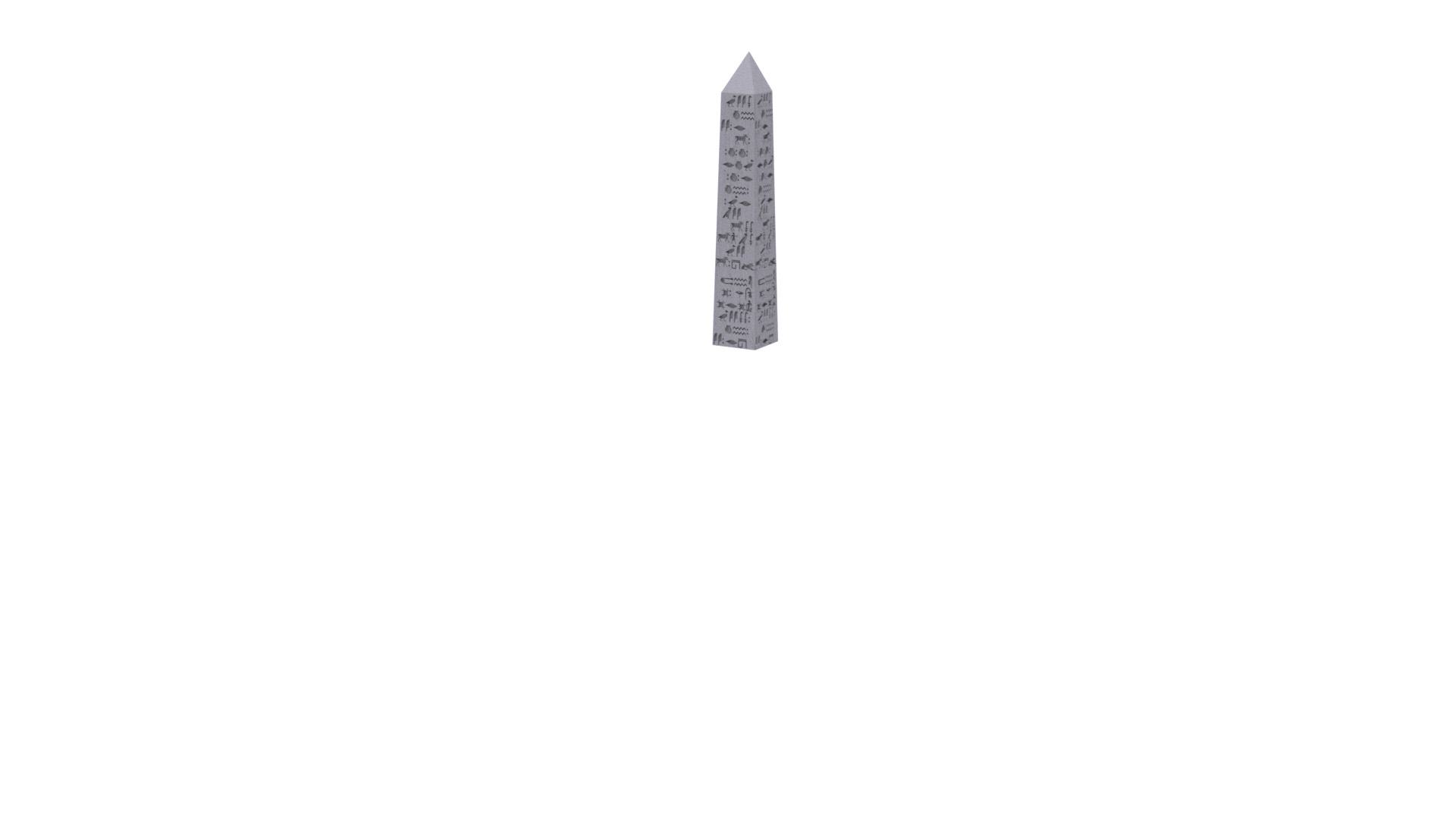}\hfill
\includegraphics[trim=700 100 700 50,clip,height=0.4\linewidth]{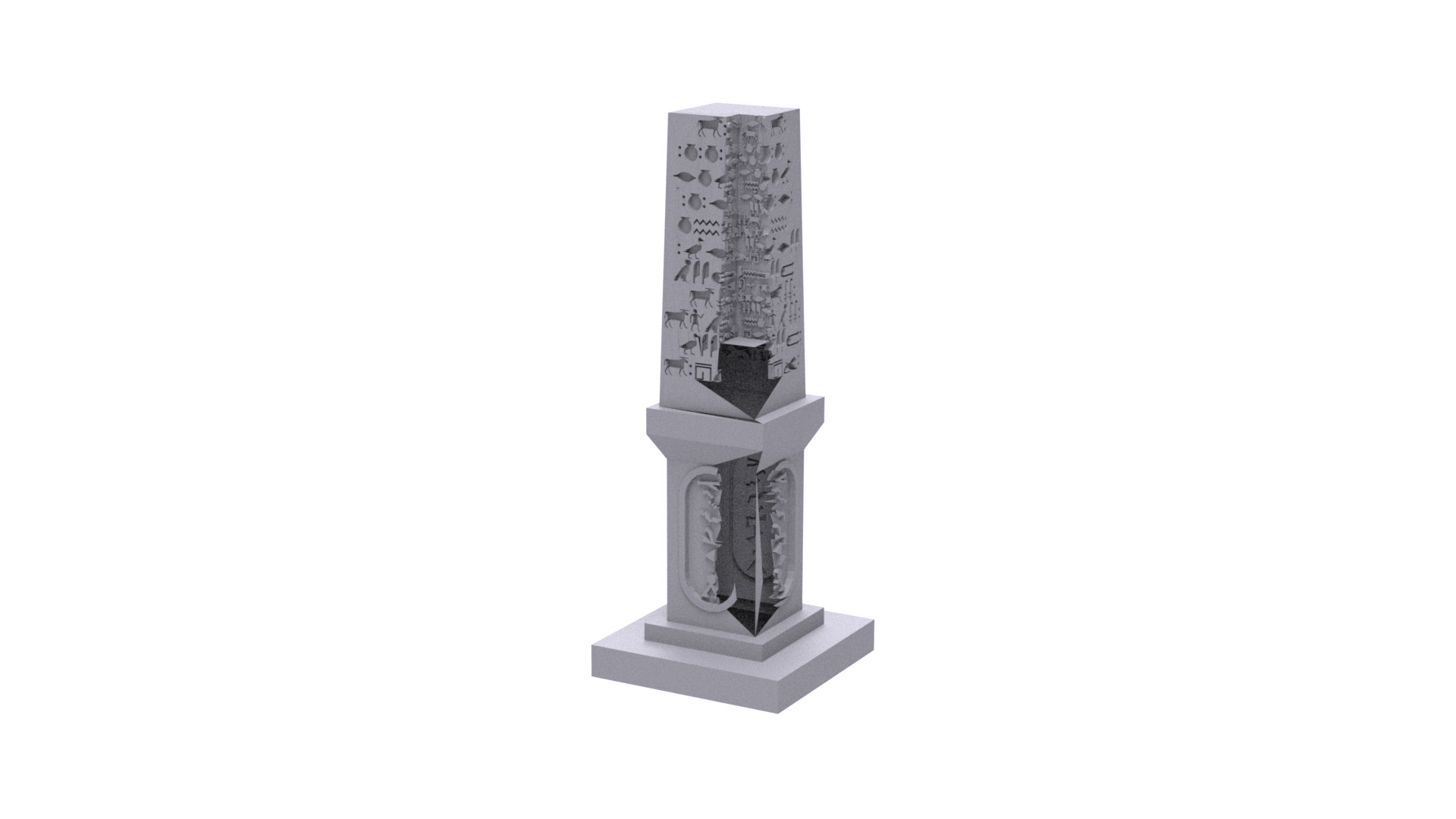}\hfill
\includegraphics[trim=700 100 700 50,clip,height=0.4\linewidth]{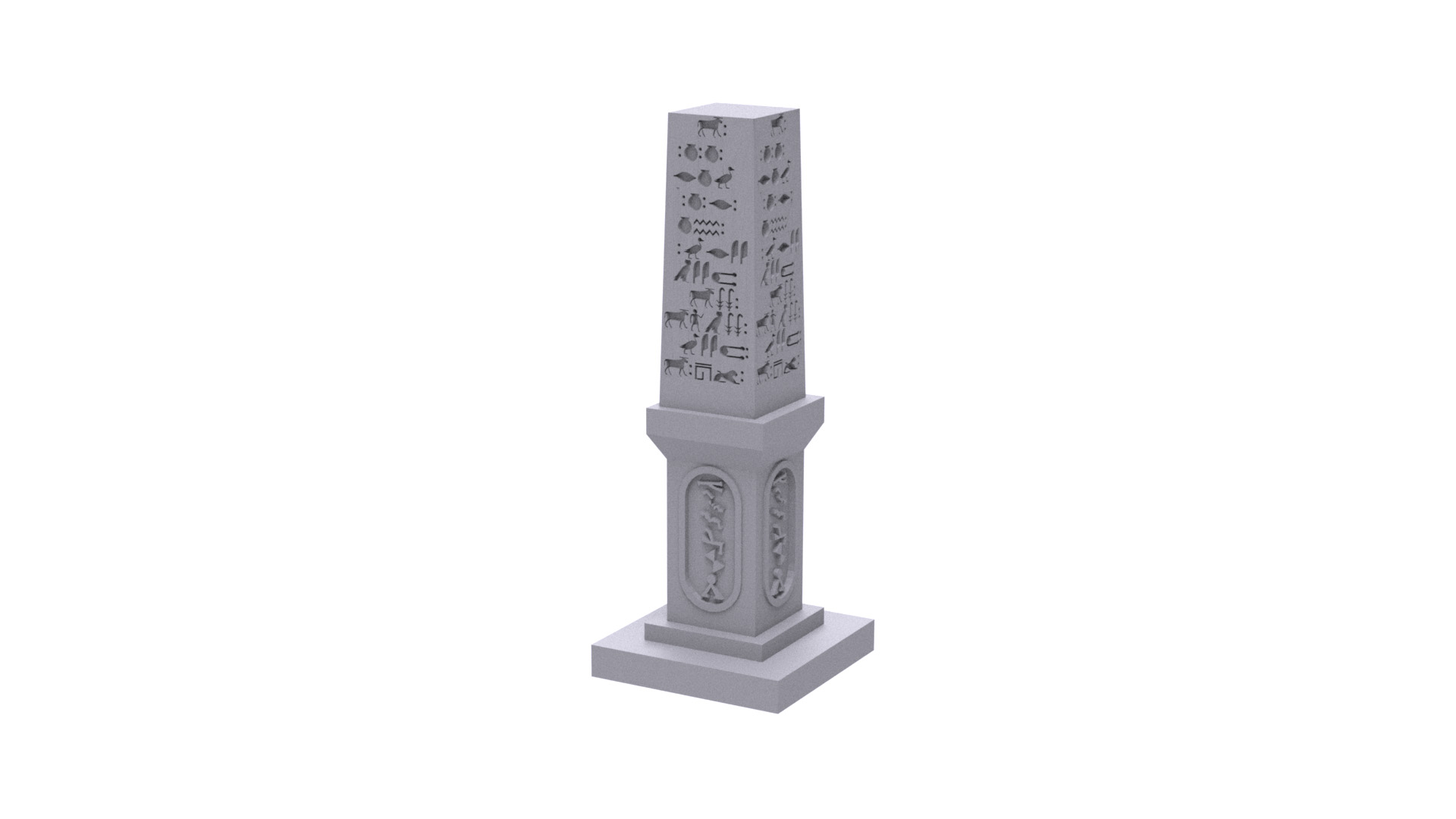}\hfill
\includegraphics[trim=900 100 900 50,clip,height=0.4\linewidth]{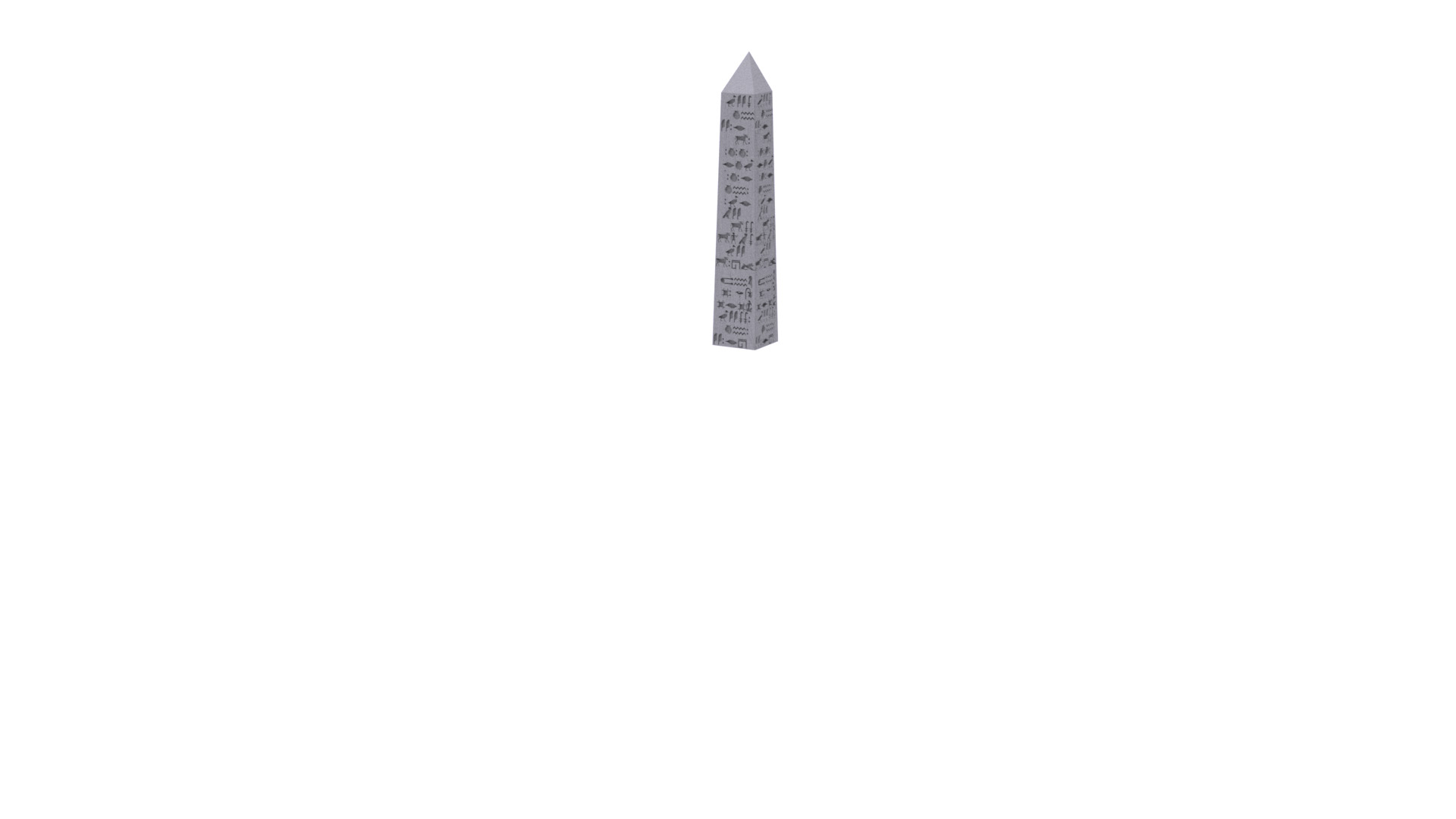}\hfill
\includegraphics[trim=700 100 700 50,clip,height=0.4\linewidth]{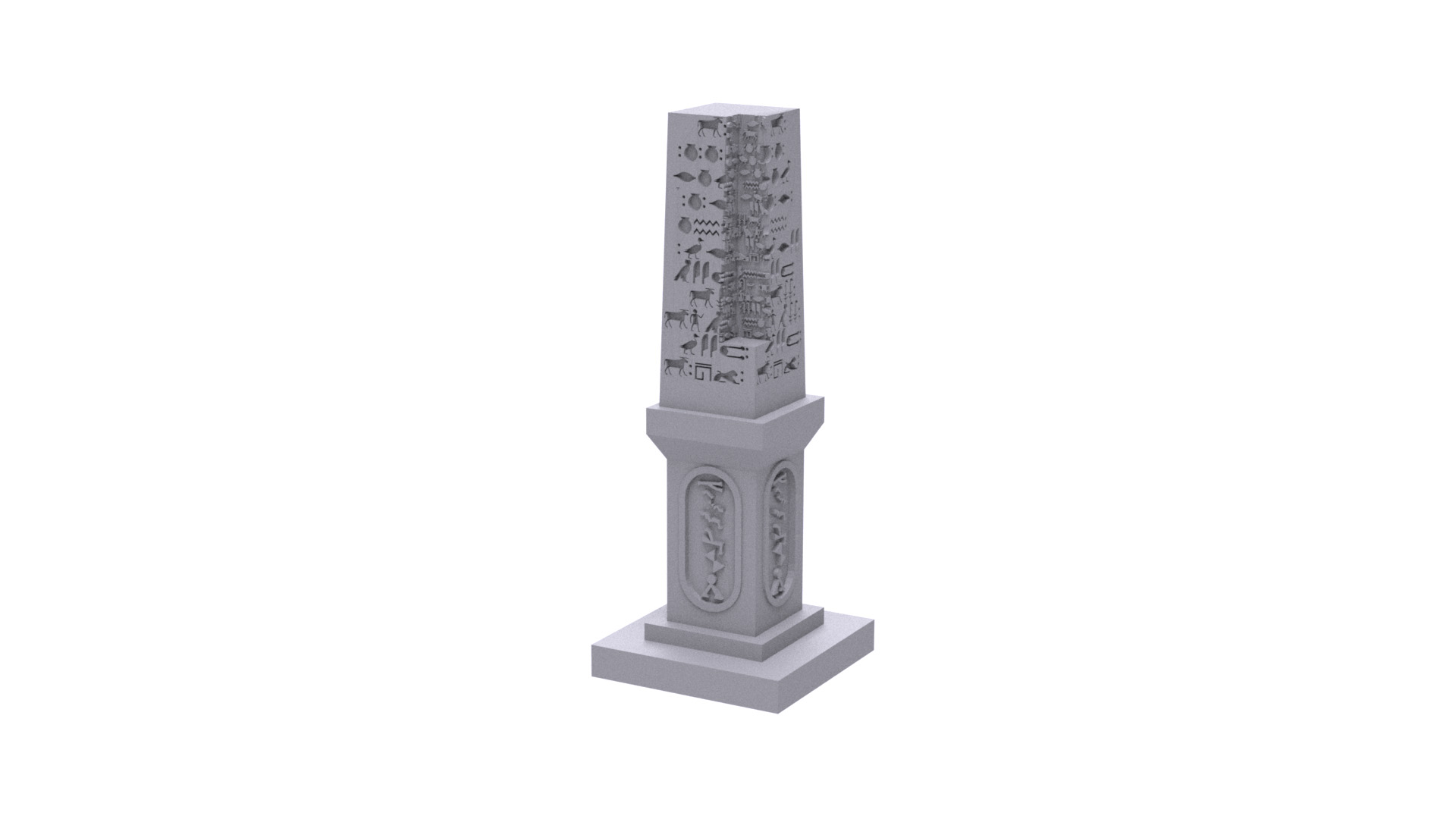}\\
\vspace{-.4in}
\begin{flushleft}{\small \hspace{6.5em} Input \hspace{12em} Ours}\end{flushleft}\vspace{-.1in}
\caption{\textbf{Example application of boolean operation:} After applying the boolean operation~\cite{zhou2016mesh} directly to the input meshes, the resulting mesh is broken due to wrong orientation at $M_1$'s bottom and tiny gaps, which may not be visible in the image. However, our method can successfully repair the mesh, enabling the boolean operation to output the correct result.}
\label{fig:boolean}
\end{figure}

\begin{figure}[ht]
\includegraphics[trim=280 120 270 70,clip,width=0.24\linewidth]{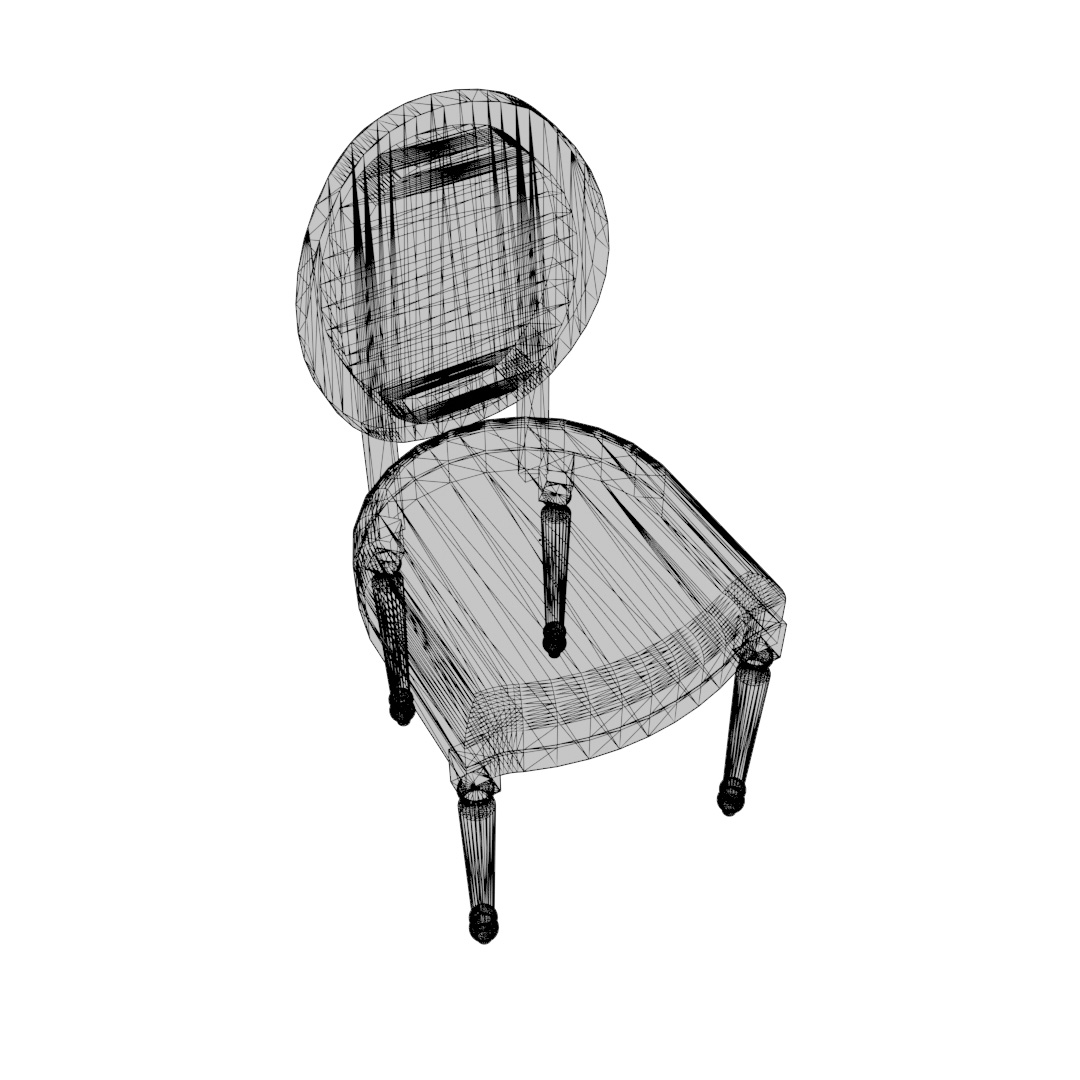}\hfill
\includegraphics[trim=280 120 270 70,clip,width=0.24\linewidth]{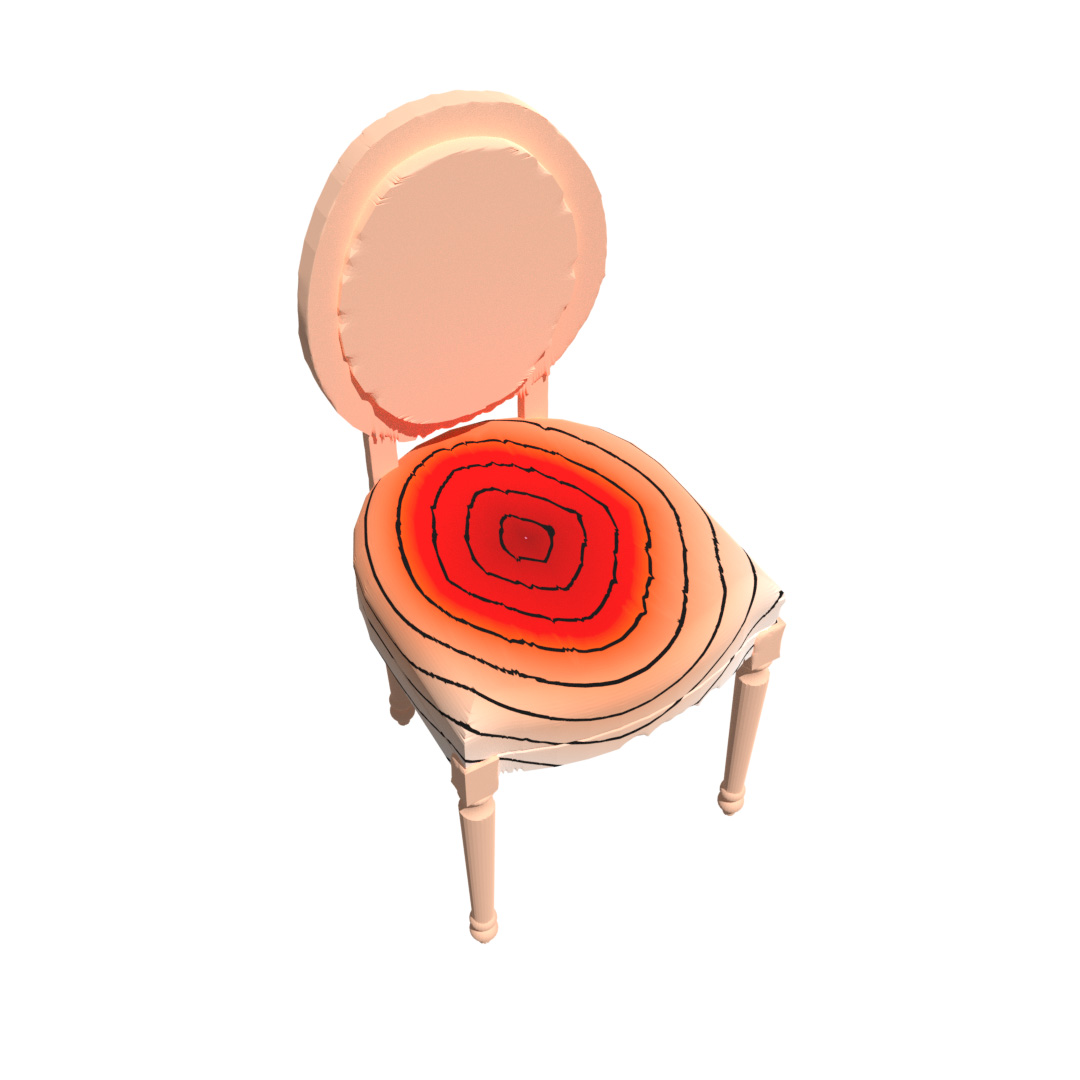}\hfill
\includegraphics[trim=280 120 270 70,clip,width=0.24\linewidth]{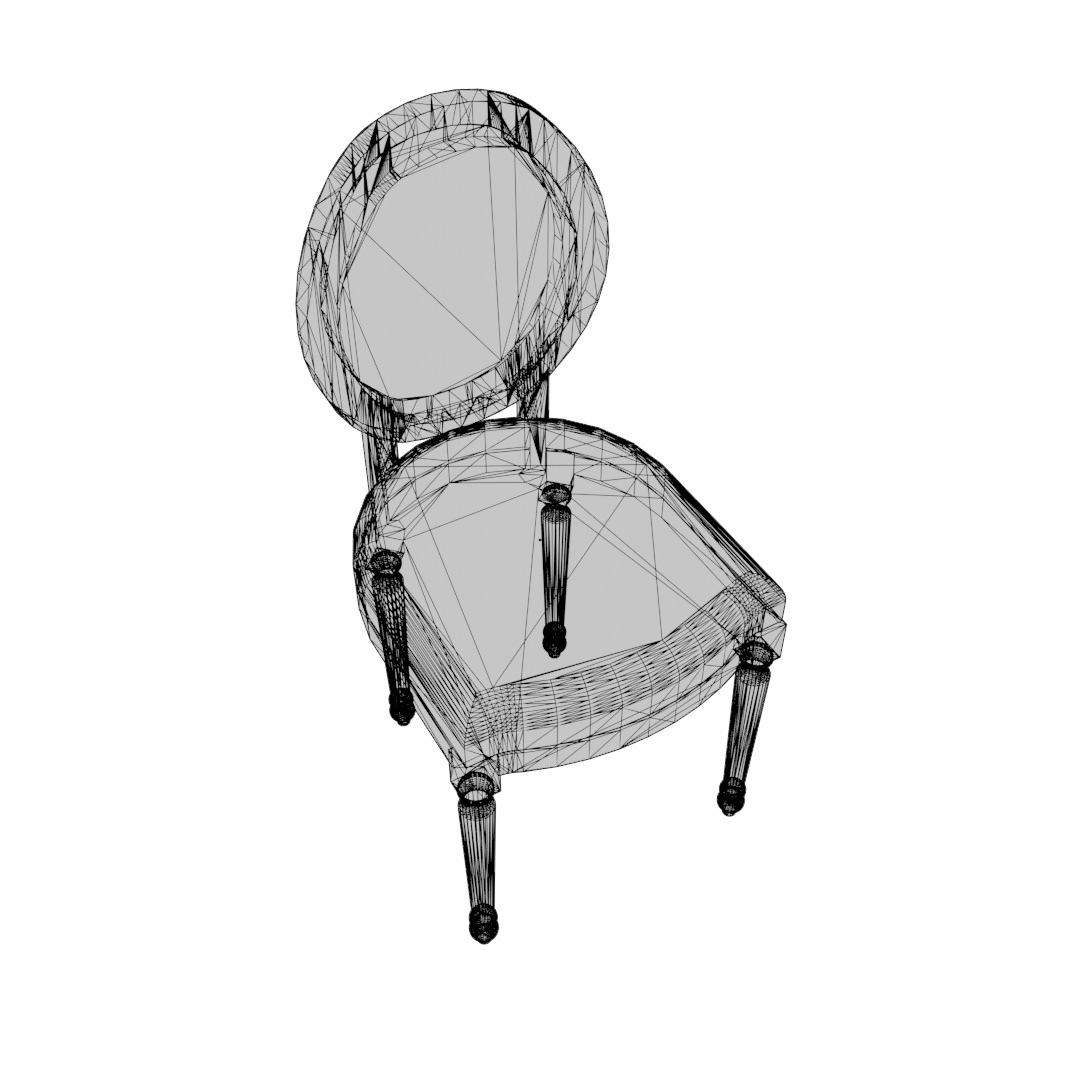}\hfill
\includegraphics[trim=280 120 270 70,clip,width=0.24\linewidth]{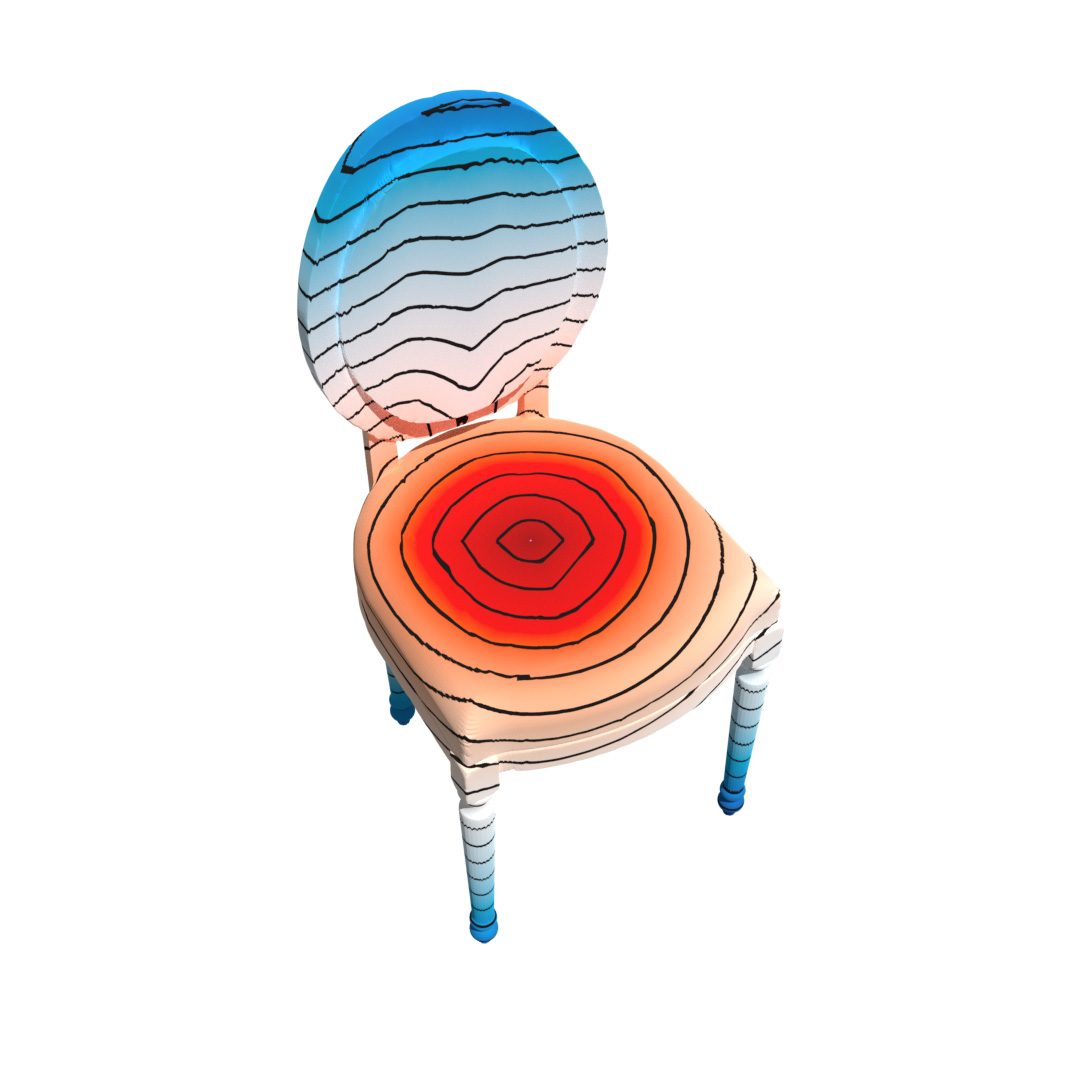}\\
\vspace{-2em}
\begin{flushleft}{\small \hspace{6.5em} Input \hspace{12em} Ours}\end{flushleft}
\vspace{-.1in}
\caption{\textbf{Example application of geodesic distance computation:} Computing geodesic distances on the input mesh through the Heat method~\cite{crane2017heat} resulted in incorrect results due to disconnectivity and inner structure. However, our method produced a more desired distance map.}
\label{fig:heatmethod}
\end{figure}   

\begin{figure}[ht]
\newcommand{\figcap}[1]{\begin{minipage}{0.49\linewidth}\centering#1\end{minipage}}
\frame{\includegraphics[trim=700 200 400 300,clip,width=0.495\linewidth]{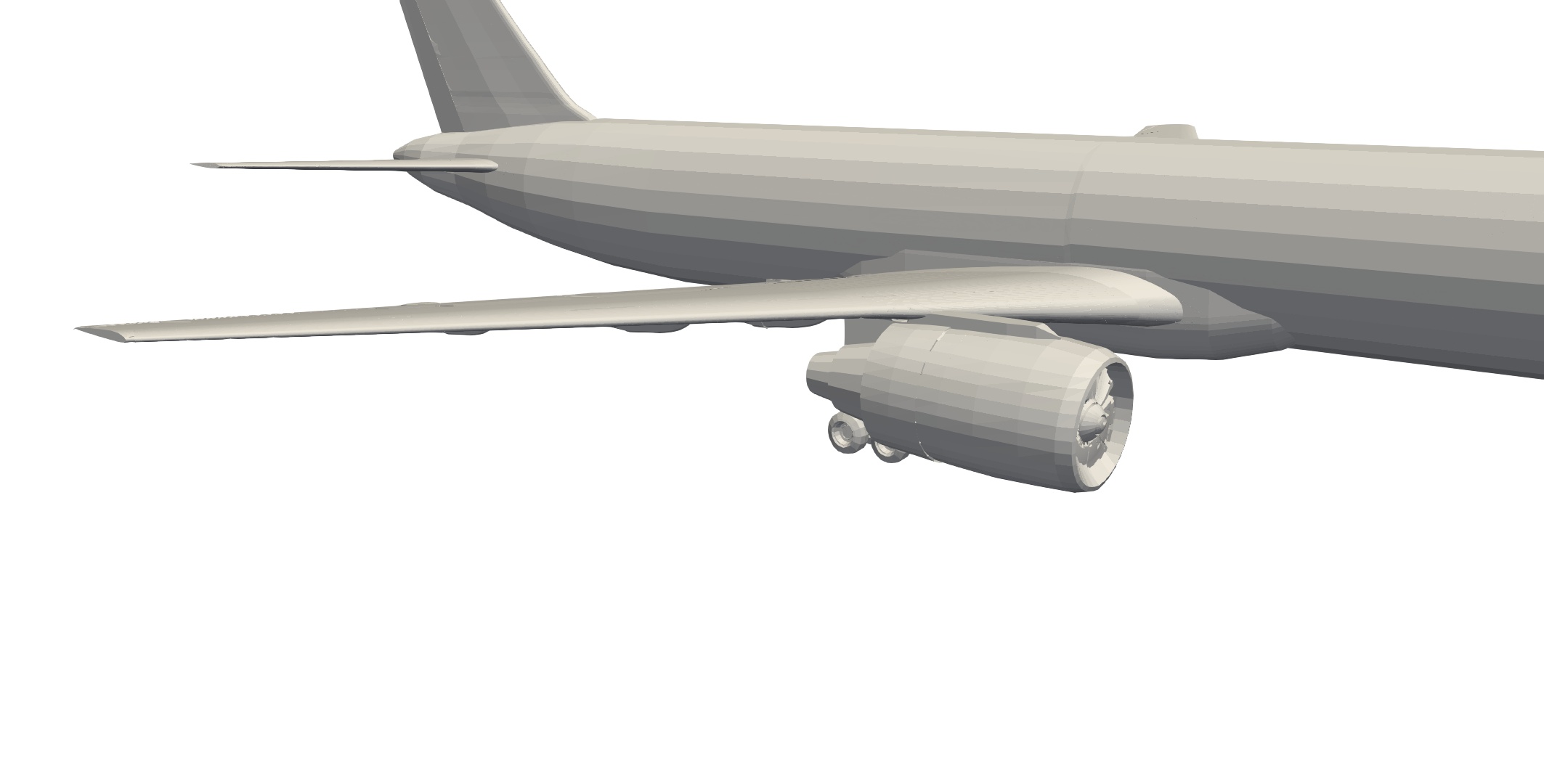}}\hfill
\frame{\includegraphics[trim=700 200 400 300,clip,width=0.495\linewidth]{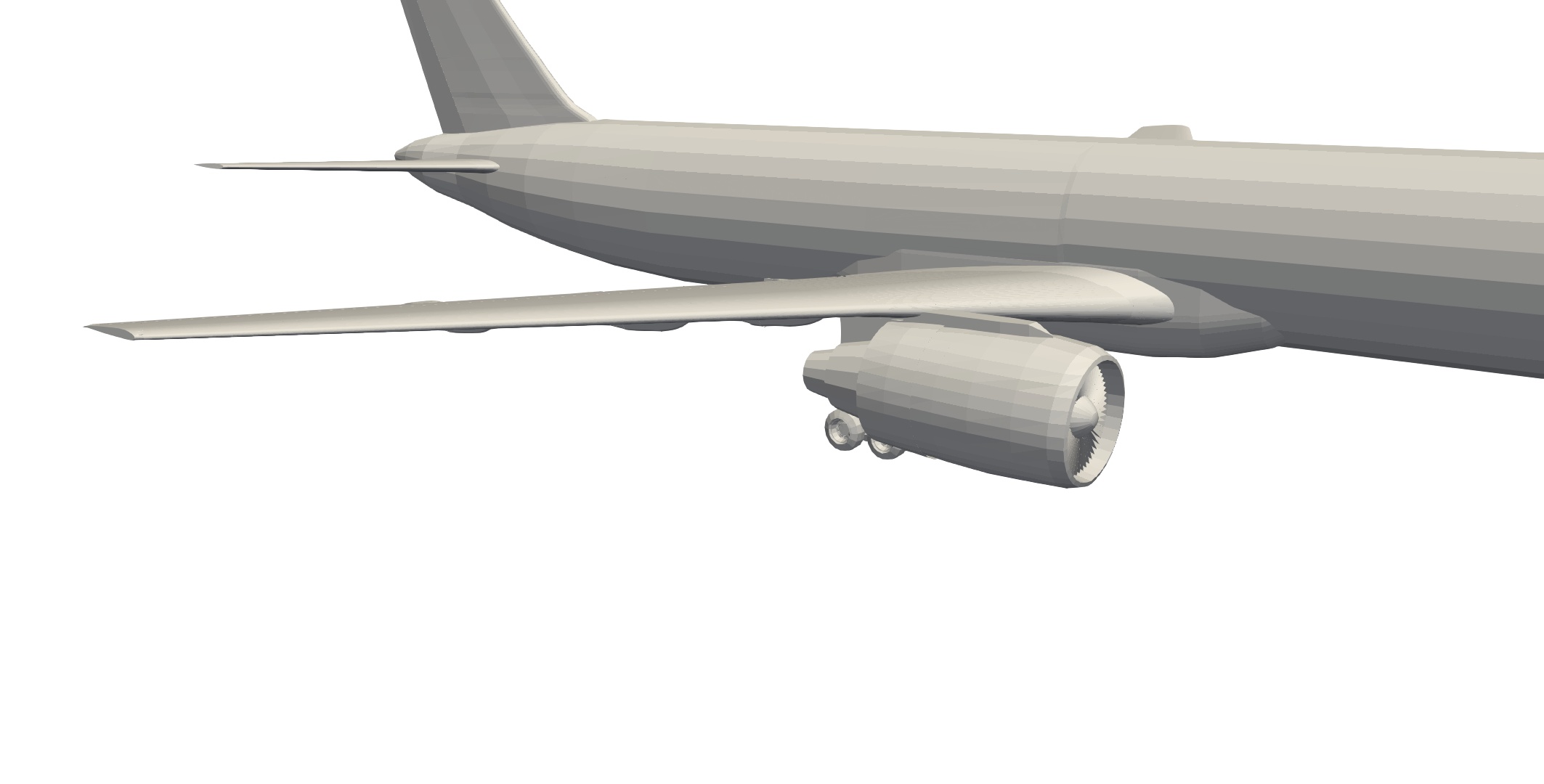}}\\
\frame{\includegraphics[trim=700 200 400 300,clip,width=0.495\linewidth]{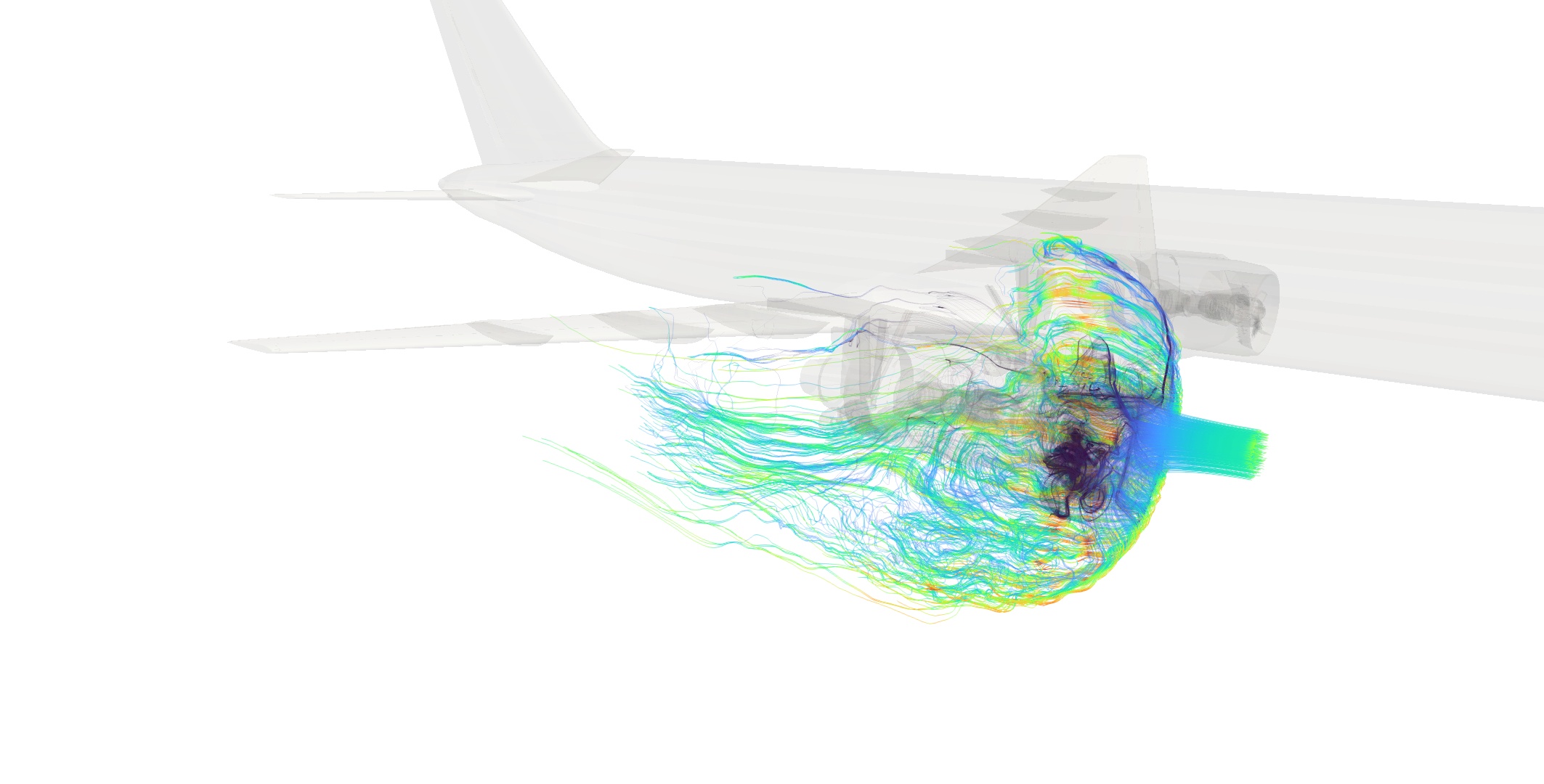}}\hfill
\frame{\includegraphics[trim=700 200 400 300,clip,width=0.495\linewidth]{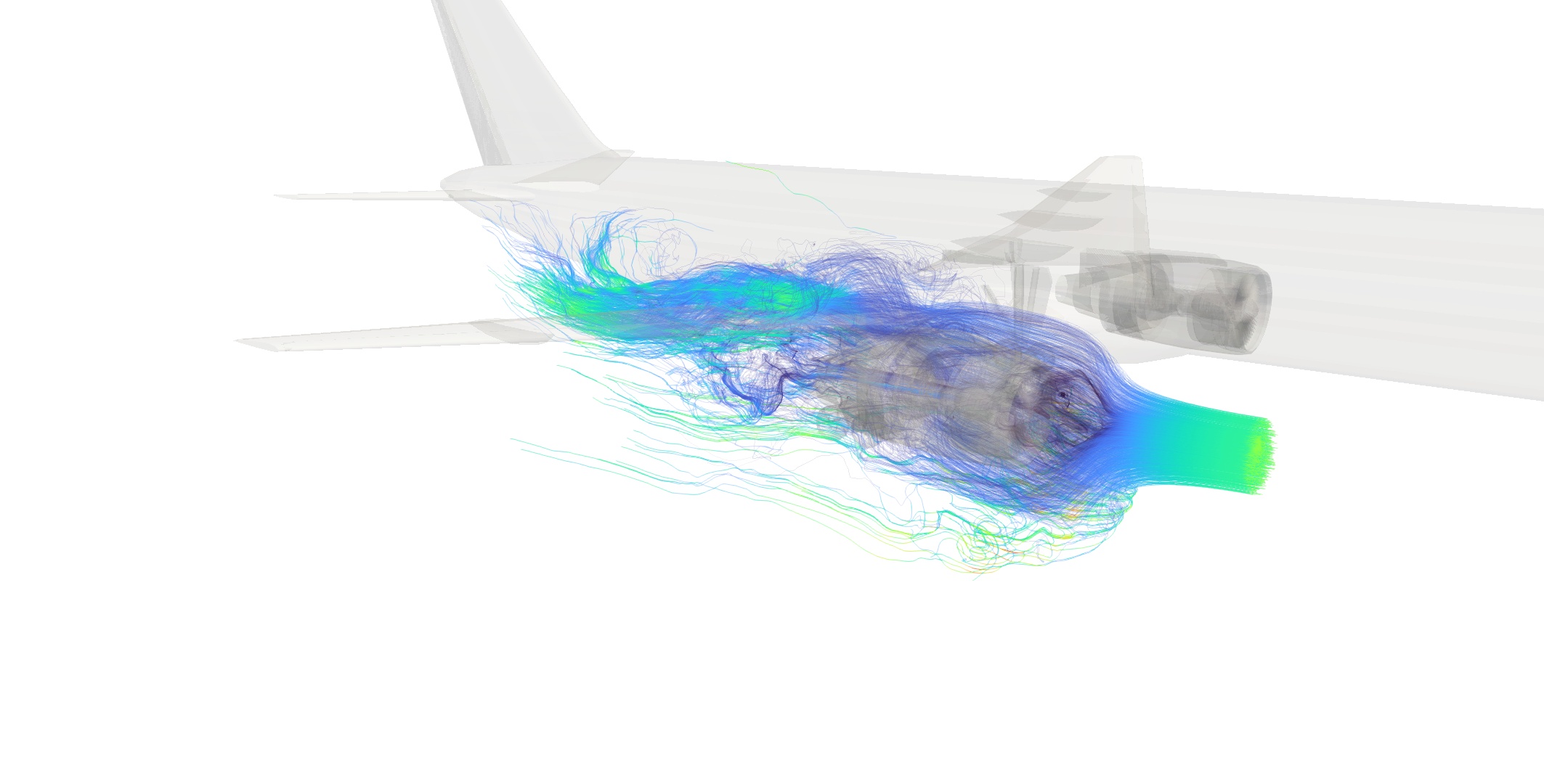}}\\
\figcap{\small VisualRepair + VolumeMesher }\hfill%
\figcap{\small Ours } \\
\vspace{-.15in}
\caption{\textbf{Example application of flow simulation:} The airplane mesh (\autoref{fig:teaser}) repaired using VisualRepair + VolumeMesher results in a closed engine model that does not allow airflow passage. In contrast, our method preserves the original structure of the engine and allows for the correct flow of fluid through the model.}
\label{fig:sim}
\end{figure}

\textbf{Applications.} We present four downstream applications of our methods, mesh simplifications (\autoref{fig:QEM}), Boolean operations on meshes (\autoref{fig:boolean}), geodesic distance computation (\autoref{fig:chair}), and fluid simulation (\autoref{fig:sim}). It is evident that meshes repaired by our method facilitate these applications, while the input meshes cannot be used due to their geometric and topological errors. On the other hand, combining the latest mesh repairing techniques, VR+VM results in the loss of original geometric structures, rendering the simulation results useless (see~\autoref{fig:sim}).

\section{Conclusion} \label{sec:conclusion}
In this work, we presented three crucial visual measures to assess visibility, orientation, and openness, and we proposed a novel framework for mesh repair that incorporates these measures into critical steps such as local open surface closing, face reorientation, and global graph cut using a visual-guided objective function. Our method was evaluated on a set of 500 models randomly selected from ShapeNet, showcasing its effectiveness and robustness in comparison to existing techniques.

\textbf{Limitations.} Nevertheless, it is important to acknowledge that our method does possess certain limitations.

Firstly, the inclusion of visual measures can enhance the resilience and efficacy of mesh repair techniques. However, evaluating structures with channel-like characteristics poses challenges in terms of efficiency, primarily due to a limited number of bounce and ray samples. For instance, even with a substantial opening at the base, the ears of the "Bunny" model do not manifest a shell-like structure. In the future, we would like to utilize differentiable rendering techniques to fill textures for newly added faces. It would also be interesting to remesh our output to improve the mesh quality, e.g., mesh aspect ratio. 

\begin{figure}[ht]
\newcommand{\figcap}[1]{\begin{minipage}{0.24\linewidth}\centering#1\end{minipage}}
\frame{\includegraphics[trim=50 300 1150 50,clip,width=0.20\linewidth]{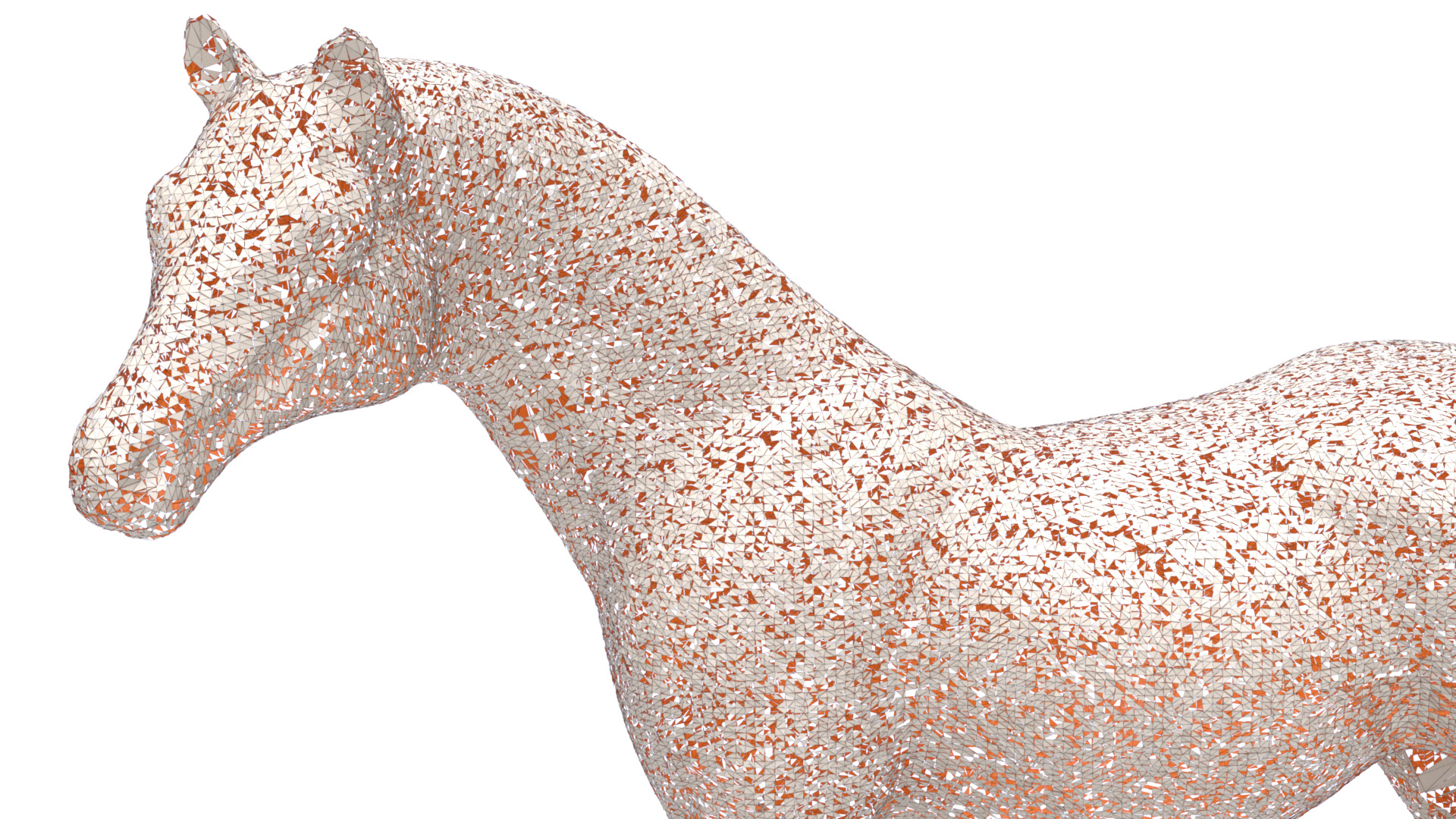}}\hfill
\includegraphics[trim=350 100 500 100,clip,width=0.28\linewidth]{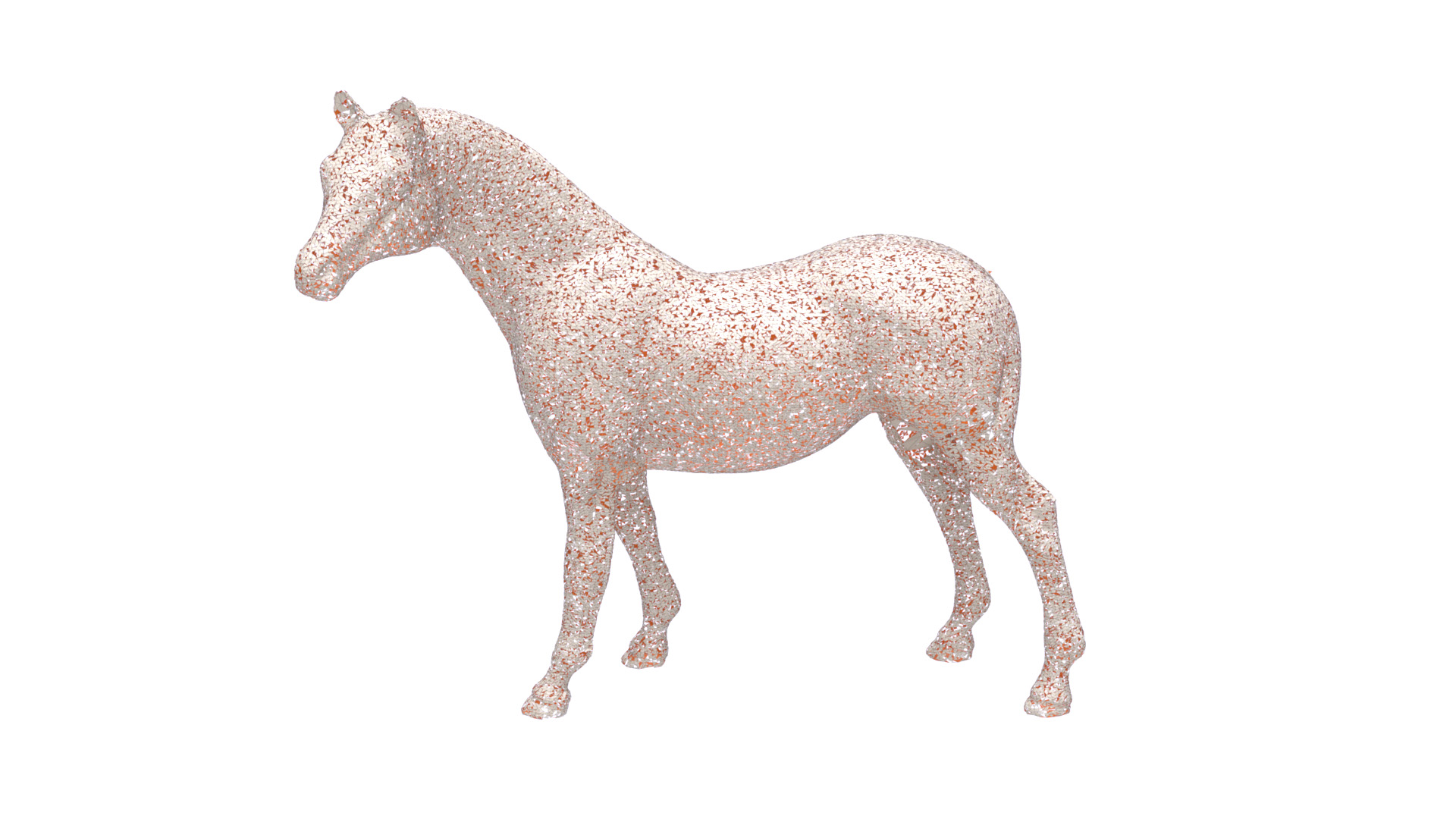}\hfill
\frame{\includegraphics[trim=50 300 1150 50,clip,width=0.20\linewidth]{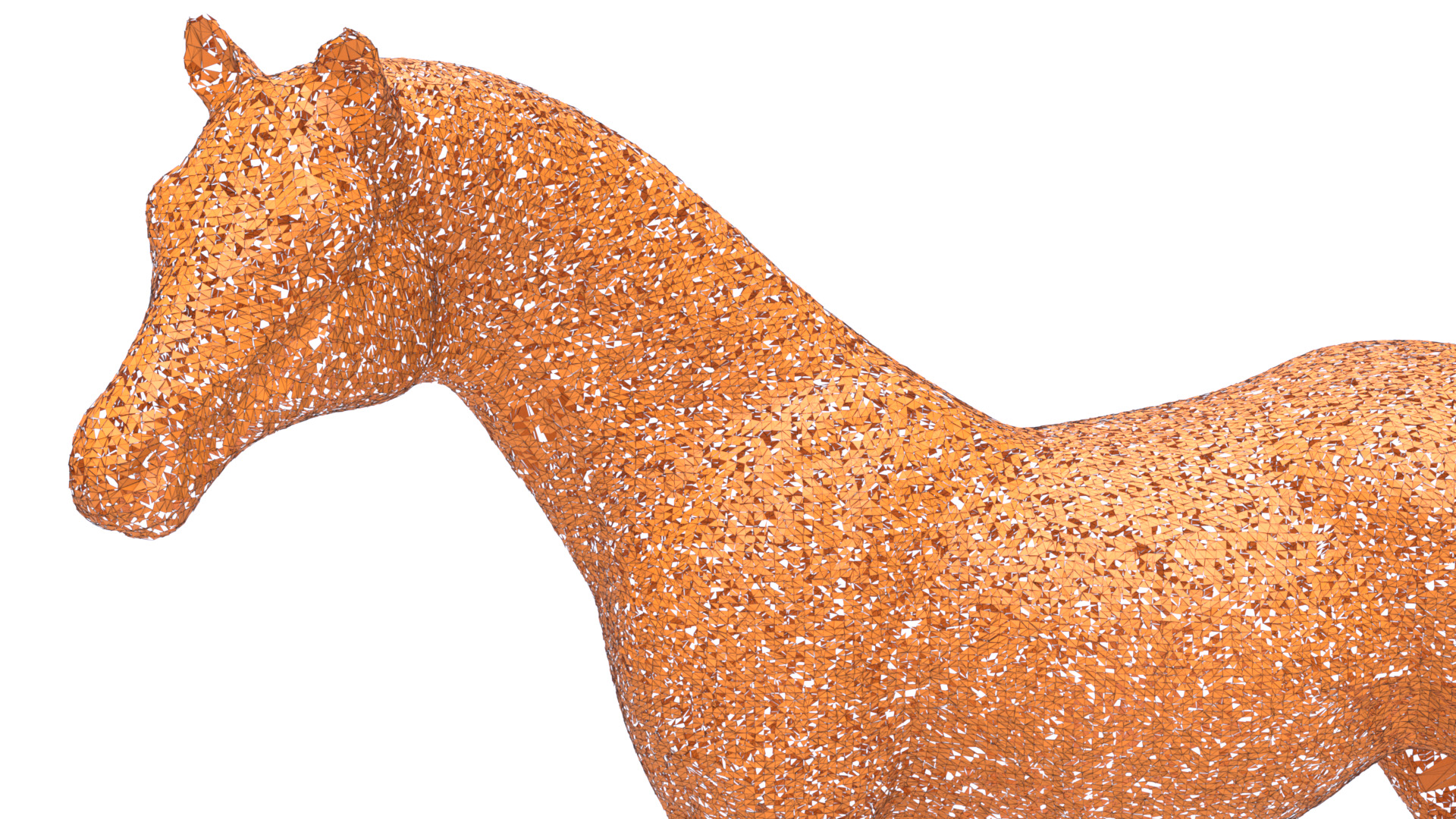}}\hfill
\includegraphics[trim=350 100 500 100,clip,width=0.28\linewidth]{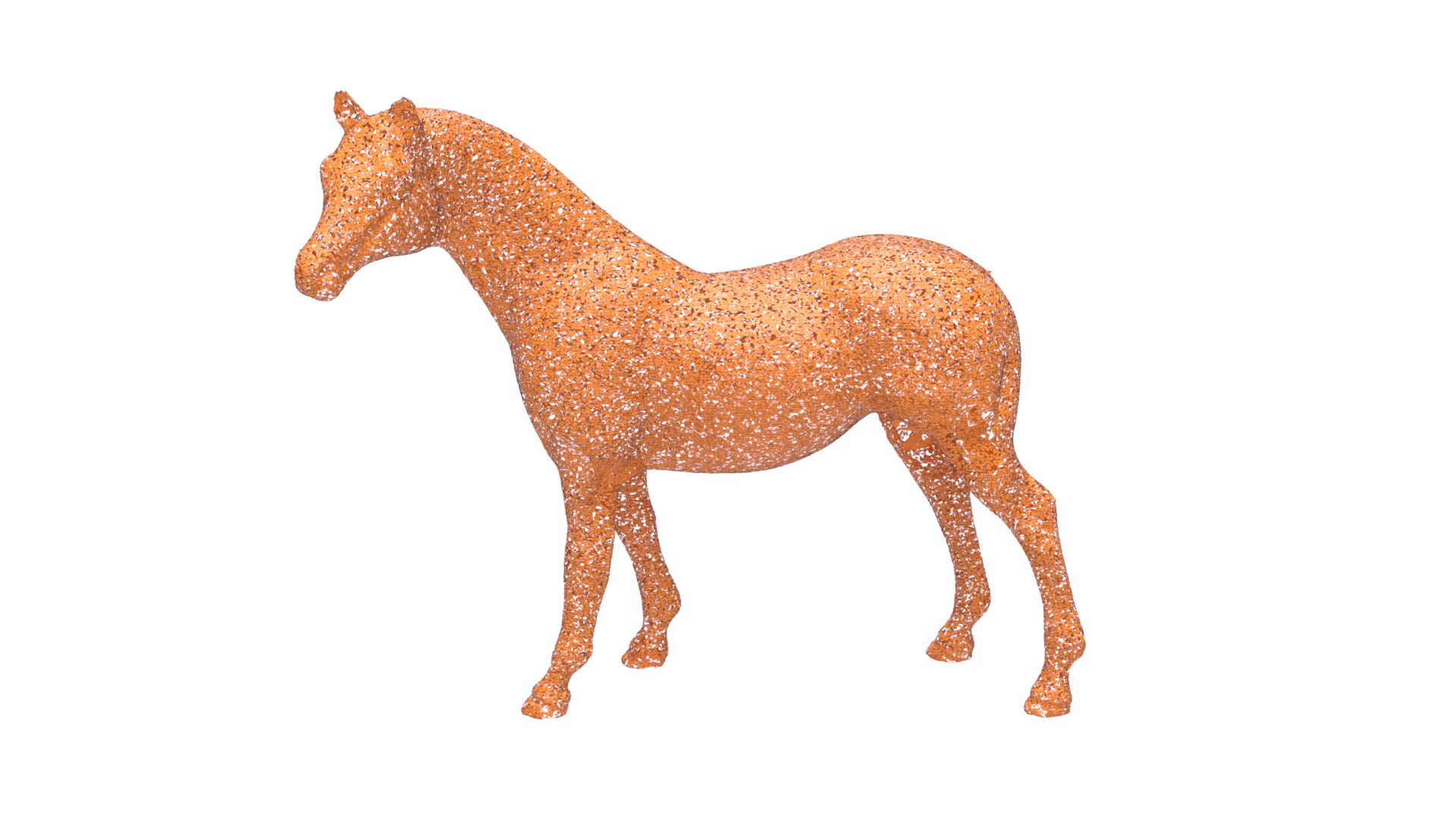}\\
\vspace{-2em}
\begin{flushleft}{\small \hspace{6.5em} Input \hspace{12.8em} Ours}\end{flushleft}\vspace{-1.5em}
\captionsetup{justification=centering}
\caption{{Failure case}}
\label{fig:horse}
\end{figure}

Besides, although our method can robustly convert any input mesh into a watertight manifold mesh, it tends to preserve the topological input features rather than closing holes and gaps to produce a single closed surface, as done in PolyMender~\cite{Ju2004polymender}. This behavior is sometimes undesired, as illustrated in~\autoref{fig:horse}.

Finally, our method guarantees topological manifoldness and watertightness, and ensures geometrically self-intersection-free and degenerate-face-free under exact arithmetic, but geometric guarantees are lost under finite floating point precision. This problem can be potentially solved by using TetWild to further process our output mesh.

\bibliographystyle{IEEEtran}
\bibliography{reference}
\end{document}